\documentclass[useAMS,usenatbib]{mn2e}
\usepackage{lscape}%
\usepackage{epstopdf}%
\usepackage{epsfig}%
\usepackage{natbib}
\usepackage{graphics}
\usepackage{longtable}

\title[Cluster classification and evolution]{X-ray/optical classification of cluster mergers and the evolution of the cluster merger fraction}
\author[Andrew W. Mann, Harald Ebeling]{Andrew W. Mann$^{1 *}$, Harald Ebeling$^{1}$\\
$^{1}$Institute for Astronomy, University of Hawai'i, 2680 Woodlawn Dr, Honolulu, HI 96822\\
$^{*}$amann@ifa.hawaii.edu}
\begin{document}

\pagerange{\pageref{firstpage}--\pageref{lastpage}} \pubyear{2011}

\maketitle

\begin{abstract} 
We present the results of a simple but robust morphological classification of a statistically complete sample of 108 of the most X-ray luminous clusters at $0.15\le z\le 0.7$ observed with Chandra. Our aims are to (a) identify the most disturbed massive clusters to be used as gravitational lenses for studies of the distant universe and as probes of particle acceleration mechanisms resulting in non-thermal radio emission, (b)  find
cluster mergers featuring subcluster trajectories that make them suitable for quantitative analyses of cluster collisions, and (c) constrain the evolution with redshift of the cluster merger fraction. Finally, (d) this paper represents the third public release of clusters from the MACS sample, adding 24 clusters to the 46 published previously. To classify clusters by degree of relaxation, we use the projected offset of the brightest cluster galaxy from the peak (or the global centroid) of the X-ray emission as a measure of the segregation between the intracluster gas and dark matter, and also perform a visual assessment of the optical and X-ray morphology on all scales. Regarding  (a), we identify ten complex systems likely to have undergone multiple merger events in the recent past.
Regarding  (b), we identify eleven systems likely to be post-collision, binary, head-on mergers (BHOMs), as well as another six mergers that are possible BHOMs but probably harder to interpret because of non-negligible impact parameters and merger axes closer to our line of sight.
Regarding (c), we find a highly significant increase with redshift in the fraction of morphologically disturbed clusters (and thus a clear decrease in the number of fully relaxed systems) starting at $z\sim 0.4$, in spite of a detection bias in our sample against very disturbed systems at high redshift.  Since our morphological diagnostics are all based on imaging data and thus sensitive to projection effects, the measured merger fractions should be considered lower limits and our list of mergers incomplete, as we are likely to miss systems forming along axes close to our line of sight. A larger sample of clusters with high-quality X-ray data in particular at high redshift will be needed to trace the evolutionary history of cluster growth and relaxation closer to the primary epoch of cluster formation $z\sim 1$.
\end{abstract}

\section{Introduction}
As the largest gravitationally bound objects in the universe, clusters of galaxies represent unique laboratories for studies of  structure formation and evolution on scales ranging from tens to thousands of kpc. In the observationally and theoretically well supported framework of hierarchical structure formation clusters  grow through a series of mergers. The resulting negative evolution with lookback time of, e.g.,  the cluster mass function has long been recognized as a powerful tool to constrain cosmological parameters \citep[e.g.,][]{1998MNRAS.298.1145E, 2001ApJ...561...13B, 2003ApJ...590...15V, 2008MNRAS.387.1179M, 2010MNRAS.406.1759M}. However, the very process underlying this evolution -- cluster merging -- is much more complex than the physics assumed by many evolutionary models. Specifically,  shocks created in cluster collisions have been demonstrated to cause temporary but dramatic increases in the immediate observables, i.e., the Comptonization parameter, the X-ray flux and luminosity, and the gas temperature \citep{2002ApJ...577..579R, 2007MNRAS.380..437P, 2008ApJ...680...17W}. Unaccounted for, such dramatic deviations from virialization are likely to bias the results of cosmological cluster studies.  This is true in the extreme for work based on measurements of the  baryon fraction in clusters that rely explicitly on the assumptions of spherical symmetry and hydrostatic equilibrium \citep{2004MNRAS.353..457A, 2008MNRAS.383..879A}.

While thus somewhat of a nuisance in the context of cosmological work, disturbed clusters are sought-after targets for investigations  of the dynamics and physics governing the interactions between the three main cluster constituents, dark matter, diffuse gas, and galaxies. The most extreme mergers, head-on collisions of massive clusters of roughly equal mass, constitute the most energetic events in the Universe, involving the conversion of up to $10^{64}$ erg of potential energy into thermal and kinetic energy. As a result, massive mergers provide us with a rare opportunity to study a wide range of astrophysical phenomena, such as galaxy evolution via ram-pressure stripping, the destruction of cool cores, or the acceleration of relativistic particles and the role of magnetic fields  in the creation of radio halos and relics.

In this paper, we attempt to facilitate the selection of massive clusters most suitable for cosmological or astrophysical applications by classifying a statistically complete sample of clusters according to relaxation state. The pronounced segregation of gaseous and (essentially) collisionless matter during a merger represents a simple and powerful diagnostic to identify systems that experienced a significant merger event fairly recently and are still far from relaxation.
Our classification scheme uses a combination of optical and X-ray diagnostics, and is applied to a sample of the most X-ray luminous clusters at redshifts between $z=0.15$ and $z=0.7$. Our goals are threefold: we aim to (a) compile a list of the most complex mergers which, although often too challenging for dynamical modeling, are likely to be extraordinary gravitational telescopes and prime targets for the detection of, e.g., extreme gas temperature or giant radio halos and relics;
(b) identify the most massive, binary head-on cluster mergers that feature sufficiently simple trajectories to allow quantitative studies of the properties of dark matter and the physics of cluster and galaxy evolution; and c) quantify the fraction of dynamically disturbed clusters as a function of redshift in order to test theoretical and numerical predictions. 

Throughout this paper we assume the concordance $\Lambda$CDM cosmology ($\Omega_M = 0.3$, $\Omega_\Lambda = 0.7$, and $H_0 = 70$ km s$^{-1}$ Mpc$^{-1}$).

\subsection{Extreme complex mergers}\label{sec:compmerg}

The most complicated mergers, i.e.\ systems that have recently undergone several mergers along different axes, are extraordinary observational targets --- and can be extraordinarily difficult to interpret. They are as far from relaxation as possible for a gravitationally bound system, feature pronounced substructure in the distribution of dark matter, gas, and galaxies, often exhibit dramatic variations in the temperature of the intra-cluster medium (ICM), and cannot be assumed to be in hydrostatic equilibrium on any scale. Clearly these systems grossly violate the assumptions made in cosmological models of the growth of structure built solely upon spherical gravitational collapse.
 
The exceptional properties of active, serial mergers render them extremely valuable in other regards though. Owing to the large regions of high mass surface density in their centres, complex mergers often make exquisite gravitational telescopes whose large Einstein radii allow studies of distant background objects over larger areas than accessible with relaxed cluster lenses.  Examples at intermediate redshift include MACS\,J0717.5+3745 \citep{2004ApJ...609L..49E, 2007ApJ...661L..33E, 2011MNRAS.410.1939Z} and MACS\,J1149.5+2223 \citep{2007ApJ...661L..33E, 2009ApJ...707L.163S}. Although often too complex to be easily modeled dynamically, some exceptionally active mergers have been used successfully also for astrophysical studies of the cluster proper, on subjects ranging from ICM heating to galaxy evolution \citep[e.g.,][]{2008ApJ...684..160M, 2009ApJ...693L..56M,2010MNRAS.406..121M}. Last, but not least, cluster mergers have been found to feature a very high incidence of radio halos and relics, making them excellent targets for studies of particle acceleration mechanisms and the role of magnetic fields in the confinement of relativistic plasma \citep[see][and references therein]{2009A&A...507..487M, 2009A&A...507.1257G}.

\subsection{Binary, head-on cluster mergers (BHOMs)}\label{sec:bhomdef}

Because of their relatively simple  geometry, binary head-on mergers (henceforth BHOMs) of massive clusters are the most promising targets for studies of the physics of mergers if  a robust physical interpretation of the observational evidence is sought. Not only does a head-on collision maximize the observable signal from all types of interactions between the cluster components, the facts that only two systems are involved and that a small or negligible impact parameter ensures a linear collision trajectory also greatly reduce systematic uncertainties and facilitate the modeling of the merger history of BHOMs.

In recent studies of cluster mergers, the different dynamical behaviour of the viscous  ICM and the essentially collisionless galaxies and dark matter has yielded compelling observational evidence of the existence of dark matter \citep{2006ApJ...648L.109C, 2010A&A...524A..95L, 2010MNRAS.404..325R} and has, in addition, been used to place upper limits on the dark-matter self-interaction cross section, $\sigma/m$,  of the putative dark-matter particle \citep{2004ApJ...606..819M,2006ApJ...648L.109C,2008ApJ...687..959B}. BHOMs involving clusters of similar mass provide the tightest constraints in this regard, and also provide the energetically most extreme environment for studies of any interaction between the constituents of the merging systems. The absence of complications caused by an unknown impact parameter makes head-on massive mergers particularly suitable also for quantitative studies of the collision of the gaseous cluster components which leads to shock fronts, collisional heating, and the gradual destruction of cool cores \citep{2007PhR...443....1M, 2010MNRAS.405.1624M, 2011MNRAS.411.1641E}. In addition, BHOMs have yielded valuable insights into the prevailing mechanisms behind accelerated galaxy evolution in dense environments \citep[e.g.][]{2010MNRAS.406..121M}.

\subsection{Redshift evolution of the merger fraction}

Large, statistically well defined cluster samples are required to constrain the evolution of cluster properties over cosmological timescales.  
Of particular interest in this context is the rate of cluster formation via mergers, a quantity that affects several key statistics, among them the cluster X-ray luminosity, temperature, and mass functions, and which is a strong function of the adopted cosmological world model. Numerical simulations predict an increase in the merger fraction with redshift \citep{2005APh....24..316C,2007MNRAS.377..317K,2008MNRAS.387..631E}, consistent with observations of a decline in the number of cool-core clusters with redshift \citep{2007hvcg.conf...48V, 2008A&A...483...35S,2011ApJ...731...31S}, as well as with expectations from essentially any world model based on hierarchical structure formation.

A low-redshift baseline for such studies is provided by work on the HIghest X-ray Flux Galaxy Cluster Sample \citep[HIFLUGCS,][]{2002ApJ...567..716R}, a sample of the X-ray brightest clusters, most of which are nearby (${<}z{>}{=}0.053$). \citet{2003ApJ...585..687K} derive a value of 17.5\% for the fraction of dynamically disturbed clusters using a subset of 40 clusters. A lower estimate of 12\% is found in a later study by \citet{2009MNRAS.398.1698S} who use all 64 clusters in the HIFLUGCS sample. For X-ray luminous clusters at somewhat higher redshifts ($z_{\rm median}{=}0.23$) Reiprich \& B\"ohringer as well as \citet{2010A&A...513A..37H} find 20\% for the same statistic; the comparison of these values is, however, complicated by the fact that the two studies differ both in  their cluster selection criteria and in the diagnostics employed to quantify dynamical state (see Section~\ref{sec:diagnostics}). 

Systematic effects also hamper studies of the evolution of the merger fraction at higher redshift, such as the investigation of \citet{2011ApJ...731...31S} who use the detection of strong [OII]$\lambda$3727 emission in the optical spectra of BCGs as a proxy for the presence of a cool core. Although these authors report a dramatic decline in the number of such line emitters as a function of redshift, their local and distant cluster samples differ significantly in their mean X-ray luminosity,  making a direct comparison difficult. Similarly, \citet{2008ApJS..174..117M} report an increase at $z>0.5$ in the number of clusters with large shifts in the centroid of the X-ray surface brightness as measured from Chandra observations of 115 clusters at $0.1 < z <1.3$, but stress that their sample is not statistically complete and their analysis subject to ``unquantifiable biases".

A census of merging clusters from a well defined and uniformly selected sample spanning a significant redshift range that allows a robust determination of the evolution of the merger fraction would thus be of considerable interest. We here attempt to perform this determination as well as to provide a list of BHOMs and complex mergers, using a statistically nearly complete sample of X-ray luminous clusters at $z=0.15-0.7$ and the diagnostics described in the following.

\section{Dynamical substructure diagnostics}\label{sec:diagnostics}

Substructure in clusters is notoriously difficult to quantify reliably. In principle, spatially resolved X-ray spectroscopy can probe the thermodynamics of the ICM in great detail, revealing merger-induced shocks and temperature variations as well as cold fronts and cool cores, all of which represent sensitive diagnostics of a system's relaxation state. In practice, however, the required exquisite spatial resolution and photon statistics can be attained only for few nearby and very X-ray bright clusters. Observationally less ``expensive" probes of substructure in the ICM use only the projected X-ray surface brightness distribution to compute, e.g., power ratios \citep{1995ApJ...452..522B} or centroid variations and axial ratios \citep{1995ApJ...447....8M}. More recent studies found the combination of data from different wavelength regimes to be particularly powerful for the purpose of allowing a crude but robust classification.  \citet{2010A&A...513A..37H} investigate the usefulness and efficiency of several diagnostics for the identification of cool-core clusters, considered dynamically fully relaxed systems. Among other criteria, the presence of a central temperature drop, short cooling time scales, significant mass deposition rate, and the presence of diffuse large-scale radio emission are all found to reliably characterise a system's dynamical state. However, Hudson and co-workers also demonstrate that an even simpler diagnostic, namely the projected separation of the locations of the brightest cluster galaxy (BCG) and of the peak of the X-ray emission from the diffuse intra-cluster gas, constitutes just as robust an indicator.  \citet{2009MNRAS.398.1698S}, in their study of the dynamical state of clusters at $0.15\,{<}\,z\,{<}\,0.3$, use two different flavours of such a BCG-X-ray offset to identify dynamically disturbed systems: one using the location of the X-ray peak, the other using the overall centroid of the diffuse X-ray emission. 

The physical reason for which the mentioned X-ray/optical offsets allow the straightforward identification of disturbed clusters lies in the different dynamical behaviour of gas and galaxies during cluster mergers that we mentioned before. On cluster scales, galaxies can be taken to be fast-moving, pointlike particles that have a very low cross section for collision. By contrast, the intracluster gas is highly collisional and, during a head-on cluster merger, will be subject to intense ram pressure and shock heating. As a result, the galaxy populations of the merging systems will proceed nearly unimpeded, slowed only by gravity through the process of dynamical friction, while the strongly self-interacting viscous cluster gas is slowed by the collision. Since the resulting offset  will diminish as the combined system gradually settles into dynamic and hydrostatic equilibrium, a pronounced spatial segregation of gas and galaxies constitutes powerful evidence of a recent or ongoing merger \citep[e.g.,][]{2005MNRAS.359..417S,  2010MNRAS.406.1134S}; conversely,  excellent alignment of the BCG  with the intra-cluster gas distribution is typical of relaxed systems which often also exhibit significant radiative cooling in the core region \citep{2009A&A...501..835M}. If the merger axis lies approximately in the plane of the sky, the apparent separation is maximised, allowing observers to deduce the direction of motion and to more readily separate the merging subclusters for detailed study.

How exactly this ``X-ray/optical offset" is quantified affects the respective diagnostic's sensitivity to the complexity of the merger (more than two merging components) and/or the inclination of the merger axis relative to the plane of the sky. Specifically, the observed offset of the BCG from the peak of the diffuse X-ray emission is most useful to derive the projected trajectory of the cores of merging clusters, whereas its projected separation from the global centroid of the cluster X-ray emission is often more reliable to identify morphologically complex systems with significant X-ray substructure. We here use both statistics in order to maximize our ability to reliably identify disturbed systems and mitigate the dependence on merger geometry and viewing angle.

Finally, we use, as a third, qualitative diagnostic of the relaxation state of all clusters the result of a visual classification of  the morphology of the X-ray contours and the degree of alignment between the BCG and the X-ray peak, using the simple classification scheme devised by \cite{2007ApJ...661L..33E}.  The assigned codes and classification criteria (from relaxed to extremely disturbed) are 1 (pronounced cool core, perfect alignment of X-ray peak and single BCG), 2 (good X-ray/optical alignment, concentric contours), 3 (non-concentric contours, obvious small-scale substructure), and 4 (poor X-ray/optical alignment, multiple peaks, ambiguous BCG).  We estimate the uncertainty in the assigned morphological code to be less than 1.  

\section{Cluster Sample}\label{sec:sample}
Our sample is selected from a master list of clusters identified in the course of the Massive Cluster Survey \cite[MACS;][]{2001ApJ...553..668E,2007ApJ...661L..33E,Ebeling:2010mz}. While MACS, by definition, is limited to the most distant systems ($z\ge 0.3$), clusters (or candidate clusters) at lower redshift were also flagged by the MACS team during their screening of thousands of X-ray sources from the Bright Source Catalogue compiled in the course of the ROSAT All-Sky Survey \citep[RASS,][]{1993Sci...260.1769T}. We here draw from this master list of RASS clusters with $f_{\rm RASS}\ge 1\times 10^{-12}$ erg s$^{-1}$ cm$^{-2}$ (0.1--2.4~keV), a database that covers the entire extragalactic sky observable from Mauna Kea, i.e., the solid angle within the boundaries defined by $|b| > 20^\circ$ and $-40^\circ < \delta(J2000) < 80^\circ$.

\begin{figure} 
   \centering
   \includegraphics[width=.5\textwidth]{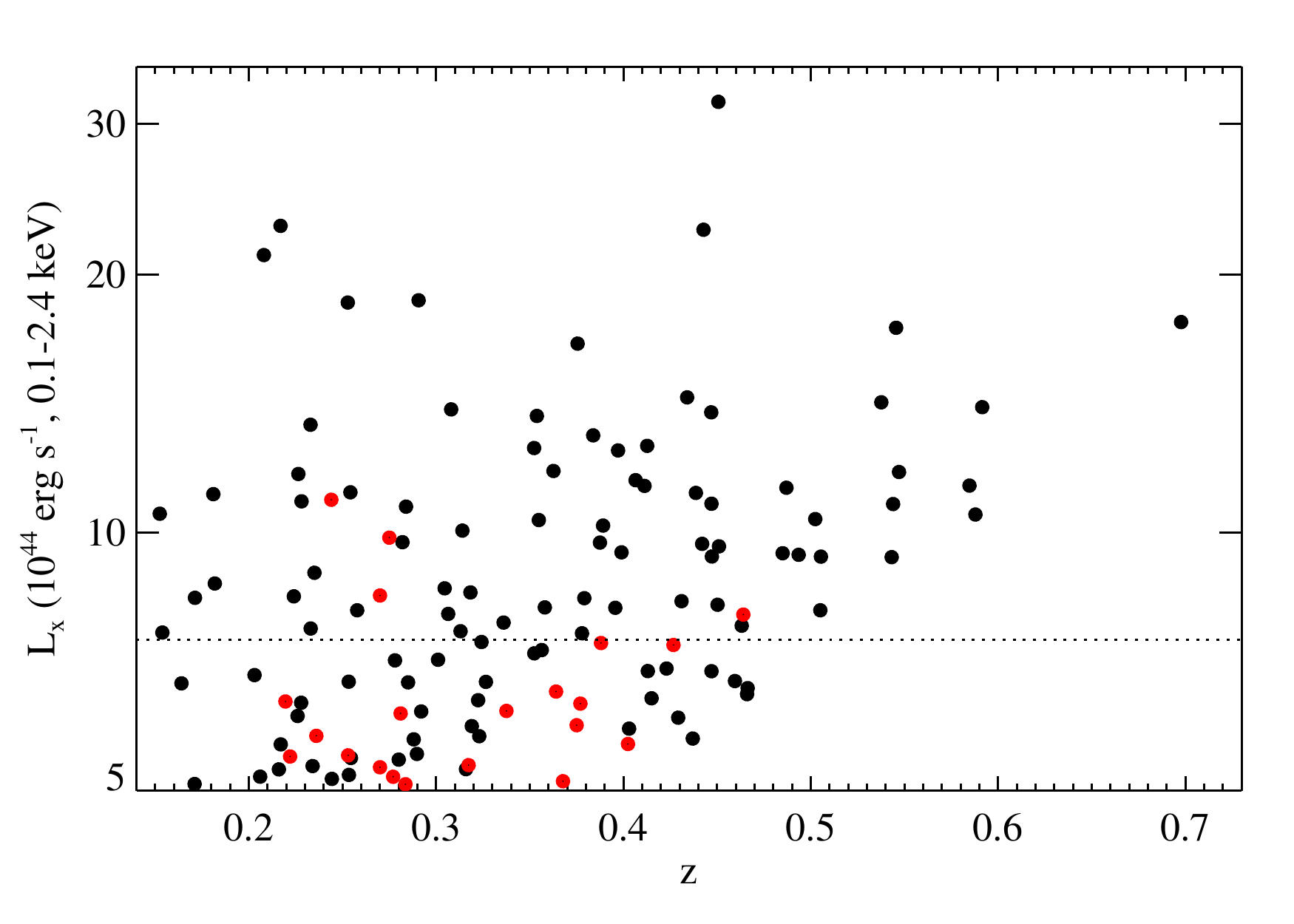} 
   \caption{ Luminosity-redshift distribution of all 129 clusters meeting our initial selection criteria (see Section \ref{sec:sample}).  Black symbols mark the 108 clusters that have been observed with CXO/ACIS and are thus included in this study, red symbols represent the 21 clusters without CXO data.  The X-ray luminosity values correspond to the flux enclosed in the RASS detect cell. The overall sample is volume limited out to $z \sim 0.4$ and becomes flux limited thereafter. Less X-ray luminous lusters are less likely to have been observed by CXO.   The dashed line marks the limiting X-ray luminosity of a nearly complete subsample for which the CXO coverage is 95\%.}
   \label{fig:sample}
\end{figure}

Application of the additional selection criteria $z\,{>}\,0.15$ and $L_{\rm X,RASS}\,{>}\, 5\times10^{44}$ erg s$^{-1}$ (0.1--2.4~keV) yields a list of 129 clusters (Fig.~\ref{fig:sample}).  The quoted lower redshift limit was imposed to ensure that the diffuse X-ray emission from all clusters is well contained within the field of view of the ACIS (Advanced CCD Imaging Spectrometer) instrument aboard the Chandra X-Ray Observatory (CXO). Carrying the only X-ray telescope that provides sub-arcsec resolution, Chandra is the observatory of choice for the unambiguous distinction between small-scale structure (especially cold cores) and X-ray point sources, an essential requirement for a robust assessment of the X-ray morphology of the clusters under study\footnote{The original detections (median, net photon count of 54) from the RASS allow only a determination of the total flux / luminosity, and even for the brightest system (574 net photons within the detect cell) even a crude morphological classification is made impossible by the poor angular resolution of the RASS.}.  The quoted luminosity cut was applied to select an unbiased sample of similarly X-ray luminous (and thus, by inference, massive) clusters across the full redshift range, and to increase the probability of the selected systems having been observed by CXO. As shown in Fig.~\ref{fig:sample} our cluster sample is volume complete out to $z \sim 0.4$ and X-ray flux limited beyond this redshift.  Of the 129 clusters, 108 have been observed with ACIS as of July 2011.  

\section{Data and Analysis}\label{sec:dataanalysis}
\subsection{X-ray Data}
Table~1 lists the exposure time for each cluster observation. We merged data from multiple observations wherever doing so increased the combined exposure time by at least 30\%. We used CIAO 3.3.0 to create images in the 0.5--7~keV band, which  we then smoothed using the adaptive-kernel  algorithm ASMOOTH \citep{2006MNRAS.368...65E}.  

The only quantitative X-ray characteristics required for our study are the locations of the peak and of the global centroid of the emission. We calculate the former from the adaptively smoothed image to mitigate the impact of fluctuations in the photon statistics, in particular for observations of short duration.   In order to measure positions to better than 5\,kpc accuracy\footnote{The pixel scale of ACIS, 0.5\arcsec, corresponds to 6 kpc at $z=0.68$, the highest redshift of all clusters in our sample.} we compute the peak position as the photon-weighted mean within a  circle of  2\arcsec\ radius around the brightest X-ray pixel.  We confirmed that the result of this process is robust and insensitive to the precise choice of radius. We also stress that, thanks to the sub-arcsec angular resolution of CXO,  X-ray point sources at or very near the location of the peak of the diffuse X-ray emission can be masked out efficiently and have no impact on the measurement of the position of the peak of the diffuse emission.  Where multiple peaks of comparable brightness are encountered (peak brightness difference of less than a factor of two) we adopt the brighter peak for the computation of the X-ray/BCG offset but flag the respective cluster for separate visual assessment.

\begin{figure*} 
   \centering
  \includegraphics[width=.49\textwidth]{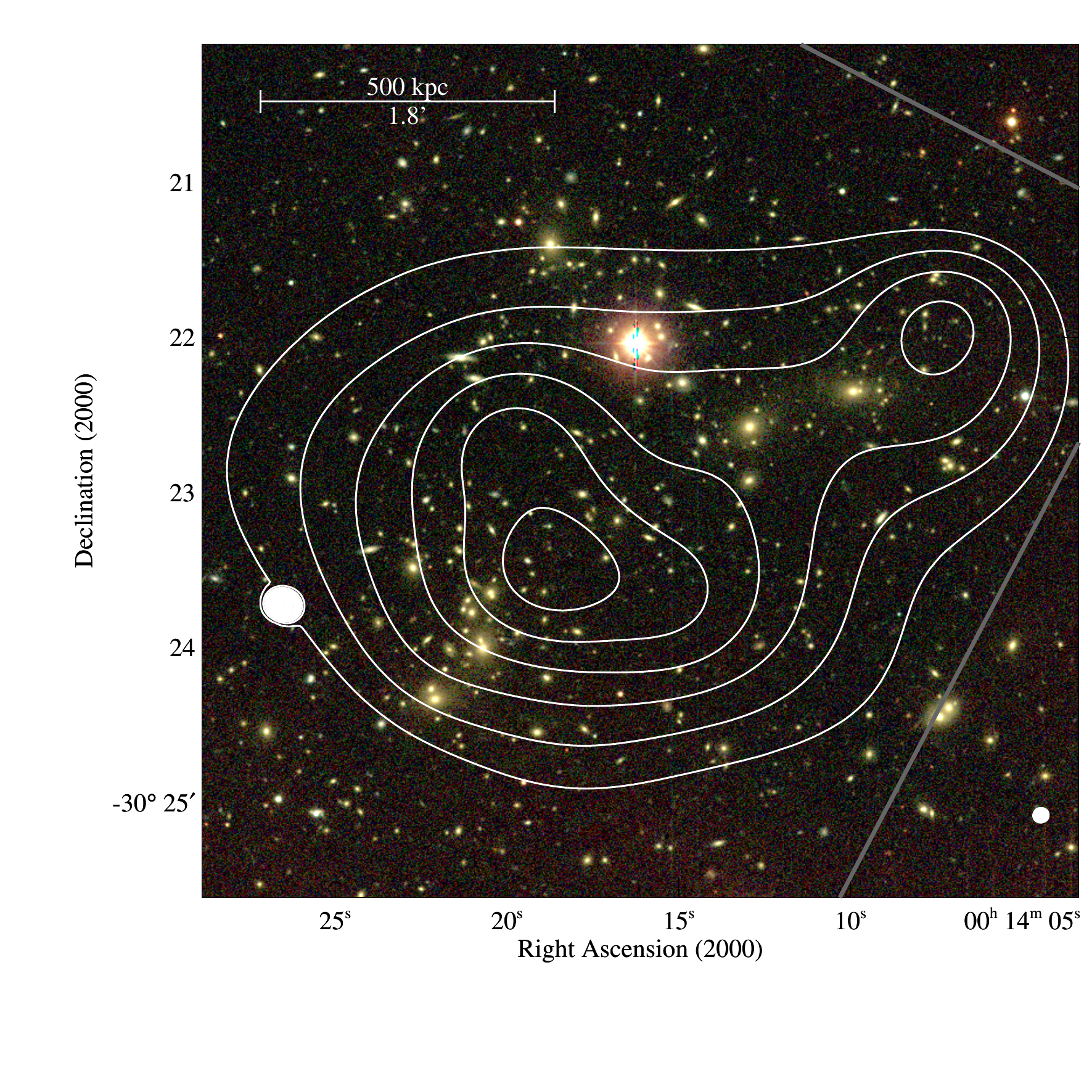} 
  \includegraphics[width=.49\textwidth]{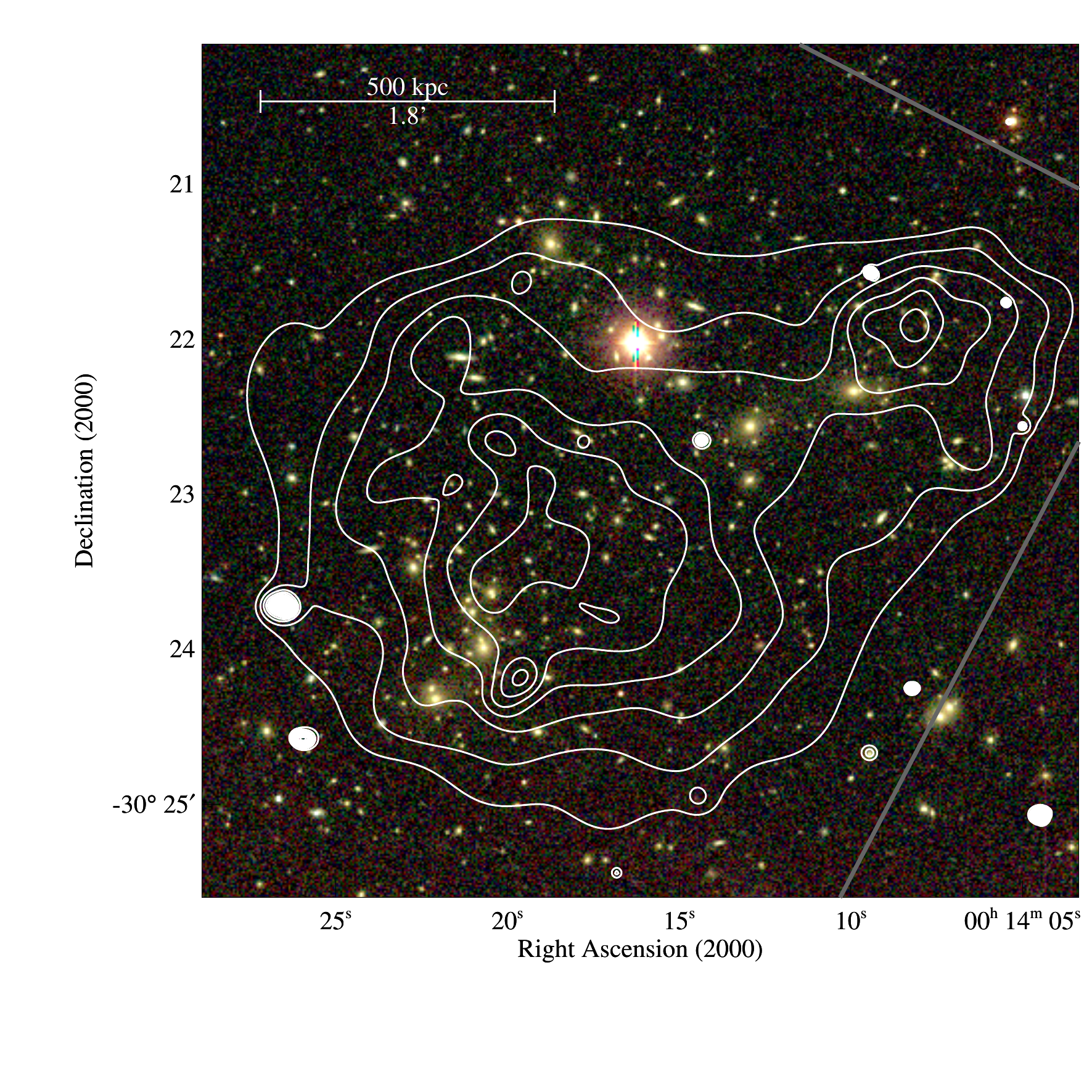} 
   \caption{ VRI image of A\,2744 obtained with the UH2.2m telescope; overlaid are logarithmically spaced isodensity contours of the adaptively smoothed X-ray surface brightness as observed with CXO/ACIS-I. The X-ray contours in this and all other X-ray/optical overlays (see figure captions for exceptions) are placed at $(1.2\times10^{-6})\times1.4^n$ counts s$^{-1}$ in the 0.5--7\,keV range.    The right panel shows a combined exposure, with a total of 74\,ks (resulting in over 36,000 counts).  The left panel is created by scaling the effective exposure time down (1.5\,ks exposure time, resulting in 1100 net counts) to match the total counts from the poorest CXO data in our sample (see Fig.~\ref{fig:countsvsz}).  The comparison of the two overlays shown here illustrates that even dramatic differences in CXO/ACIS exposure time have little impact on our quantitative and qualitative diagnostics (BCG-X-ray peak, and BCG-X-ray-centroid separation, as well as overall morphological classification). Note, however, that in this particular case the location of the peak of the X-ray emission shifts significantly from the approximate centroid of the southeastern (main) cluster to the cold core almost due south of it \citep{2011ApJ...728...27O}. }
   \label{fig:exptests}
\end{figure*}

The second X-ray characteristic required for this work is the location of the overall centroid of the diffuse X-ray emission. We compute this position as the photon-weighted mean of the coordinates of all pixels exceeding a surface-brightness threshold of $1.2\times 10^{-6}$ counts s$^{-1}$ for ACIS-I data, and $1.6\times 10^{-6}$ counts s$^{-1}$ for ACIS-S, thereby accounting for the higher instrumental background of the back-illuminated CCD at the ACIS-S aim point. The quoted surface-brightness thresholds also mark the count rate levels of the lowest contours in the X-ray/optical overlays used to assess the overall cluster morphology (see Fig.~\ref{fig:exptests} for an example).

\begin{figure} 
   \centering
   \includegraphics[width=.5\textwidth]{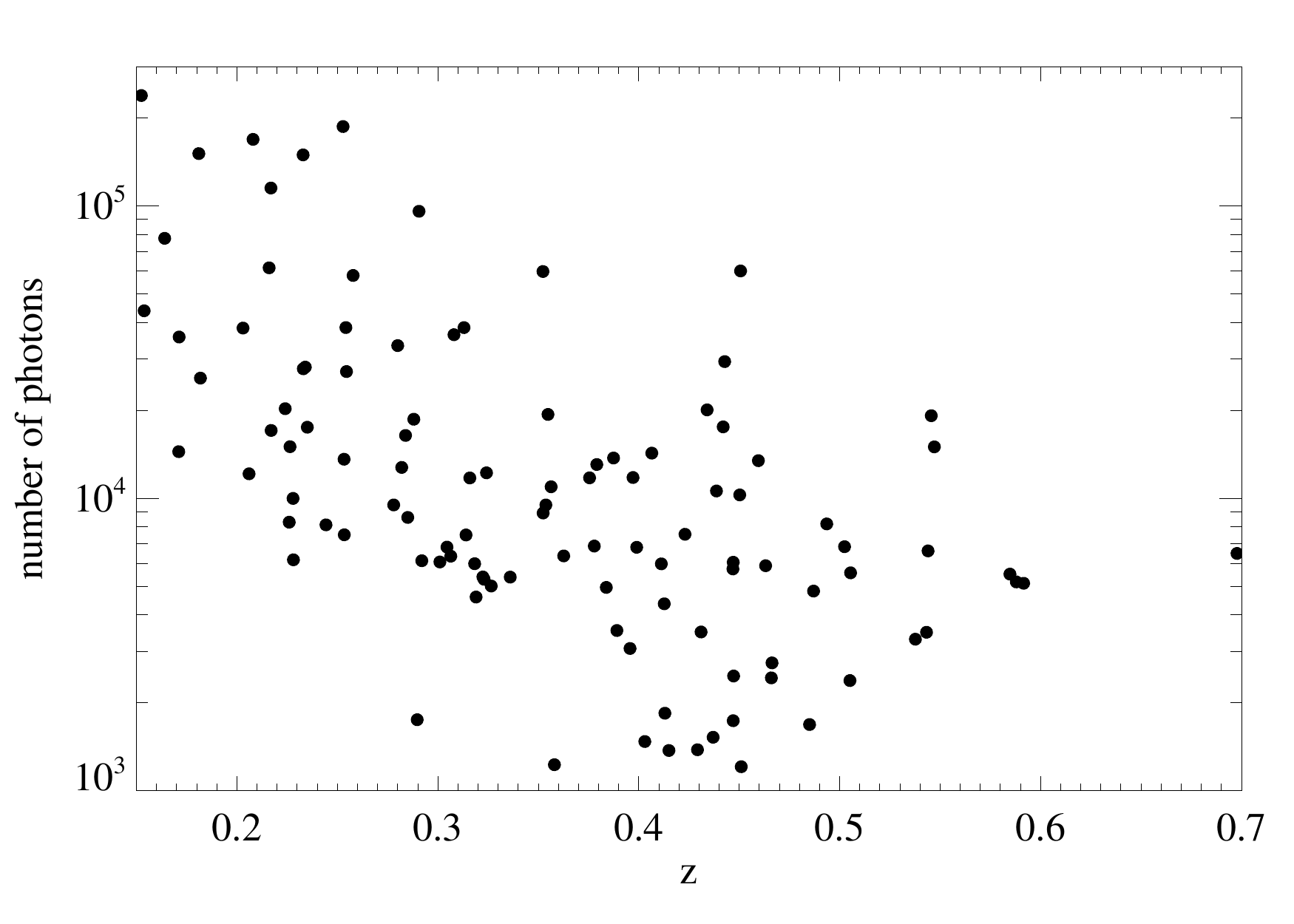} 
   \caption{ Total number of X-ray photons (after background subtraction) for each cluster within the lowest X-ray contour shown in the respective X-ray/optical overlay as a function of redshift.}
   \label{fig:countsvsz}
\end{figure}

Our ability to discern multiple X-ray peaks (or, for that matter, to accurately measure the position of any X-ray peak)  depends on the depth of the respective X-ray image,  as well as on the X-ray brightness of the cluster and its X-ray morphology, all of which vary significantly between clusters. Figure~\ref{fig:exptests} illustrates this point by showing, side by side, the adaptively smoothed images of A\,2744 (MACS\.J0014.3--3022) from the  74\,ks observation obtained with ACIS-I, as well as from just the first 1.5\,ks of the same data set. As is apparent from the comparison, the global centroid of the X-ray emission is hardly affected by the depth of the observation used; the exact location of the X-ray peak(s), however, can change significantly as a function of photon statistics.
To assess whether variations in the photon statistics within our sample affect our ability to obtain a robust morphological classification at all redshifts we show in Fig.~\ref{fig:countsvsz} the photon count (net of background) for all clusters as a function of redshift. As a result of observer bias\footnote{The most distant X-ray luminous systems receive special attention (and thus more CXO exposure time) owing to their importance for cosmological studies.} photon statistics are poorest at roughly the midpoint of the redshift range of our sample. Although still sufficient for reliable determinations of the X-ray centroids, the CXO photon statistics for some clusters at $z{\sim}0.45$ may, for certain morphologies and viewing angles, be too poor to allow accurate measurements of the X-ray peak location.

\subsection{Optical Data}

In addition to X-ray images of good angular resolution (for the computation of the X-ray peak and centroid positions), we need optical images that allow the unambiguous identification of the brightest cluster galaxy (BCG). High-quality colour images of all clusters at $z>0.3$ in our sample were obtained with the UH2.2m telescope in the course of the compilation of the MACS sample. At lower redshift, we performed similar observations for a subset of our sample and rely on SDSS \citep[Sloan Digital Sky Survey,][]{2002AJ....123..485S} and DSS (Digitized Sky Survey) data for the remainder. A brief description of each of these imaging datasets follows.

\emph{UH2.2m Data:} Imaging with the  UH2.2m telescope was performed for 77 of the 108 clusters in our sample between 1999 and 2010, using the Tek2048 CCD (pixel size $0.22\arcsec$) and 240 second dithered integrations in the $V$, $R$ and $I$ filters. Obtained in photometric conditions and sub-arcsec seeing, these images allow the determination of the BCG centre with negligibly small errors. 

\emph{SDSS Data:} For 23 clusters SDSS colour images were produced from data taken through the $g$, $r$, and $i$ filters \citep{1996AJ....111.1748F}.  Although not as deep as the UH2.2m images and also featuring worse angular resolution \citep[median seeing $\sim 1.4\arcsec$ and using $0.5\arcsec$ pixels; ][]{2009ApJS..182..543A}, the SDSS images still yield BCG positions with negligibly small errors. 

\emph{HST Data:}  Three clusters (A\,2146, A\,2163, and A\,2667) located outside the SDSS footprint and without UH2.2m images have been observed with the Hubble Space Telescope (HST) allowing highly accurate measurements of the BCG position.

\emph{DSS Data:}  For the remaining five clusters, the only readily available optical data hail from the DSS, which is  much shallower\footnote{For instance, the limiting magnitude of the DSS is $g \sim 21.5$  \citep{2004AJ....128.3082G}  compared to $g \sim 23$ for the SDSS \citep{2000AJ....120.1579Y}.}, of poorer angular resolution\footnote{The median seeing is $\sim 1.5\arcsec$ and the pixel size $1\arcsec$.}, and of inferior cosmetic quality\footnote{Much of the DSS data was obtained from scans of photographic plates, a process that, in addition to objects of astrophysical interest, also faithfully recorded fingerprints, scratches, and other artefacts.} than the aforementioned imaging databases. We use infrared, red, and blue images from the DSS to create colour images. The relatively poor quality of the DSS data leads to significant uncertainties in the determination of the BCG centre. 

\subsection{Astrometric Corrections}\label{sec:astrometriccorrections}

The primary diagnostics used by us to characterise the dynamical state of clusters and to identify merging systems are the offsets between the location of the BCG and that of the X-ray peak or global X-ray centroid, respectively. Either diagnostic thus mandates that our optical and X-ray images be astrometrically aligned as accurately as possible. The magnitude of the required astrometric corrections can be estimated from the absolute astrometric accuracy of the datasets involved. On the X-ray side, CXO provides astrometric solutions that are accurate to better than $0.8\arcsec$ within 3 arcmin of the ACIS aimpoint\footnote{See http://cxc.harvard.edu/cal/ASPECT/celmon/}. For our optical data, we can expect astrometric uncertainties for our UH2.2m images of less than $0.5\arcsec$ after processing with WCSTools \citep{2006ASPC..351..204M}, less than $0.2\arcsec$ for the SDSS \citep{2003AJ....125.1559P}, and less than $0.4\arcsec$ for the DSS images used \citep{2008AJ....136..735L}. The resulting error in the {\it relative}\/ astrometry can thus reach, and occasionally even exceed, $1\arcsec$, which corresponds to 5~kpc at the median redshift ($z = 0.35$) of our sample -- a significant contribution to the overall error  budget.

\begin{figure} 
   \centering
   \includegraphics[width=.5\textwidth]{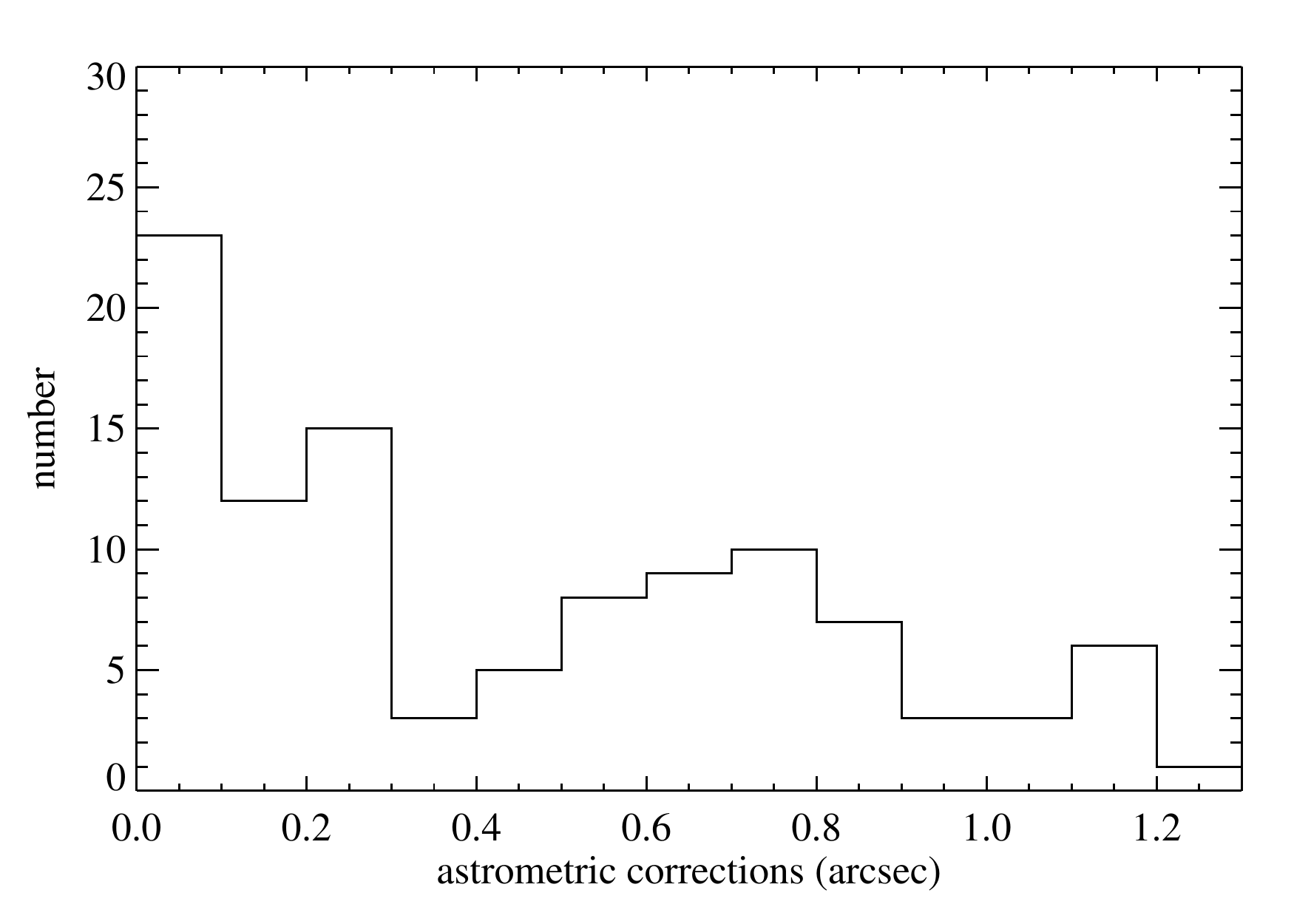} 
   \caption{ Histogram of astrometric corrections applied to the imaging data for our 108 clusters.  The median correction is 0.36\arcsec\ for all optical sources.  Optical data from the UH2.2m telescope require considerably larger corrections (median $\sim$0.8\arcsec) than data from the SDSS and DPOSS2.  Three anomalous cluster fields that require astrometric corrections in excess of 1.2\arcsec\ are not shown.}
   \label{fig:corrections}
\end{figure}

In order to eliminate, or at least minimise, this source of measurement error, we compute corrections in the {\it relative} astrometry between the optical and X-ray images by identifying the optical counterparts to the brightest X-ray point sources within the field of view common to both images. Typically, over half a dozen matches with X-ray/optical offsets that are consistent to within $0.3\arcsec$ are found for each cluster.  Figure~\ref{fig:corrections} shows a histogram of the astrometric corrections applied; the distribution is consistent with the expectations discussed above.

\subsection{Merger classification criteria}\label{sec:mergclass}

While the morphological diagnostics discussed in Section~\ref{sec:diagnostics} are well suited to allow us to select the most disturbed systems, several selection biases enter.

Most importantly, morphological classification based solely on imaging data is subject to projection effects. Given that all clusters selected for this study are intrinsically X-ray luminous, projections involving physically unrelated fore- or background structure are improbable. Hence, overestimates of the complexity of a system, and thus of the merger fraction, are negligible. The opposite effect, however, is bound to be present, as mergers proceeding along or close to our line of sight are likely to go unrecognized. We shall return to this important point when discussing the results of our study. An additional complication arises from the fact that we are interested only in mergers observed {\it after}\/ the primary collision, thereby rendering them useful laboratories for in-depth study of the interactions of all three cluster components (Sections~\ref{sec:compmerg},~\ref{sec:bhomdef}). Visual scrutiny of the combined X-ray/optical appearance of a cluster is usually sufficient to eliminate close double clusters or pre-collision mergers, but -- alas -- again not for prolate systems forming along an axis close to our line of sight. Finally,  the subdivision of disturbed clusters into complex mergers and BHOMs is not always possible and necessarily subjective to some extent.

These challenges notwithstanding, we attempt to identify complex mergers as well as BHOM candidates, all of them observed {\it after}\/ the primary collision, by using the following criteria:

\begin{description}
\item[{\bf Morphological class:}] Any ongoing or recent merger along an axis sufficiently misaligned with our line of sight will have a value of 3 or 4 for the morphological code described in Section~\ref{sec:diagnostics}. A high value for the morphological class is thus a necessary but not a sufficient criterion for the selection of both complex mergers and post-collision BHOMs;
\item[{\bf X-ray-centroid/BCG separation:}] A large offset between the global centroid of the X-ray emission and the location of the BCG {\em may} be an indicator of a post-collision merger in which the ICM of the merging clusters have coalesced into a single central gas cloud; however, large offsets are also observed for pre-collision mergers. Conversely, viewing angle and merger phase may conspire to yield a small X-ray-centroid/BCG separation even for very disturbed systems. A high value for this diagnostic is thus  (with the noted caveat concerning pre-collision mergers)  a sufficient but not a necessary selection criterion;
\item[{\bf X-ray-peak/BCG separation:}] A large separation between the X-ray peak and the location of the BCG is an unambiguous sign of an ongoing merger. Correctly interpreting a small separation can, however, be a difficult task since this metric depends on both the stage of the merger process (BCG and surviving cool core will align again approximately at turnaround) and the impact parameter (in particular for mergers of clusters of significantly different mass). A high value for the X-ray-peak/BCG separation is thus again a sufficient but not a necessary selection criterion for our purposes.
\end{description}

In view of the complexity and enormous variety in the projected geometry and dynamics of real-life cluster mergers,  we therefore require likely post-collision mergers to a) meet our morphological selection criterion (class 3 or 4) and b) feature a large offset of the BCG from either the X-ray centroid or the X-ray peak.

Noticeably absent from the discussion above is a criterion that clearly distinguishes between complex mergers and BHOMs, i.e., a criterion that assesses whether two, and only two, clusters are involved and clearly discernible in the optical image of a BHOM candidate. Ideally we would hence like to add a fourth criterion akin to:
\begin{description}
\item[{\bf BCG-BCG separation:}] Unless it proceeds along an axis very close to our line of sight, a two-cluster merger observed at any stage other than core passage should feature two well separated BCGs, one for each of the two participating clusters;  for a post-collision BHOM, the X-ray centroid will be located between them.  
\end{description}
Unfortunately, we found this  BHOM selection criterion (intuitive and straightforward as it seems) to be almost impossible to implement in practice, the primary reason being that, for mergers of very different mass, the two brightest galaxies within the lowest  X-ray contour are frequently both members of the more massive merger component. This is true, for instance, for the Bullet Cluster: the two brightest galaxies are both members of the (unrelaxed) main cluster. While this criterion can thus not be applied without visual scrutiny of the respective system, it is useful in principle, as illustrated by the case of the highly complex merger A\,2744 which features no fewer than {\it five} galaxies of essentially equal brightness (within the measurement uncertainties, Fig.~\ref{fig:exptests}). 

In recognition of the difficulties inherent in the subclassification of merger types we therefore apply the first three criteria to select extreme mergers of all kinds, and then resort to visual inspection of all  candidates in order to distinguish between complex mergers (Section~\ref{sec:complex}) and likely post-collision BHOMs (Section~\ref{sec:bhom}).

\section{Results}\label{sec:results}
\subsection{X-ray/optical separation and morphological code}
Table~1 lists the values of all three quantitative diagnostics used in this study, i.e., the assessed global morphology described in Section \ref{sec:diagnostics}, as well as the projected metric separations between the BCG and the X-ray peak and centroid, respectively,  for each of our 108 clusters. Also listed are the $1 \sigma$ errors of the latter two measurements.  We find the distribution to be roughly log-normal for either type of separation (Fig.~\ref{fig:logsepall}), and to be centred at 11.5 and 21.2 kpc for the offset of the BCG from the X-ray peak and from the X-ray centroid, respectively.

15\% (16 out of 108) of the clusters in our sample have projected X-ray-peak/BCG separations of at least 50~kpc, slightly higher than, but (within the uncertainties) consistent with, the percentage of 12\% (8 out of 64) reported by \cite{2010A&A...513A..37H} for the HIFLUGCS sample.  Application of the same cut (${>}50$~kpc) to the X-ray-centroid/BCG separations selects 30\% (32 out of 108) compared to a value of 20\% (13 out of 65) found for the LoCuSS sample   \citep{2009MNRAS.398.1698S}. Both of these comparisons are, however, problematic because of the fact that HIFLUGCS is X-ray flux limited, and LoCuSS uses Galactic $n_{\rm H}$ as a selection criterion, whereas the sample used in this work is primarily X-ray luminosity limited. In addition, the redshift ranges covered by the three cluster samples differ substantially (HIFLUGCS: $\textless z\textgreater = 0.05$; LoCuSS: $\textless z\textgreater = 0.23$; this study: $\textless z\textgreater = 0.35$).

\subsection{Extreme and active mergers}

\begin{figure} 
   \centering
   \includegraphics[width=.5\textwidth]{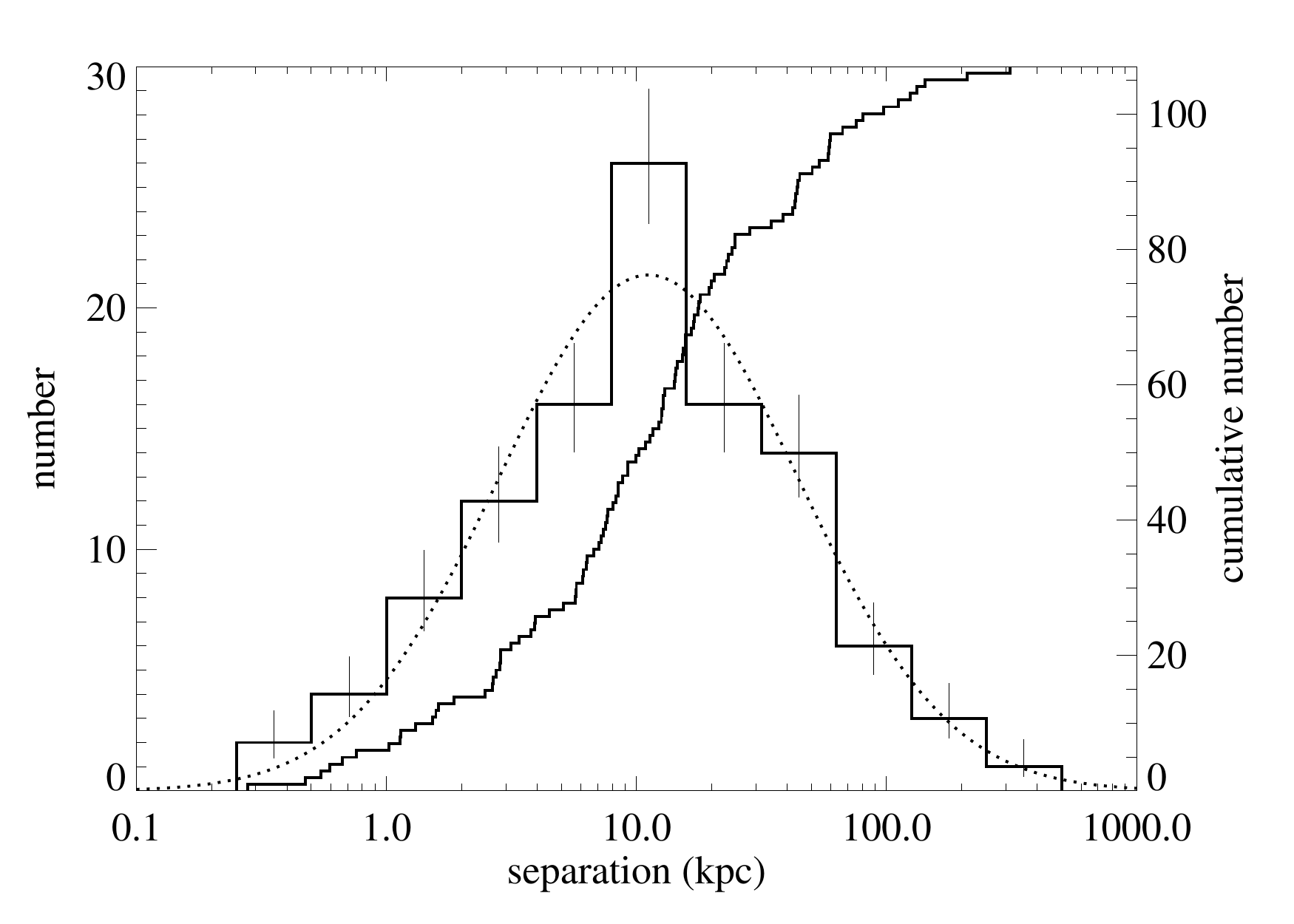} 
   \includegraphics[width=.5\textwidth]{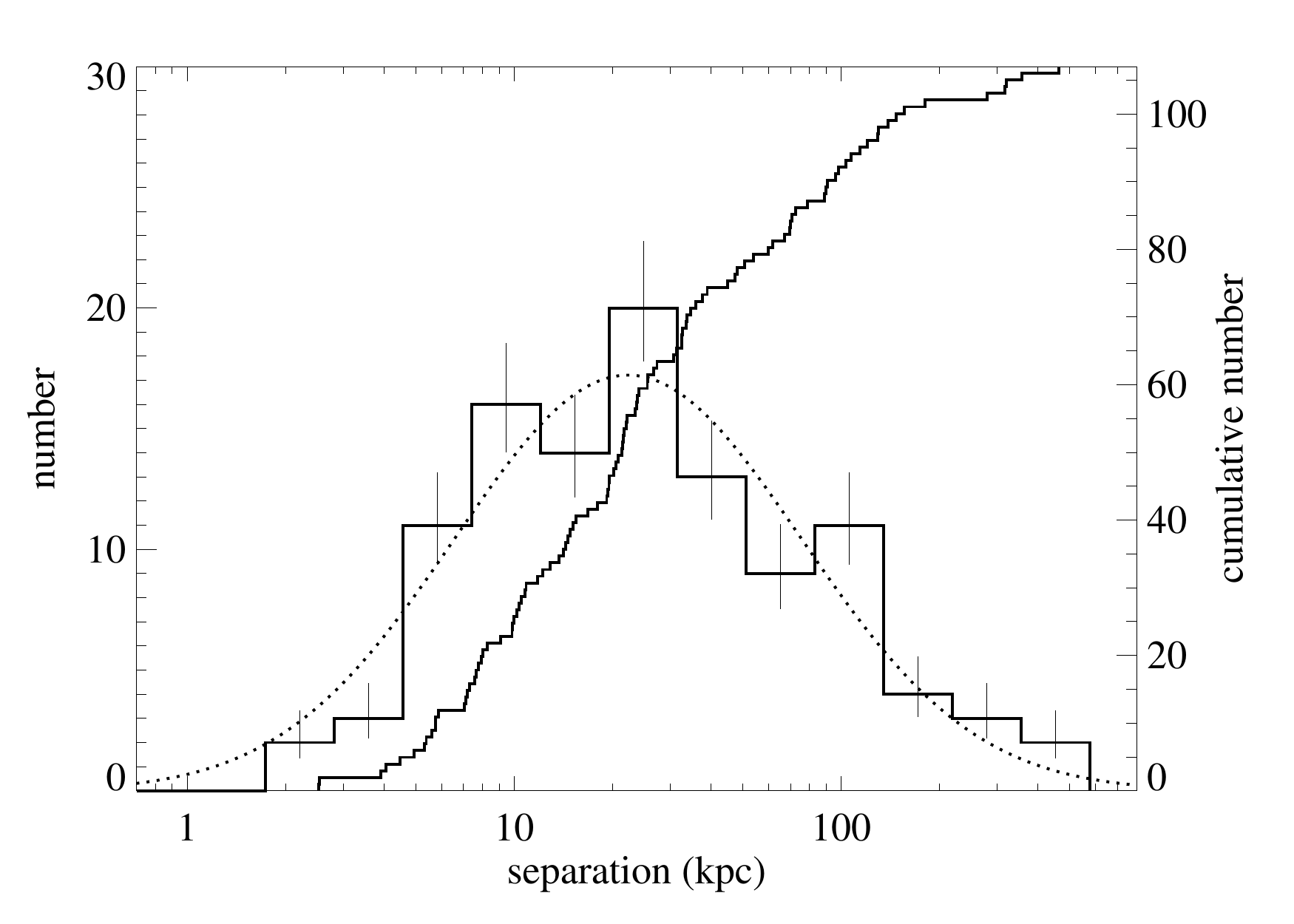} 
   \caption{ Histogram of the offsets of the BCG from the X-ray peak (top) and X-ray centroid (bottom) for all 108 clusters in our sample.  The distributions are lognormal centered at $11.5^{+30}_{-9}$ and $21.2^{+50}_{-16}$ for the peak and centroid respectively.  Poisson errors are shown, as are the cumulative distributions of offsets.}
   \label{fig:logsepall}
\end{figure}

A total of 29 clusters in our sample meet the criteria discussed in Section~\ref{sec:mergclass}, i.e., they have been assigned a morphological code of 3 or 4  {\em and}\/ feature a large offset of the BCG from the X-ray centroid  or the X-ray peak. We implement the latter criterion by requiring the respective offset to exceed a threshold value of 71 or 42\,kpc, i.e., the mean of the respective distribution of separations plus one standard deviation, rounded to 1\,kpc precision
(Fig.~\ref{fig:logsepall}).

All systems thus selected are far from relaxed and thus prime candidates for studies of the physics of cluster mergers; however, not all of them are  BHOMs, and some may not meet the important requirement that we observe them after the primary core passage. As discussed in Section~\ref{sec:mergclass}, it is difficult to devise an objective, quantitative criterion for either charactistic. We therefore screen all candidates visually, focusing on the systems' appearance in the optical and X-ray waveband. Our assessment then uses primarily the merger's morphological complexity as well as the direction of motion of the merger participants inferred from the BCG-X-ray-peak offsets. 

As a first result of this screening process we eliminate MACS\,J0159.0--3412 and ZwCl\,1459.4+4240 because, in both cases, the cluster emission falls so close to the ACIS-I chip gaps as to raise doubts about the reliability of the X-ray morphological classification listed in Table~1.

\subsubsection{The most complex massive mergers}\label{sec:complex}

Scrutiny of the optical and X-ray appearance of the 27 initial candidates leads to the classification of 10 systems as complex mergers that have recently undergone, or are still undergoing, multiple mergers along differing axes. We briefly discuss them individually below.

\noindent

%{\it MACS\,J0014.3-3022}
\noindent
{\it A\,2744}, a well studied cluster, features multiple galaxies of very similar brightness, rendering the identification of a BCG difficult.  Although A\,2744 could be mistaken for a BHOM when only its X-ray appearance is considered, it features X-ray/optical offsets that are inconsistent with the linear trajectories expected in such a scenario. Indeed, detailed analyses at optical and X-ray wavelengths showed the system to be the result of a recent line-of-sight primary merger combined with a secondary merger proceeding at high impact parameter \citep{2004MNRAS.349..385K,2011ApJ...728...27O,Merten:2011qy}. Consistent with such an active dynamical history, A\,2744 is also known to host a radio halo \citep{2006A&A...449..461B}.

\noindent
{\it A\,520} is a complex system whose unusual morphology led to claims of the presence of a ``dark core" \citep{2007ApJ...668..806M}. A dynamical analysis of the galaxies in the field suggests a multiple merger at the intersection of three large-scale filaments \citep{2008A&A...491..379G}. Evidence of past and ongoing merging is found in the form of a bow shock (one of less than a handful of supersonic shocks detected so far in cluster mergers) and a complex radio halo \citep{2005ApJ...627..733M}.

\noindent
{\it MACS\,J0717.5+3745}\/ is the most disturbed massive cluster known at $z{>}0.5$ \citep{2007ApJ...661L..33E}, accreting matter along a 6\,Mpc long filament \citep{2004ApJ...609L..49E}. A detailed study of the ICM and the cluster galaxy population yielded compelling evidence of a triple merger, resulting in ICM temperatures as high as 20 keV and galaxy transformations caused by ram pressure \citep{2008ApJ...684..160M,2009ApJ...693L..56M,2011MNRAS.410.2593M}. Shocks triggered by a series of mergers are likely responsible for the acceleration of electrons in the ICM to relativistic energies, making  MACS\,J0717.5+3745 a source of extreme non-thermal emission at radio wavelengths \citep{2003MNRAS.339..913E,2009A&A...503..707B,2009A&A...505..991V}. The system was also successfully targeted in Sunyaev-Zel'dovich observations \citep{2006ApJ...652..917L}.

\noindent
{\it A\,665} lacks optical evidence of multiple cluster cores in spite of a very disturbed morphology. A joint X-ray/optical analysis suggests a head-on merger along the line of sight, observed close to core passage  \citep{2000ApJ...540..726G}, a scenario that would explain the system's morphology as viewed in projection as well as the observed unimodal, but very broad ($\sigma{=}1400$ km s$^{-1}$) distribution of radial velocities for the cluster galaxy population. A\,665 is thus an example of a likely BHOM (see Section~\ref{sec:bhom}) missed by the set of diagnostics used by us here because the merger axis is nearly aligned with our line of sight. This merger too is giving rise to non-thermal emission in the form of a well studied synchrotron radio halo \citep{2001ApJ...563...95M, 2004A&A...423..111F,2010A&A...514A..71V}.

\noindent
%{\it MACS\,J1131.8--1955} 
{A\,1300} is a highly disturbed cluster whose optical complexity rivals that of A\,2744 in that it features no fewer than five galaxies vying for the title of BCG. The cluster's complex X-ray morphology suggests a series of recent mergers, as well as an ongoing merger reflected by the potential presence of a cold front. Hosting both a giant radio halo and a radio relic, this system features frequently in studies of the connection between cluster mergers and diffuse radio emission \citep{1999MNRAS.302..571R,2008A&A...480..687C,2010ApJ...721L..82C} but has, to the best of our knowledge, yet to attract similar attention at other wavelengths.

\noindent
{\it MACS\,J1149.5+2223} is part of the complete subsample of MACS clusters at $z>0.5$. Deep follow-up observations with HST/ACS revealed spectacular strong-lensing features which allowed detailed modelling of the cluster mass distribution. Comprising at least four large-scale dark-matter halos,  MACS\,J1149.5+2223 is one of the most complex clusters presently known \citep{2009ApJ...707L.163S}. The system's extreme velocity dispersion of 1800 km s$^{-1}$ \citep{2007ApJ...661L..33E} strongly suggests merger activity along our line of sight.

\noindent
{\it MACS\,J2129.4--0741} is the third cluster from the $z>0.5$ subset of the MACS sample to make our list of complex mergers and, much like A\,665, does so only by virtue of a significant offset of the BCG from the large-scale centroid of the X-ray emission recorded by Chandra. Featuring the same velocity dispersion of 1400\,km s$^{-1}$, MACS\,J2129.4--0741 may be dynamically similar to A\,665 also in terms of its recent merger history.

\noindent
%{\it MACS\,J2228.5+2036}
{\it RX\,J2228.5+2036} was discovered during the ROSAT Brightest Cluster Survey \citep[][BCS,]{1998MNRAS.301..881E} and followed up in pointed X-ray observations with the ROSAT HRI and XMM-Newton. A high mass in excess of $10^{15}\;h^{-1}$ M$_\odot$ was estimated from X-ray and Sunyaev-Zel'dovich effect observations under the assumptions of sphericity and hydrostatic equilibrium \citep{2002A&A...387...56P,2008A&A...489....1J}, both of which are unlikely to apply given the cluster's X-ray/optical appearance \citep[Fig.~\ref{fig:complexmerg}; see also][]{2008ApJS..174..117M}. Although our classification of this system as a merger is seemingly called into question by the absence of diffuse radio emission \citep{2008A&A...484..327V}, the strong correlation between mergers and non-thermal radio emission is not without exceptions, as demonstrated by \,\citet{Russell:2011fk} whose GMRT observations of A\,2146 (see Section~\ref{sec:bhom}) also failed to detect any sign of diffuse emission. 

\noindent
{\it MACS\,J2243.3--0935} is a highly disturbed cluster according to all three of our diagnostics. The merger axis suggested by its highly elongated X-ray emission is significantly misaligned with the axis implied by the two main galaxy concentrations, thus all but ruling out a BHOM scenario. The system's high velocity dispersion of   over 1500\,km s$^{-1}$ \citep{Ebeling:2010mz} may be reflective not only of this cluster's high mass but  also of line-of-sight merger events in its recent dynamical history.

\noindent
{\it A\,2631} exhibits an X-ray/optical morphology similar to MACS\,J2243.3--0935 but at much lower galaxy surface density. Although clearly highly unrelaxed, the cluster is listed as a non-detection by \citet{2008A&A...484..327V} who conducted a search for diffuse radio emission in 50 X-ray selected clusters. The absence of non-thermal radio emission from this and several other obvious mergers presents a formidable challenge for theories of particle acceleration during massive cluster mergers.

\subsubsection{BHOMs}\label{sec:bhom}

The remaining 17 clusters are all potential or probable early post-collision BHOMs, and are discussed individually below (in order of right ascension).

\begin{center}
{\it Primary candidates}
\end{center}

\noindent
{\it MACS\,J0025.4--1222}\/ is a textbook case of the optical and X-ray morphology expected for a post-collision BHOM. It is also one of only two systems studied already in depth in attempts to constrain the collisional properties of dark matter (the other one being the Bullet Cluster). The simple geometry of the merger and the wide separation of its components made  MACS\,J0025.4--1222 a rewarding target also for studies of the impact of mergers on galaxy evolution. We refer to the literature for a detailed discussion and results \citep{2007ApJ...661L..33E, 2008ApJ...687..959B, 2010MNRAS.406..121M}.

\noindent
{\it MACS\,J0140.0--0555}\/ is very similar in appearance to MACS\,J0025.4--1222, suggesting a post-collision BHOM involving two clusters of comparable mass. Although the metric separation of the two components of about 250 kpc may pose challenges for the separation of the corresponding dark-matter halos in a weak-lensing study, the likely high mass of either cluster makes this BHOM an excellent candidate for constraining dark-matter properties following the approach taken for MACS\,J0025.4--1222 \citep{2008ApJ...687..959B}. A first characterization of the system based on all existing optical and X-ray data is given by Ho et al.\ (2011, in preparation).

\noindent
{\it MACS\,J0417.5--1154}\/ has a comet-like appearance in X-rays, with an extremely X-ray bright and compact core that is coincident with the optical core of one of the two merging clusters, trailed by diffuse emission towards, and encompassing, the second cluster. The system's X-ray morphology strongly suggests a high-velocity encounter akin to the scenario observed in A\,2146 or the Bullet Cluster. The absence of a noticeable offset between the locations of the BCG and X-ray peak of the compact core suggest that the merger is observed either close to turnaround and / or proceeds along a merger axis that is significantly tilted in the direction of our line of sight. A detailed analysis of the apparent shock front to the SW of the compact core (cf.~Fig.~\ref{fig:bhomstier1}), as well as of the mass distribution of both components as revealed by strong- and weak-lensing observations, is underway.

%MACS\,J0454.1--1014
\noindent
{\it A\,521}\/ deserves special mention: although the main cluster in this merger appears morphologically unremarkable\footnote{Note that the X-ray appearance of this system in Fig.~\ref{fig:bhomstier1} is misleading due to the ACIS-I chip gaps truncating the very extended emission from the presumably more massive southern merger component.}, its extremely low surface density, both in X-rays and in terms of projected galaxy numbers, make it stand out among clusters of similar X-ray luminosity at this redshift ($z{=}0.253$). The presence of an extended radio source \citep{2006NewA...11..437G, 2009ApJ...699.1288D} as well as of multiple components in the redshift distribution of cluster galaxies \citep{2003A&A...399..813F} strongly suggests a long history of merger events, possibly along different axes. In spite of the complexity of the primary cluster component, the ongoing merger that appears to occur close to the plane of the sky makes {\it A\,521} a promising target for quantitative studies of the physics of cluster mergers \citep{2006A&A...446..417F}.

\noindent
{\it MACS\,J0553.4--3342}\/ is again a BHOM that is clearly observed past core passage and that appears to involve at least two similarly massive clusters. The large metric separation of the involved clusters makes it likely that the merger axis lies approximately in the plane of the sky. Whether a third, less massive cluster component is involved (as suggested by both the ICM and galaxy distributions), and whether the merger is viewed sufficiently shortly after the initial collision to allow the participating dark-matter halos to still be discernibly  separated will be revealed by an ongoing X-ray/lensing study using deep Chandra and HST/ACS observations.

\noindent
{\it MACS\,J1006.9+3200}\/ resembles A\,1682 (see below) in its X-ray/optical morphology. Clearly observed after the primary collision (possibly one of several recent merger events), the system may be near turnaround, as suggested by the wide separation of the two subclusters and the direction of the BCG/X-ray peak offset of the minority component. To the best of our knowledge, no in-depth study of this merger has been conducted to date.

\noindent
%{\it MACS\,J1306.8+4633}
{\it A\,1682}\/ appears at first glance like a likely pre-collision merger of two systems of very different mass (Fig.~\ref{fig:bhomstier1}). A pair of BCGs in the primary cluster and diffuse X-ray emission between the merging systems, however, constitutes convincing evidence of prior interaction. This interpretation is supported by the detection of very complex diffuse radio emission, interpreted as a halo and two relics by \citet{2008A&A...484..327V}.

\noindent
{\it A\,1758}\/ is an obvious post-collision BHOM (prominent X-ray emission between the two participating clusters) in which both merger components appear to have retained some of their original cold cores, visible as compact X-ray peaks near the respective BCG. An existing weak-lensing study of  A\,1758 based on ground based optical data \citep{2008PASJ...60..345O} finds two well separated mass concentrations and assigns the system a total mass of $(5\pm 1.5)\times 10^{14}h_{\rm 0}^{-1}\; {\rm M}_{\odot}$ and a velocity dispersion of $(670\pm 110)$ km s$^{-1}$. Although thus likely less massive than MACS\,J0025.4--1222, A\,1758 may be another good candidate for a measurement of $\sigma/m$. The system is of interest also as a future merger on much larger scales, given the presence of a second cluster, of lower X-ray luminosity but at the same redshift,  about 2 Mpc south of the system discussed here \citep{1998MNRAS.301..328R,2004ApJ...613..831D}. Observations conducted with the Spitzer observatory revealed an unusually high rate of star formation, primarily from the northern, main cluster, with much of the presumably merger-induced activity originating in obscured galaxies near the cluster core and at the extreme outskirts \citep{2009MNRAS.396.1297H}.

\noindent
%{\it MACS\,J1426.0+3749}
{\it A\,1914}\/ is a well known merger with a morphologically very complex core, but qualifies for ``post-collision BHOM" status by virtue of the presence of two distinct (if close) optical cluster cores on either side of both the peak and the overall centroid of the X-ray emission. However, the noticeable misalignment between the elongated X-ray contours in the core and the merger axis suggested by the line connecting the BCGs of the two components raises the possibility that the collision occurred with a significant impact parameter, rather than head-on. In addition, the large extent and near-sphericity of the X-ray emission on larger scales may be the result of the merger axis being distinctly inclined with respect to the plane of the sky. This interpretation is supported by a large discrepancy between mass estimates based on X-ray and weak-lensing analyses \citep{2008PASJ...60..345O,2010ApJ...711.1033Z}. Featuring a powerful radio halo, A\,1914 is also part of a larger cluster sample for which \citet{2004ApJ...605..695G} investigated the relation between variations in the ICM temperature and the presence and location of diffuse radio emission. 

\noindent
%{\it MACS\,J1556.1+6621}
{\it A\,2146}\/ is almost a clone of {\it MACS\,J0417.5--1154} in terms of optical and X-ray morphology. A recent in-depth study of a deep CXO/ACIS-S observation of this system \citep{Russell:2010gf} finds two pronounced shock fronts (with Mach numbers of 2.2 and 1.7, respectively), rendering this merger the closest rival yet to the Bullet Cluster. Although the estimated velocity of the bullet in A\,2146 is lower than the one measured for 1E\,0657--56, the ongoing analysis of this BHOM is likely to yield competitive constraints on $\sigma/m$. The system is also a prime candidate for a search for evidence of accelerated galaxy evolution as a result of the collision. Surprisingly for such an extreme merger, and so far inexplicably, A\,2146 shows no signs of diffuse radio emission \citep{Russell:2011fk}.
 
\noindent
{\it MACS\,J1731.6+2252}\/ exhibits an X-ray/optical morphology similar to that of A\,2744 or the Bullet Cluster (although at low fidelity, owing to the short exposure time of the Chandra observation), in that a merger of two clusters of very different mass is viewed well after collision. The main component of MACS\,J1731.6+2252 shows no significant X-ray/optical offset though, suggesting either a collision at large impact parameter and/or a significant inclination of the merger axis with respect to the plane of the sky. Like A\,2744, this system may also be a merger of more than two components. A much longer Chandra observation would allow a better estimate of the three-dimensional merger trajectories to be obtained.

\begin{center}
{\it Secondary candidates}
\end{center}

The following systems show compelling evidence of being post-collision BHOMs but feature (projected) separations between the BCGs of the merging clusters of less than 200~kpc, either because the respective mergers proceed along axes close to our line of sight, or because the merging components are not yet (or no longer) sufficiently far apart. If the apparent close separation of the latter is indeed due to a greatly inclined merger axis, projection effects and the degeneracy between peculiar velocity and Hubble flow are likely to impede if not prevent the determination of the true three-dimensional merger geometry and the physical interpretation of the observables. If, on the other hand, these mergers proceed approximately in the plane of the sky and simply happen to be observed by us at a time of small separation of the components, all of the systems listed below will still be of great interest for studies of all merger-induced physical effects, with the likely exception of measurements of $\sigma/m$.

\noindent
{\it A\,2813} appears moderately relaxed on large scales but exhibits a double core in both optical and X-ray images \footnote{We note that the XMM-Newton study by \citet{2006A&A...456...55Z} does not classify the system as a merger, underlining again the importance of high-resolution X-ray data.}. Its morphology is most easily explained by a merger proceeding along an axis that is highly inclined with respect to the plane of the sky; if so, it is conceivable that we observe the system before the primary collision.

\noindent
{\it MACS\,J0358.8--2955}\/ was only recently  recognized as a very X-ray luminous cluster at $z=0.434$ \citep{Ebeling:2010mz} after \citet{2004A&A...425..367B} misidentified the system as A\,3192 at $z=0.169$. Both at X-ray and optical wavelengths two cluster components are clearly separated, but as evident are signs of previous interactions. The combination of a small projected separation of the cluster cores and an X-ray/optical offset of the more compact component that is not aligned with the apparent merger axis suggests a collision at significant impact parameter and possibly a line-of-sight merger. A detailed study with Chandra and HST is underway in an attempt to better constrain the merger geometry.

\noindent
{\it MACS\,J0404.6+1109}\/ features a very interesting morphology with an X-ray peak located between two almost equally bright BCGs. The very small separation of the two BCGs of only about 100\,kpc suggests either a line-of-sight merger or a very recent collision. A deeper Chandra observation is needed to improve upon the presently poor photon statistics and allow a better assessment of the relative distributions of collisional and collisionless matter in the cluster core.

\noindent
{\it MACS\,J0416.1--2403} would easily make our list of primary BHOM candidates if  the separation of the system's BCGs did not fall just shy of the 200\,kpc threshold. Illustrating an element of subjectivity of our classification, this merger (clearly observed after the primary collision, and probably even after turnaround) should make an excellent target for studies of the physics of cluster mergers.

\noindent
{\it ZwCl 0847.2+3617}, aka RX\,J0850.1+3604, is an X-ray discovery from the Brightest Cluster Sample \citep{1998MNRAS.301..881E} that exhibits the classical signature of a post-collision BHOM: a single X-ray peak between two galaxy concentrations. Its other morphological characteristics, however, raise doubts about the system's usefulness for quantitative physical investigations. Specifically, the elongation of its X-ray emission on large scale runs at right angles to the presumed merger axis suggests a complex merger history. In addition, the small separation of the mentioned two galaxy concentration relative to the extent of the X-ray emission would be most easily explained by a merger axis close to the line of sight. Diffuse radio emission, as expected from such a massive merger, was detected by \citet{2009ApJ...697.1341R} who, however, attribute the emission to a radio galaxy.

\noindent
{\it MACS\,J1115.2+5320}\/ resembles MACS\,J0140.0--0555 (and thus also MACS\,J0025.4--1222) but features a smaller separation of the two BCGs. While a partial alignment with our line of sight is likely to contribute to the small separation observed in projection (\citet{Ebeling:2010mz} report a high velocity dispersion of 1300 km s$^{-1}$ for this cluster), we may also simply be catching this merger at a later stage than MACS\,J0025.4--1222. Several spectacular gravitational arcs visible in our UH2.2m images were also independently noted by \citet{2008AJ....135..664H} and \citep{2010MNRAS.406.1318H}. An in-depth study of the mass distribution in the cluster core is underway based on strong lensing constraints derived from HST imaging data and Keck spectroscopy.

\subsection{Redshift evolution of the merger fraction} \label{sec:evolution}

Theories of hierarchical structure formation  predict an increase in the fraction of unrelaxed clusters as we approach the epoch of  formation for massive clusters at $z \sim 1$.  Although the redshift range of our sample ($0.15\,{<}\,z\,{<}\,0.7$) falls short of probing this era we may still expect to see the imprint of hierarchical merging in the form of  an increase with redshift in the fraction of dynamically disturbed clusters, as evidenced by higher X-ray/optical separations and by a preference for higher values in the assigned morphological code\footnote{Note that our inability to reliably identify mergers proceeding close to our line of sight does not affect our measurement of the evolution of the merger fraction since the distribution of merger axes does not correlate with redshift.}.

In order to ensure that any observed trends with redshift are not caused by selection bias (not all clusters in our original list have Chandra data), all results presented and discussed in this section are based on the analysis of a subset of only 75 clusters, defined by a more aggressive X-ray luminosity cut of $L_{\rm X, RASS} > 7.5\times10^{44}$ erg s$^{-1}$ (and shown as a dashed line in Fig.~\ref{fig:sample}). Above this threshold 95\% of the clusters in our sample have CXO data, rendering selection bias negligible.

\begin{figure}
   \centering
   \includegraphics[width=.5\textwidth]{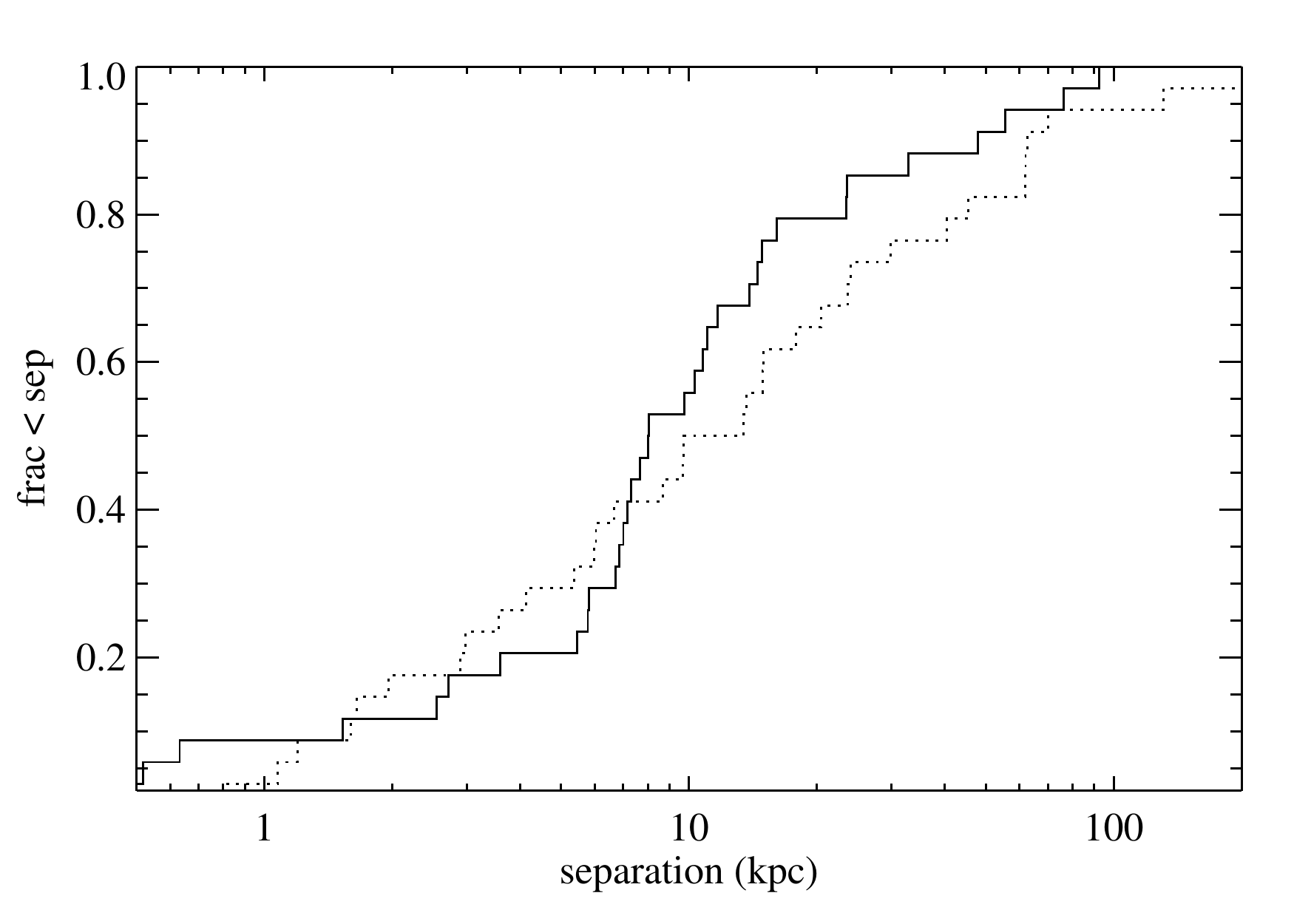} 
   \includegraphics[width=.5\textwidth]{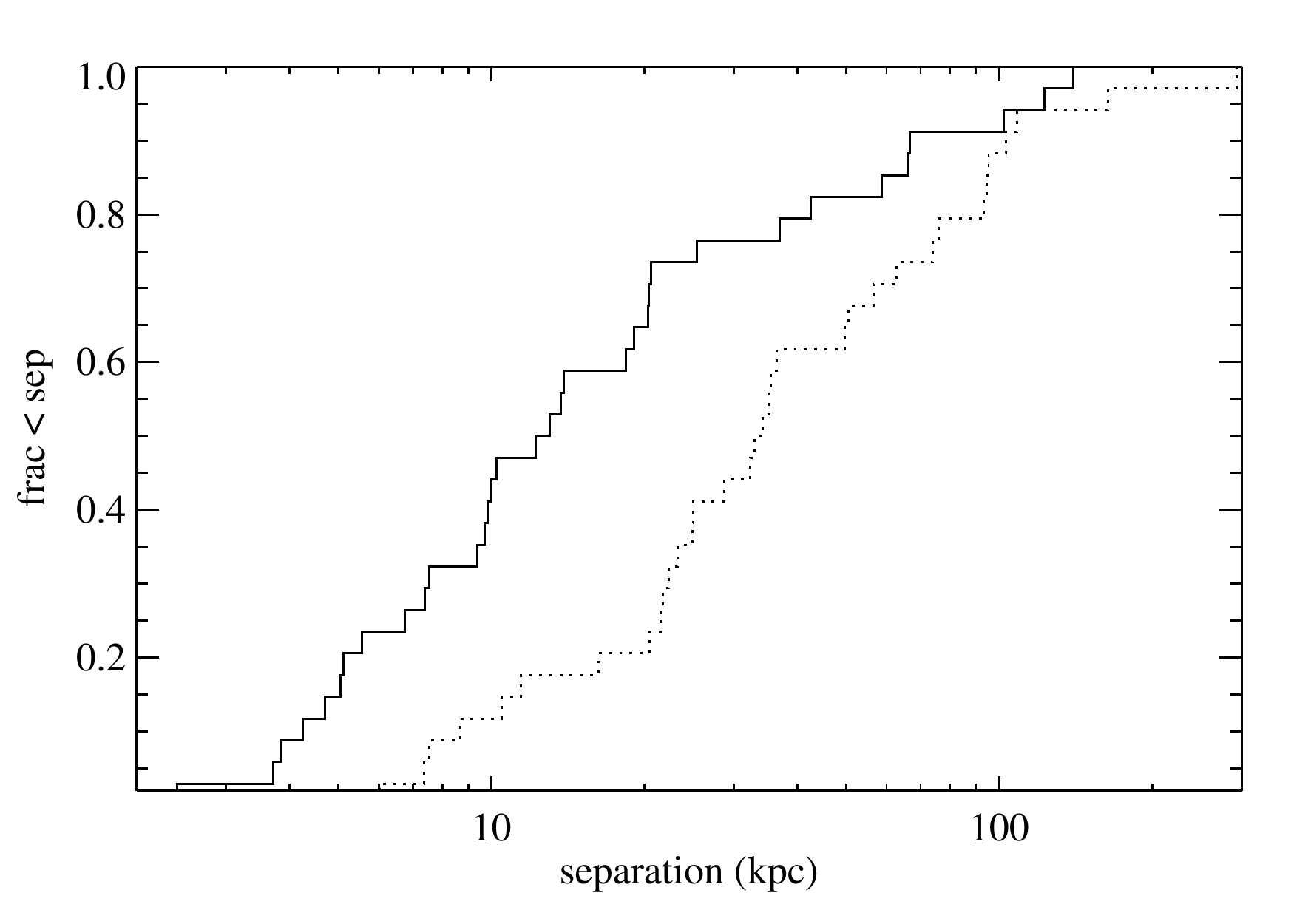} 
   \caption{ Cumulative distribution of the offsets of the BCG from  the X-ray peak (top) and X-ray centroid (bottom)  for two subsets of our high-completeness sample, comprising clusters with redshifts below and above the median redshift of $z{=}0.387$ (solid and dotted lines, respectively).  While the distribution of offsets between BCG and X-ray peak (left)  changes little between subsamples generated by this particular redshift cut, we see a pronounced shift to larger offsets of the BCG from the X-ray centroid for the more distant clusters (right). A Kolmogorov-Smirnov test yields probabilities of 44 (0.3)\% for the low- and high-redshift distributions of the observed BCG-X-ray-peak (BCG-X-ray-centroid) values being drawn from the same parent distribution.}
   \label{fig:cumredshift}
\end{figure}

\begin{figure}
   \centering
   \includegraphics[width=.5\textwidth]{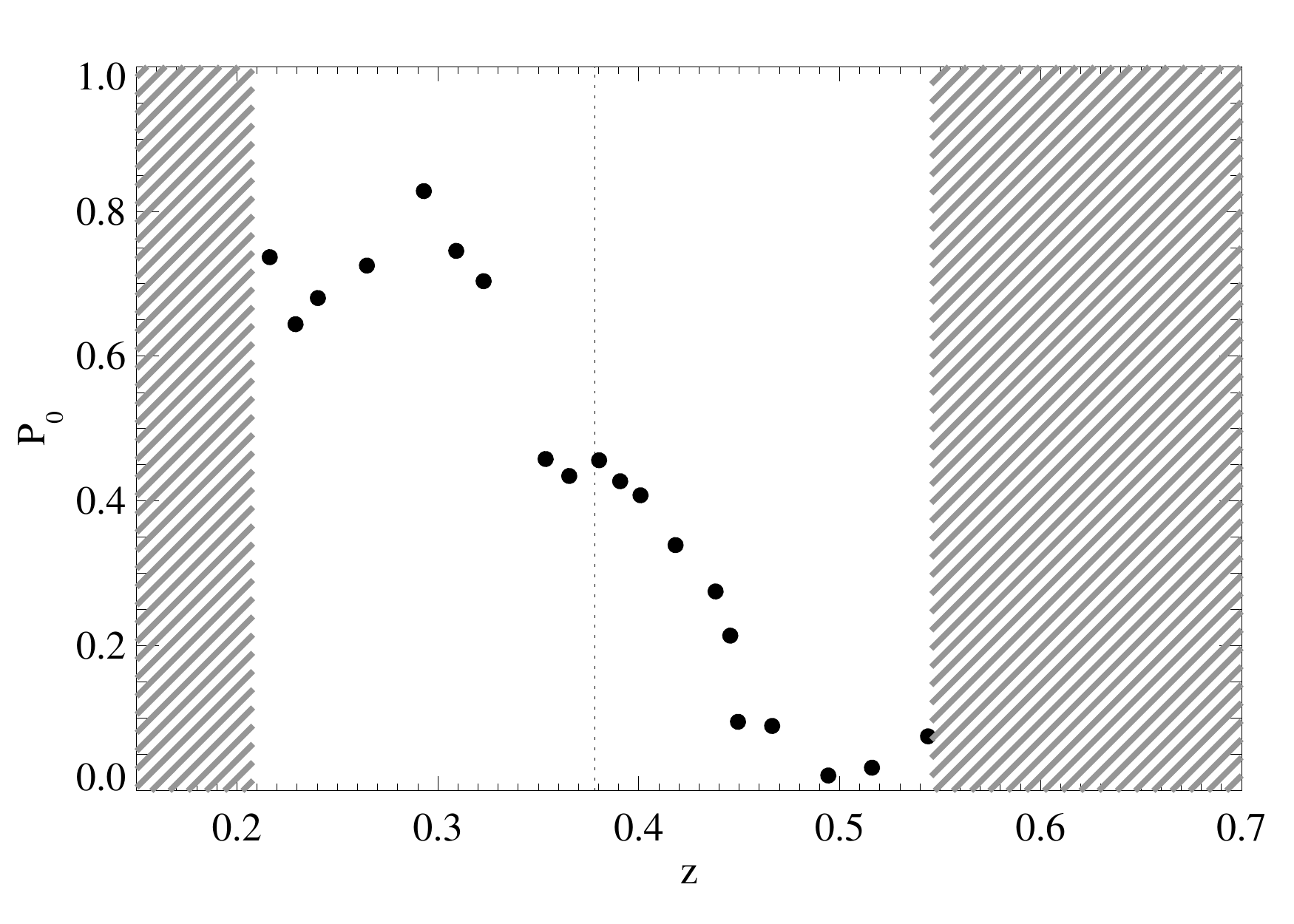} 
   \includegraphics[width=.5\textwidth]{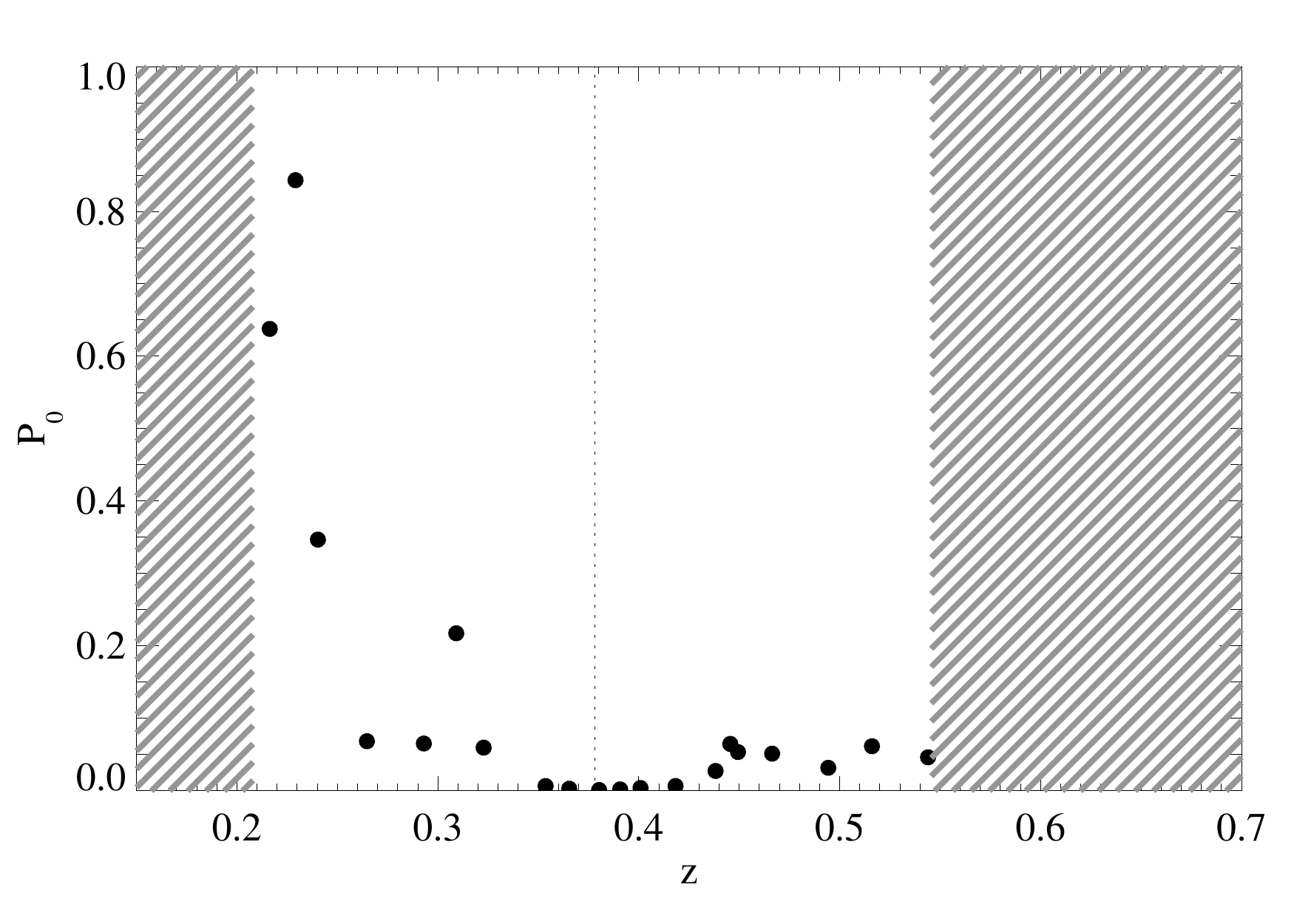} 
   \caption{ In order to quantify any systematic redshift dependence of the X-ray/optical offsets, we split our high-completeness sample at a given redshift and compare the offset distributions of the two  subsamples. Shown here are the Kolmogorov-Smirnov probabilities for the resulting distributions of BCG/X-ray peak offsets (top) and of BCG/X-ray centroid offsets (bottom) as a function of the redshift used to divide the sample. We use a running median of the Kolmogorov-Smirnov probability to reduce statistical variations (although they are still present), and require at least 7 clusters in either subsample, which leads to the effective exclusion of the shaded redshift range.  In each panel the dashed line marks the median redshift used to create the subsets compared in Fig.~\ref{fig:cumredshift}.}
   \label{fig:ksprobs}
\end{figure}

Figure \ref{fig:cumredshift} shows the cumulative distribution of the offset of the BCG from the X-ray peak (or centroid) for clusters above and below the median redshift of this high-completeness sample.  A two-sided Kolmogorov-Smirnov (KS) test finds the probability of the two data sets to be drawn from the same parent population to be 44\% and 0.3\% for the BCG/X-ray peak and BCG/X-ray centroid separation, respectively. Fig.~\ref{fig:ksprobs} shows the KS probabilities for the same comparison but as a function of the redshift at which we divide the high-completeness sample.  Both panels of  Fig.~\ref{fig:ksprobs} show a clear increase in the fraction of disturbed clusters with redshift, indicating a higher incidence of large offsets for the more distant clusters. However, the redshift at which this trend becomes significant depends markedly on the diagnostic used ($z\sim 0.45$ for the offset between BCG and X-ray peak; $z\sim 0.35$ for the offset between BCG and X-ray centroid). Such a systematic difference does not come unexpected though, since the two kinds of offsets probe different spatial (and temporal) scales. In a post-merger system, a cool core close to  either BCG will either have survived the collision, or will form as soon as partial relaxation allows radiative cooling of a stable region of high gas density around the BCG. As a result, X-ray/optical alignment will be restored quickly on small scales. By contrast, realignment between BCG and X-ray centroid after a merger will only occur much later, when relaxation has progressed to a global scale.  

\begin{figure} 
   \centering
   \includegraphics[width=8cm]{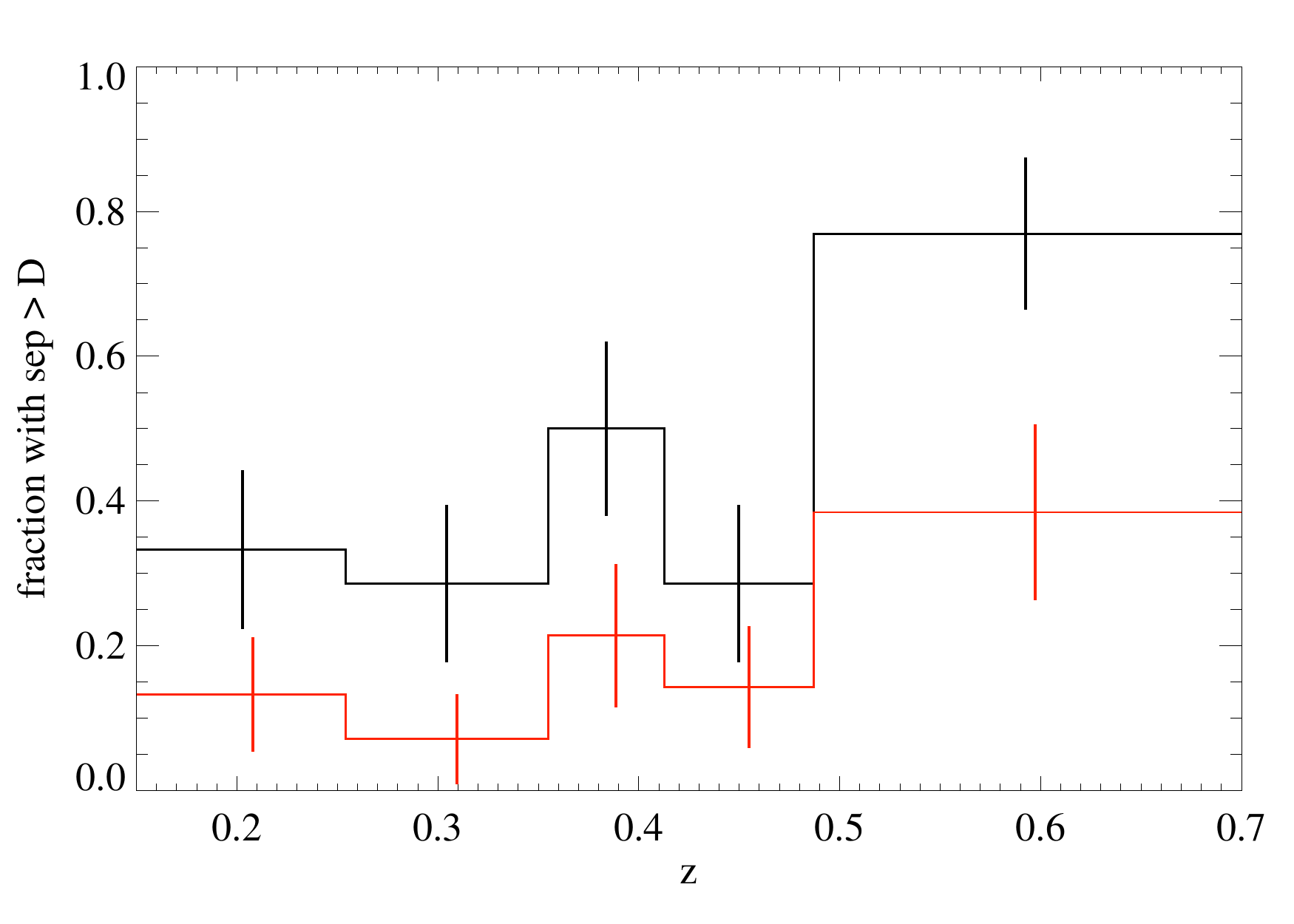} 
   \includegraphics[width=8cm]{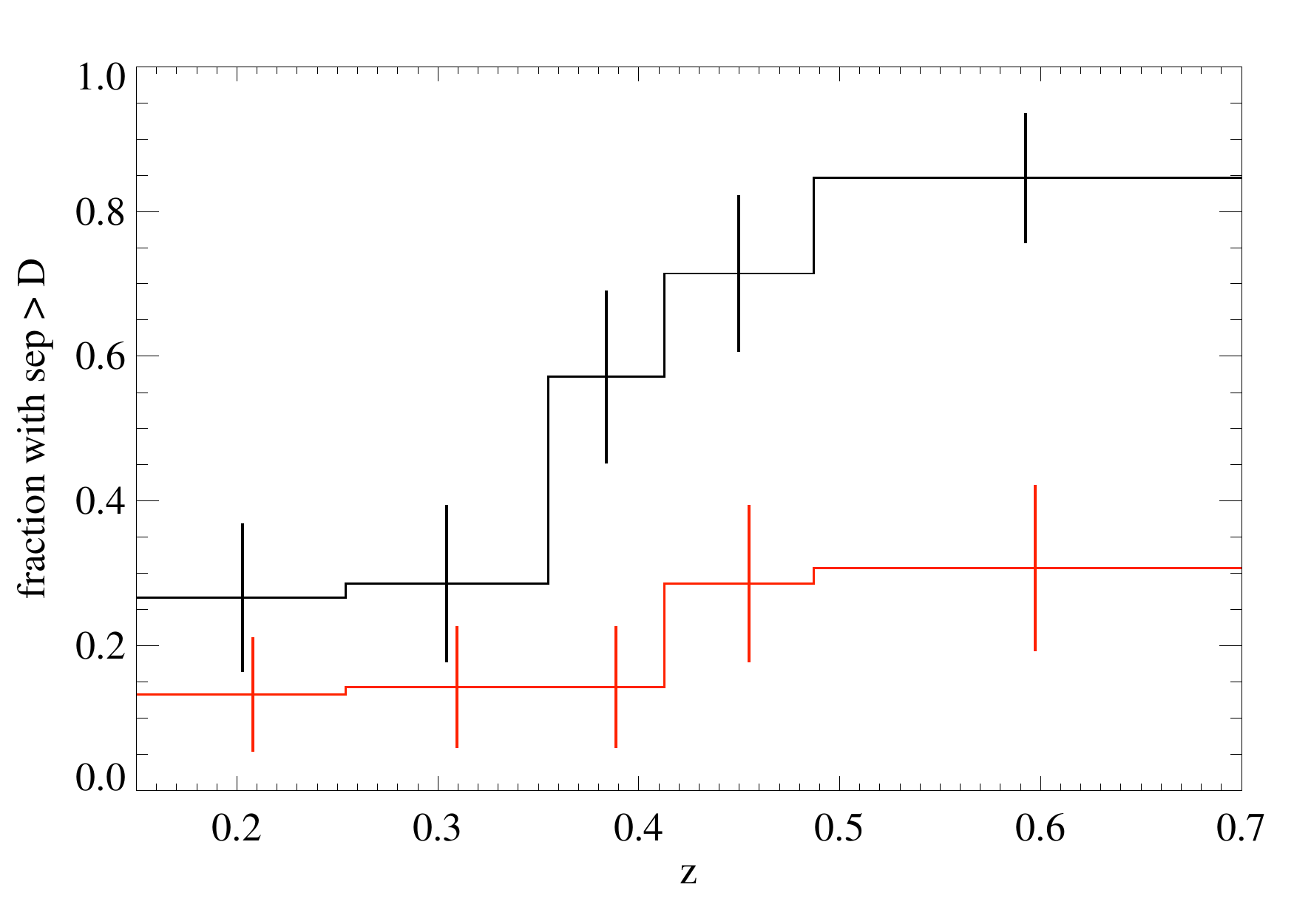} 
      \caption{ Top: The fraction of clusters with offsets of the BCG from the X-ray peak  in excess of 12\,kpc (the mean of the distribution of offsets; black), and 42\,kpc (mean plus one standard deviation; red) as a function of redshift. The evolution visible in the black histogram is significant at the 2.7$\sigma$ confidence level. Bottom: Same as left panel, but for the offsets of the BCG from the X-ray centroid, and using thresholds of  21 (mean) and  71\,kpc (mean plus $1\sigma$). The evolution observed for the larger sample is significant at the 4.5$\sigma$ confidence level. For both panels the bin boundaries are chosen such that each bin contains 15 clusters; errors are computed following \citet{2004MNRAS.351..125D}.  The evolutionary trend apparent in either panel and the approximate redshift at which the merger fraction increases significantly are fully consistent with the results shown in Fig.~\ref{fig:ksprobs} based on unbinned data.}
   \label{fig:sepz2small}
\end{figure}

\begin{figure} 
   \centering
   \includegraphics[width=8cm]{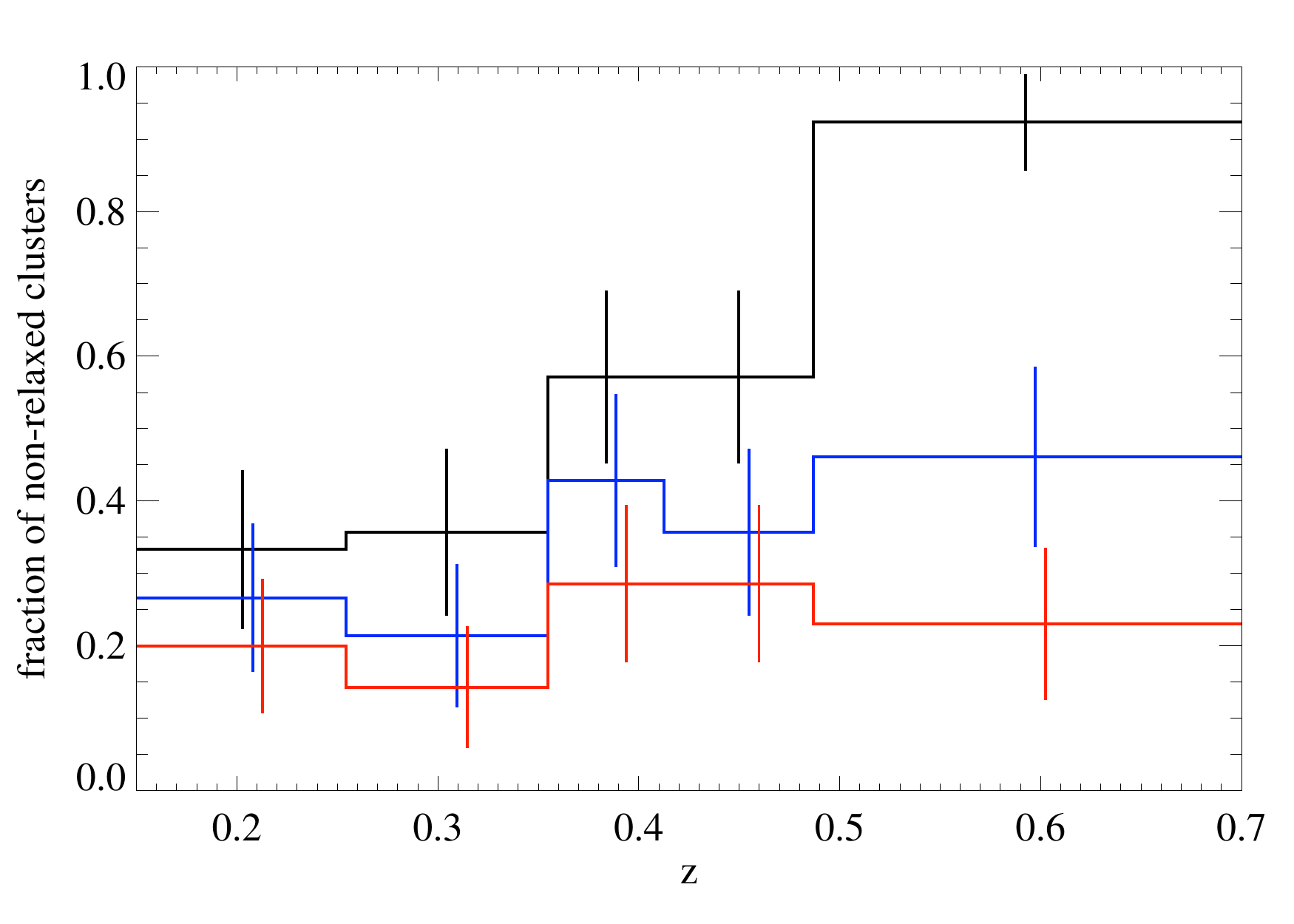} 
      \caption{ The fraction of clusters with morphological classifications $\ge 2$ (black), $\ge 3$ (blue), and $=4$ (red).  Again the fraction of disturbed clusters (all systems except the ones classified as fully relaxed) clearly increases with redshift (significant at the 5.0$\sigma$ confidence level), consistent with the trends apparent from Figs.~\ref{fig:ksprobs} and \ref{fig:sepz2small}. See text for further discussion.}
   \label{fig:morphcodes}
\end{figure}

The same trend is apparent in Fig.~\ref{fig:sepz2small}, which shows the redshift dependence of the fraction of clusters with offsets in excess of two threshold values, namely the mean, and the mean plus one standard deviation, of the X-ray/optical separation. Similarly, Fig.~\ref{fig:morphcodes} probes any variations with redshift in the fraction of systems with morphological classifications $\ge$ 2, 3, and 4. A highly significant increase with redshift is observed in the fraction of disturbed clusters, regardless of the diagnostic used. 

The systems classified by us as the most extreme (i.e., the ones featuring the highest X-ray/optical separations or a morphological code of 4), however, exhibit little, if any, evolution with redshift. This flattening of the redshift distribution of the most obvious mergers is likely to be due to two effects. The first one is the fact that our classification is based on projected, not three-dimensional separations. Hence, systems merging approximately in the plane of the sky will be ranked as the most extreme (by all of our metrics), while mergers featuring entirely comparable physical separations (or morphological complexity), but proceeding along axes closer to our line of sight, may still be recognised as disturbed, but will be assigned lower values for the (projected) X-ray/optical separation and morphological class. This effect causes the redshift distributions of clusters of differing degrees of relaxation to scale approximately by factors that correspond to different effective inclination angles of the merger axis. A second effect is clearly at work though in Figs.~\ref{fig:sepz2small} (right panel) and \ref{fig:morphcodes} which show a flatter redshift distribution for the most extreme mergers than could be explained by such scaling.  Although the still relatively small size of our sample prevents us from quantifying the effect, we suspect that the low X-ray surface brightness of the most disturbed systems caused them to be detected less reliably near the flux limit of the RASS, resulting in a bias against very extended mergers (class 4 clusters) at the high-redshift end of our study. The ultimate cause of such a bias, namely the combination of poor photon statistics (at $z>0.4$ the median net photon count in the RASS for our sample is 30) with the use of a detection algorithm optimised for the detection of point sources, has been identified before. \citet{2000ApJ...534..133E}, investigating the failure of the Einstein Medium Sensitivity Survey to detect Cl\,J0152.7--1357, a BHOM at $z=0.83$, in spite of sufficient photon statistics in the Einstein raw data, find that complex morphology causes the significance of sources near the detection limit to be underestimated, resulting in a bias against mergers with pronounced substructure. 

\section{Summary}\label{sec:summary}

In order to identify the most extreme cluster mergers and characterize their evolution with redshift, we have applied a set of simple X-ray/optical diagnostics to a statistically complete sample of 108 clusters at $0.15 < z < 0.70$ and at declinations north of $-40^\circ$, featuring X-ray luminosities of $L_{\rm X} > 5\times 10^{44}$ erg s$^{-1}$ (0.1--2.4 keV), and observed with the ACIS imaging spectrometer aboard Chandra. We determine the projected (2D) separation between the X-ray peak or the X-ray centroid and the BCG, and perform a visual morphological classification following the criteria of \citet{2007ApJ...661L..33E}.  

We identify 27 clusters as extreme and active mergers by requiring such systems to feature a morphological code of 3 or 4, and values of at least 71\,kpc for the offset between BCG and X-ray centroid, or 42\,kpc for the separation between BCG and X-ray peak. Note that, given the nature of our diagnostics, these criteria  will favour the selection of mergers proceeding along an axis close to the plane of the sky. 

Of those 27, 17 exhibit X-ray/optical characteristics suggestive of a binary, head-on merger (BHOM), making them promising targets for quantitative studies of, e.g., galaxy-gas interactions, ICM shocks, or the collisional properties of dark matter. This list includes MACS\,J0025.4--1222 and A\,2146, both of which have already proven extremely valuable for such work, and also A\,1758 which is the target of an ongoing multi-wavelength follow-up campaign. Importantly, more than half of this list of potential BHOMs are, however, yet to be studied in depth. 

The remainder of our initial sample of 27 extreme mergers are likely to be too complex to be easily modeled, rendering the physical interpretation of the observables (all of them accessible only in projection) difficult at best. Such challenges notwithstanding, our list of 10 complex mergers still holds great promise for, e.g., statistical studies of the properties of gas and galaxies in cluster mergers, as well as for investigations of the origin of diffuse, non-thermal radio emission which is indeed ubiquitous among the mergers selected here \citep{2010ApJ...721L..82C}. Among this second subsample to emerge from this study are several famous clusters, such as A\,2744 and the most disturbed cluster presently known at $z>0.5$, MACS\,J0717.5+3745. Although many systems on this list have been known to be complex mergers before, about one third are new identifications resulting from this work. 

In order to probe the evolution of the merger fraction on cosmological timescales, we examine the redshift distribution of the clusters in our sample identified as unrelaxed by each of our three selection criteria. We find strong evidence for such evolution regardless of the merger criterion and regardless of the statistic used. The increase with redshift of the merger fraction starts approximately at $z=0.4$; the observed  dependence of this value on the kind of X-ray/optical separation used (relative to the peak or the centroid of the X-ray emission) is consistent with the difference in relaxation scales (both spatial and temporal) probed by the two diagnostics. The fraction of fully relaxed clusters decreases dramatically with lookback time even within the limited redshift range of our sample, lending strong support to previous claims of a pronounced dearth of cool-core clusters at high redshift \citep{2007hvcg.conf...48V, 2008A&A...483...35S,2011ApJ...731...31S}. Since our analysis is based solely on imaging data and thus on quantities observed in projection, mergers proceeding at angles close to our line of sight have a high probability of being missed. As a result, the true evolution of the merger fraction with redshift is almost certainly even stronger than reported here.

Finally, 24 clusters are highlighted in Table~1 as new releases from the Massive Cluster Survey, bringing the total to 70. We note that, contrary to the previous MACS releases \citep{2007ApJ...661L..33E,Ebeling:2010mz}, this set of 24 does not represent a statistically complete subsample. 

\section{Outlook}

Deeper Chandra observations of several clusters in our sample with currently poor photon statistics will allow a more robust determination of the merger fraction around $z=0.4$ and thus a more accurate measurement of  cluster evolution at this dynamically important juncture. In addition, extending this sample of X-ray luminous clusters to higher redshift could provide significant constraints closer to the era of cluster formation, especially with regard to the evolution of cool-core clusters. Finally, we aim to overcome the primary limitation of the study discussed here (an unavoidable bias against mergers proceeding along an axis close to our line of sight) by including dynamical information (ICM gas temperatures and radial velocity dispersions of the cluster galaxy population) that will allow us to extend the identification of cluster mergers to three dimensions.

\section{Acknowledgements}
We thank an anonymous referee for careful reading, several corrections, and helpful suggestions.
HE gratefully acknowledges financial support from NASA grant  NAG 5-8253 as well as from SAO grants GO2-3168X, GO3-4168X, GO5-6133X, GO9-0146X, GO0-11137X, GO1-12172X, and GO1-12153X. We are grateful to the University of Hawai'i telescope time allocation committee for their unwavering support of optical follow-up observations of MACS clusters. This research has made use of data obtained from the Chandra Data Archive and the Chandra Source Catalog, and software provided by the Chandra X-ray Center (CXC) through the application package CIAO.

Funding for the SDSS and SDSS-II has been provided by the Alfred P. Sloan Foundation, the Participating Institutions, the National Science Foundation, the U.S. Department of Energy, the National Aeronautics and Space Administration, the Japanese Monbukagakusho, the Max Planck Society, and the Higher Education Funding Council for England. The SDSS Web Site is http://www.sdss.org/.

\bibliography{/Users/amann/Dropbox/fullbiblio}

\clearpage

\begin{landscape}
\begin{table}
 \caption{Data for full sample of clusters}
 \label{tab:alldata}
 \begin{tabular}{@{}llllllllllllllllllllllllllllllllllllllllllllllllllllllllllllll}
Name&$\alpha$&$\delta$&Optical&Obs ID&CXO $t_{exp}$&$L_{x}$ ($10^{45}$)&z&Peak Sep&$\sigma_{peak}$&Cent Sep&$\sigma_{cent}$&Morph&$z$\\
&&&Source&&(ks)&erg~s$^{-1}$&&(kpc)&(kpc)&(kpc)&(kpc)&Code&ref\\
\hline
\hline
Abell 2744$^{a}$&00:14:19&-30:22:15&UH2.2&8477, 8557&45.9+27.8& 17.6&0.31&   80.3&  5.4&  357.3& 5.11&4&(18)\\
MACS\,J0025.4--1222$^{a}$&00:25:29&-12:22:47&UH2.2&       10413&    75.6& 13.7&0.58&  210.0&  6.8&  279.3& 3.43&4&(8)\\
Abell 2813$^{a}$&00:43:24&-20:37:28&UH2.2&        9409&    19.9&  7.9&0.29&   45.0&  4.3&   50.7& 1.55&4&(3)\\
MACS\,J0140.0--0555$^{a}$$^{b}$&01:40:01&-05:55:13&UH2.2&5013, 12243&10.2+19.3& 11.8&0.45&   59.8&  5.0&   23.7& 2.00&4&(1)\\
MACS\,J0159.0--3412$^{b}$&01:59:04&-34:12:50&UH2.2&        5818&     9.4&  8.5&0.41&  142.7&  6.4&  130.3& 9.44&4&(1)\\
MACS\,J0358.8--2955$^{a}$&03:58:51&-29:55:21&UH2.2&12300, 13194, 11719&29.7+20.0+9.6& 17.6&0.43&    9.2&  4.7&   98.1& 1.67&4&(9)\\
MACS\,J0404.6+1109$^{a}$&04:04:38&+11:09:20&UH2.2&        3269&    21.8& 10.2&0.36&   58.6&  4.9&   69.7& 3.17&4&(9)\\
MACS\,J0416.1--2403$^{a}$$^{b}$&04:16:09&-24:03:58&UH2.2&       10446&    15.8& 10.1&0.40&    5.7&  4.5&   72.5& 2.04&4&(1)\\
MACS\,J0451.9+0006$^{b}$&04:51:54&00:06:14&UH2.2&        5815&    10.2&  7.5&0.43&   23.2&  4.7&   32.7& 9.76&4&(1)\\
Abell 520$^{a}$&04:54:08&+02:55:05&UH2.2&        4215&    66.3&  9.2&0.20&   75.9&  5.7&  463.1& 1.91&4&(14)\\
Abell 521$^{a}$&04:54:09&-10:14:31&UH2.2&         901&    38.6&  8.7&0.25&   19.9&  4.1&  138.8&12.43&4&(19)\\
MACS\,J0553.4--3342$^{a}$$^{b}$&05:53:26&-33:42:36&UH2.2&        5813&     9.9& 10.2&0.43&   28.4&  4.5&   90.8& 1.69&4&(1)\\
MACS\,J0717.5+3745$^{a}$&07:17:33&+37:45:19&UH2.2&        4200&    59.1& 20.9&0.55&  312.8&  9.3&  316.5& 4.61&4&(8)\\
Abell 665$^{a}$&08:30:56&+65:51:04&SDSS&        3586&    29.7& 10.2&0.18&   50.5&  6.0&  147.5& 2.52&4&(14)\\
ZwCl 0847.2+3617$^{a}$$^{b}$&08:50:11&+36:04:21&UH2.2&        1659&    24.9&  9.5&0.38&  111.9&  7.2&   61.8& 0.75&4&(6)\\
Abell 773&09:17:51&+51:43:27&SDSS&5006, 533&19.8+11.3&  7.5&0.22&   12.6&  6.1&   23.5& 4.92&4&(6)\\
MACS\,J1006.9+3200$^{a}$$^{b}$&10:06:55&+32:00:57&UH2.2&        5819&    10.9&  7.3&0.40&   43.1&  5.1&  180.1& 6.11&4&(1)\\
Abell 1300$^{a}$&11:31:53&-19:55:43&UH2.2&        3276&    13.9& 10.2&0.31&    7.7&  3.6&  107.3& 5.36&4&(9)\\
MACS\,J1149.5+2223$^{a}$&11:49:34&+22:23:42&UH2.2&1656, 3589&18.5+20.0& 13.0&0.54&   59.0&  5.3&   88.9& 4.25&4&(8)\\
Abell 1682$^{a}$&13:06:53&+46:33:13&SDSS&11725, 3244&19.9+9.8&  7.2&0.23&    6.7&  3.3&   96.1& 5.18&4&(2)\\
Abell 1758$^{a}$&13:32:43&+50:32:56&SDSS&        2213&    58.3&  6.9&0.28&   54.0&  3.1&  319.7& 2.21&4&(16)\\
Abell 1914$^{a}$&14:26:01&+37:49:35&SDSS&        3593&    18.9& 11.6&0.17&   97.2&  6.3&  129.1& 0.92&4&(12)\\
Abell 2219&16:40:22&+46:42:15&SDSS&        7892&     5.1& 14.3&0.23&   16.9&  2.7&   44.8& 4.54&4&(2)\\
MACS\,J1731.6+2252$^{a}$&17:31:40&+22:52:39&UH2.2&        3281&    20.5& 12.6&0.39&   66.7&  4.9&   31.4& 3.18&4&(9)\\
MACS\,J2243.3--0935$^{a}$&22:43:22&-09:35:49&UH2.2&        3260&    20.5& 16.9&0.45&  124.8&  6.0&  155.8& 4.57&4&(9)\\
Abell 2631$^{a}$&23:37:39&00:16:03&SDSS&11728, 3248&16.8+9.2&  9.1&0.28&  132.6&  4.3&   79.0& 6.22&4&(4)\\
\hline
CL 0016+1609&00:18:31&+16:26:34&UH2.2&         520&    67.4& 14.2&0.55&   38.6&  5.3&   21.3& 4.94&3&(17)\\
MACS\,J0035.4--2015&00:35:26&-20:15:37&UH2.2&        3262&    21.4&  9.0&0.35&   15.7&  3.9&   11.8& 0.28&3&(9)\\
RBS 436&01:52:35&-28:52:47&UH2.2&        3264&    17.5& 14.5&0.41&    1.1&  2.7&   53.9& 1.41&3&(9)\\
MACS\,J0417.5--1154$^{a}$&04:17:33&-11:54:12&UH2.2&       11759&    51.4& 27.6&0.44&   17.0&  5.2&  103.3& 2.27&3&(9)\\
Abell 907&05:28:55&-39:28:02&DSS&        4994&    22.5& 13.7&0.28&   24.8&  2.2&   38.9& 3.48&3&(18)\\
MACS\,J0911.2+1746&09:11:12&+17:46:34&UH2.2&3587, 5012&17.9+23.8&  9.8&0.51&   22.6&  5.5&   47.3& 4.00&3&(8)\\
MACS\,J1108.8+0906$^{b}$&11:08:53&+09:06:05&UH2.2&        5009&    24.5&  8.0&0.47&   17.8&  4.7&   15.1& 1.46&3&(1)\\
MACS\,J1115.2+5320$^{a}$$^{b}$&11:15:14&+53:20:13&UH2.2&        5008&    18.0&  7.9&0.47&   42.0&  4.6&   24.1& 2.34&3&(1)\\
Abell 1763&13:35:17&+40:59:58&SDSS&        3591&    19.6&  8.3&0.23&    1.1&  2.2&   35.9& 6.85&3&(12)\\
ZwCl 1459.4+4240&15:01:23&+42:21:06&SDSS&        7899&    13.0&  6.9&0.29&   44.0&  3.8&   67.3& 1.54&3&(7)\\
Abell 2146$^{a}$&15:56:11&+66:21:23&DSS&       10464&    35.7&  7.0&0.23&   20.5&  7.1&  119.9& 2.68&3&(2)\\
Abell 2163&16:15:46&-06:08:41&HST&        1653&    71.1& 24.3&0.21&   34.6&  3.9&   70.1& 3.88&3&(18)\\
MACS\,J2049.9--3217&20:49:55&-32:17:10&UH2.2&        3283&    23.8&  7.3&0.32&   18.0&  4.0&   25.5& 6.53&3&(9)\\
MACS\,J2129.4--0741$^{a}$&21:29:26&-07:41:28&UH2.2&3199, 3595&19.9+19.9& 12.6&0.59&   43.3&  5.6&   48.1& 6.70&3&(8)\\
RX J2228.6+2037$^{a}$&22:28:34&+20:36:47&UH2.2&        3285&    19.9& 13.9&0.41&   19.5&  4.9&   89.9& 9.74&3&(9)\\
\hline
\end{tabular}
\end{table}
\end{landscape}

\setcounter{table}{0}

\begin{landscape}
\begin{table}
 \caption{Data for full sample of clusters}
 \begin{tabular}{@{}llllllllllllllllllllllllllllllllllllllllllllllllllllllllllllll}
Name&$\alpha$&$\delta$&Optical&Obs ID&CXO $t_{exp}$&$L_{x}$ ($10^{45}$)&z&Peak Sep&$\sigma_{peak}$&Cent Sep&$\sigma_{cent}$&Morph&$z$\\
&&&Source&&(ks)&erg~s$^{-1}$&&(kpc)&(kpc)&(kpc)&(kpc)&Code&ref\\
\hline
\hline
MACS\,J0111.5+0855$^{b}$&01:11:34&+08:55:52&UH2.2&        3256&    19.4& 11.5&0.49&    9.2&  4.8&    8.3& 8.77&2&(1)\\
MACS\,J0257.1--2325&02:57:09&-23:25:48&UH2.2&1654, 3581&19.8+18.5& 11.4&0.51&    1.0&  3.5&   30.8& 0.76&2&(8)\\
Abell 402&02:57:41&-22:09:01&UH2.2&        3267&    20.5&  8.0&0.32&    2.6&  3.7&   18.0& 0.46&2&(15)\\
MACS\,J0308.9+2645&03:08:55&+26:45:39&UH2.2&        3268&    24.5&  9.1&0.36&   16.5&  4.0&    9.1& 1.06&2&(9)\\
RX J0437.1+0043&04:37:09&00:43:37&UH2.2&       11729&    30.5&  8.5&0.28&   17.6&  3.4&    5.7& 1.36&2&(7)\\
MS 0451.6--0305&04:54:10&-03:00:51&UH2.2&         529&    13.9& 17.1&0.54&   59.1&  5.2&   70.5& 4.03&2&(16)\\
MACS\,J0455.2+0657$^{b}$&04:55:17&+06:57:55&UH2.2&        5812&     9.9& 13.2&0.45&    6.3&  4.8&   23.8& 0.39&2&(1)\\
MACS\,J0520.7--1328&05:20:42&-13:28:54&UH2.2&        3272&    19.2&  9.8&0.34&    8.4&  3.7&   10.2& 0.39&2&(9)\\
MACS\,J0647.7+7015&06:47:45&+70:15:03&UH2.2&3196, 3584&19.3+20.0& 16.8&0.59&   13.0&  6.6&   33.5& 2.19&2&(8)\\
MACS\,J0744.8+3927&07:44:51&+39:27:33&UH2.2&        6111&    49.5& 21.0&0.70&    1.9&  4.4&   59.9& 4.97&2&(8)\\
Abell 697&08:42:57&+36:21:41&SDSS&        4217&    19.5& 12.4&0.28&    8.5&  2.7&   19.4& 1.47&2&(5)\\
ZwCl 0947.2+1723&09:49:53&+17:08:03&UH2.2&        3274&    14.3& 16.1&0.38&   15.7&  4.7&   10.5& 4.51&2&(9)\\
MACS\,J1105.7--1014$^{b}$&11:05:45&-10:14:29&UH2.2&        5817&    10.3&  7.9&0.41&   43.9&  4.9&  114.0& 2.54&2&(1)\\
MACS\,J1206.2--0847&12:06:13&-08:47:42&UH2.2&        3277&    23.5& 13.6&0.44&    8.3&  5.2&   33.8& 0.59&2&(9)\\
MACS\,J1226.8+2153$^{b}$&12:26:49&+21:53:01&UH2.2&        3590&    19.0&  7.0&0.44&    6.3&  4.5&   16.8& 0.92&2&(1)\\
MACS\,J1311.0--0310$^{b}$&13:11:00&-03:10:31&UH2.2&        6110&    63.2& 11.4&0.49&   14.2&  4.7&   32.5& 0.82&2&(1)\\
Abell 1722&13:19:59&+70:03:34&UH2.2&        3278&    21.6&  8.4&0.33&    7.6&  4.3&   37.7& 2.24&2&(9)\\
RX J2129.6+0005&21:29:40&00:05:44&SDSS&        9370&    29.6& 11.8&0.23&    2.9&  2.7&   21.5& 1.81&2&(6)\\
MACS\,J2211.7--0349&22:11:44&-03:49:46&UH2.2&        3284&    17.7& 15.4&0.40&   14.3&  4.3&   34.7& 3.95&2&(9)\\
MACS\,J2214.9--1359&22:14:57&-13:59:53&UH2.2&3259, 5011&19.5+18.5& 12.6&0.50&   22.9&  5.2&   15.5& 1.35&2&(8)\\
\hline
MACS\,J0011.7--1523&00:11:42&-15:23:11&UH2.2&3261, 6105&21.6+37.3& 10.4&0.38&    3.8&  4.4&    7.1& 1.56&1&(9)\\
ZwCl 0104+0048&01:06:49&+01:03:16&SDSS&       10465&    48.9&  7.1&0.25&    2.5&  2.6&    2.5& 0.90&1&(7)\\
Abell 209&01:31:52&-13:36:50&UH2.2&3579, 522&10.0+10.0&  6.9&0.21&   12.6&  3.1&   14.1& 3.62&1&(18)\\
MACS\,J0159.8--0849&01:59:49&-08:49:48&UH2.2&        6106&    35.3& 14.2&0.41&    2.8&  3.6&    7.2& 2.93&1&(9)\\
MACS\,J0242.6--2132&02:42:35&-21:32:15&UH2.2&        3266&    11.9& 12.7&0.31&    7.1&  3.8&    4.1& 0.19&1&(20)\\
Abell 3088&03:07:02&-28:40:03&UH2.2&        9414&    18.9&  6.8&0.25&   15.4&  5.4&   14.8& 0.90&1&(16)\\
MACS\,J0326.8--0043$^{b}$&03:26:49&-00:43:41&UH2.2&        5810&     9.9& 11.5&0.45&    2.7&  4.0&    5.7& 1.49&1&(1)\\
MACS\,J0329.6--0211$^{b}$&03:29:40&-02:11:51&UH2.2&        6108&    39.6& 10.1&0.45&    3.9&  5.1&   20.5& 0.86&1&(1)\\
MACS\,J0429.6--0253&04:29:36&-02:53:01&UH2.2&        3271&    23.2& 11.7&0.40&   12.8&  4.6&   19.5& 2.31&1&(9)\\
RX J0439.0+0715&04:39:01&+07:16:15&UH2.2&        3583&    19.2&  6.7&0.24&    9.9&  3.7&   25.6& 0.30&1&(6)\\
MACS\,J0547.0--3904&05:47:02&-39:04:25&UH2.2&        3273&    21.7&  7.5&0.32&    0.6&  2.7&    9.9& 0.98&1&(9)\\
ZwCl 0735.7+7421&07:41:44&+74:14:44&DSS&       10470&   142.0&  7.0&0.22&    4.5&  3.2&    5.6& 3.02&1&(10)\\
Abell 661&08:00:59&+36:03:09&SDSS&        3194&    36.1&  7.3&0.29&    0.5&  2.1&   22.0& 1.60&1&(18)\\
MACS\,J0913.7+4056$^{b}$&09:13:45&+40:56:20&UH2.2&       10445&    76.2& 11.9&0.44&    5.1&  3.7&   10.9& 0.79&1&(1)\\
RBS 0797&09:47:13&+76:23:17&UH2.2&        2202&    11.7& 17.1&0.35&   11.6&  5.8&   14.4& 0.22&1&(9)\\
ZwCl 1021.0+0426&10:23:39&+04:11:17&SDSS&909, 9371&46.0+40.2& 23.8&0.29&   12.3&  3.1&   21.4& 0.27&1&(3)\\
Abell 3444&10:23:50&-27:15:22&DSS&        9400&    36.7& 14.4&0.25&   14.6&  4.7&    9.9& 1.67&1&(18)\\
MACS\,J1115.8+0129&11:15:53&+01:29:47&UH2.2&        9375&    39.6& 12.9&0.35&    6.1&  4.0&    5.4& 0.32&1&(9)\\
Abell 1689&13:11:29&-01:20:17&SDSS&        6930&    76.1& 15.2&0.18&    7.2&  3.0&   14.6& 2.85&1&(18)\\
RX J1347.5--1145&13:47:30&-11:44:55&UH2.2&        3592&    57.7& 38.8&0.45&    5.7&  5.4&   20.8& 1.74&1&(9)\\
MACS\,J1359.1--1929$^{b}$&13:59:09&-19:29:07&UH2.2&        9378&    49.4&  8.4&0.45&    2.9&  4.2&   19.2& 1.66&1&(1)\\
Abell 1835&14:01:02&+02:52:49&SDSS&        6880&   117.9& 24.0&0.25&   11.3&  2.7&   20.1& 0.50&1&(2)\\
MACS\,J1411.3+5212$^{b}$&14:11:21&+52:12:50&UH2.2&        2254&    91.0&  8.2&0.46&    1.3&  3.9&    7.6& 1.76&1&(1)\\
MACS\,J1423.8+2404&14:23:48&+24:04:47&UH2.2&        1657&    18.5& 11.3&0.54&    3.4&  5.1&    7.0& 1.12&1&(8)\\
\end{tabular}
\end{table}
\end{landscape}

\setcounter{table}{0}

\begin{landscape}
\begin{table}
 \caption{Data for full sample of clusters}
 \begin{tabular}{@{}llllllllllllllllllllllllllllllllllllllllllllllllllllllllllllll}
Name&$\alpha$&$\delta$&Optical&Obs ID&CXO $t_{exp}$&$L_{x}$ ($10^{45}$)&z&Peak Sep&$\sigma_{peak}$&Cent Sep&$\sigma_{cent}$&Morph&$z$\\
&&&Source&&(ks)&erg~s$^{-1}$&&(kpc)&(kpc)&(kpc)&(kpc)&Code&ref\\
\hline
\hline
MACS\,J1427.2+4407$^{b}$&14:27:15&+44:07:26&SDSS&        9380&    25.8& 13.7&0.49&    0.8&  2.1&   27.3& 0.98&1&(1)\\
MACS\,J1427.6--2521&14:27:40&-25:21:06&UH2.2&3279, 9373&16.9+28.4&  9.7&0.32&    0.7&  2.7&    2.5& 0.08&1&(9)\\
MACS\,J1447.4+0827$^{b}$&14:47:25&+08:27:32&SDSS&       10481&    12.0& 20.6&0.38&    0.3&  3.2&    4.5& 0.51&1&(1)\\
Abell 1995$^{b}$&14:52:55&+58:02:51&UH2.2&        7021&    48.5&  6.7&0.32&   14.4&  4.0&   32.6& 1.67&1&(17)\\
ZwCl 1454.8+2233&14:57:15&+22:20:25&SDSS&        4192&    91.9& 10.5&0.26&    7.4&  2.7&   10.3& 0.94&1&(2)\\
Abell S780&14:59:28&-18:10:34&DSS&        9428&    39.6& 10.1&0.23&   10.2&  2.9&    5.9& 1.32&1&(18)\\
RBS 1460&15:04:07&-02:48:20&SDSS&        5793&    39.2& 30.3&0.22&    7.5&  2.5&    4.9& 0.37&1&(3)\\
RX J1532.9+3021&15:32:53&+30:21:03&UH2.2&        1665&    10.0& 14.7&0.36&   10.8&  4.4&    5.3& 0.22&1&(9)\\
MACS\,J1621.3+3810$^{b}$&16:21:24&+38:10:02&UH2.2&        6109&    37.5&  9.5&0.46&    1.6&  6.1&   10.0& 0.58&1&(1)\\
Abell 2204&16:32:46&+05:34:23&DSS&        7940&    77.1& 15.0&0.15&    2.7&  1.9&   10.8& 2.06&1&(16)\\
RXC J1720.1+2637&17:20:09&+26:37:27&SDSS&3224, 4361&23.8+25.7&  9.3&0.16&    3.1&  1.7&   12.2& 0.63&1&(10)\\
MACS\,J1720.2+3536&17:20:15&+35:36:21&UH2.2&3280, 6107&20.8+33.9& 12.0&0.39&    1.5&  4.2&   22.2& 2.14&1&(9)\\
Abell 2261&17:22:26&+32:07:52&SDSS&        5007&    24.3& 11.1&0.22&    6.1&  2.2&   12.9& 0.22&1&(4)\\
MACS\,J1931.8--2634&19:31:50&-26:34:36&UH2.2&        9382&    98.9& 15.6&0.35&   15.3&  4.8&    3.9& 0.61&1&(9)\\
RXC J2014.8--2430&20:14:49&-24:30:43&UH2.2&       11757&    19.9& 10.9&0.15&    8.1&  3.5&    7.9& 1.45&1&(3)\\
MACS\,J2046.0--3430$^{b}$&20:46:01&-34:30:04&UH2.2&        9377&    39.2&  8.5&0.42&   12.8&  4.9&    7.7& 0.52&1&(1)\\
MS 2137--2353&21:40:15&-23:39:45&UH2.2&        4974&    57.4&  9.7&0.31&    0.5&  2.1&    7.8& 1.10&1&(17)\\
Abell 2390&21:53:36&+17:41:11&UH2.2&        4193&    95.1& 15.5&0.23&    1.6&  3.8&   13.7& 3.88&1&(13)\\
MACS\,J2229.7--2755&22:29:44&-27:55:38&UH2.2&3286, 9374&16.4+14.8&  9.4&0.32&    3.9&  4.4&    7.3& 0.61&1&(9)\\
MACS\,J2245.0+2637&22:45:05&+26:37:58&UH2.2&        3287&    16.9&  9.0&0.30&    8.8&  3.9&    8.0& 2.40&1&(9)\\
Abell 2552&23:11:35&+03:38:24&UH2.2&       11730&    22.7& 10.9&0.30&   24.7&  3.7&   26.7& 1.21&1&(9)\\
Abell 2667&23:51:40&-26:05:02&HST&        2214&     9.6& 15.4&0.23&    5.7&  4.0&   21.7& 1.83&1&(15)\\
Abell 586&07:32:21&+31:37:50&SDSS&11723, 530&9.9+10.0&  7.1&0.17&   24.1&  0.0&   19.5& 0.42&1&(16)\\
\end{tabular}
\end{table}
\end{landscape}
\noindent
$^{a}$ BHOM or complex merger; see Figs.~\ref{fig:complexmerg},\ref{fig:bhomstier1},\ref{fig:bhomstier2}

\noindent
$^{b}$ Cluster detected by the Massive Cluster Survey (MACS).  Previous releases of MACS clusters are presented and discussed in \citet{2007ApJ...661L..33E,Ebeling:2010mz}

\noindent
Note.--All luminosities as measured from ROSAT All-Sky Survey data within the nominal detect cell.  Morphological classification codes are explained in Section~\ref{sec:diagnostics}.\\

References for Table~1
(1) This work, (2) \citet{1992MNRAS.259...67A}, (3) \citet{2004A&A...425..367B}, (4) \citet{1995MNRAS.274...75C}, (5) \citet{1999MNRAS.306..857C}, (6) \citet{1998MNRAS.301..881E}, (7) \citet{2000MNRAS.318..333E}, (8) \citet{2007ApJ...661L..33E}, (9) \citet{Ebeling:2010mz}, (10) \citet{1992ApJS...80..257E}, (11) \citet{1994ApJS...94..583G}, (12) \citet{1991A&AS...88..133L}, (13) \citet{1991ApJ...376...46O}, (14) \citet{1994Natur.372...75R}, (15) \citet{1982A&A...108L...7S}, (16) \citet{1991ApJS...76..813S}, (17) \citet{1999ApJS..125...35S}, (18) \citet{2000MNRAS.312..663W}, (19) \citet{1983MNRAS.205..793W}.

\clearpage

\newcommand{\otheroverlaysize}{5.4cm}
\newcommand{\overlaysize}{5.4cm}

\begin{figure*} 
   \centering
   \includegraphics[width=\otheroverlaysize]{MACS-J00143-3022uh88.pdf} %A2744
   \includegraphics[width=\otheroverlaysize]{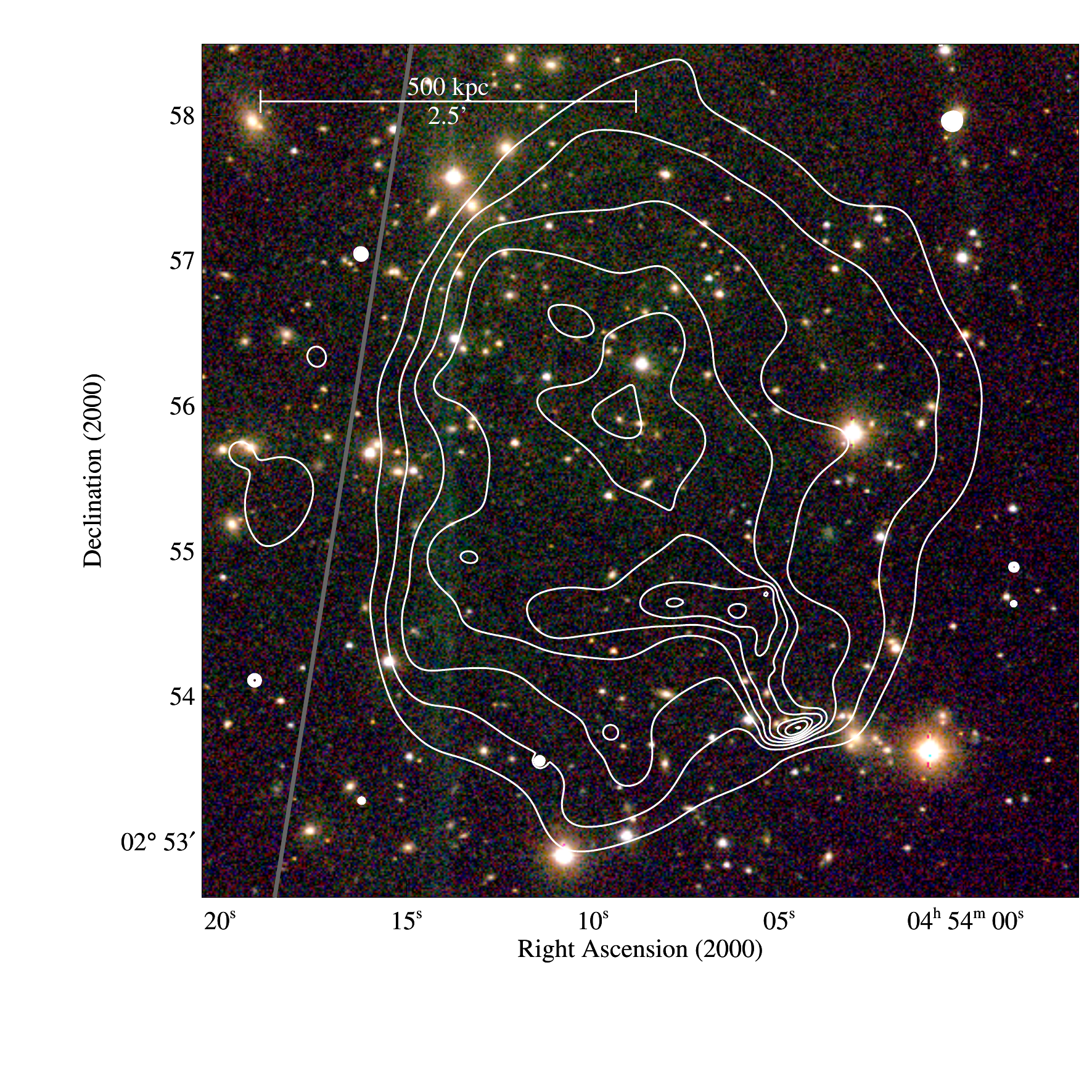} %A520
   \includegraphics[width=\otheroverlaysize]{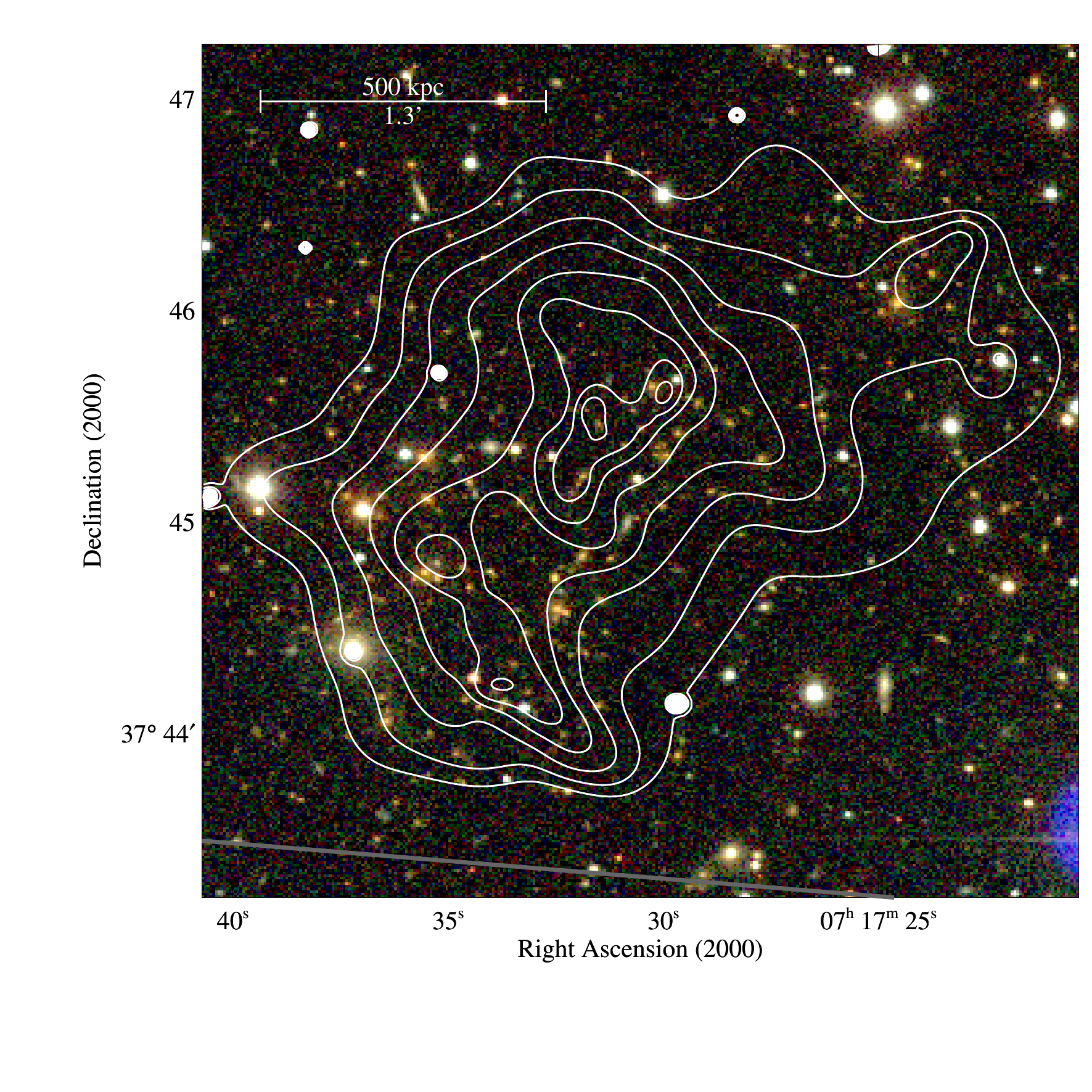} 
   \includegraphics[width=\otheroverlaysize]{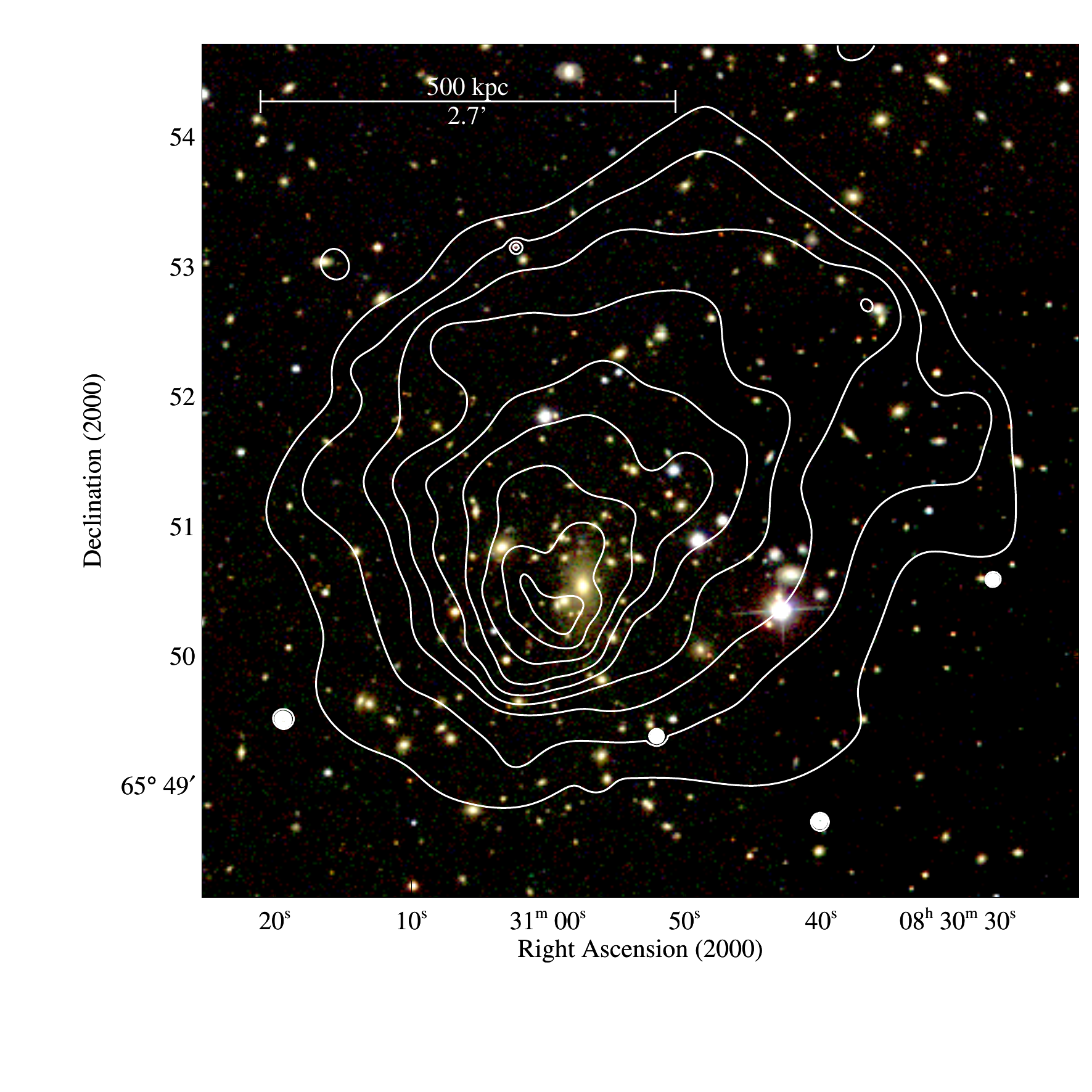} %A665
   \includegraphics[width=\otheroverlaysize]{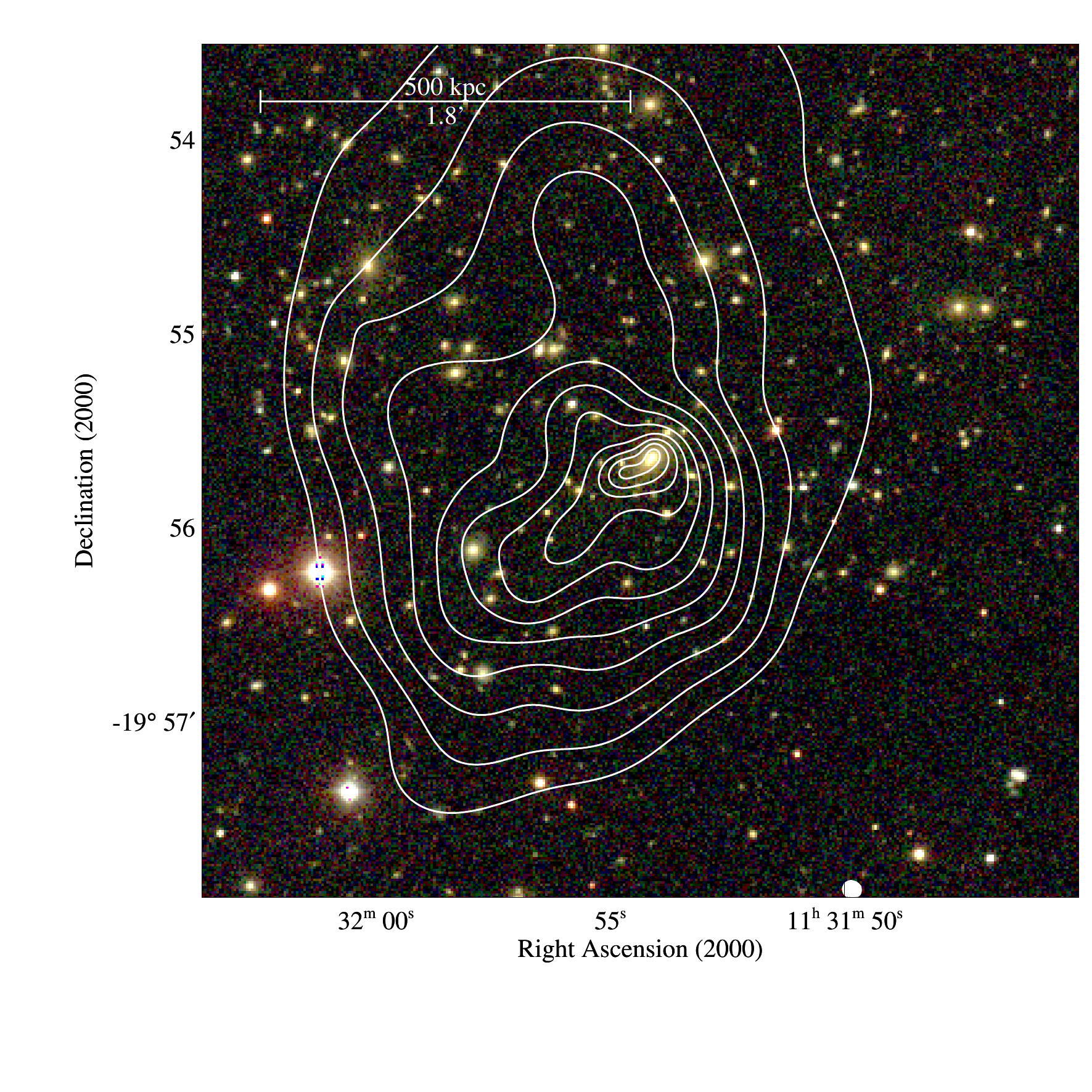} %A1300
   \includegraphics[width=\otheroverlaysize]{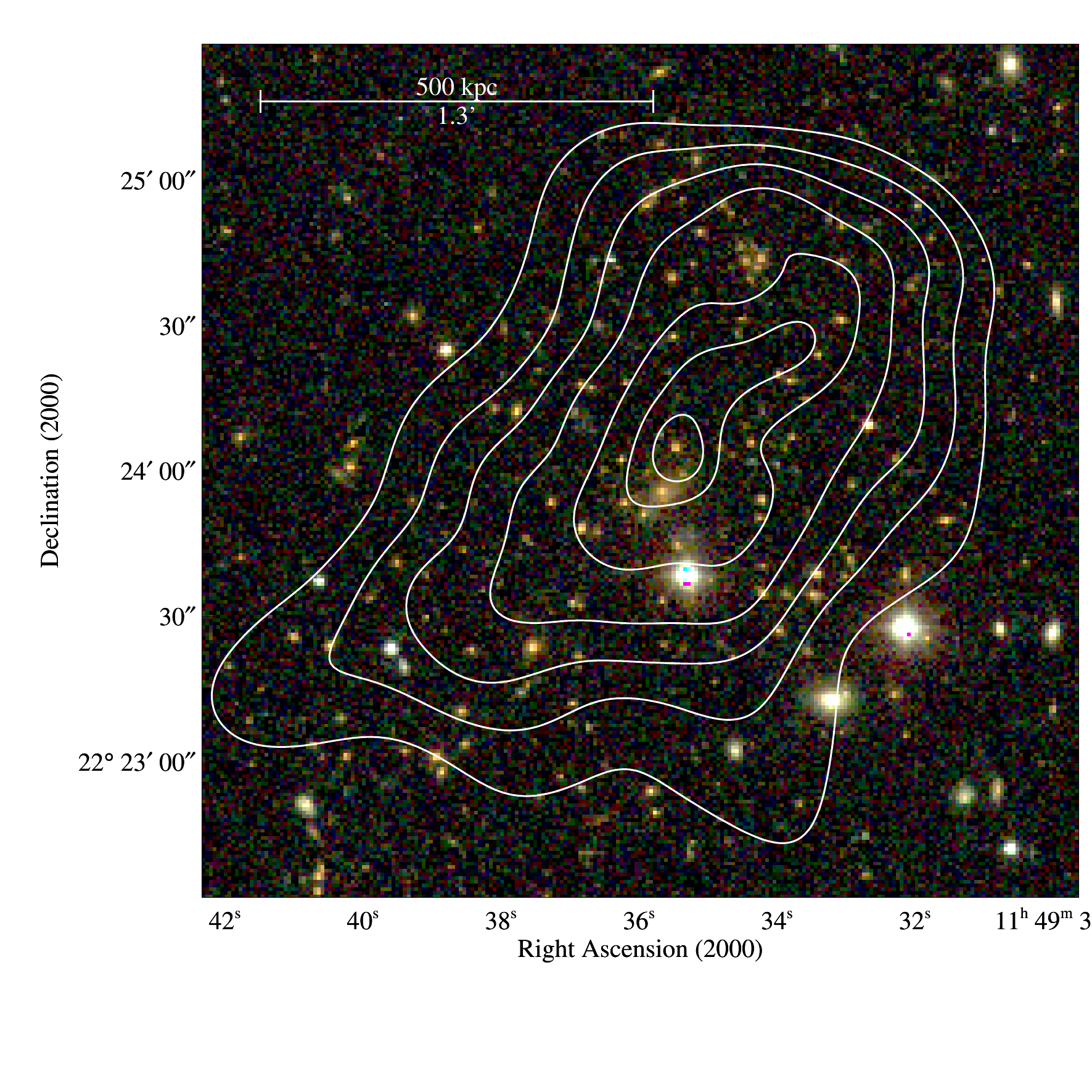}
   \includegraphics[width=\otheroverlaysize]{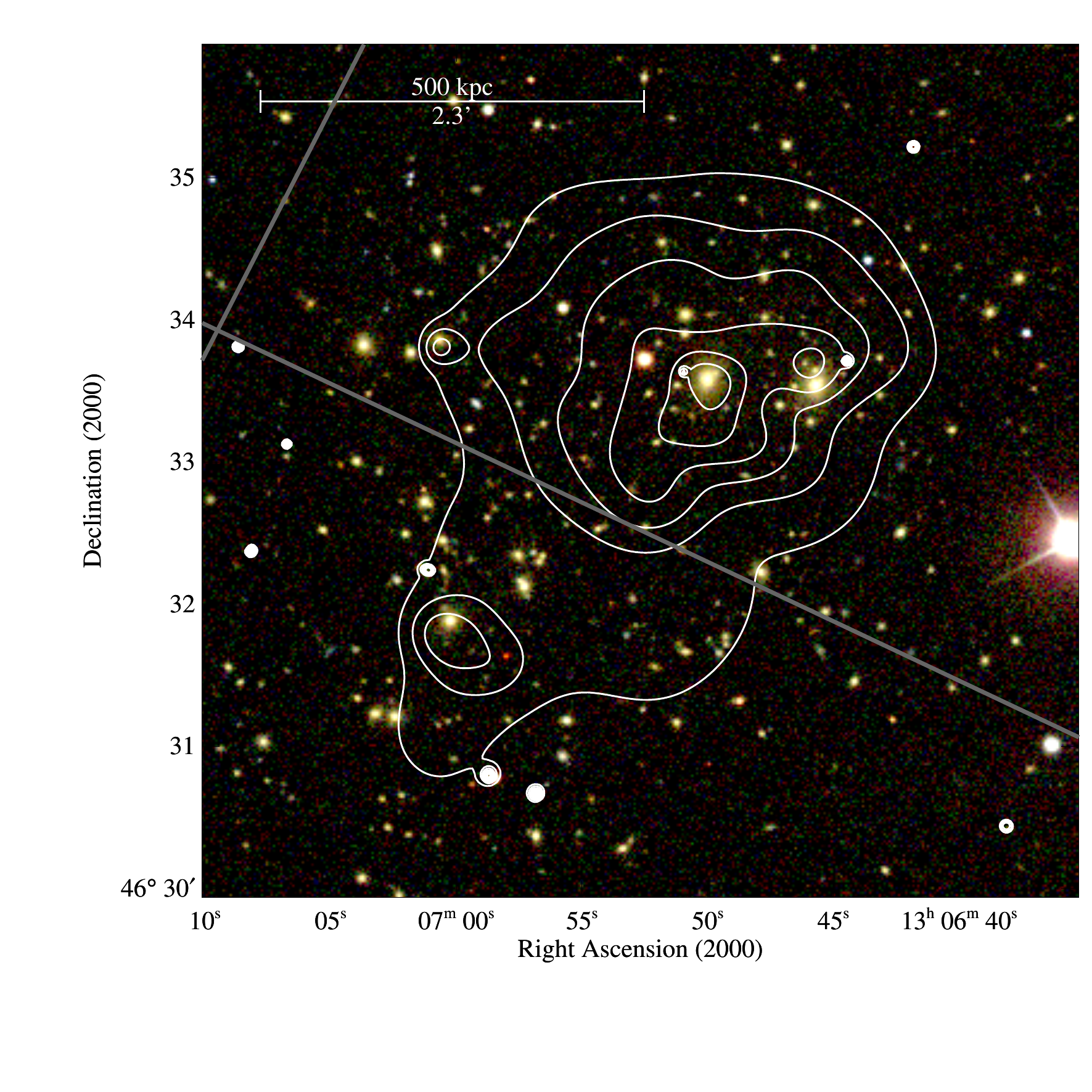} %A1682
   \includegraphics[width=\otheroverlaysize]{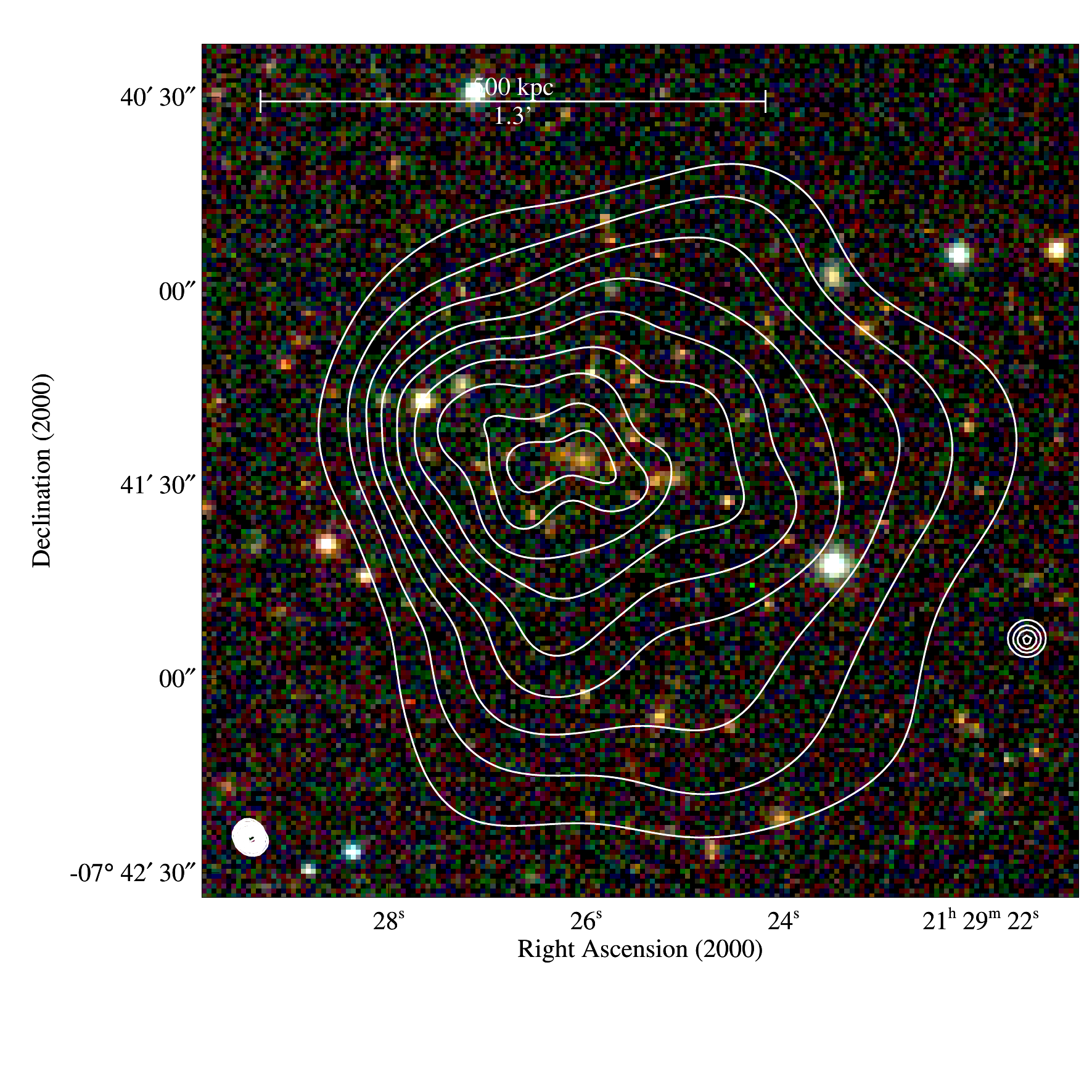}
   \includegraphics[width=\otheroverlaysize]{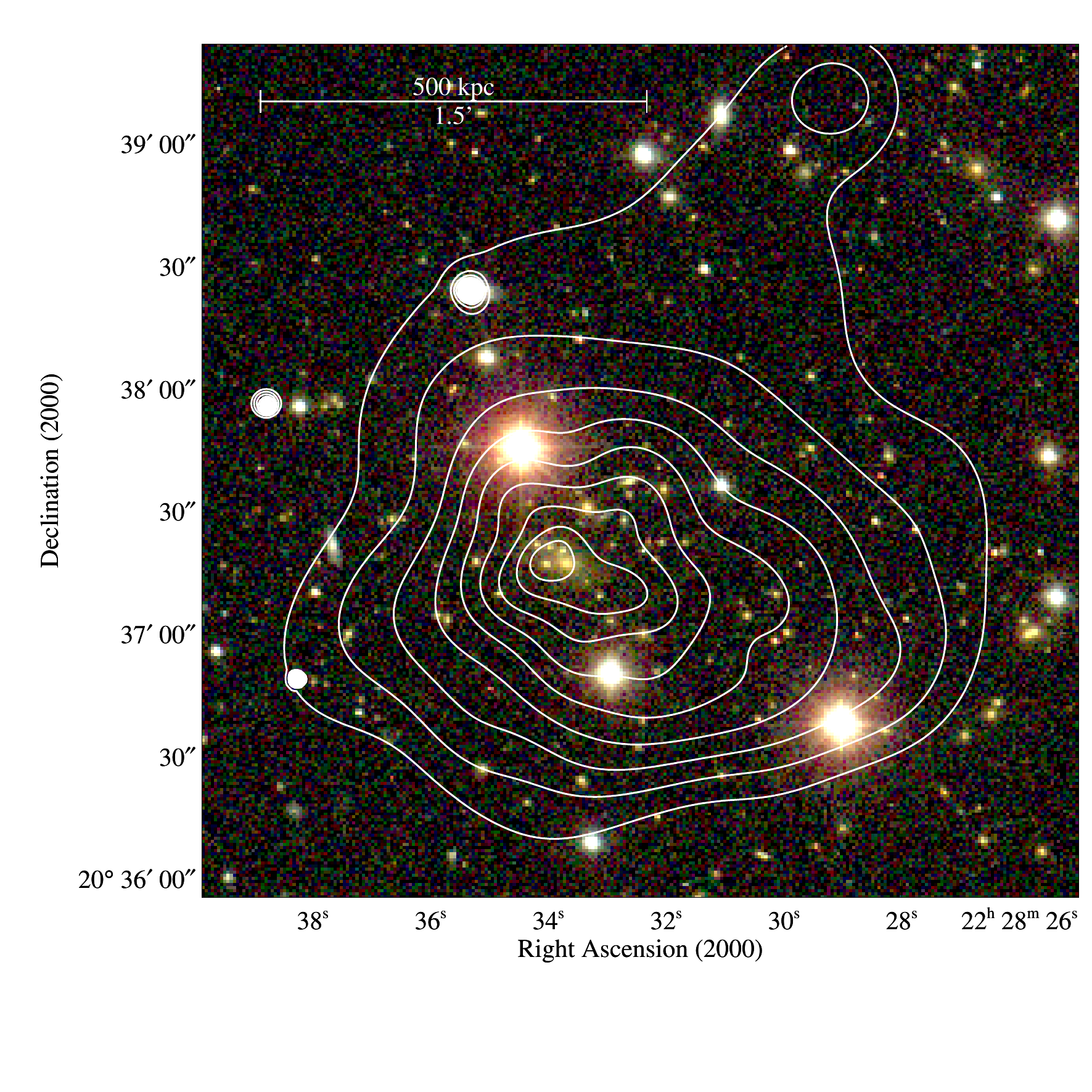} %RX J2228.5+2036
   \includegraphics[width=\otheroverlaysize]{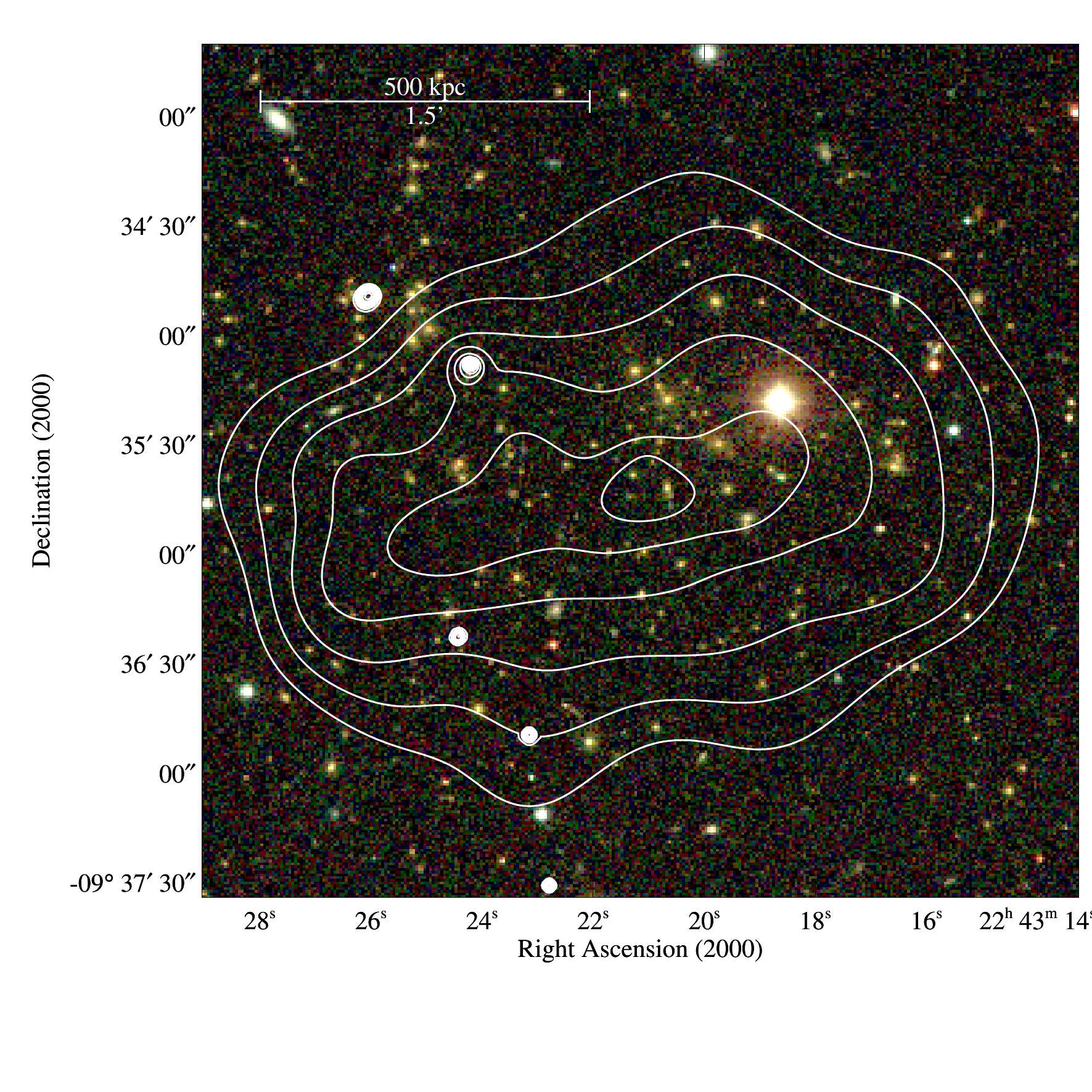}    
   \includegraphics[width=\otheroverlaysize]{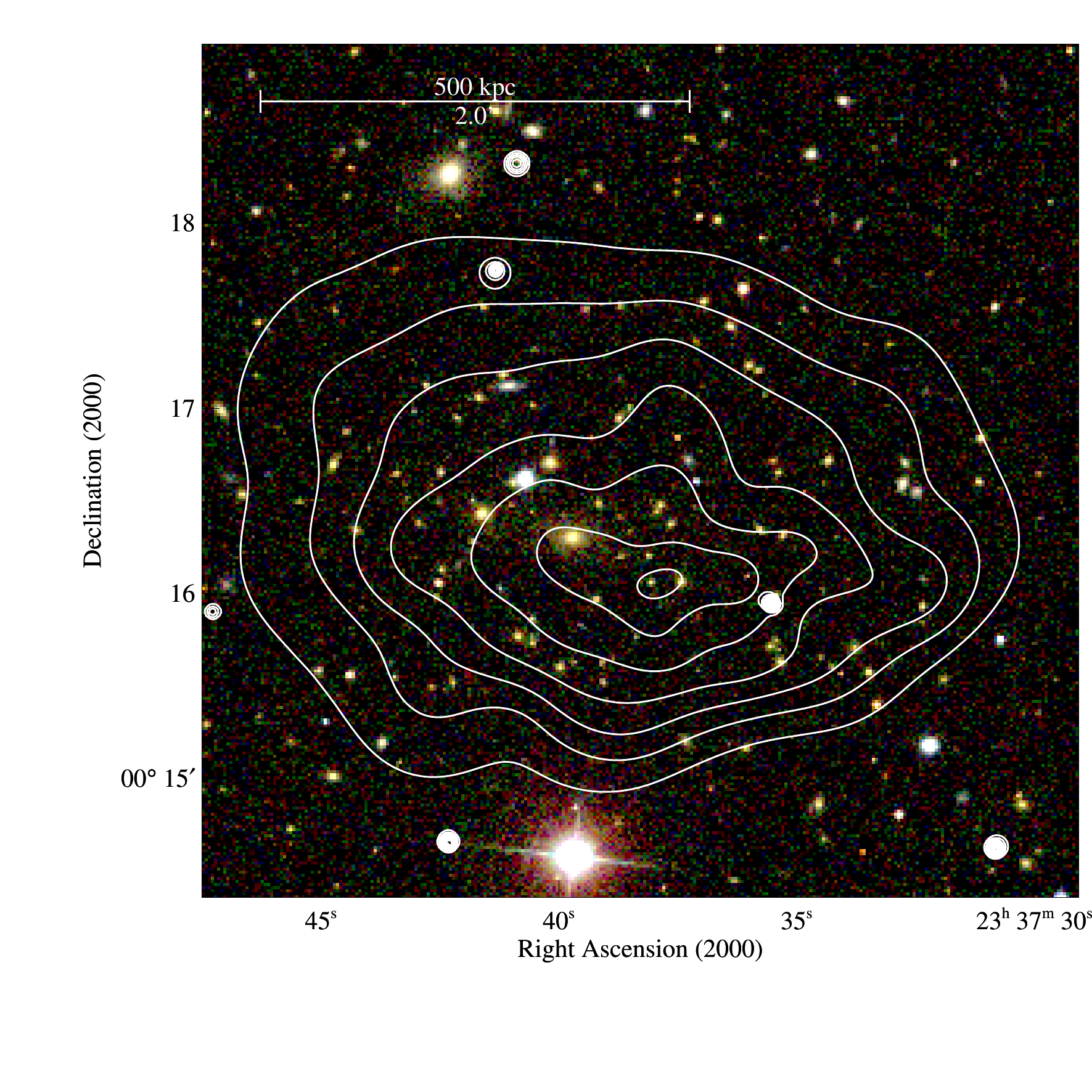} %A2631
   \caption{Logarithmically spaced isodensity contours of the adaptively smoothed X-ray surface brightness as observed with Chandra overlaid on optical images of the clusters identified as complex mergers (see Section~\ref{sec:complex}).  The units on the X-ray contours are $(1.2\times10^{-6})\times1.4^n$ photons s$^{-1}$ in the 0.5--7~keV range. Shown are (in R.A.\ order) {\it A\,2744}, {\it A\,520}, {\it MACS\,J0717.5+3745}, {\it A\,665}, {\it A\,1300}, {\it MACS\,J1149.5+2223},  {\it A\,1682}, {\it MACS\,J2129.4--0741}, {\it RX\,J2228.5+2036}, {\it MACS\,J2243.3--0935}, and {\it A\,2631}.  High resolution versions of each overlay is available at http://ifa.hawaii.edu/~amann/MNRAS2011highres/}
   \label{fig:complexmerg}
\end{figure*}

\clearpage

\begin{figure*} 
   \centering
   \includegraphics[width=\otheroverlaysize]{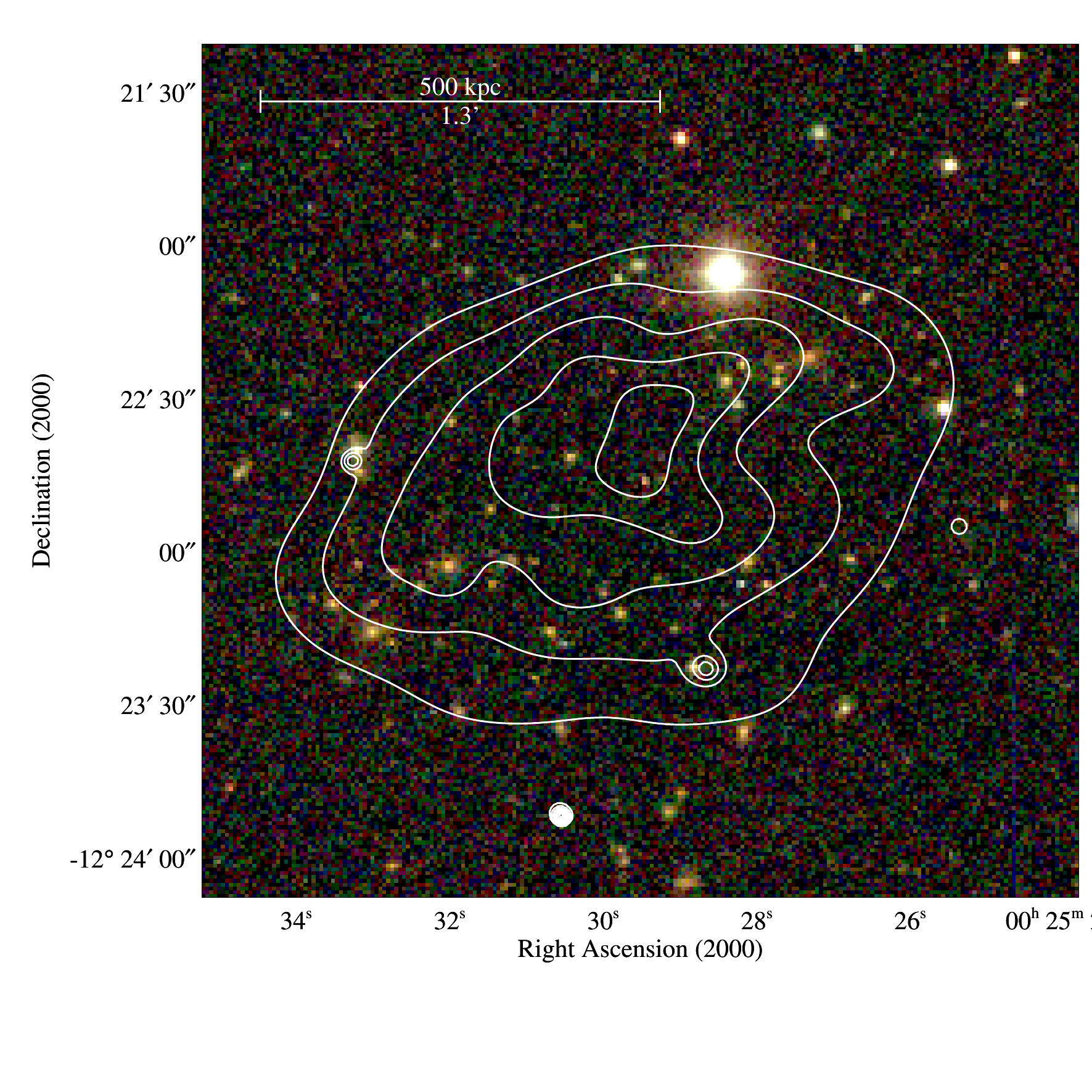} 
   \includegraphics[width=\otheroverlaysize]{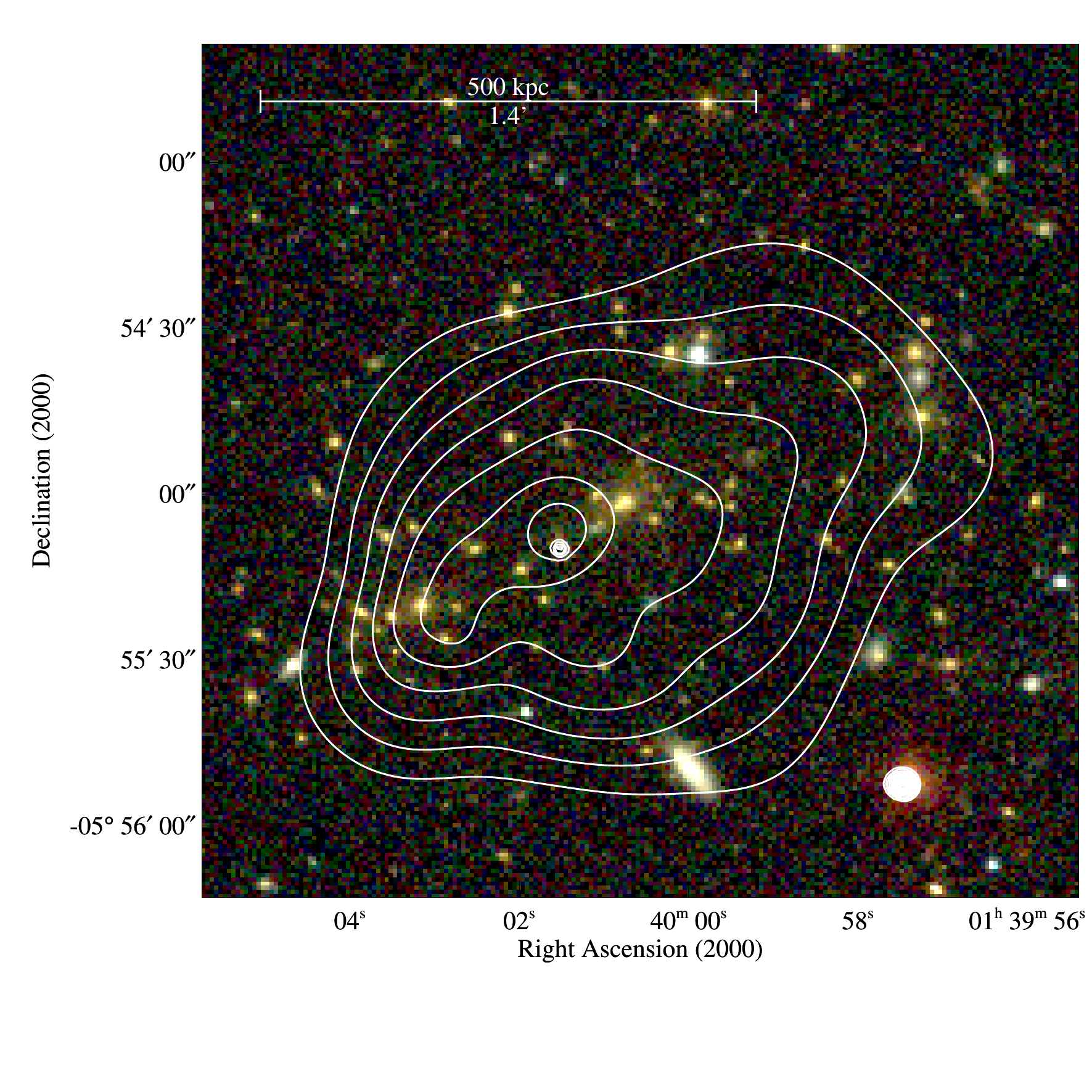} 
   \includegraphics[width=\otheroverlaysize]{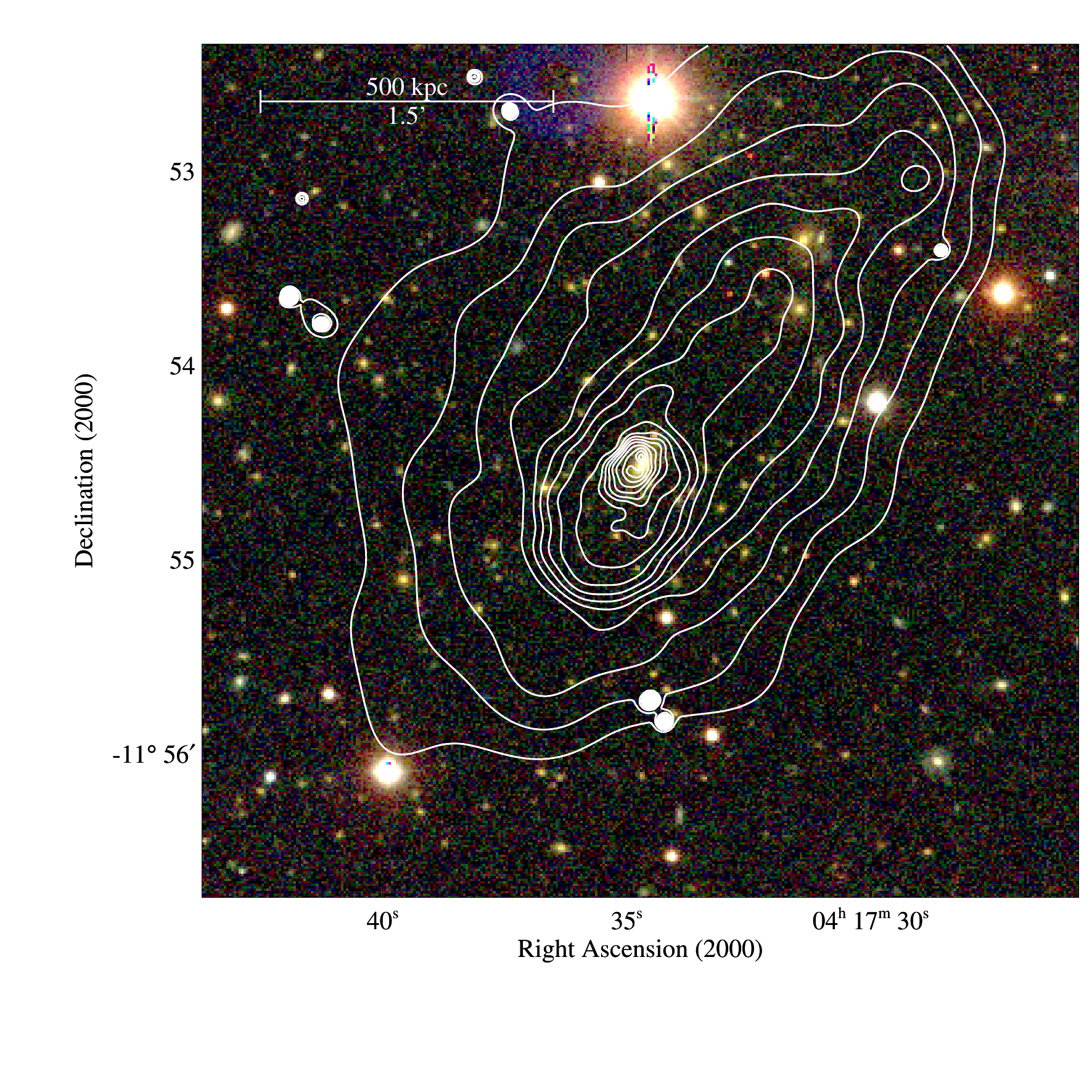}
   \includegraphics[width=\otheroverlaysize]{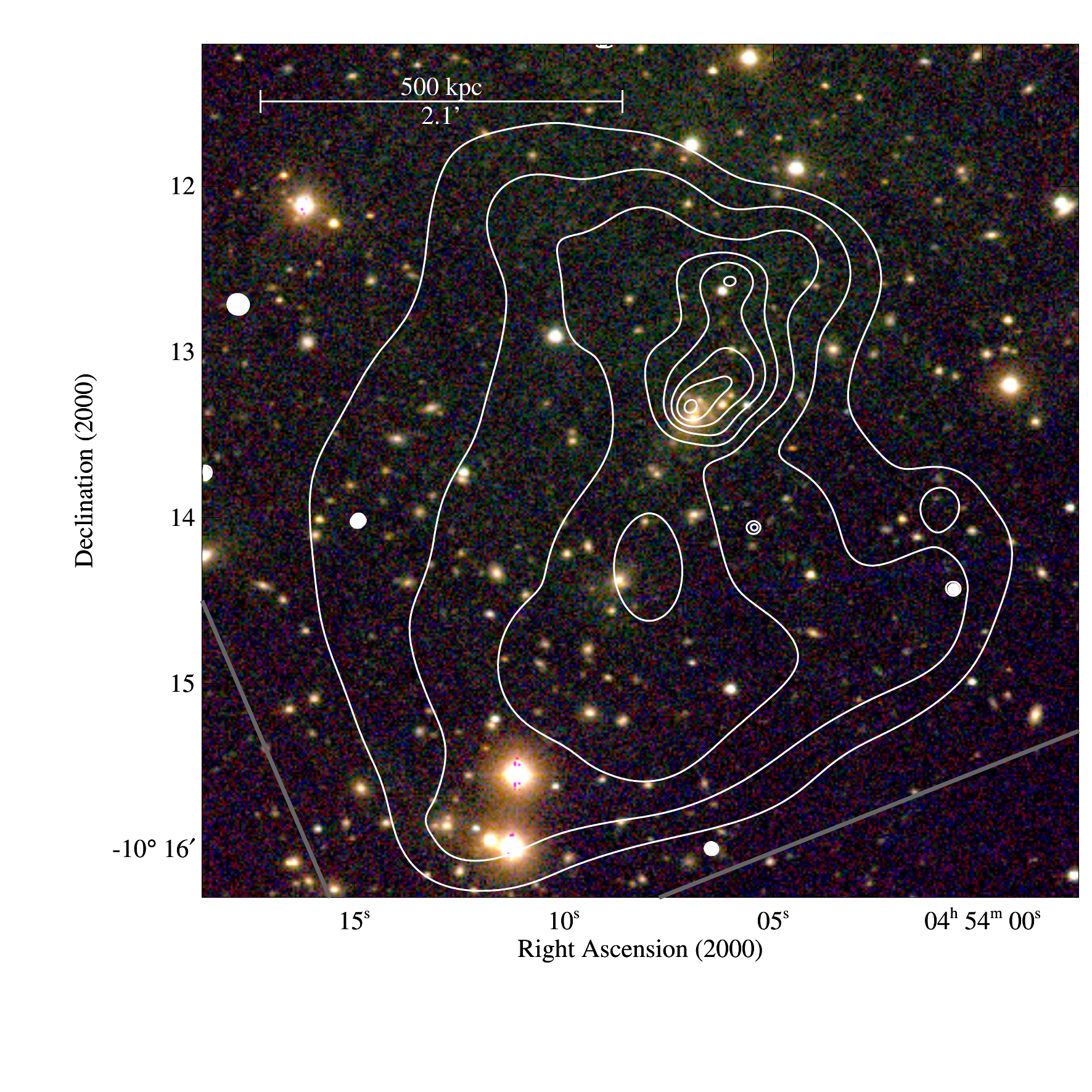} % A520
   \includegraphics[width= \otheroverlaysize]{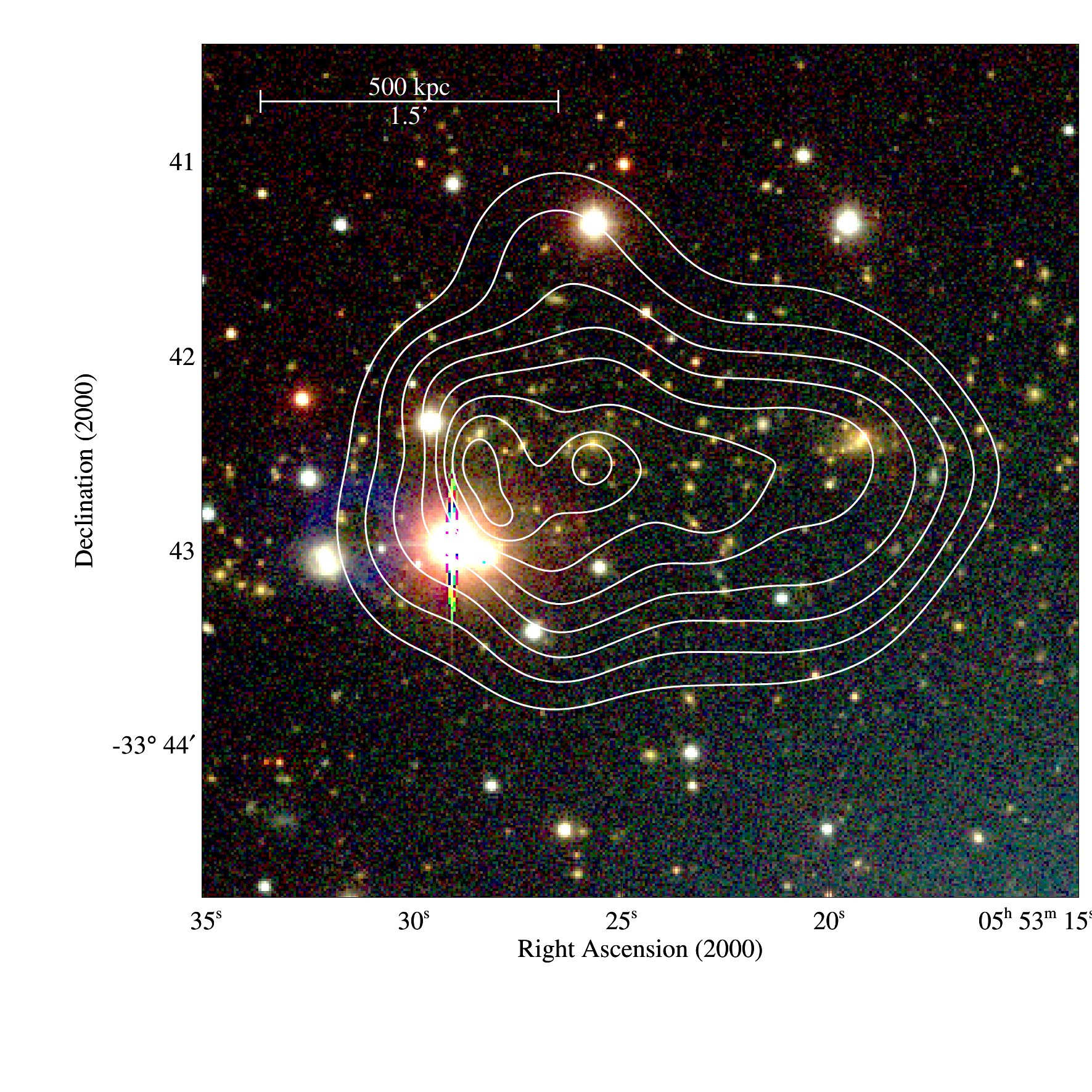} 
   \includegraphics[width=\otheroverlaysize]{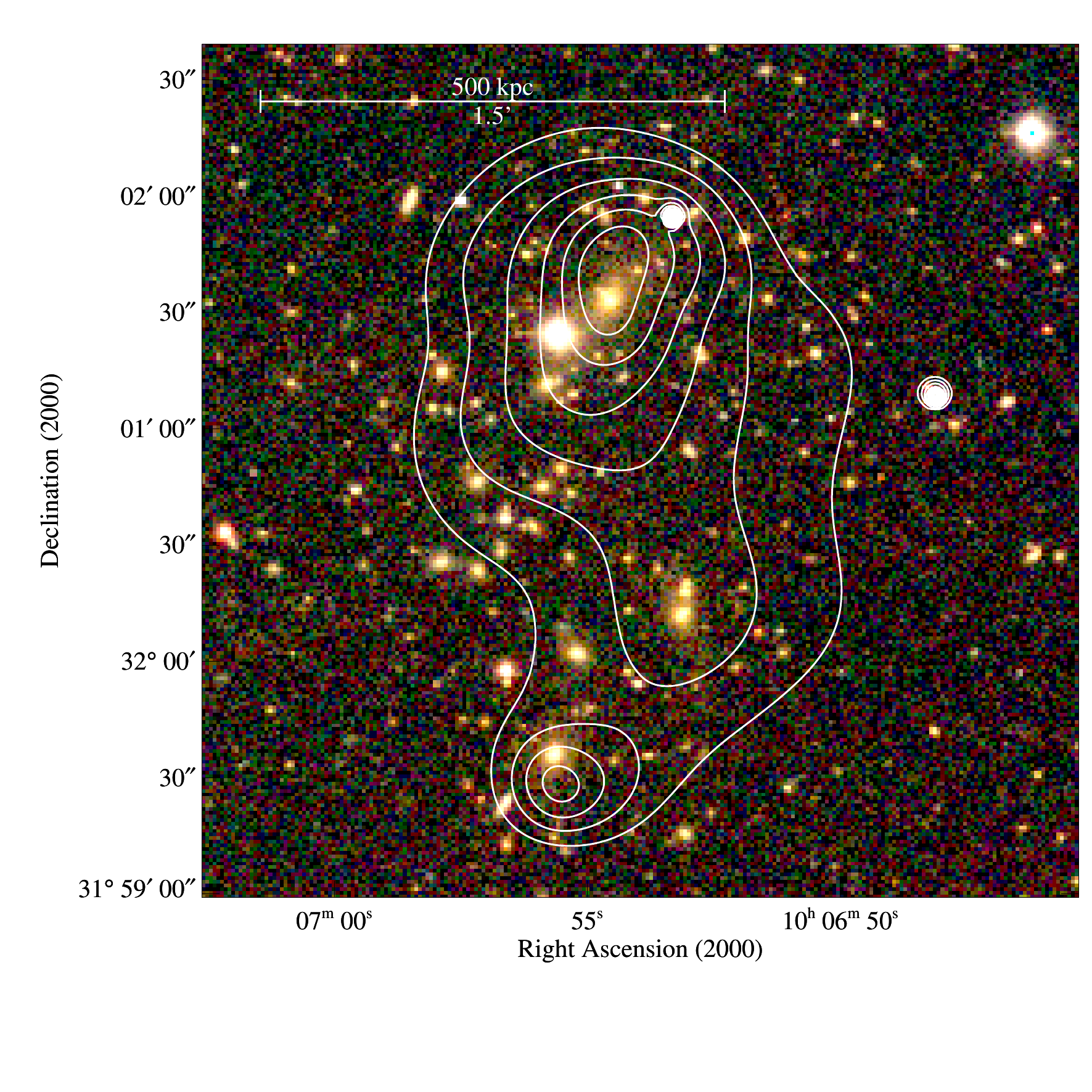}
   \includegraphics[width= \otheroverlaysize]{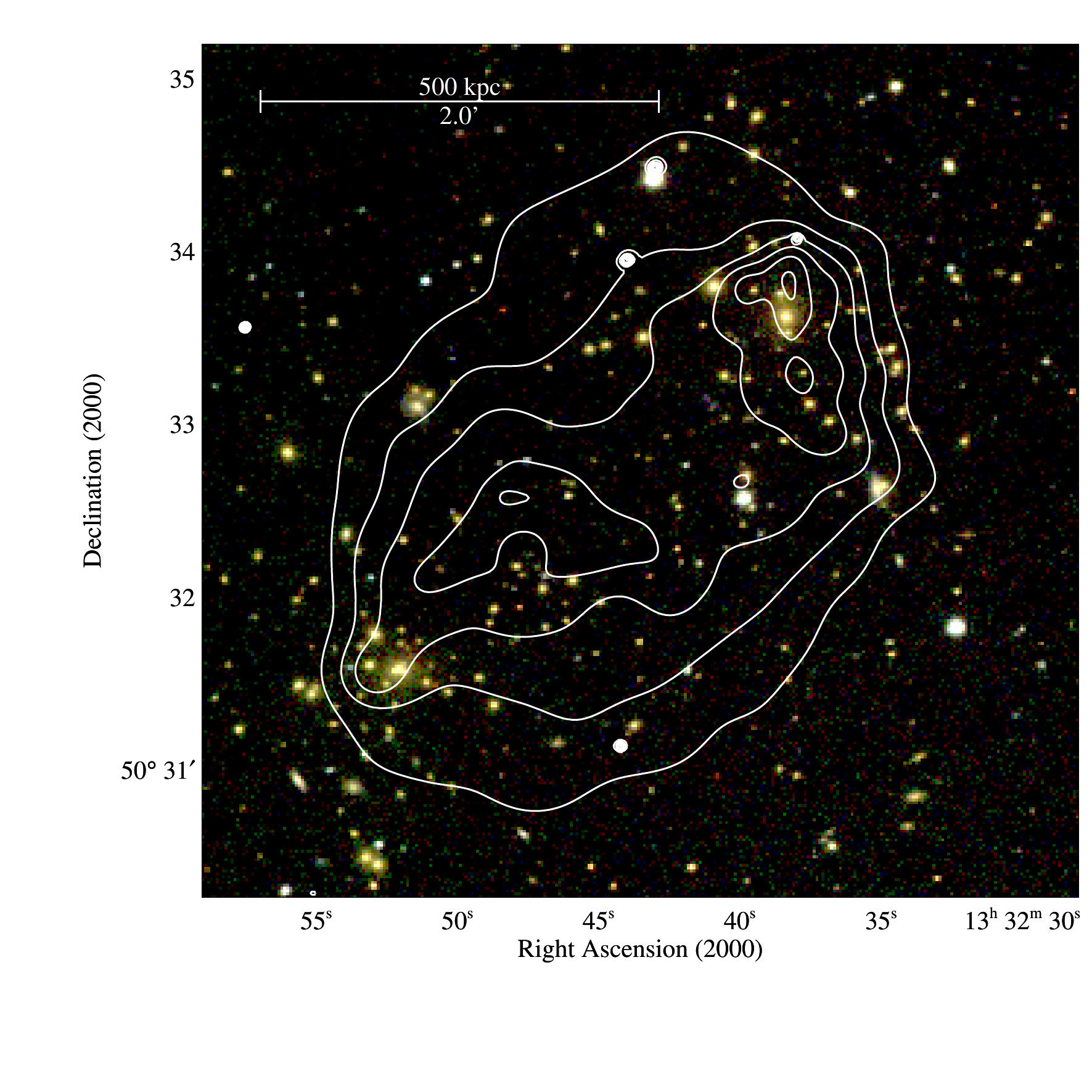} %A1758
   \includegraphics[width=\otheroverlaysize]{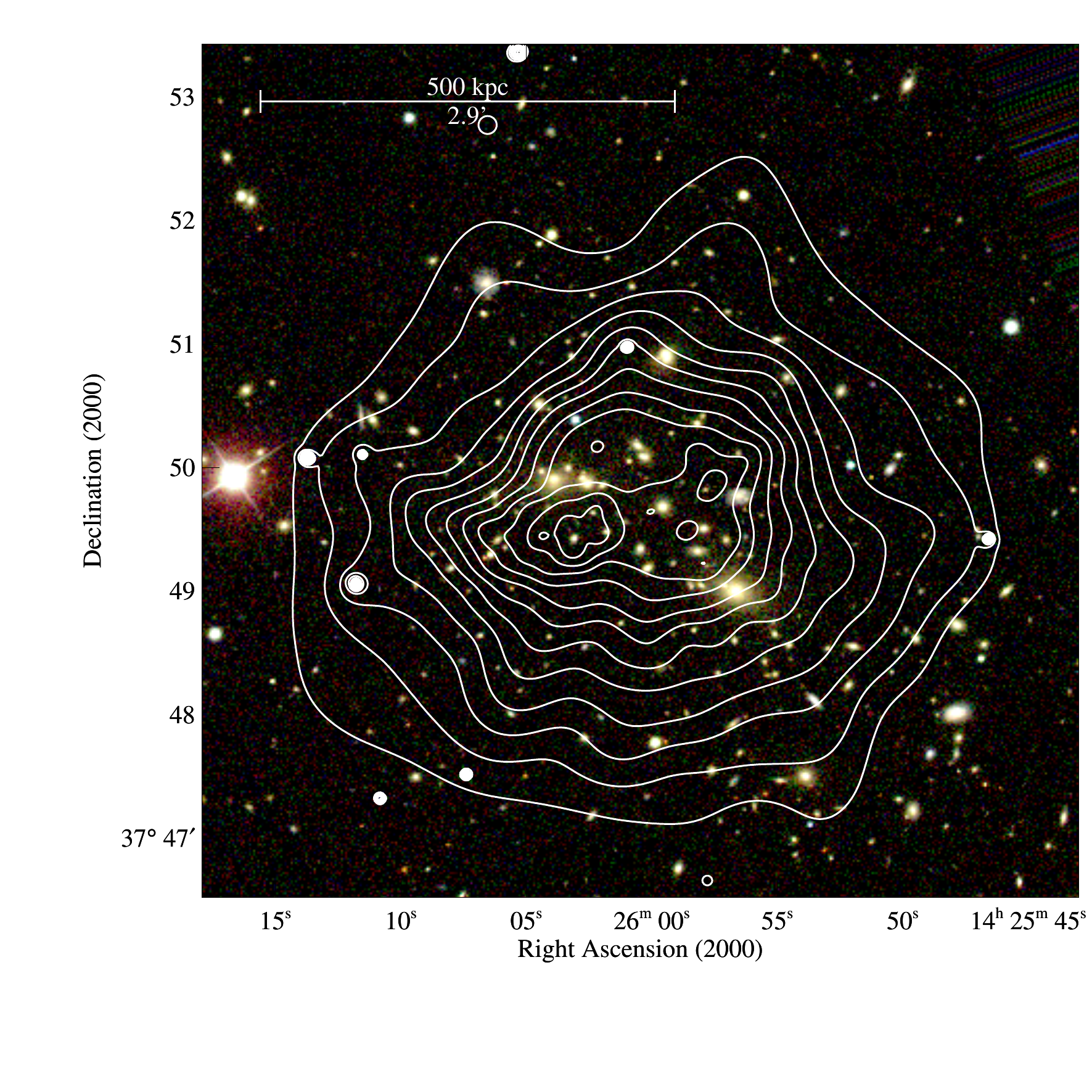} %A1914
   \includegraphics[width=\otheroverlaysize]{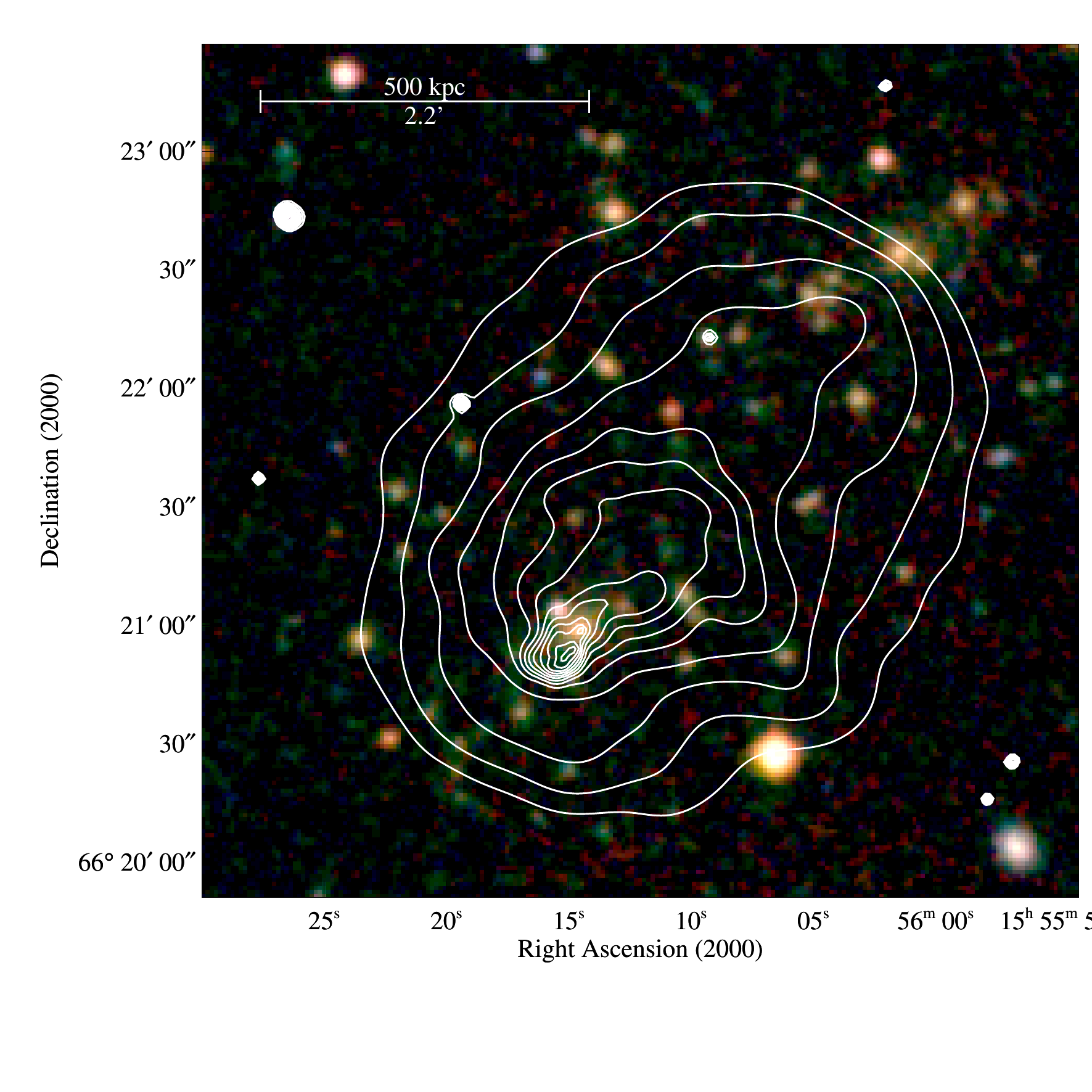} %A2416
   \includegraphics[width=\otheroverlaysize]{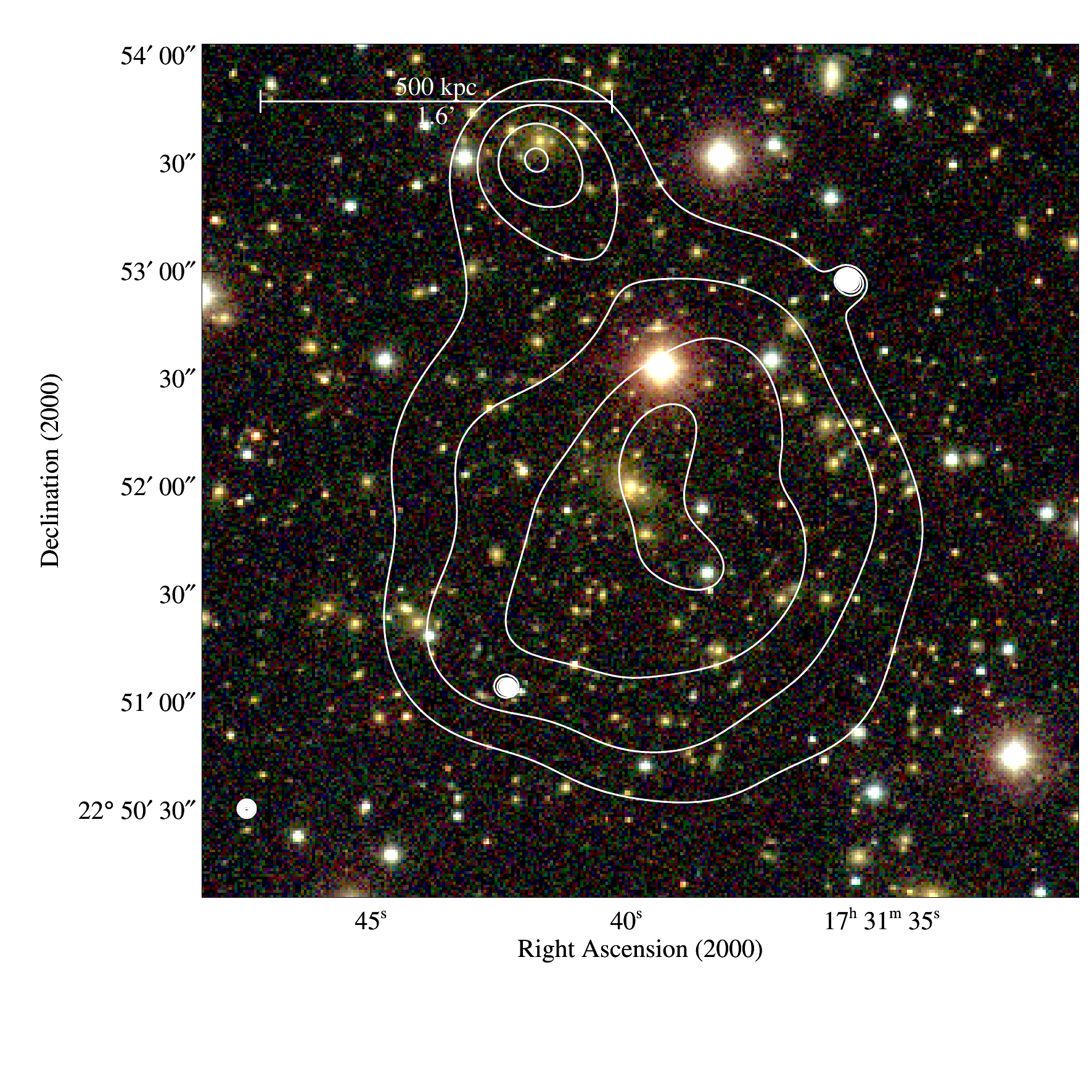} 
   \caption{Same as Fig.~\ref{fig:complexmerg} for the primary BHOM candidates discussed in Section~\ref{sec:bhom}.  Shown are (in R.A.\ order) {\it MACS\,J0025.4--1222}, {\it MACS\,J0140.0--0555}, {\it MACS\,J0417.5--1154}, {\it A\,520}, {\it MACS\,J0553.4--3342}, {\it MACS\,J1006.9+3200}, {\it A\,1758}, {\it A\,1914}, {\it A\,2146}, and {\it MACS\,J1731.6+2252}.}
   \label{fig:bhomstier1}
\end{figure*}

\clearpage

\begin{figure*} 
   \centering
   \includegraphics[width=\otheroverlaysize]{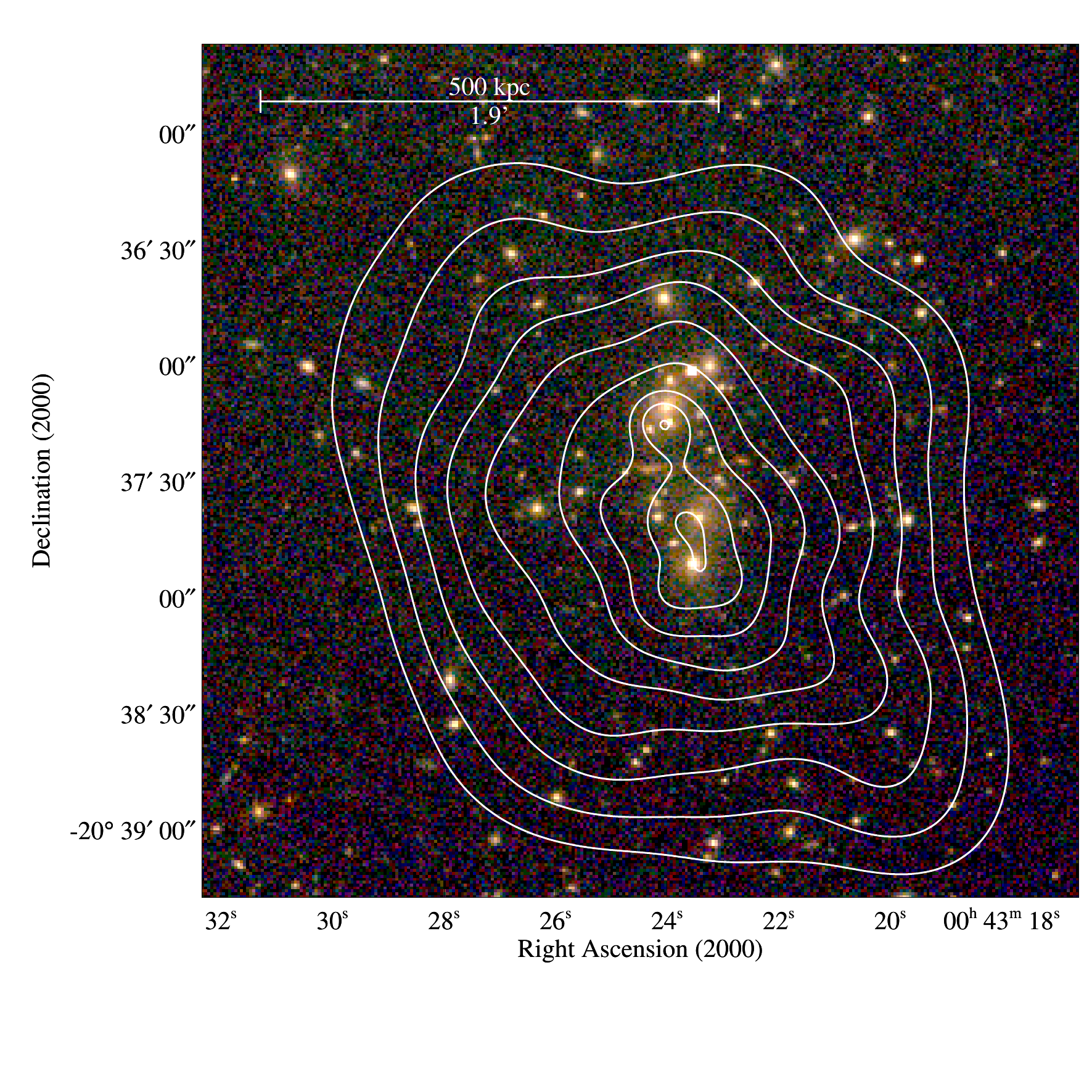} %A2813
   \includegraphics[width=\otheroverlaysize]{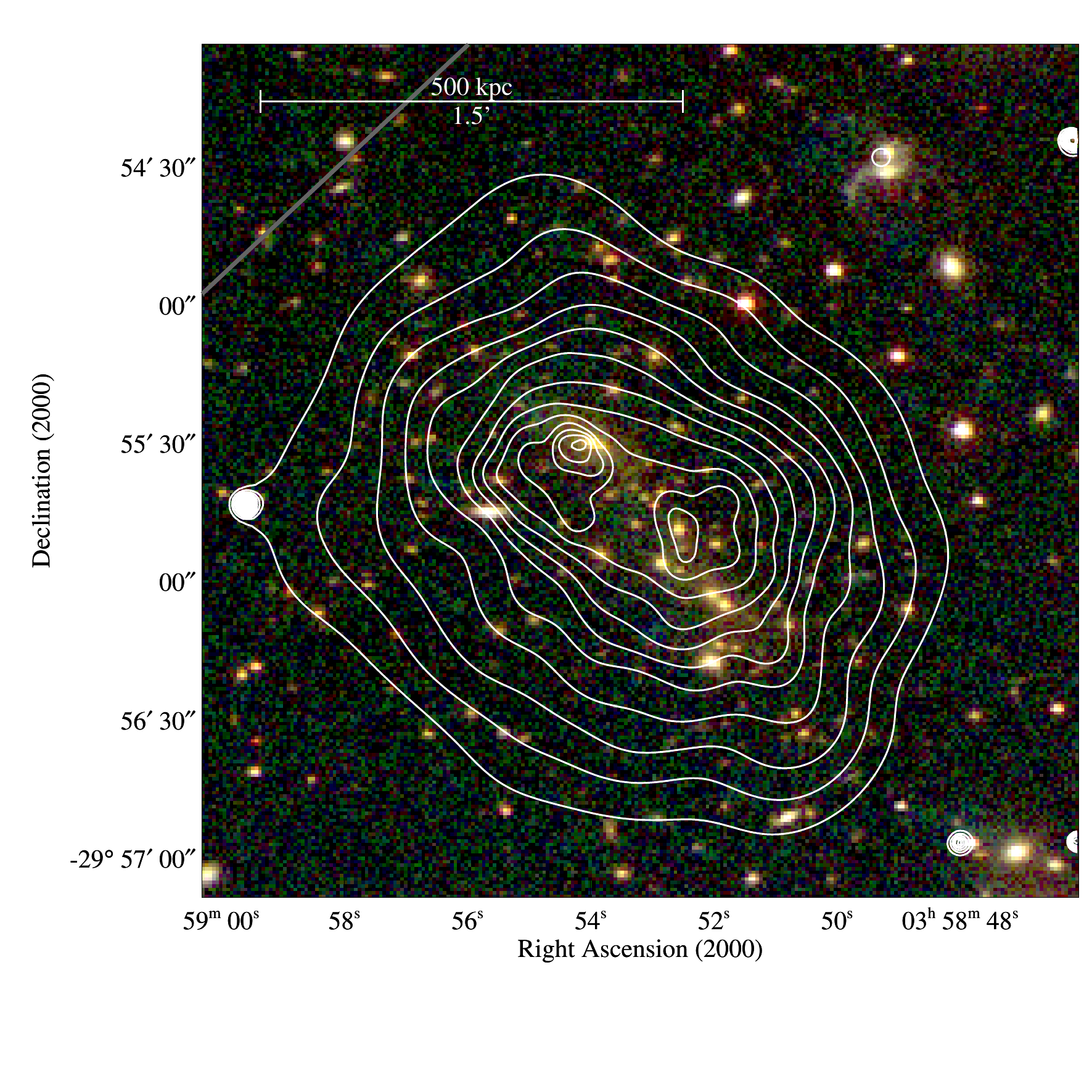} 
   \includegraphics[width=\otheroverlaysize]{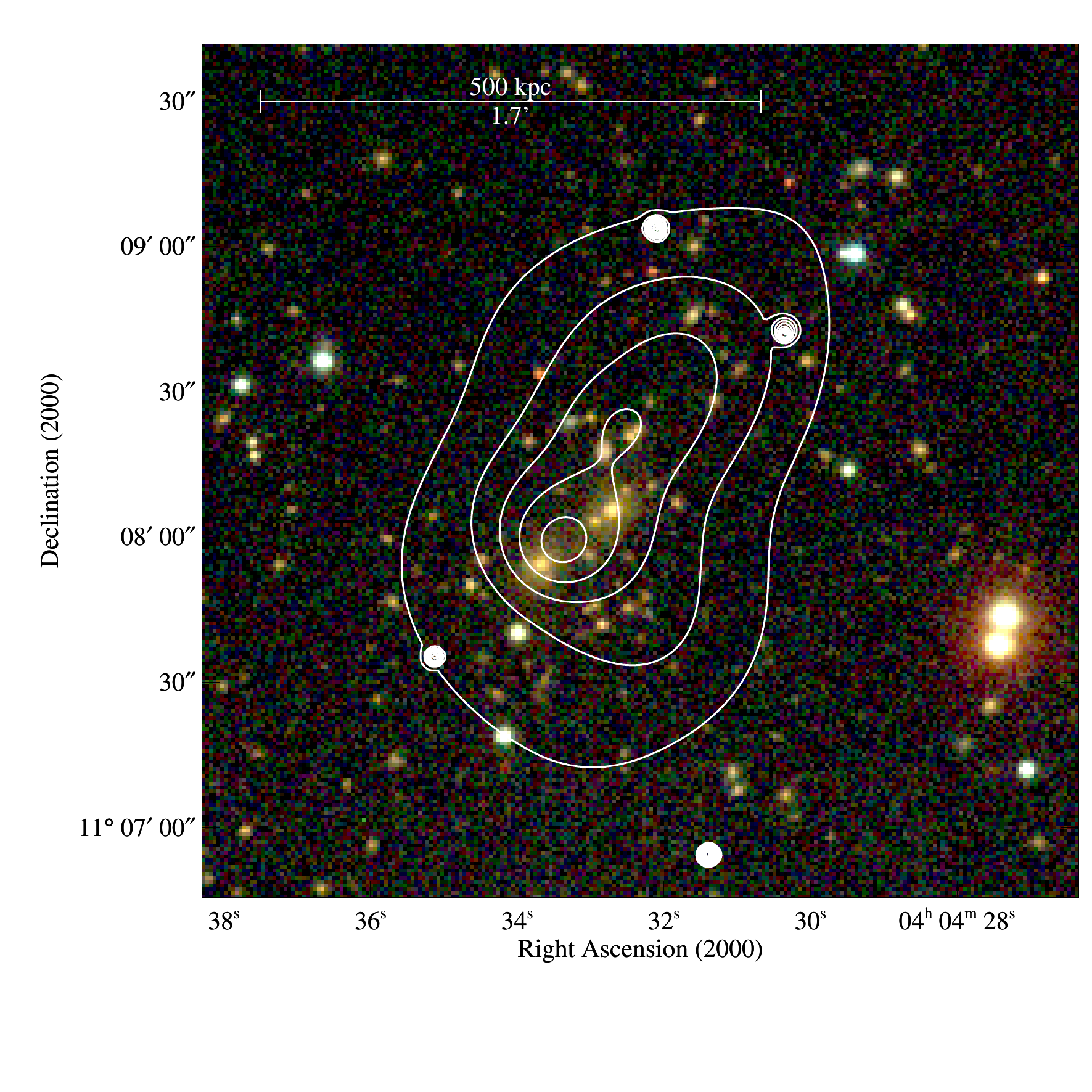}
   \includegraphics[width=\otheroverlaysize]{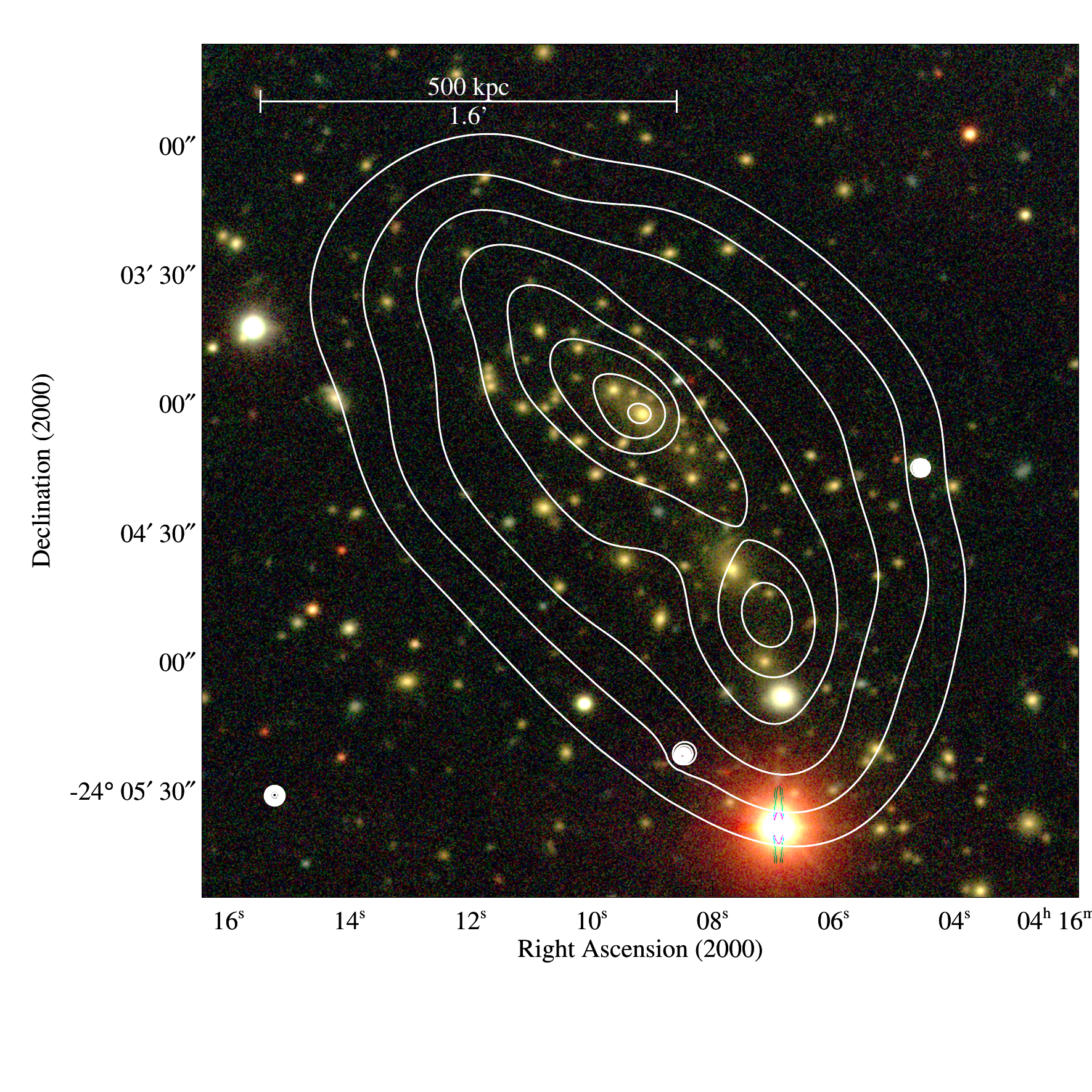}
   \includegraphics[width=\otheroverlaysize]{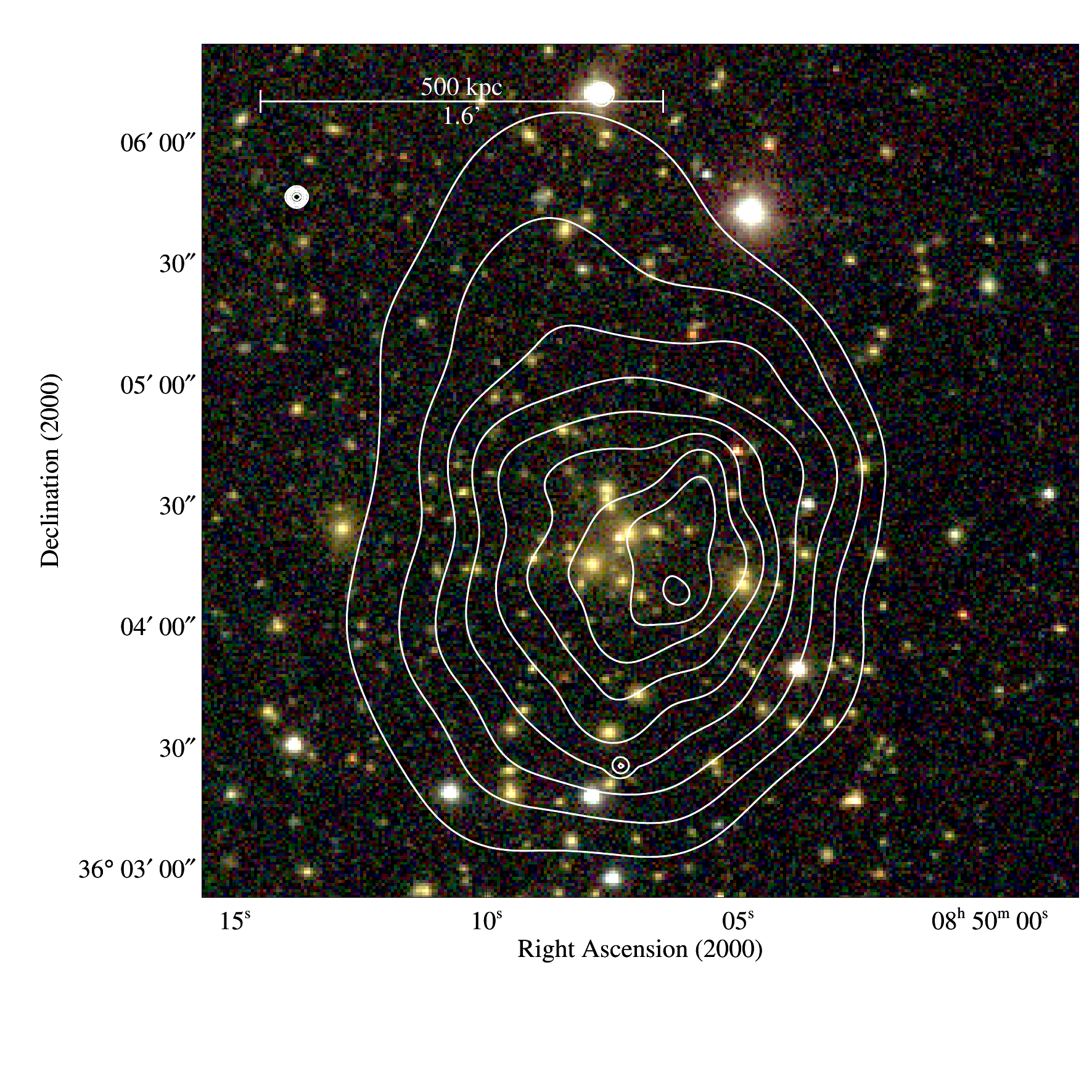}  %Z1953
   \includegraphics[width=\otheroverlaysize]{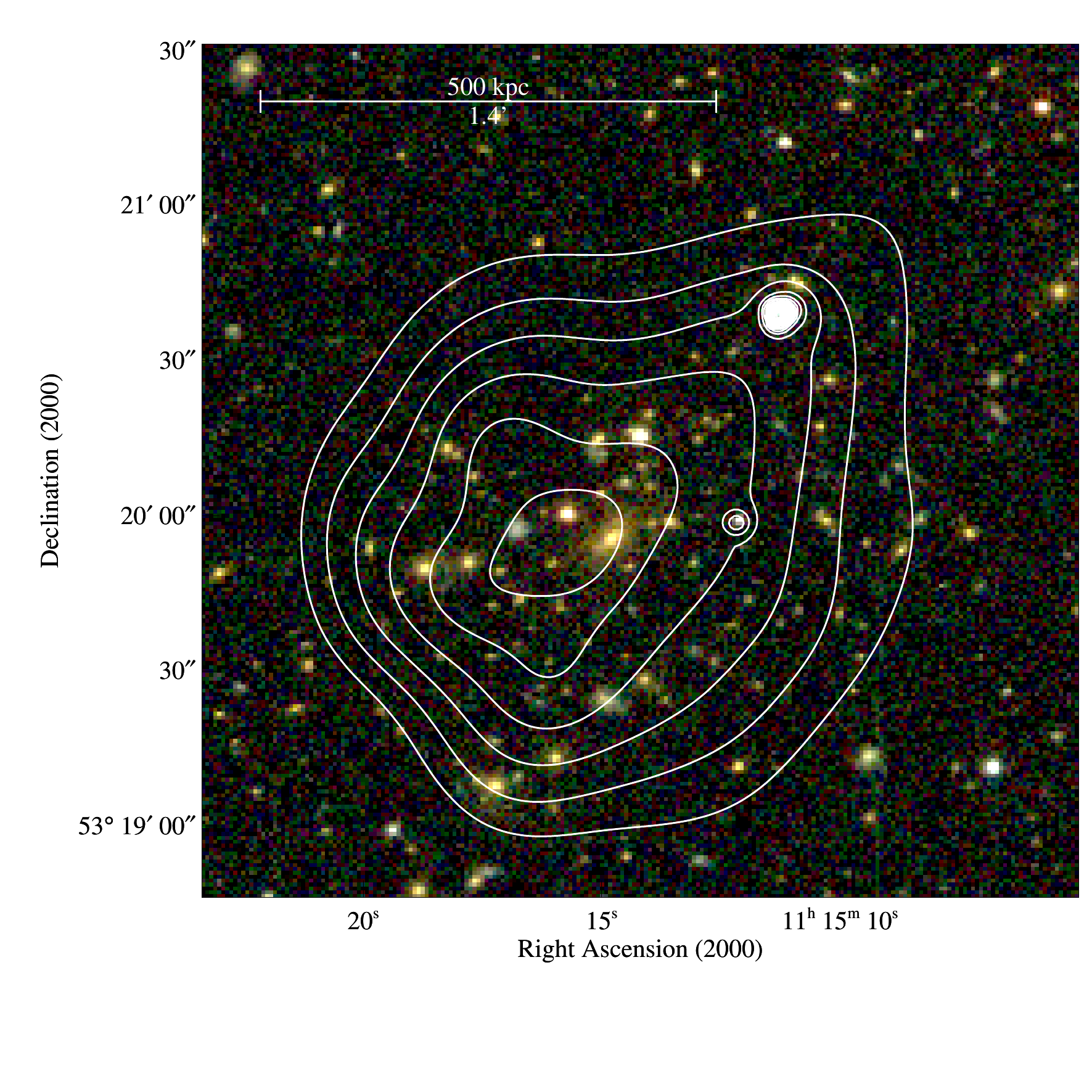}
   \caption{Same as Fig.~\ref{fig:complexmerg} for the secondary BHOM candidates.  Shown are (in R.A.\ order) {\it A\,2813}, {\it MACS\,J0358.8--2955}, {\it MACS\,J0404.6+1109}, {\it MACS\,J0416.1--2403}, {\it ZwCl 0847.2+3617}, {\it MACS\,J1115.2+5320}.}
   \label{fig:bhomstier2}
\end{figure*}

\clearpage

\appendix \label{app:allclusters}
\section{Overlays of all clusters}
\begin{figure*} 
   \centering
\includegraphics[width=\overlaysize]{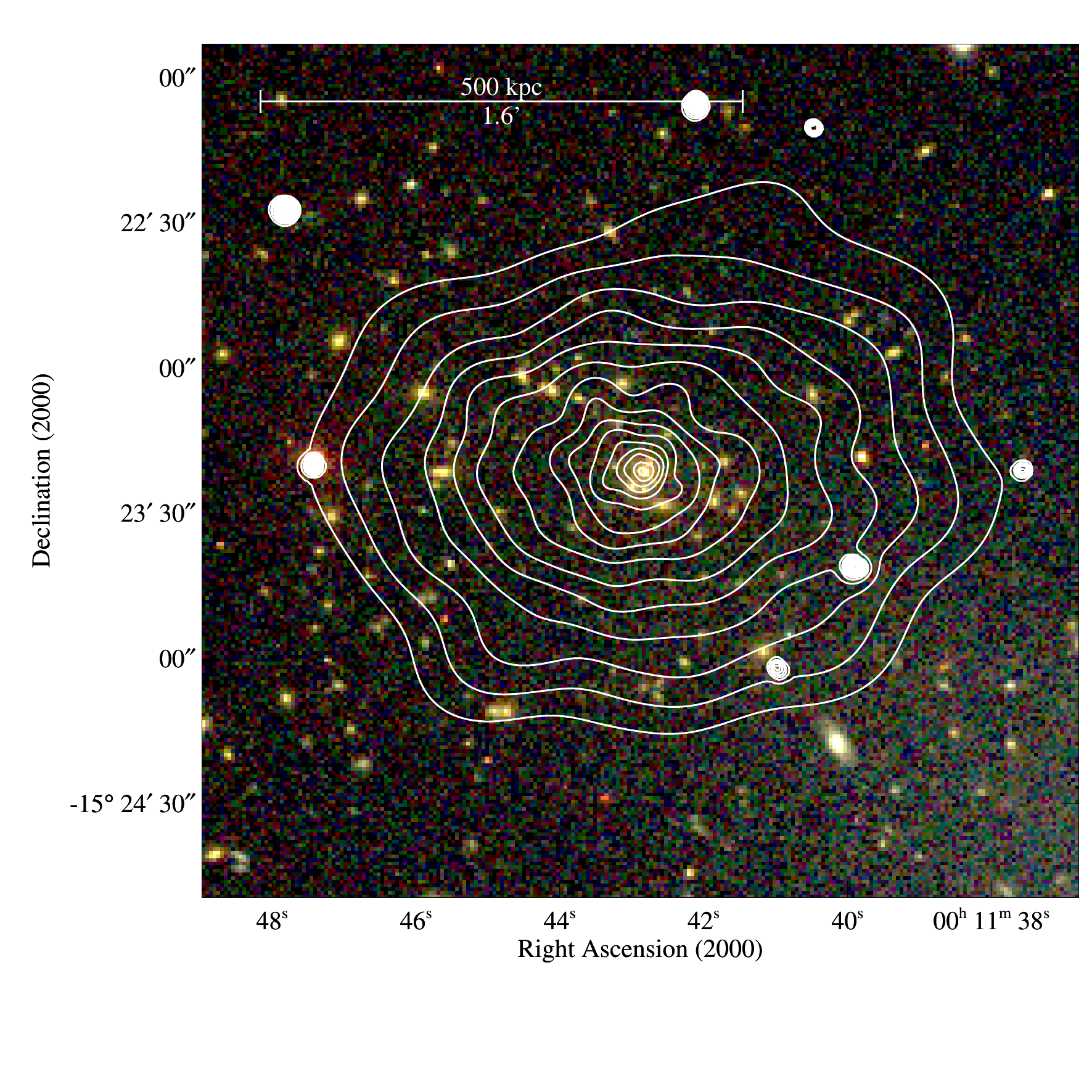}
\includegraphics[width=\overlaysize]{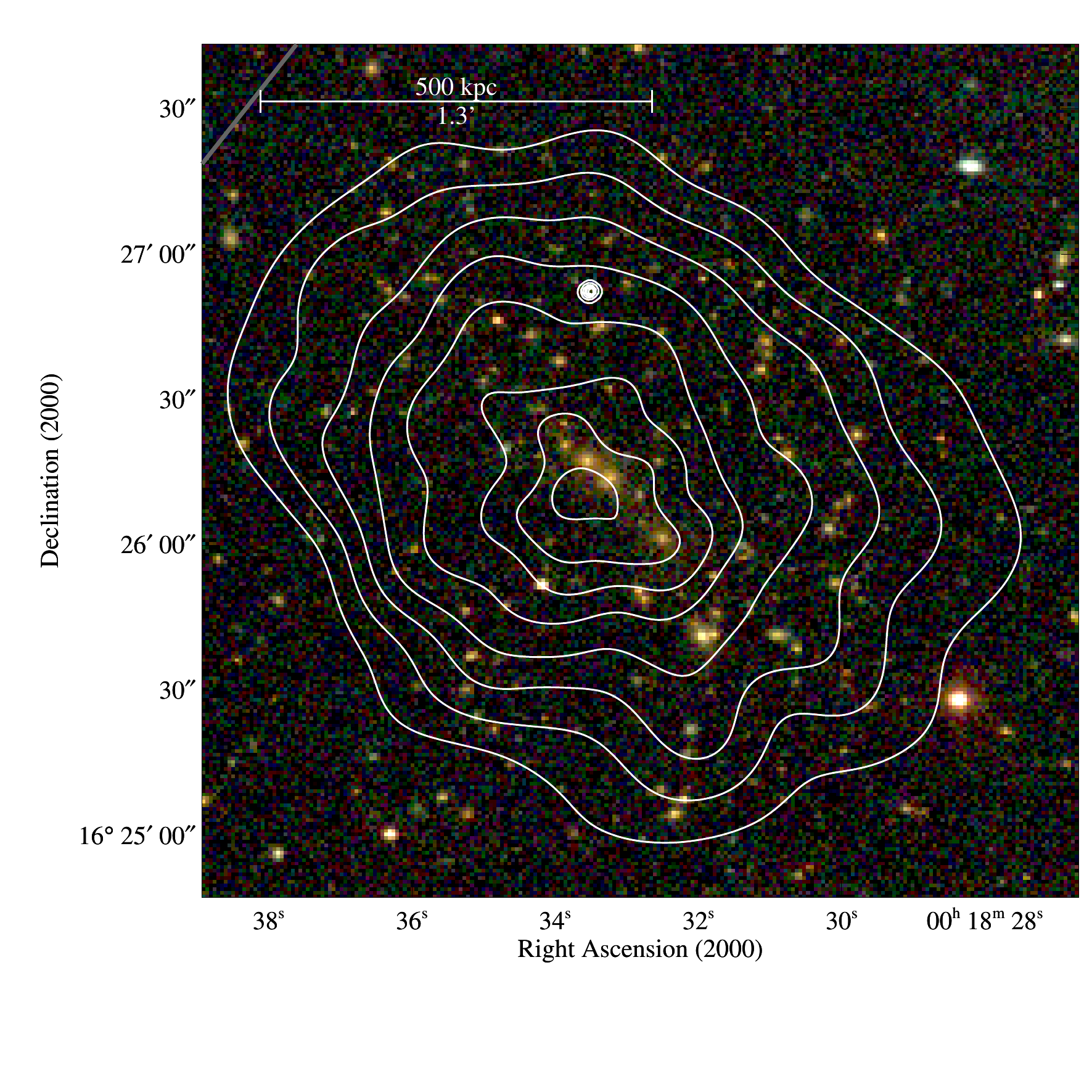}
\includegraphics[width=\overlaysize]{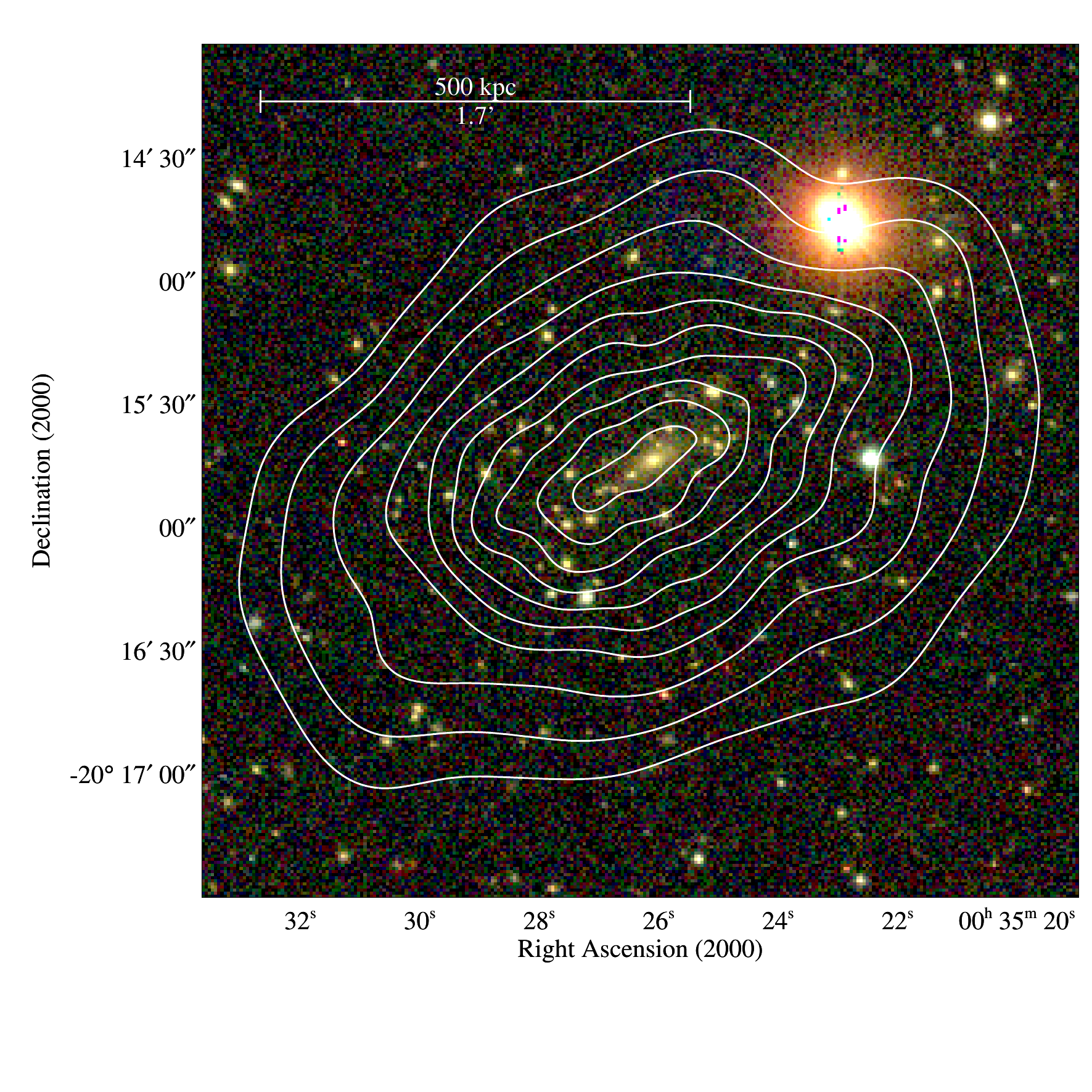}
\includegraphics[width=\overlaysize]{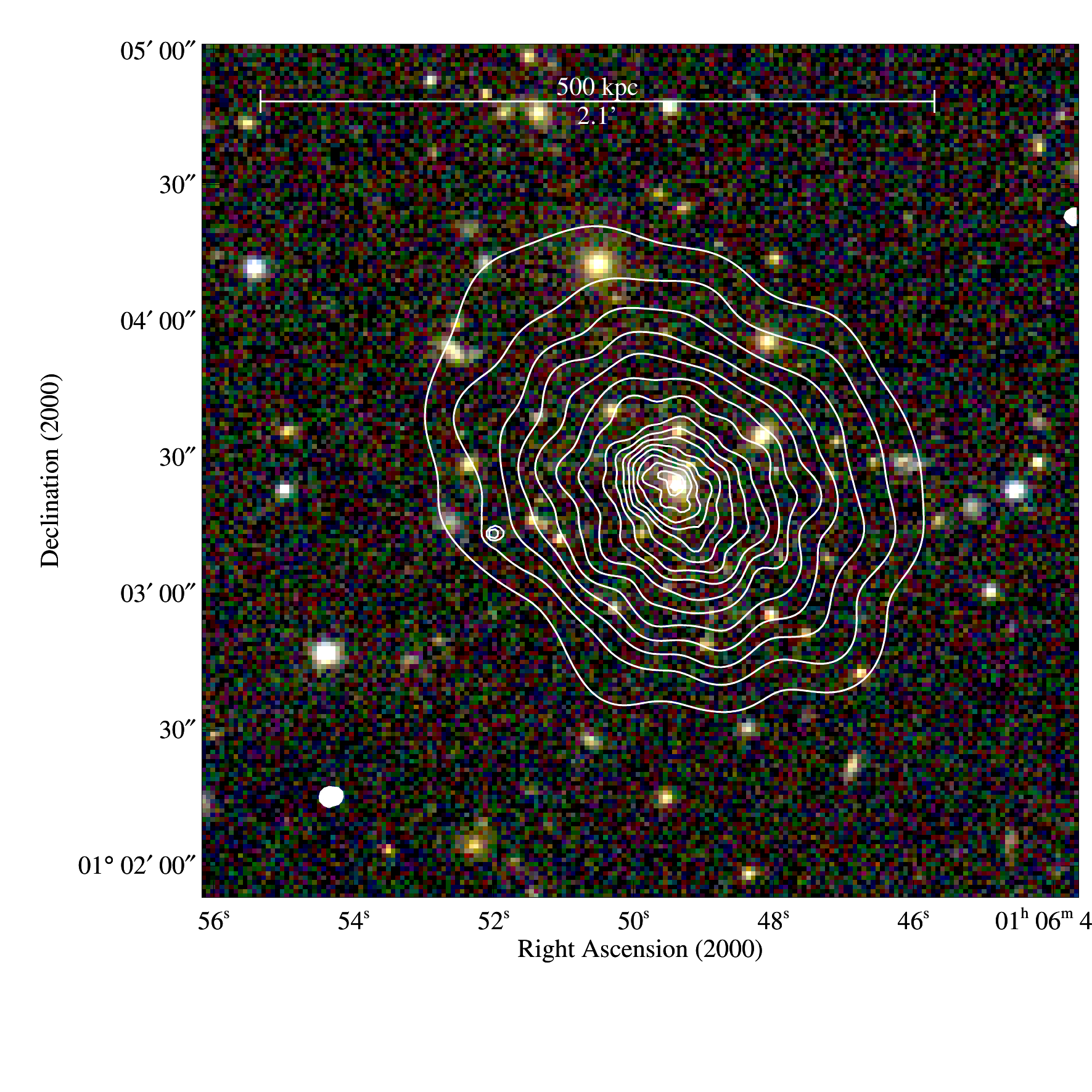}
\includegraphics[width=\overlaysize]{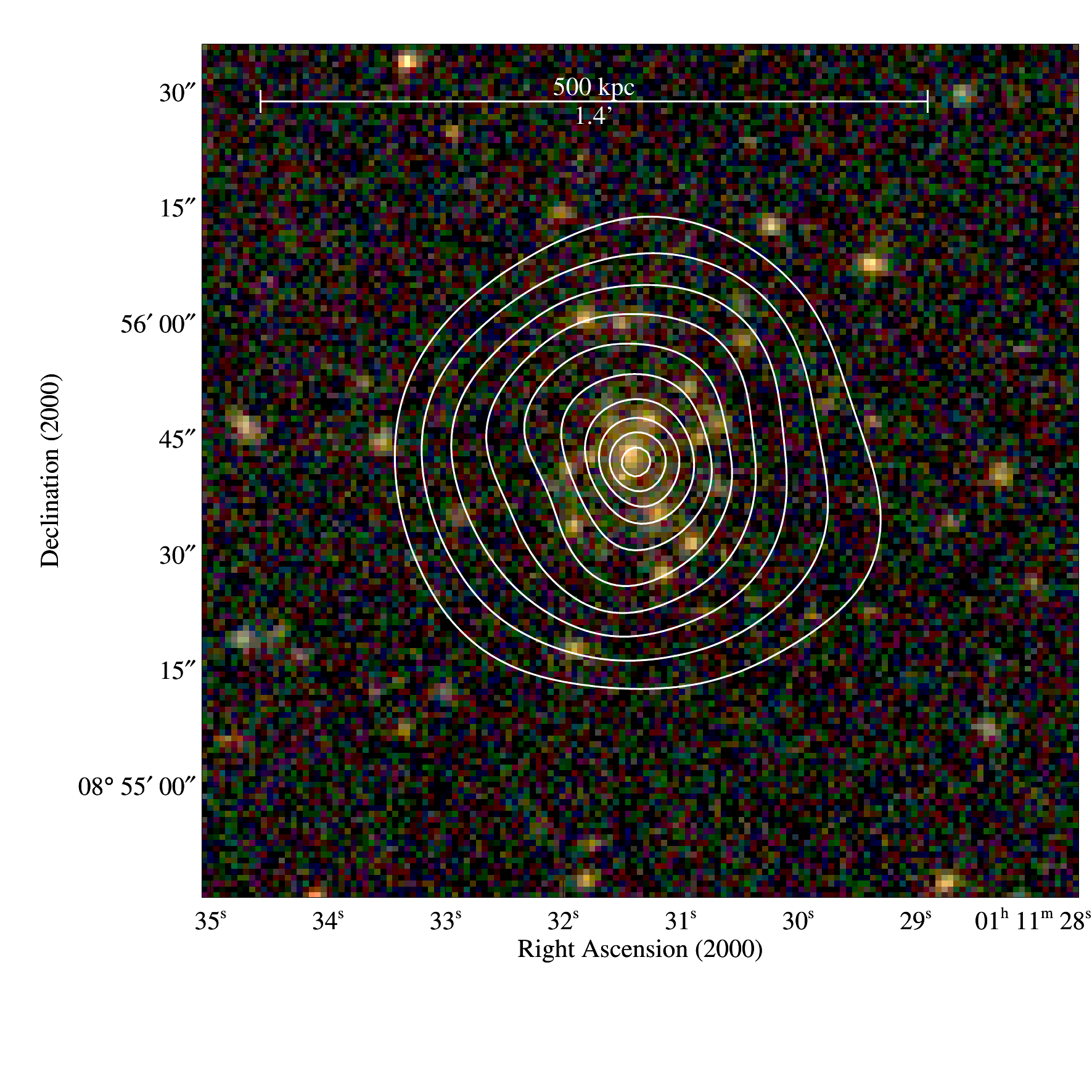}
\includegraphics[width=\overlaysize]{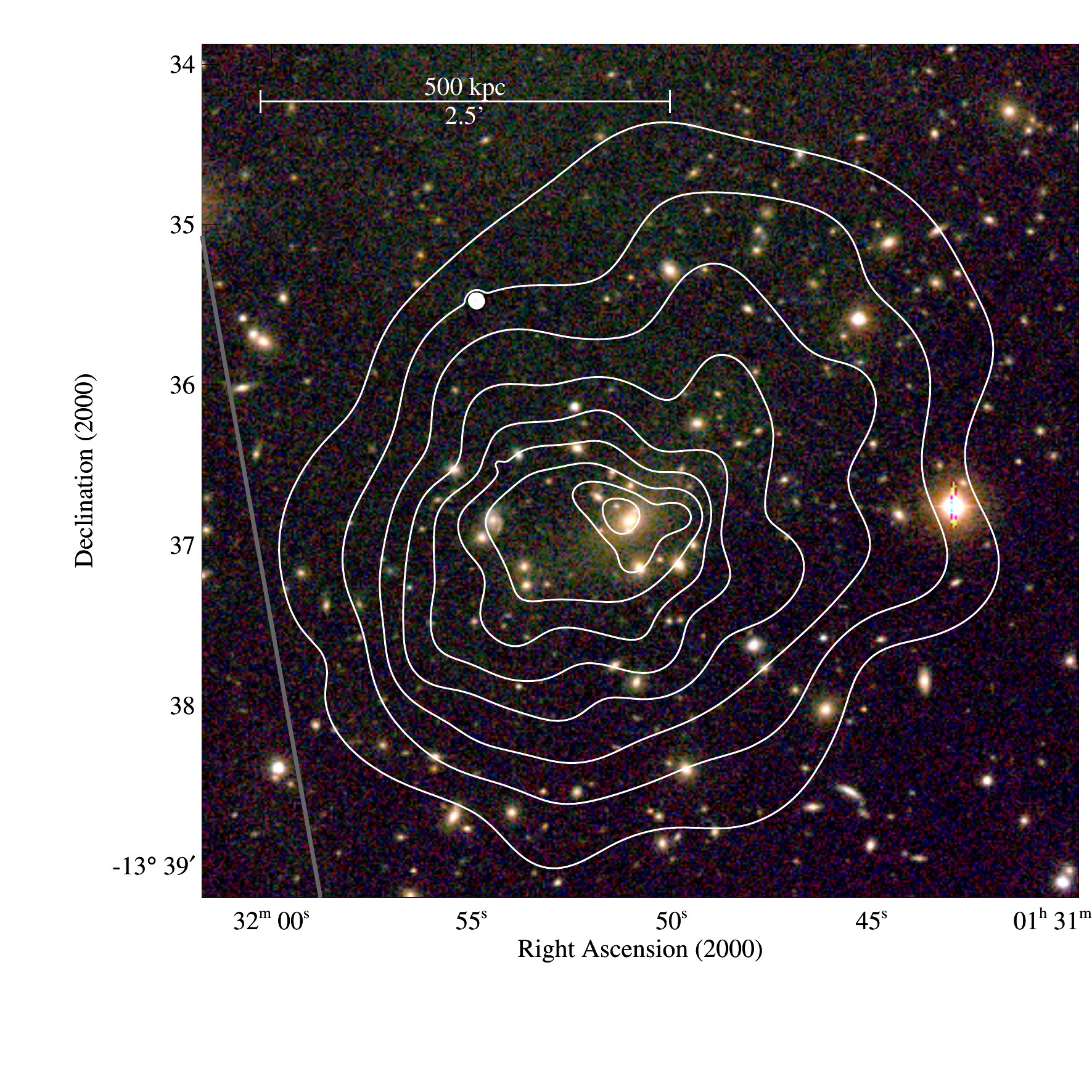}
\includegraphics[width=\overlaysize]{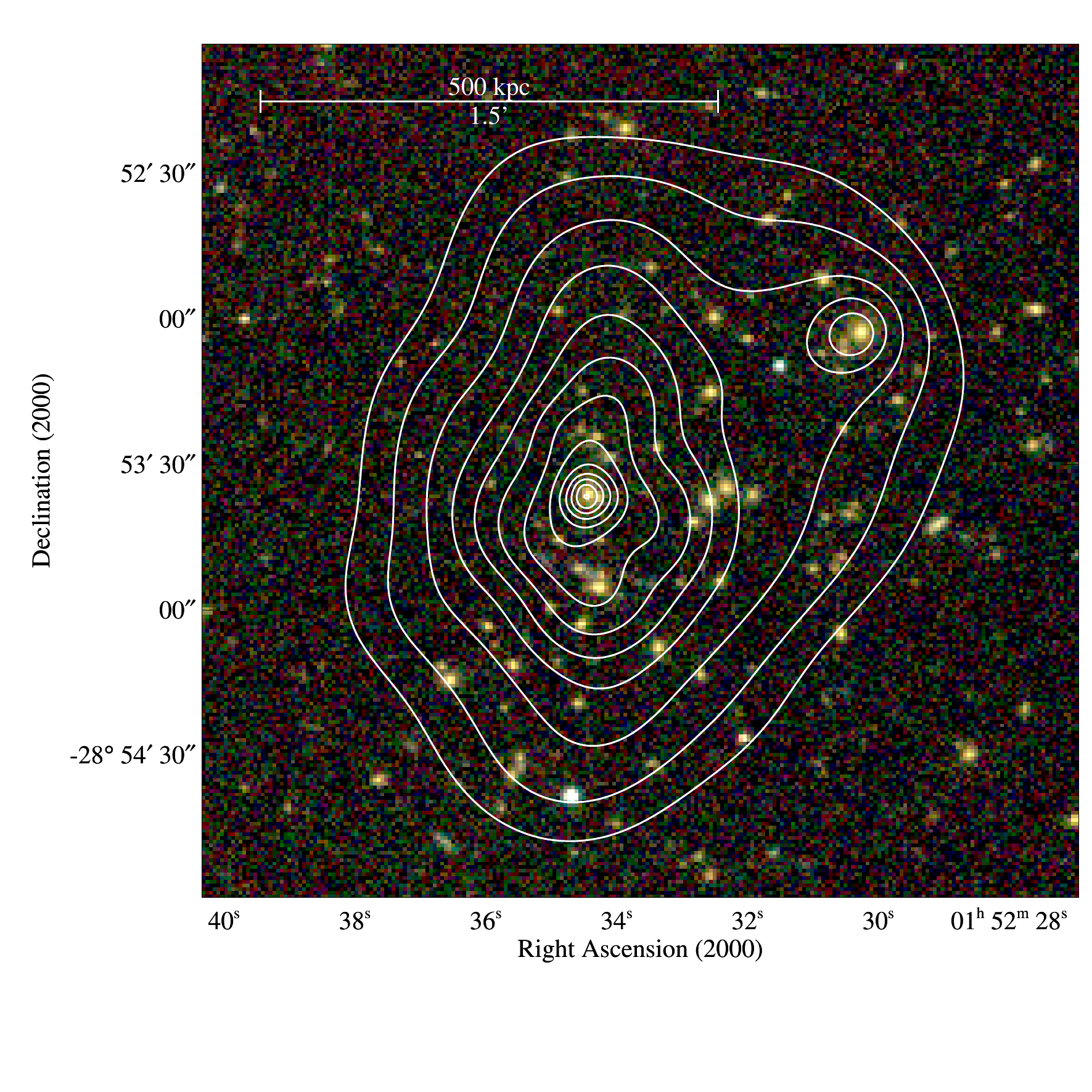} 
\includegraphics[width=\overlaysize]{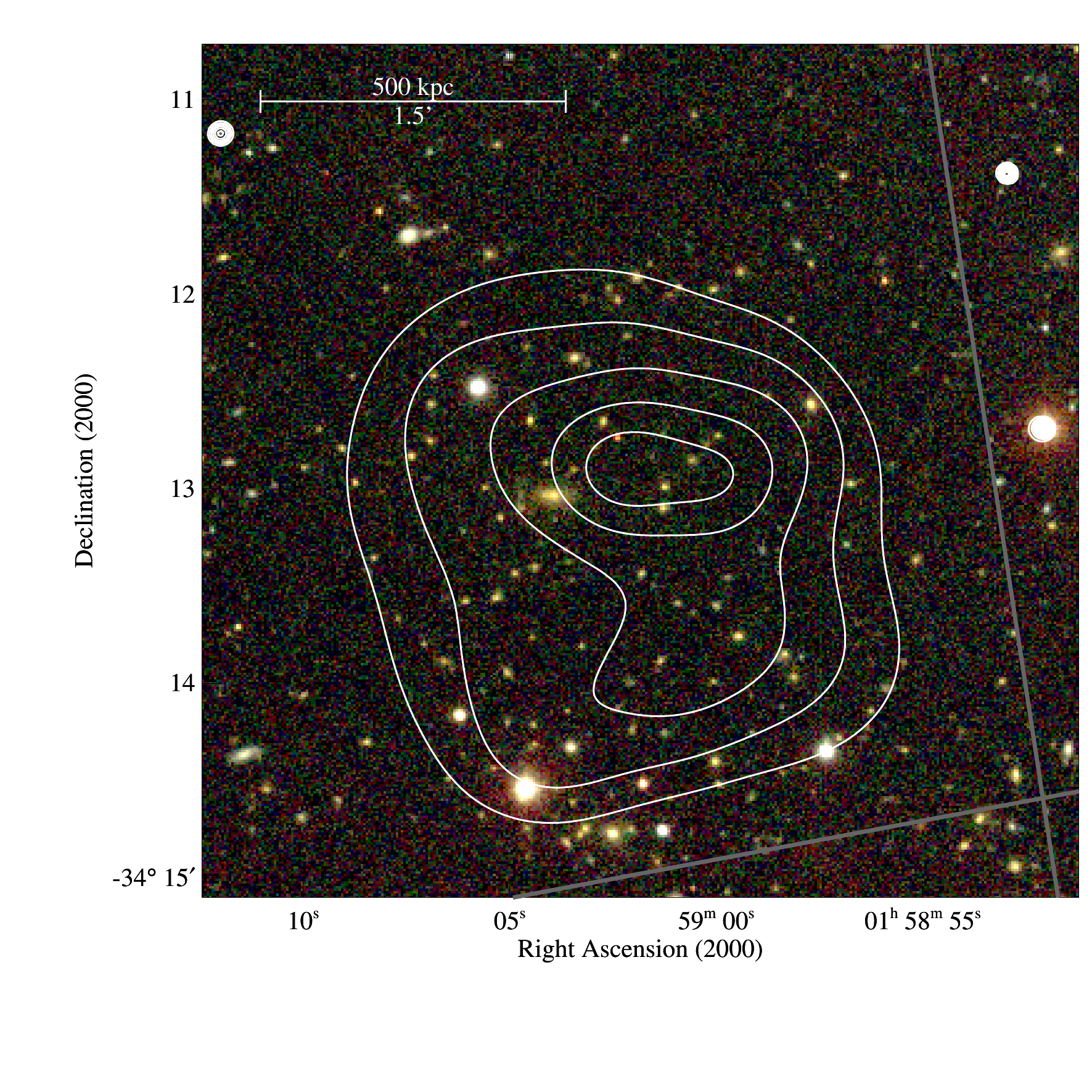} 
\includegraphics[width=\overlaysize]{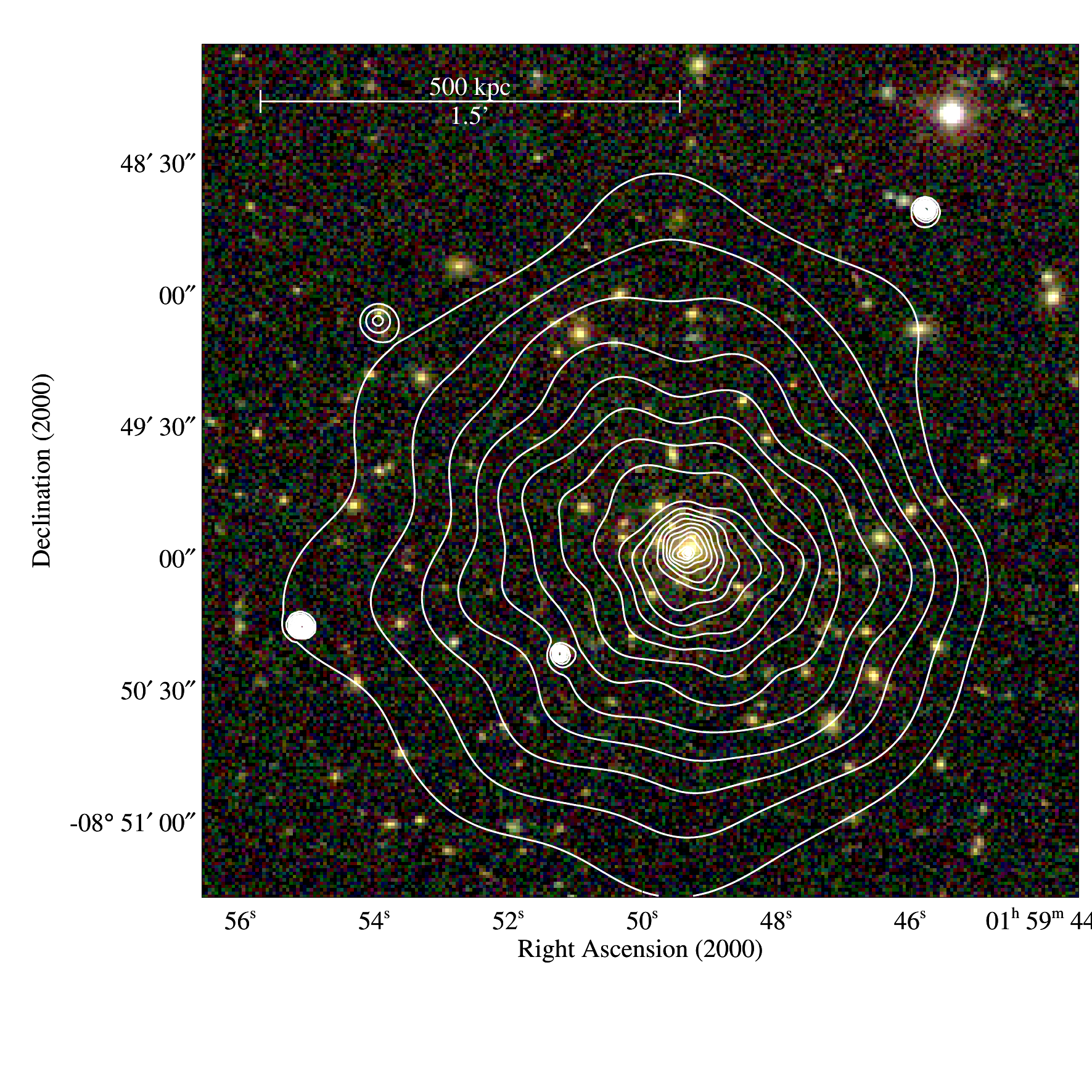}
\includegraphics[width=\overlaysize]{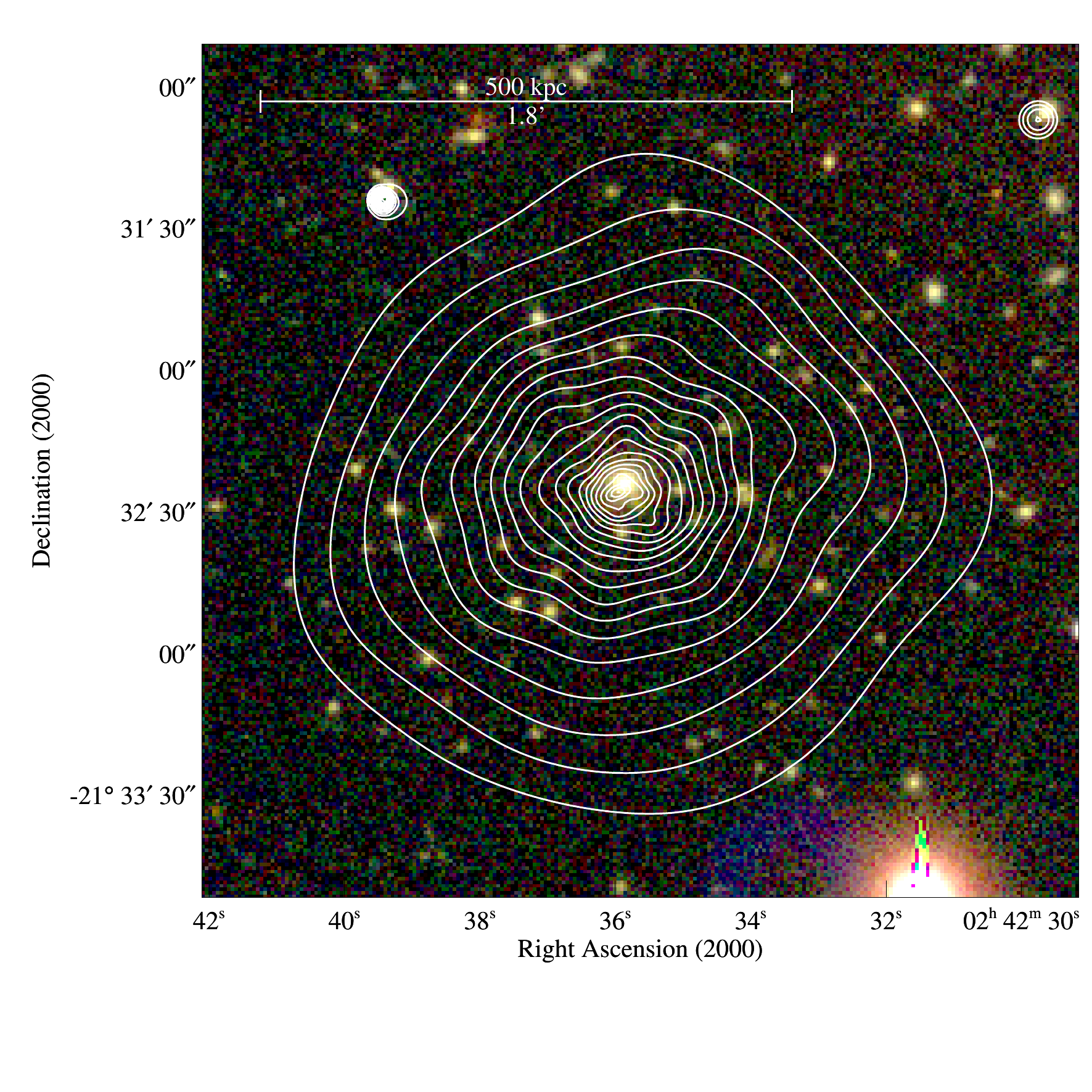}
\includegraphics[width=\overlaysize]{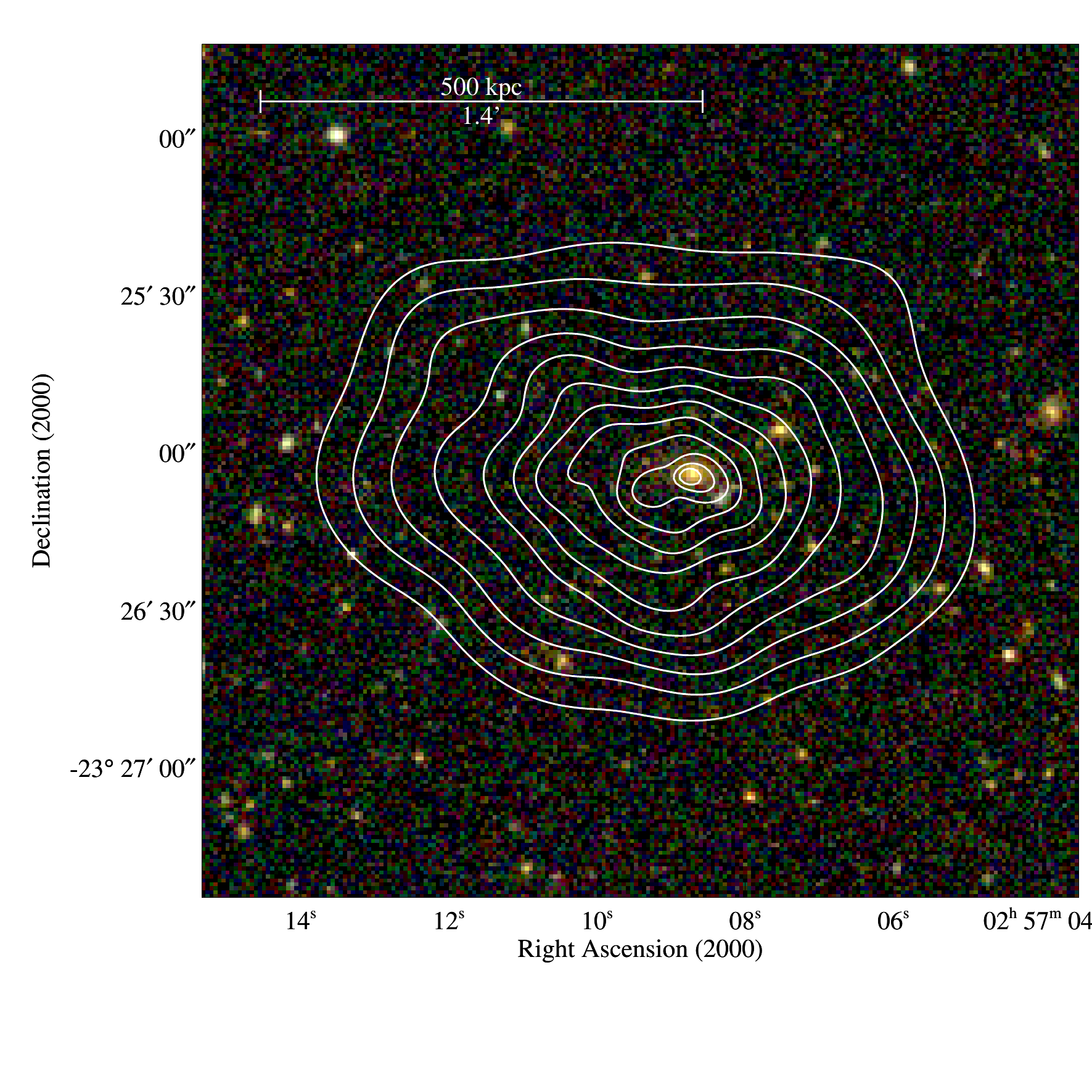}
\includegraphics[width=\overlaysize]{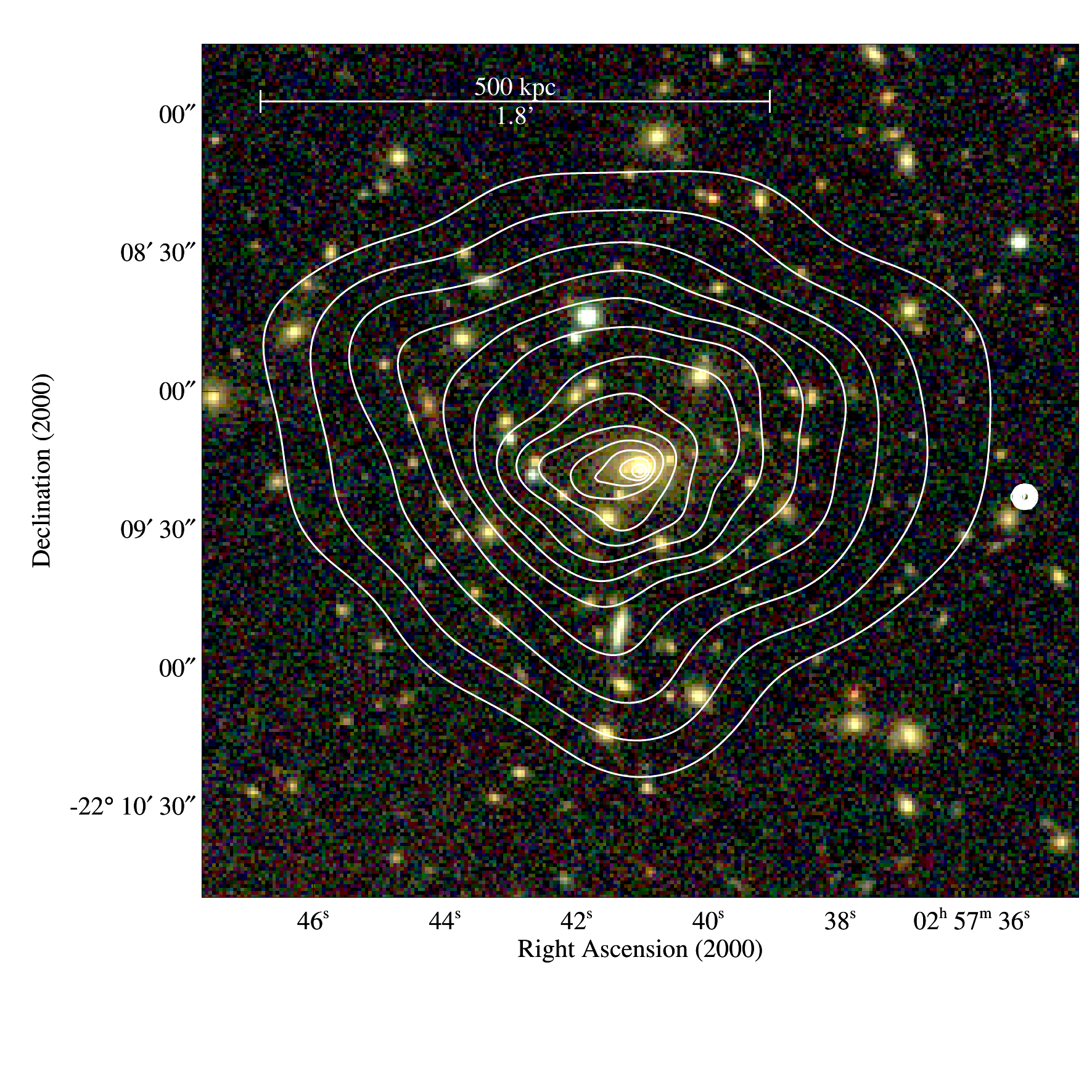}
   \caption{(Continued over next 6 pages).  Same as Fig.~\ref{fig:complexmerg} for the all clusters in our sample not shown in Fig.~\ref{fig:complexmerg}, \ref{fig:bhomstier1}, or \ref{fig:bhomstier2}.  Shown are (in R.A.\ order) {\it MACS\,0011.7--1523}, {\it CL\,0016+1609}, {\it MACS\,J0035.4--2015}, {\it ZwCl\,0104+0048}, {\it MACS\,J0111.5+0855}, {\it A\,209}, {\it RBS\,436}, {\it MACS\,J0159.0--3412}, {\it MACS\,J0159.8--0849}, {\it MACS\,J0242.6--2132}, {\it MACS\,J0257.1--2325}, and {\it A\,402}.  High resolution versions of each overlay is available at http://ifa.hawaii.edu/~amann/MNRAS2011highres/}
   \label{fig:appendix}
\end{figure*}

\setcounter{figure}{0}

\begin{figure*} 
   \centering
\includegraphics[width=\overlaysize]{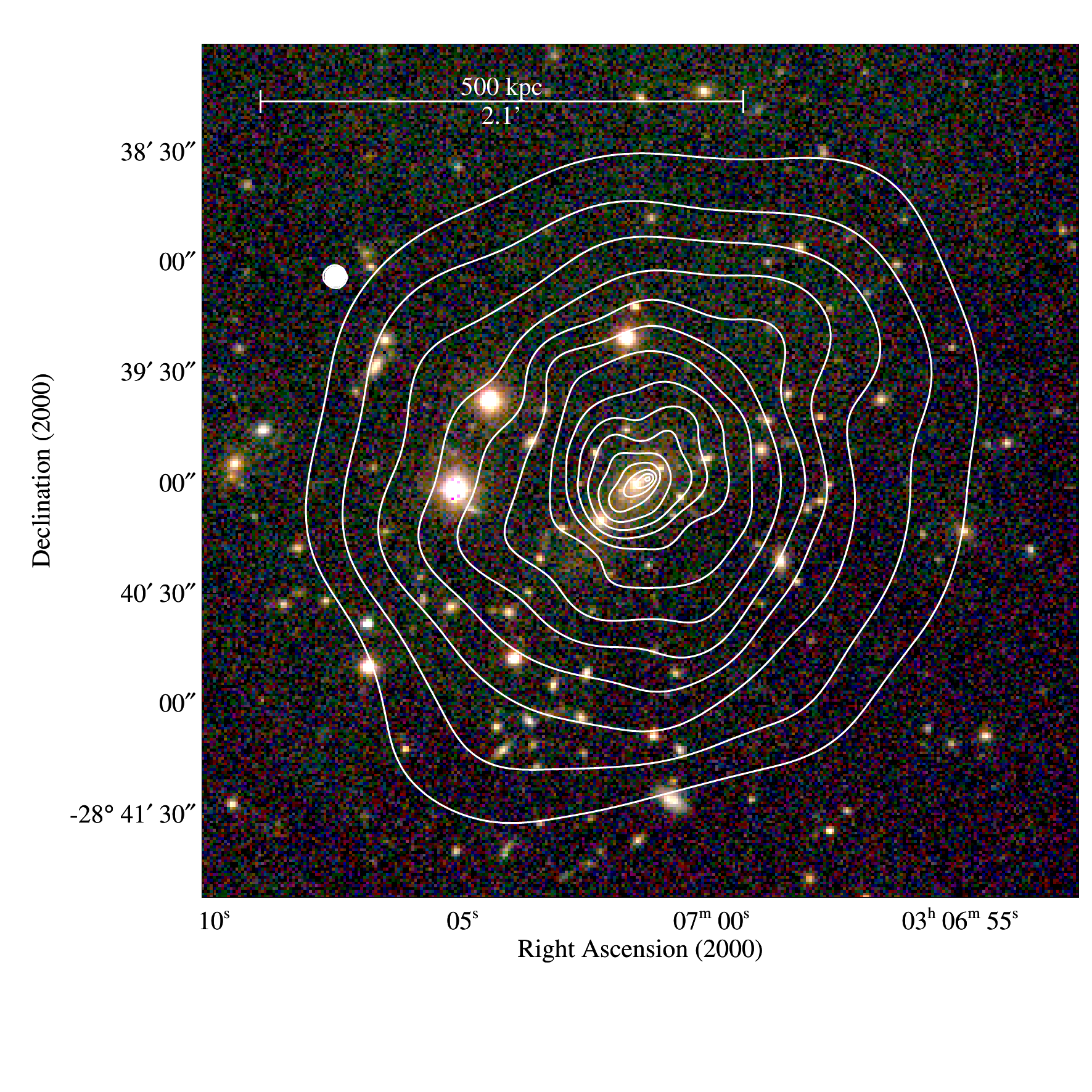}
\includegraphics[width=\overlaysize]{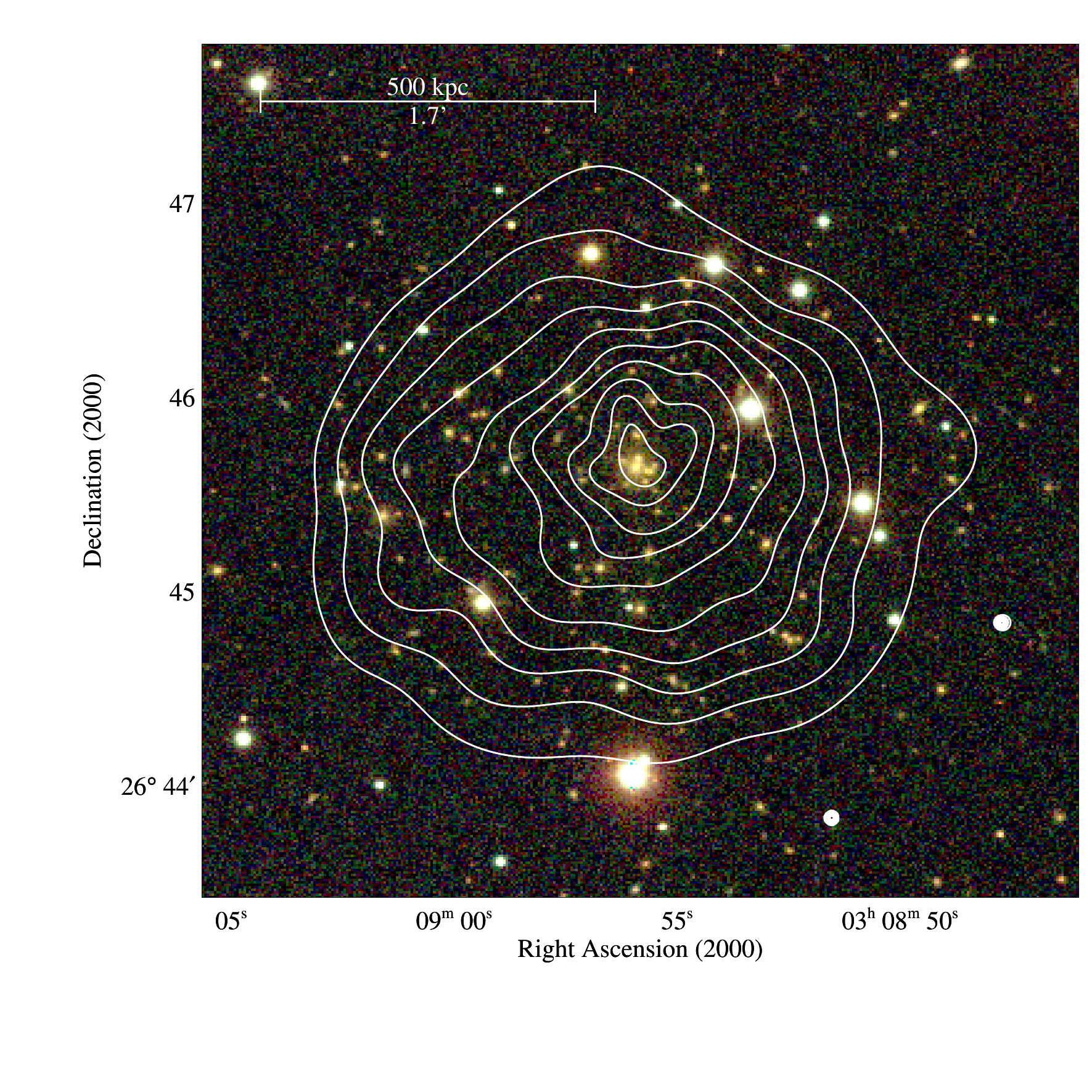}
\includegraphics[width=\overlaysize]{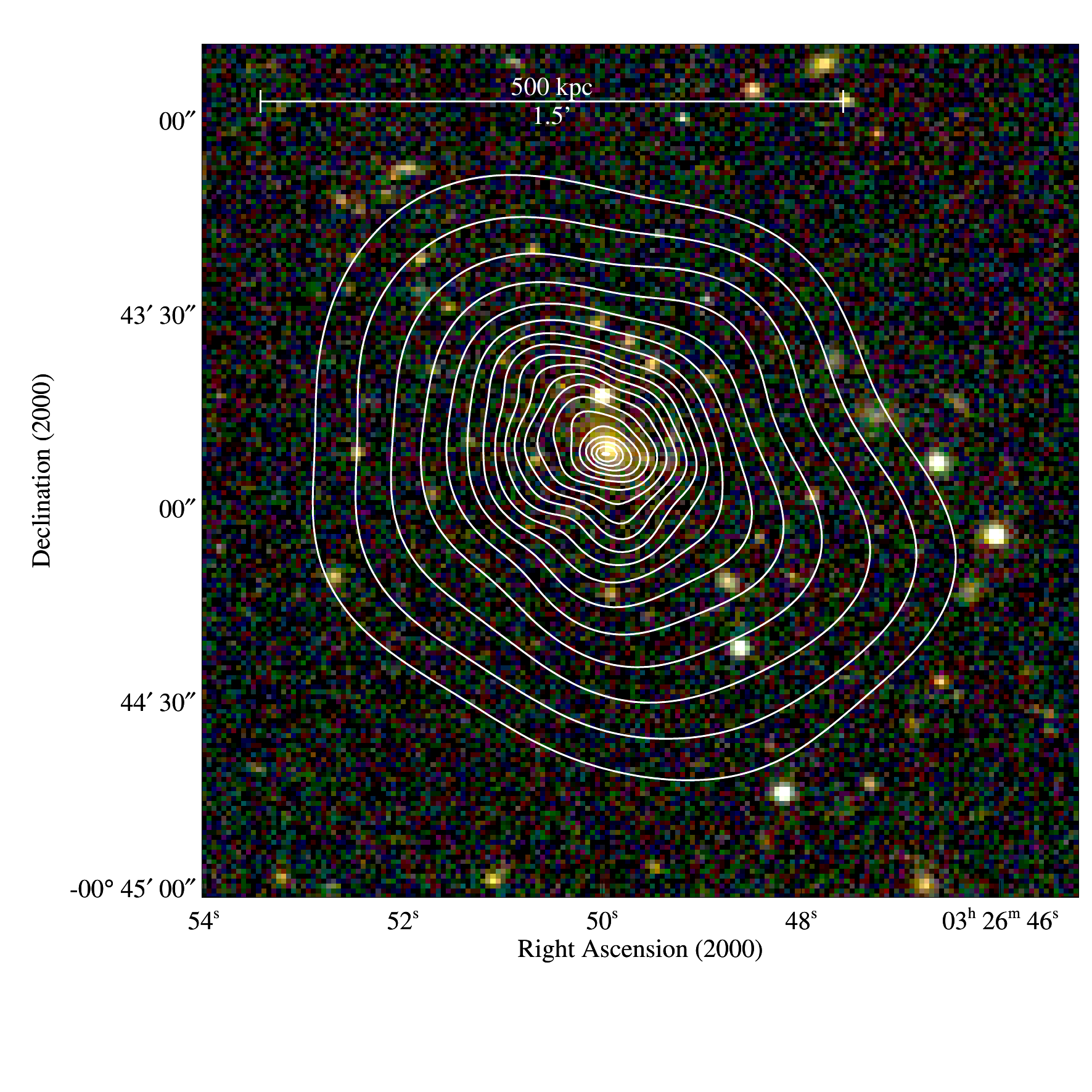}
\includegraphics[width=\overlaysize]{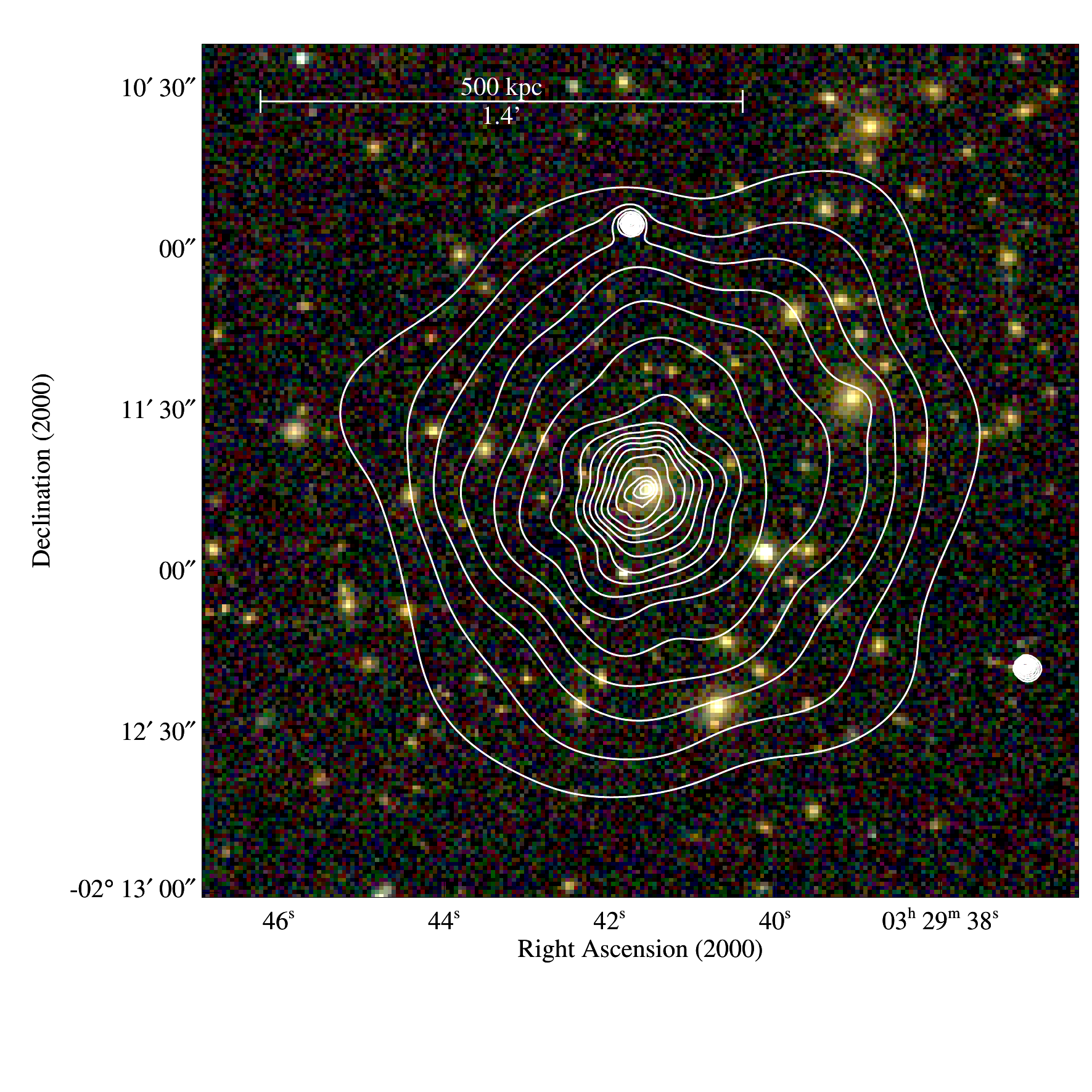}
\includegraphics[width=\overlaysize]{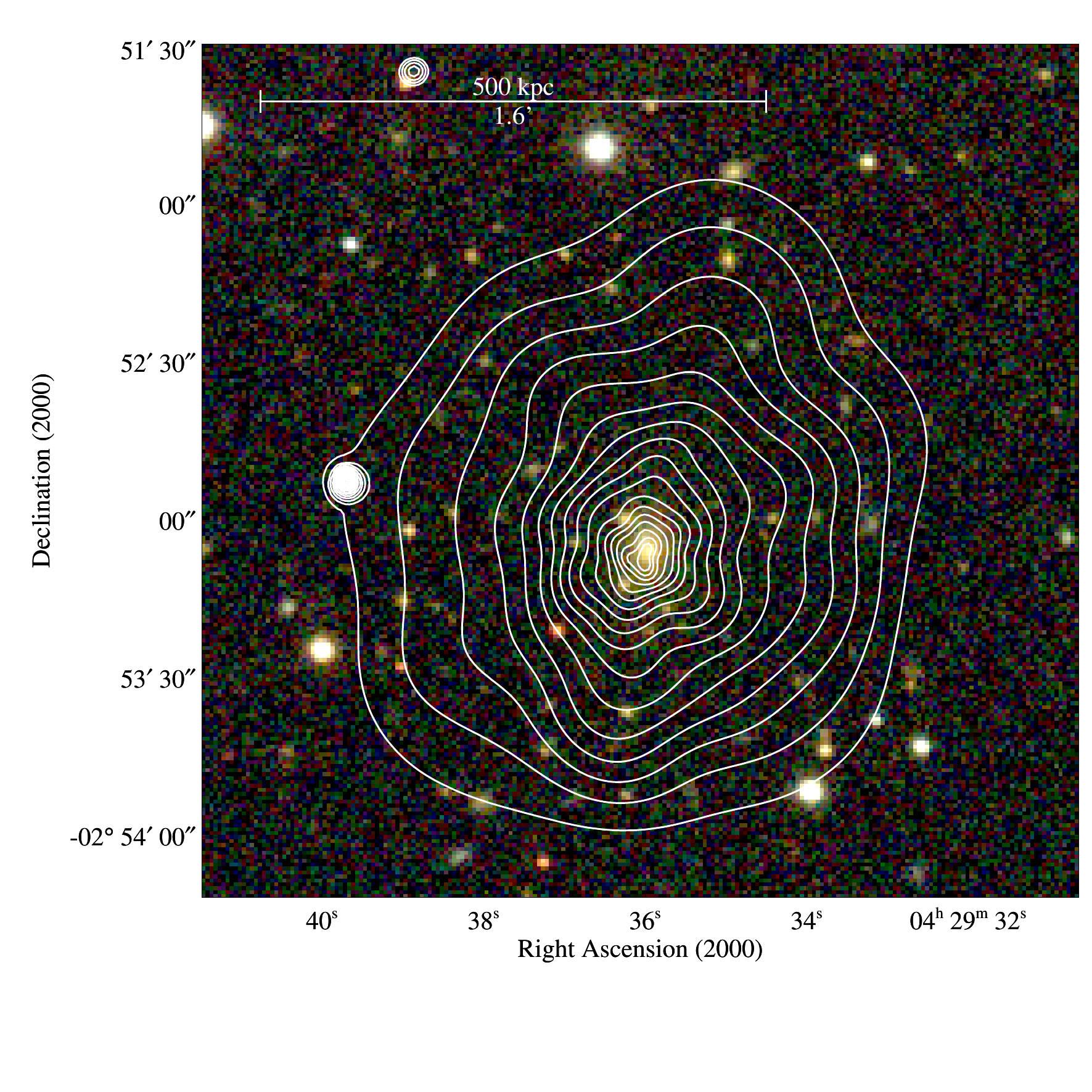}
\includegraphics[width=\overlaysize]{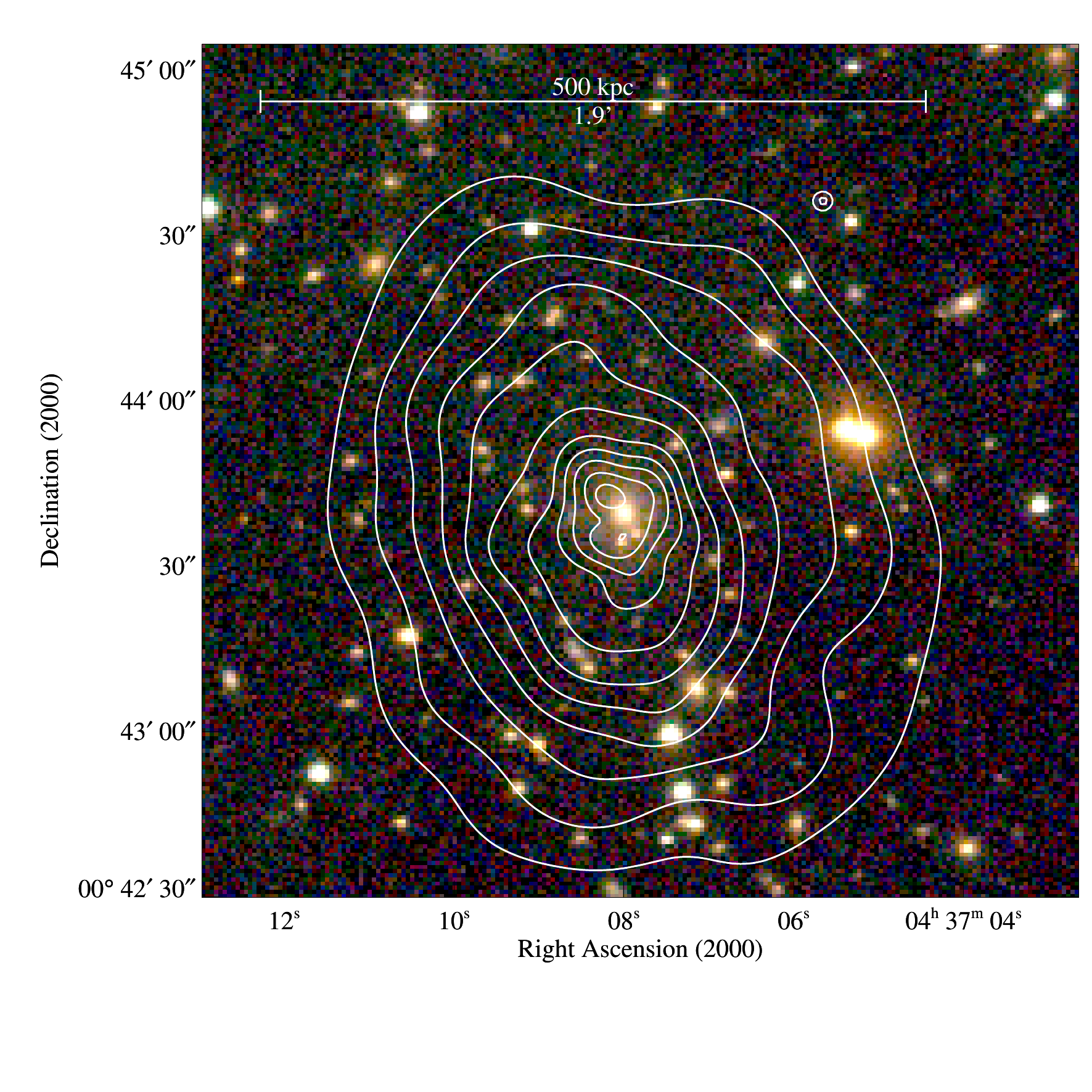}
\includegraphics[width=\overlaysize]{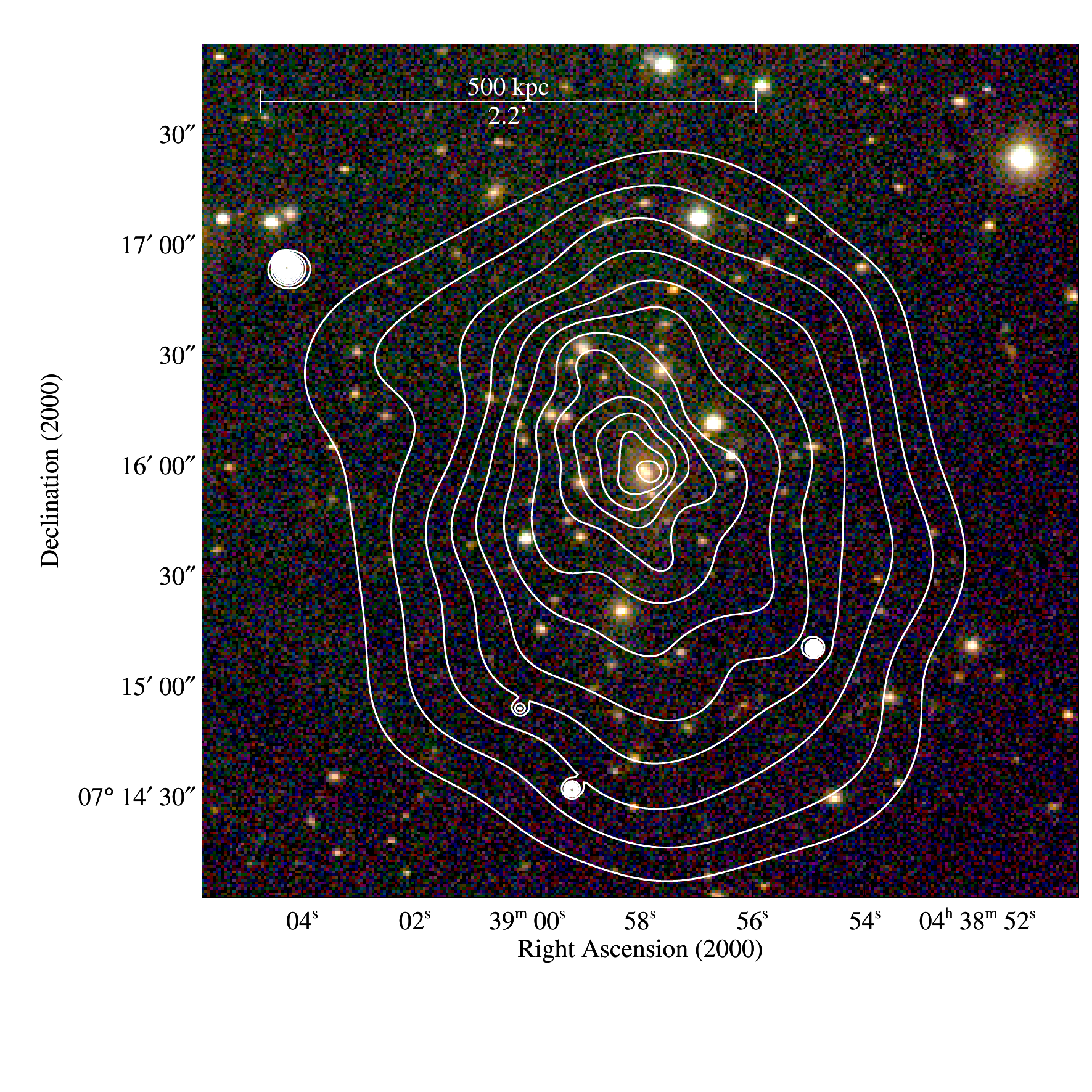}
\includegraphics[width=\overlaysize]{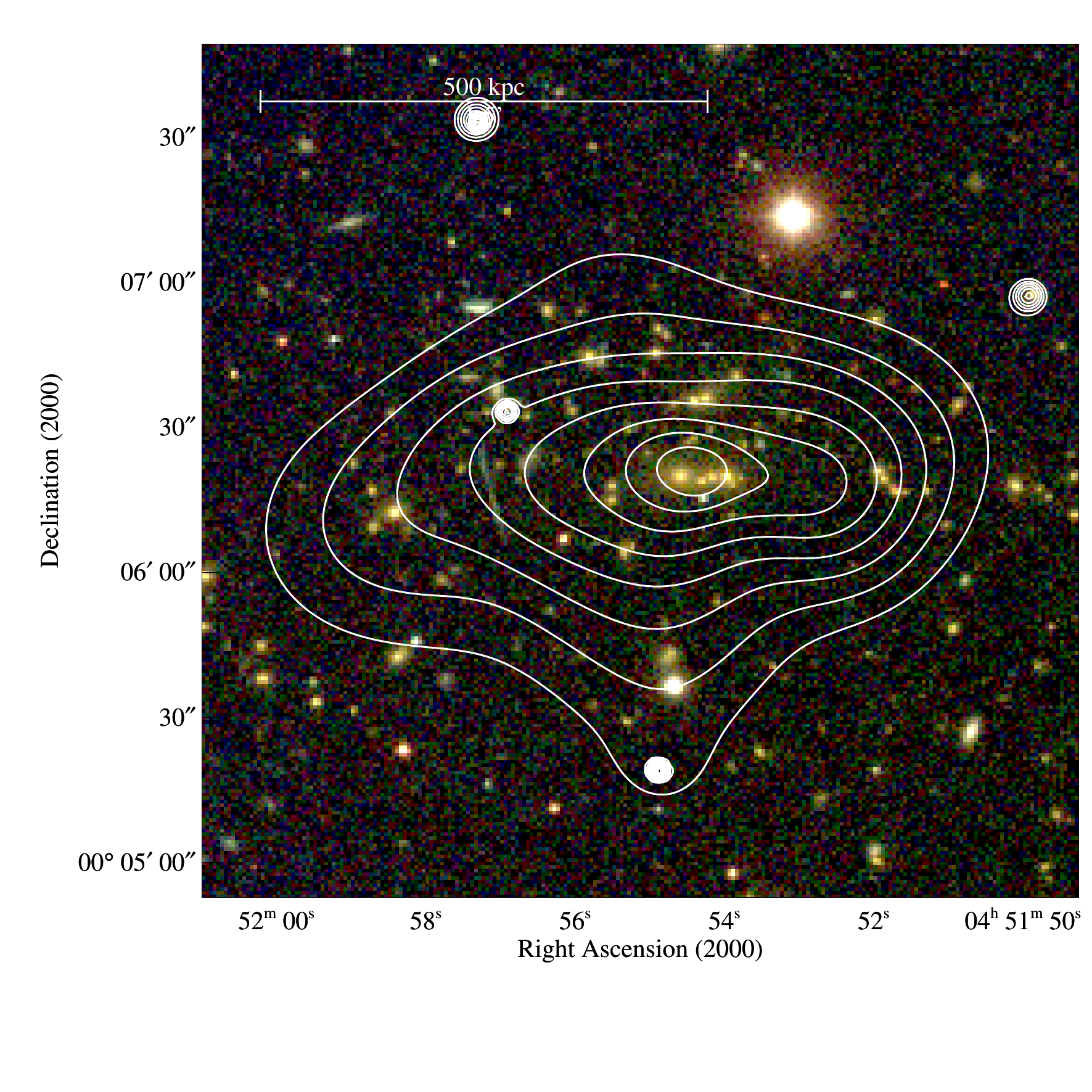}
\includegraphics[width=\overlaysize]{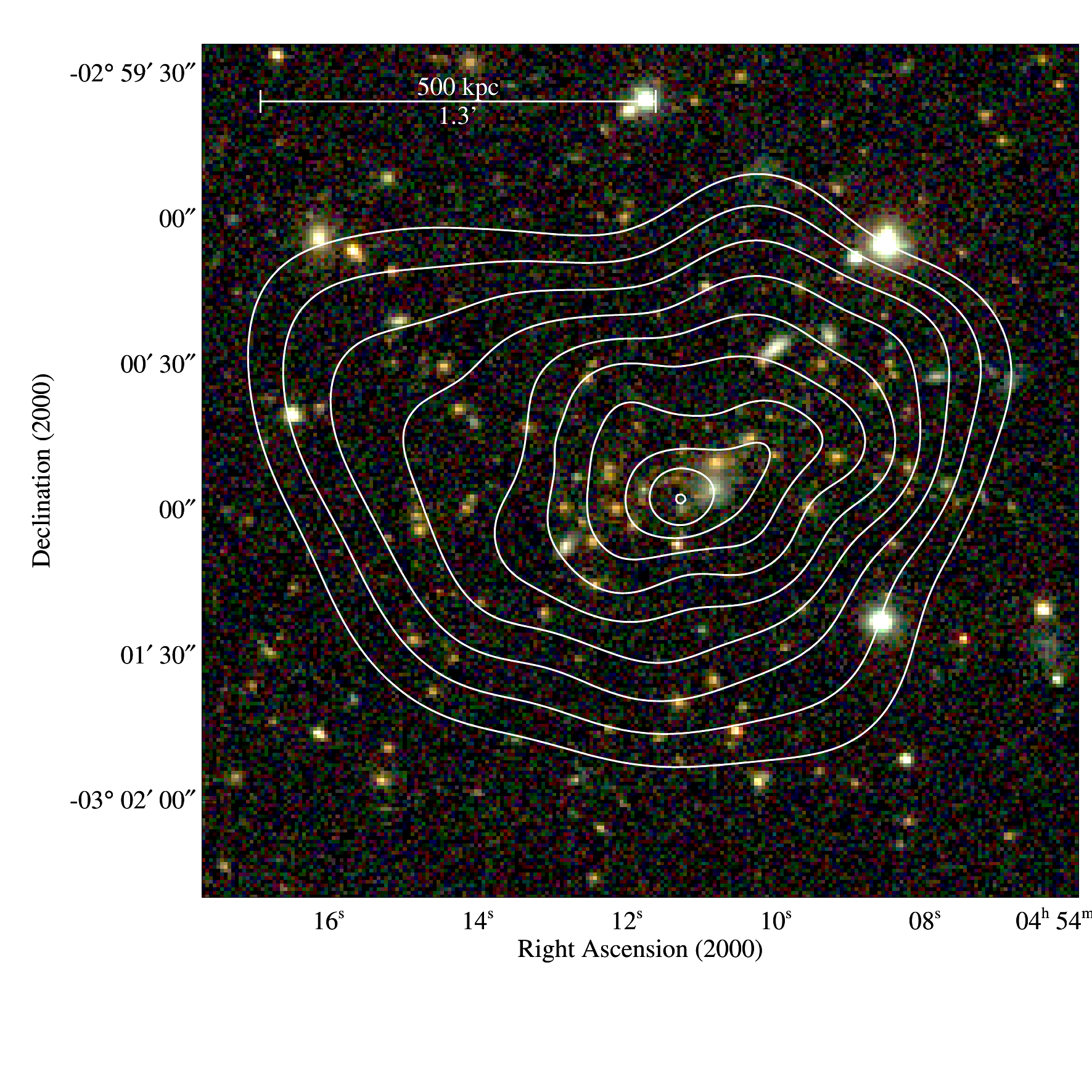}
\includegraphics[width=\overlaysize]{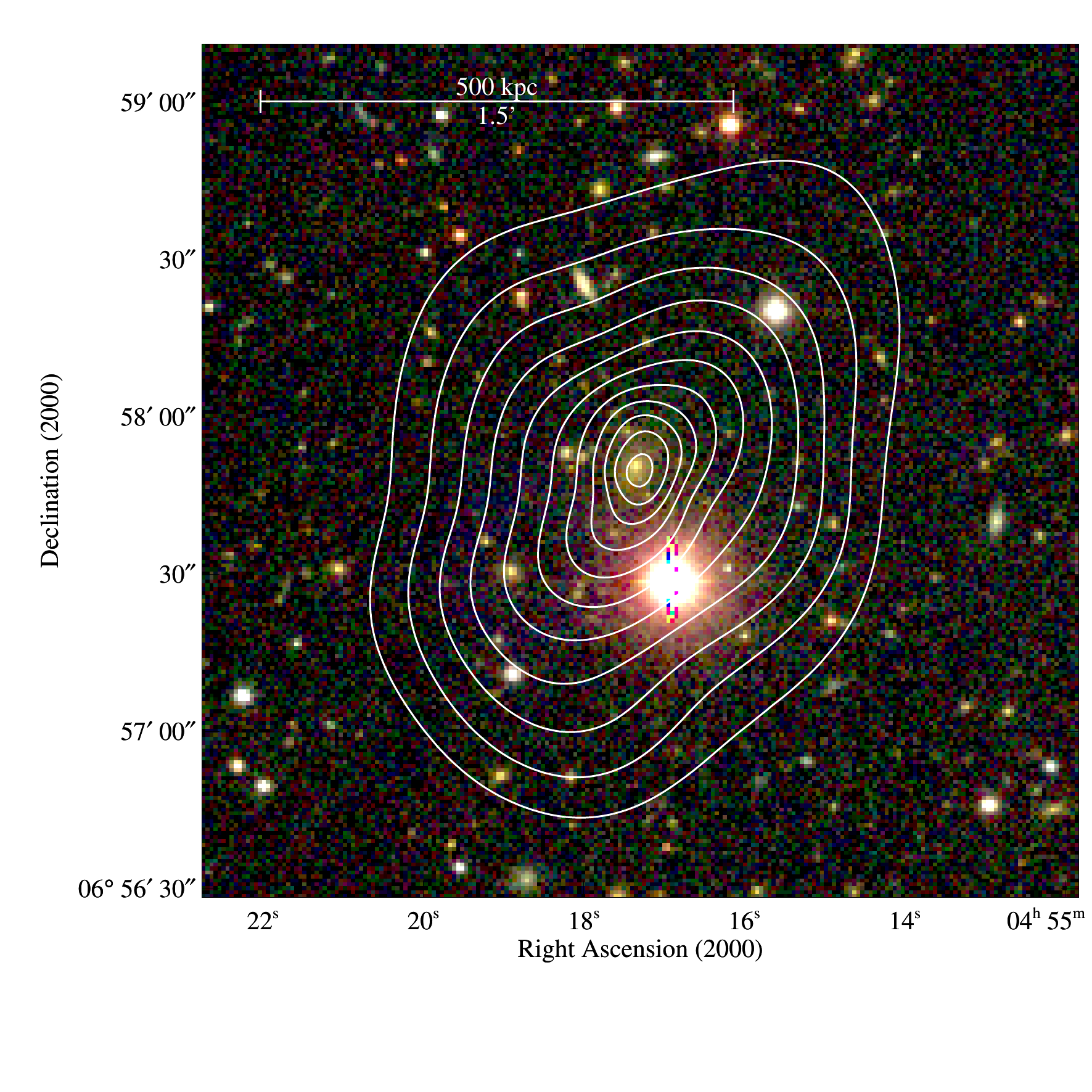}
\includegraphics[width=\overlaysize]{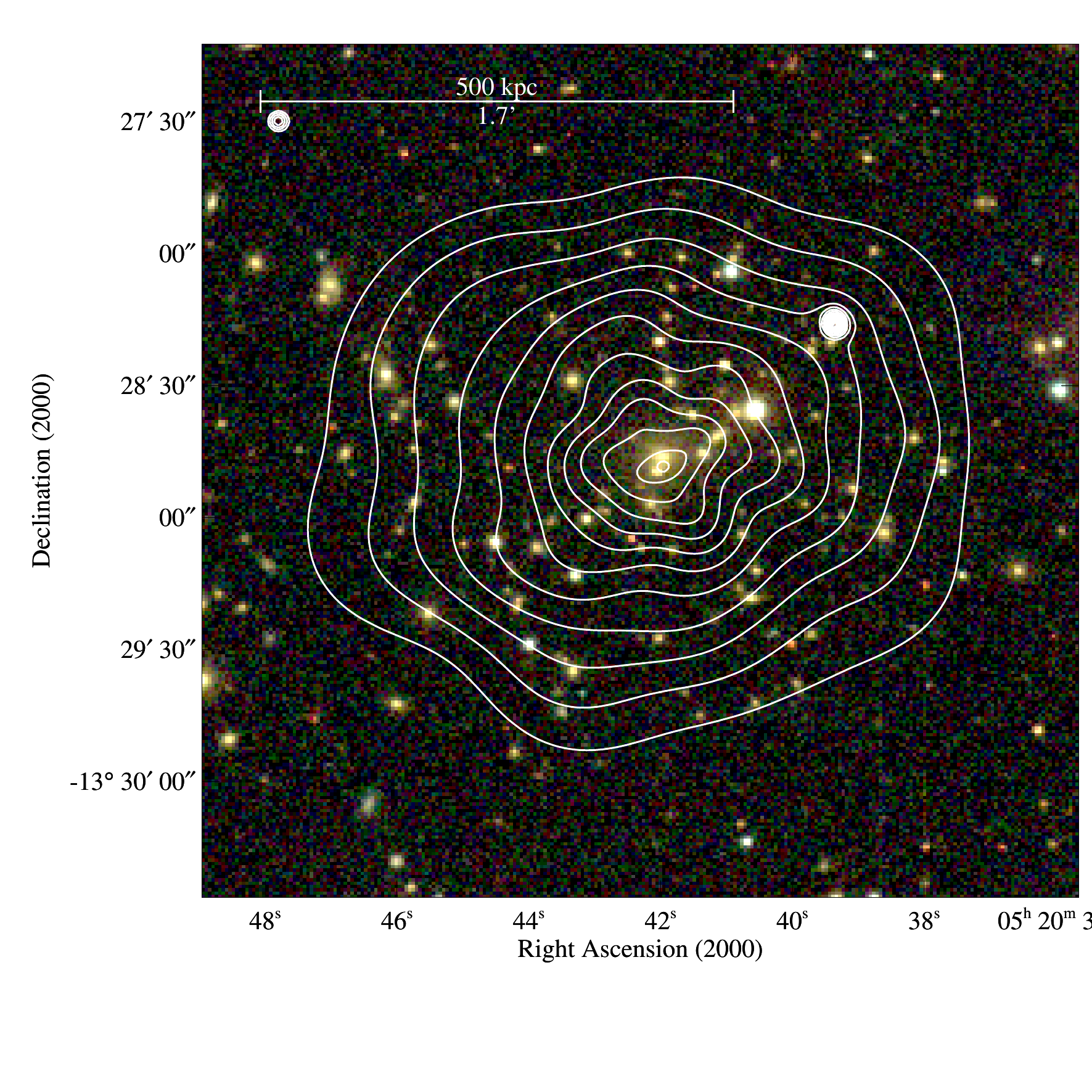}
\includegraphics[width=\overlaysize]{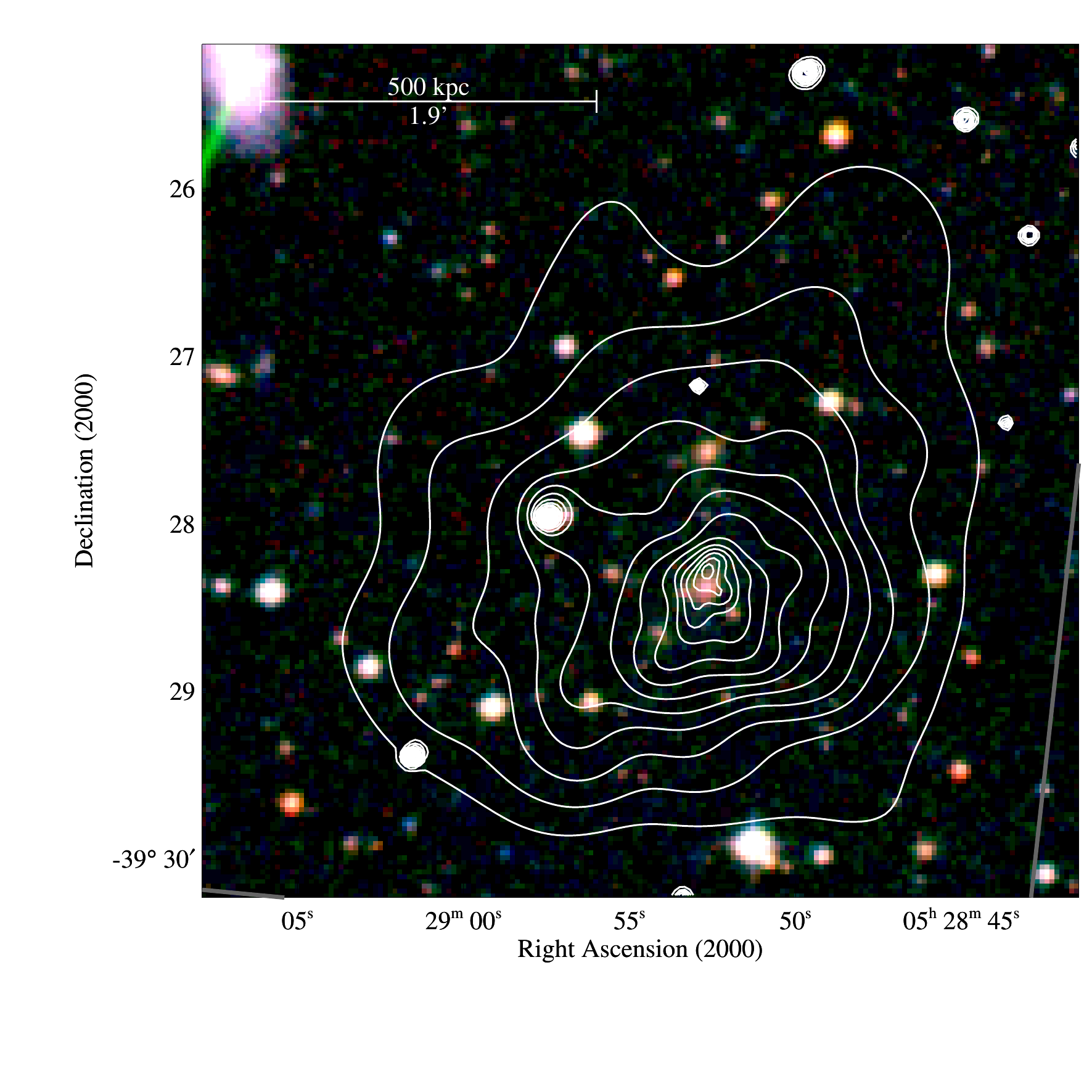}
   \caption{\emph{continued} Shown are (in R.A.\ order) {\it A\,3088}, {\it MACS\,J0308.9+2645}, {\it MACS\,J0326.8--0043}, {\it MACS\,J0329.6--0211}, {\it MACS\,J0429.6--0253}, {\it RX\,J0437.1+0043}, {\it RX\,J0439.0+0715}, {\it MACS\,J0451.9+0006}, {\it MS\,0451.6-0305}, {\it MACS\,J0455.2+0657}, and {\it MACS\,J0520.7--1328}, and {\it A\,907}.}
   \label{fig:appendix}
\end{figure*}

\setcounter{figure}{0}

\begin{figure*} 
   \centering
\includegraphics[width=\overlaysize]{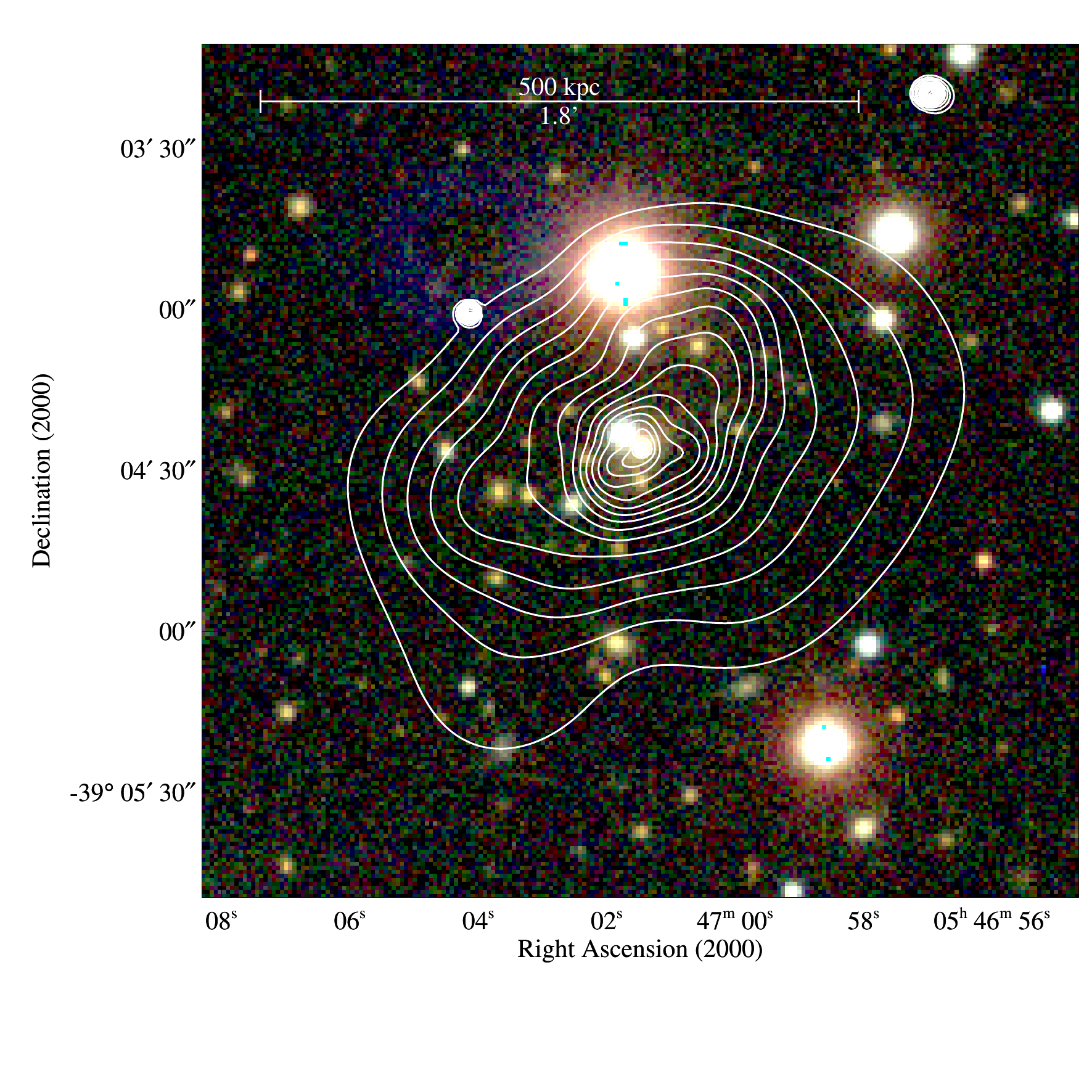}
\includegraphics[width=\overlaysize]{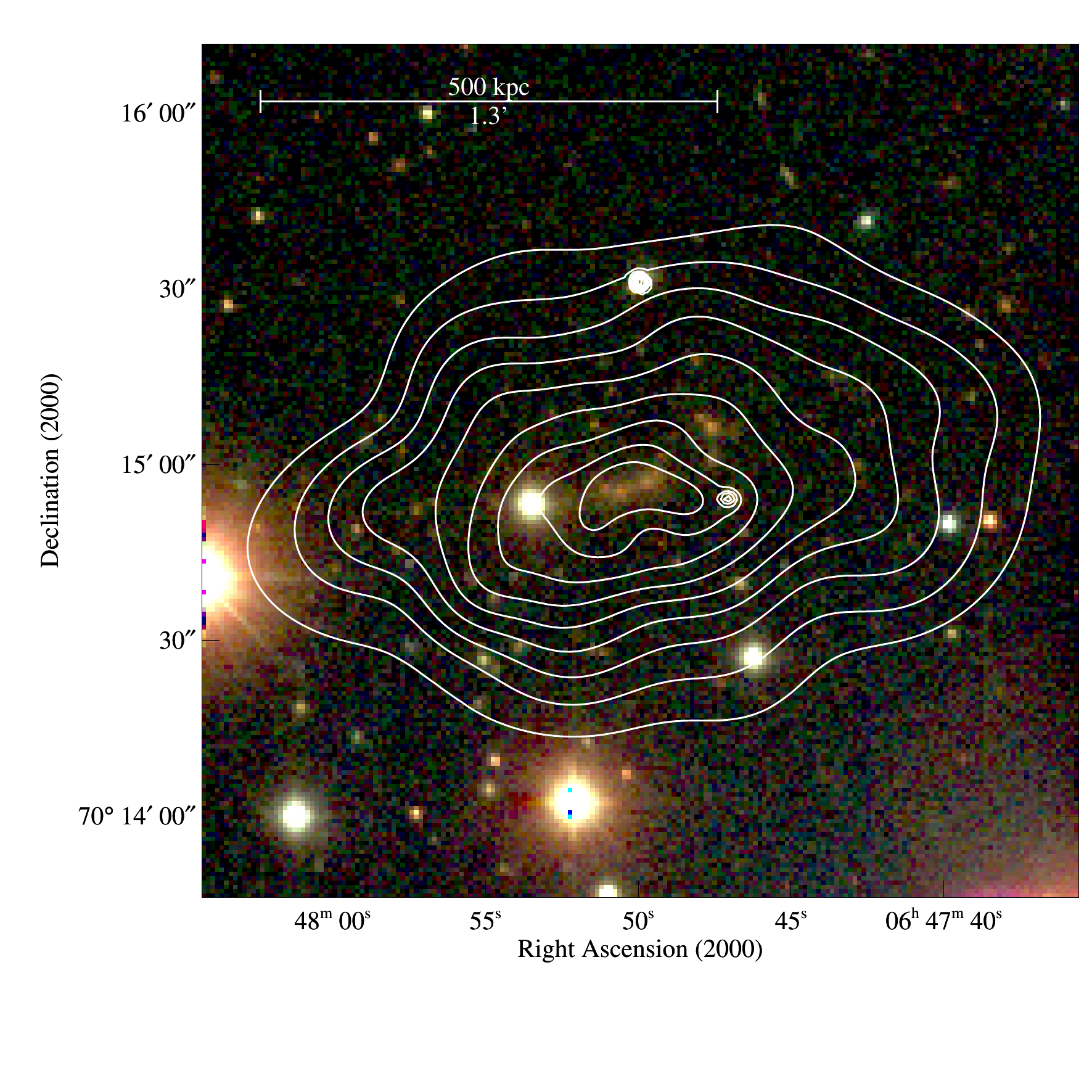}
\includegraphics[width=\overlaysize]{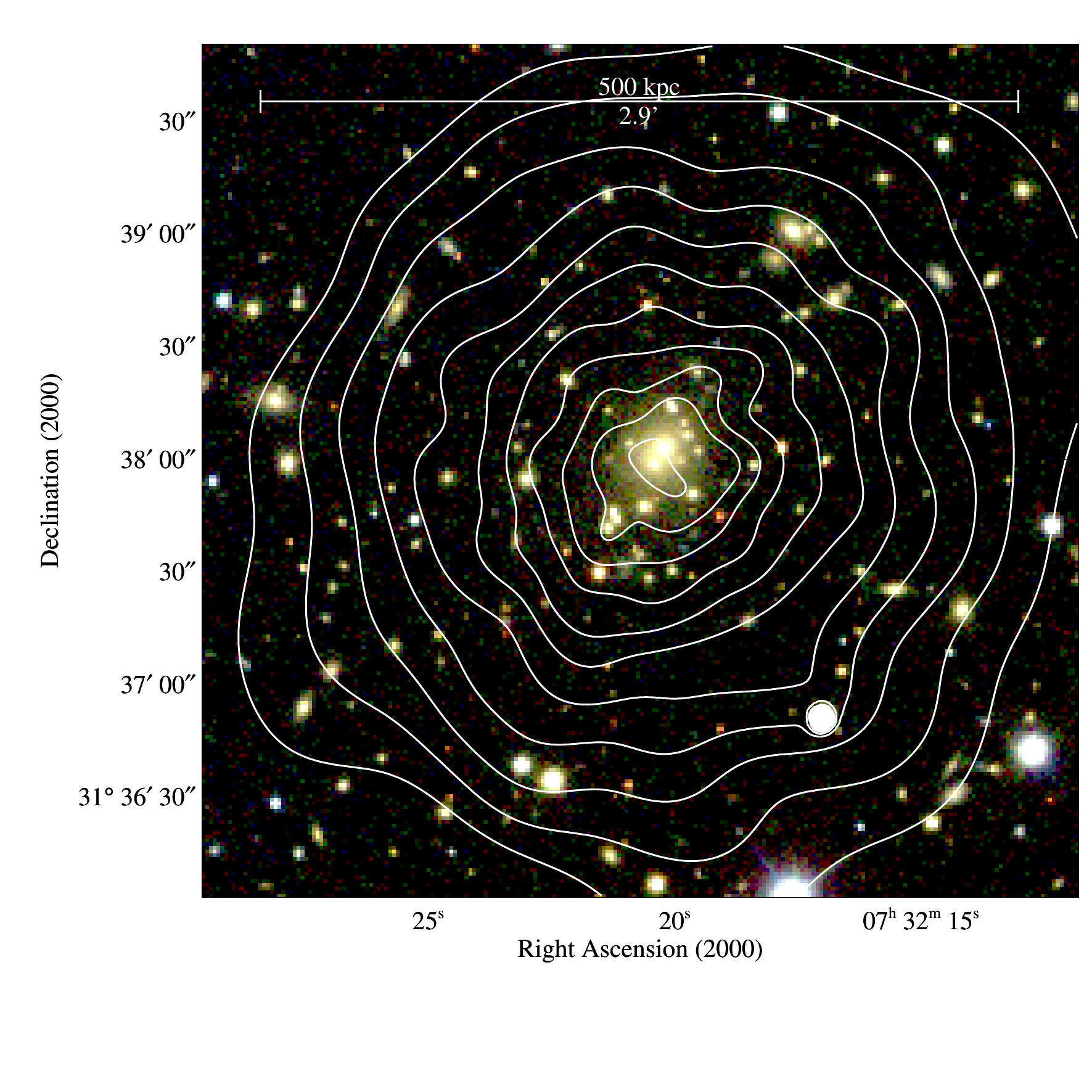}
\includegraphics[width=\overlaysize]{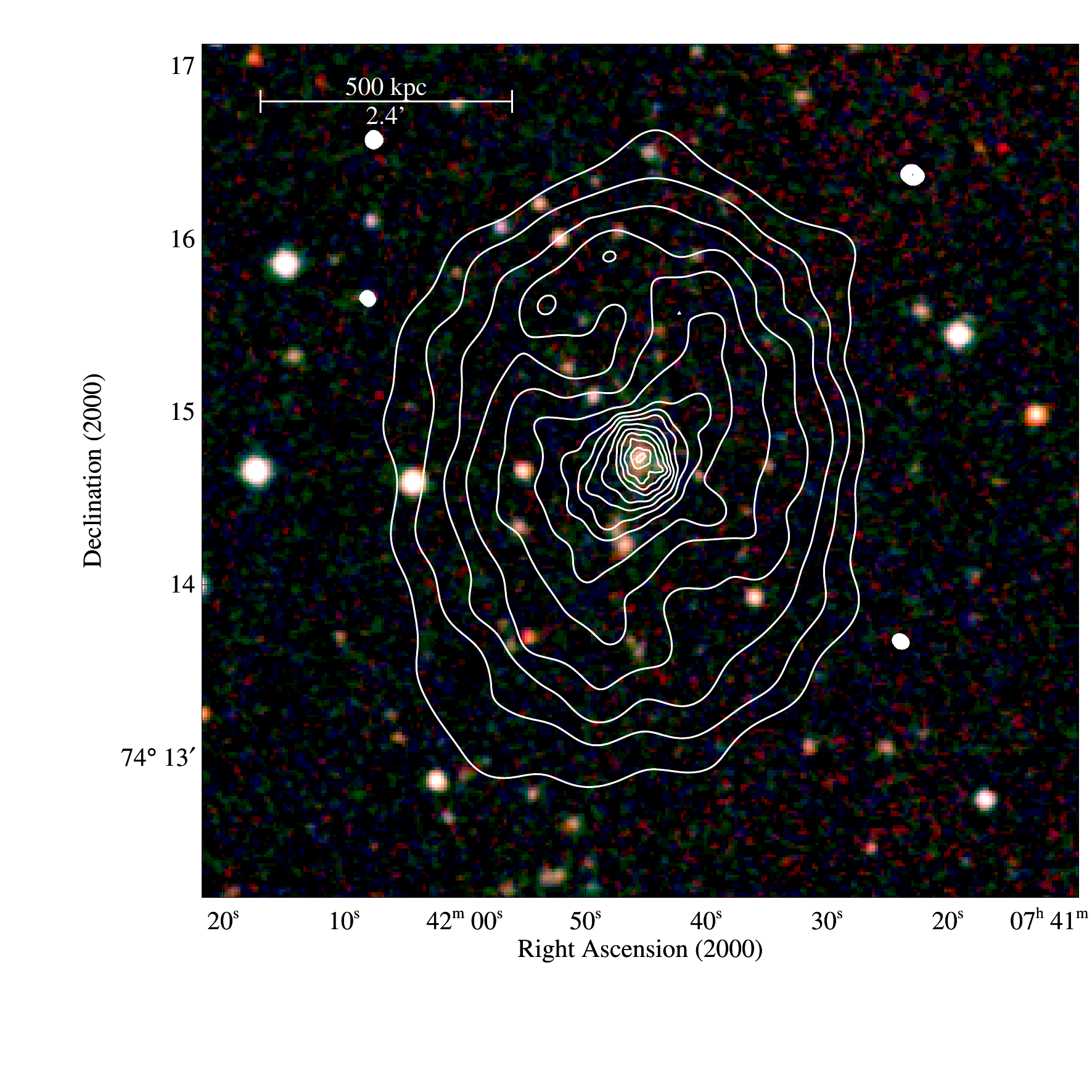}
\includegraphics[width=\overlaysize]{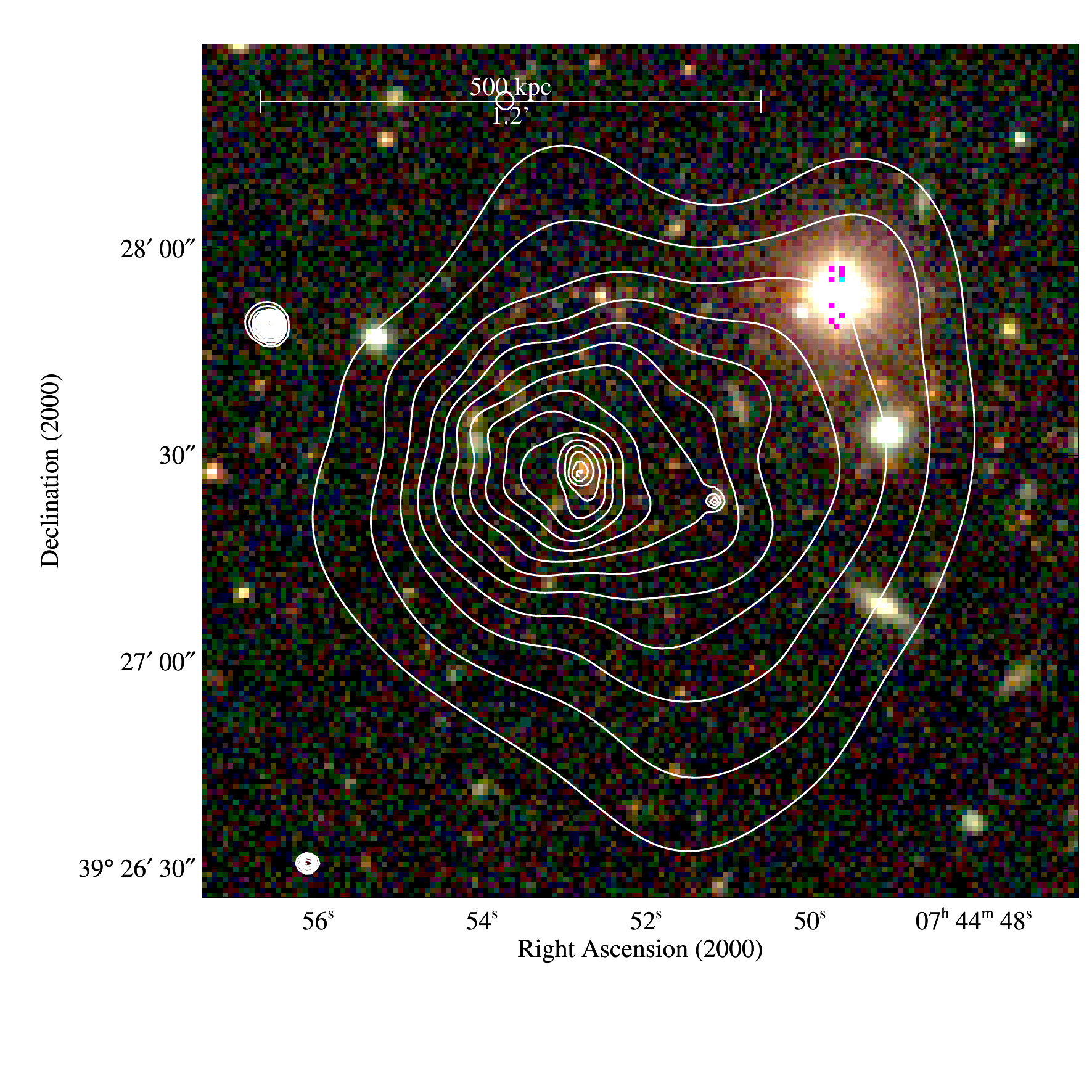}
\includegraphics[width=\overlaysize]{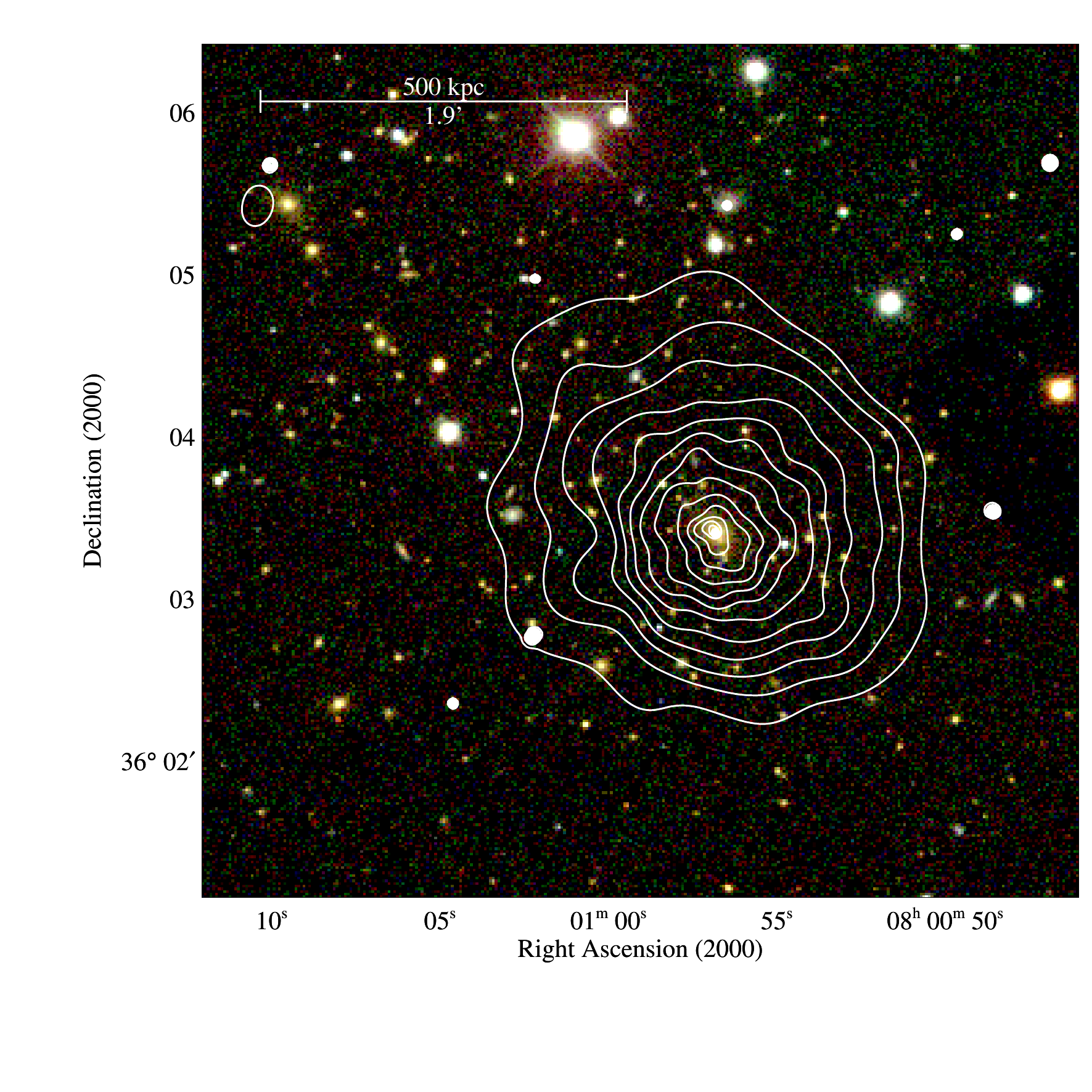}
\includegraphics[width=\overlaysize]{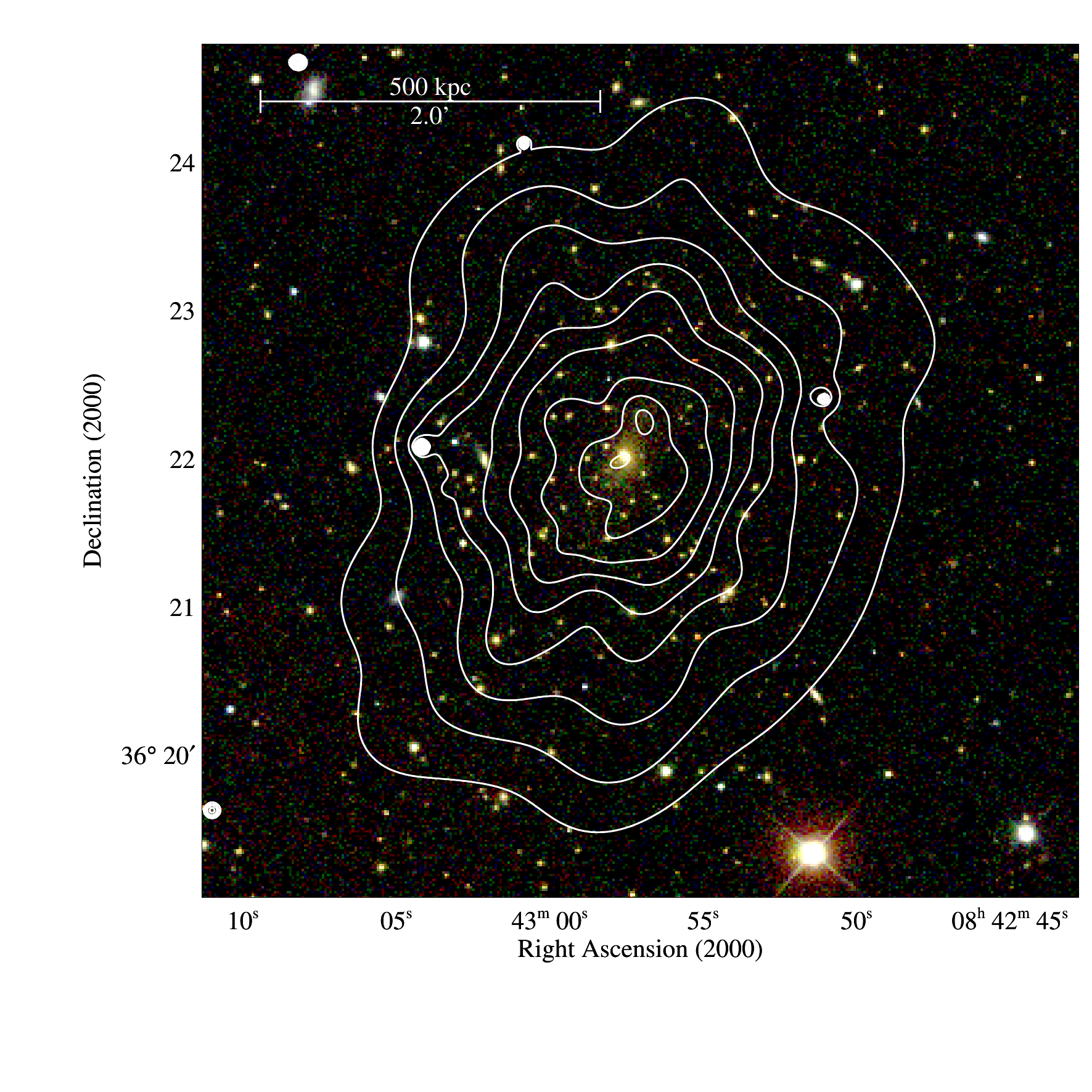}
\includegraphics[width=\overlaysize]{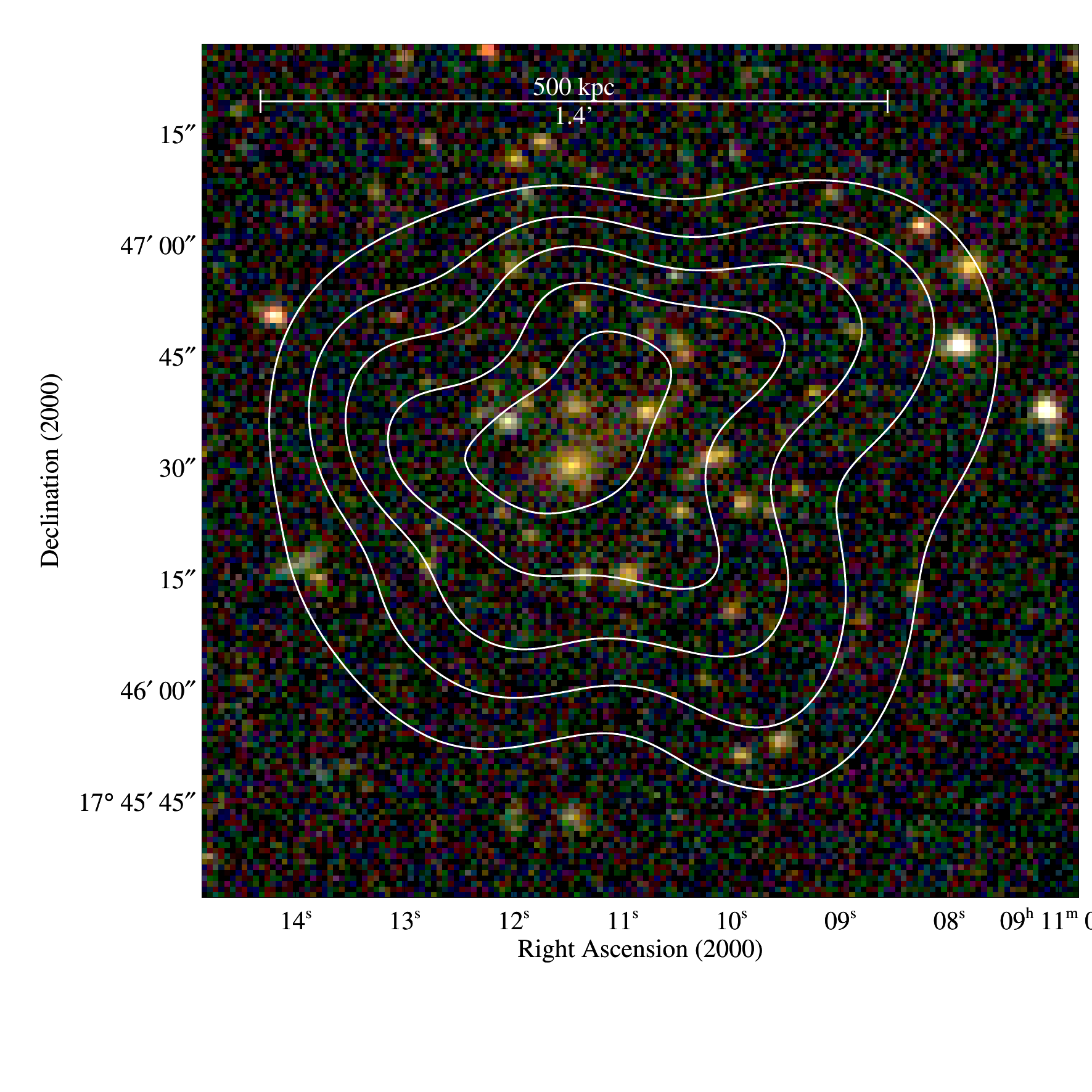}
\includegraphics[width=\overlaysize]{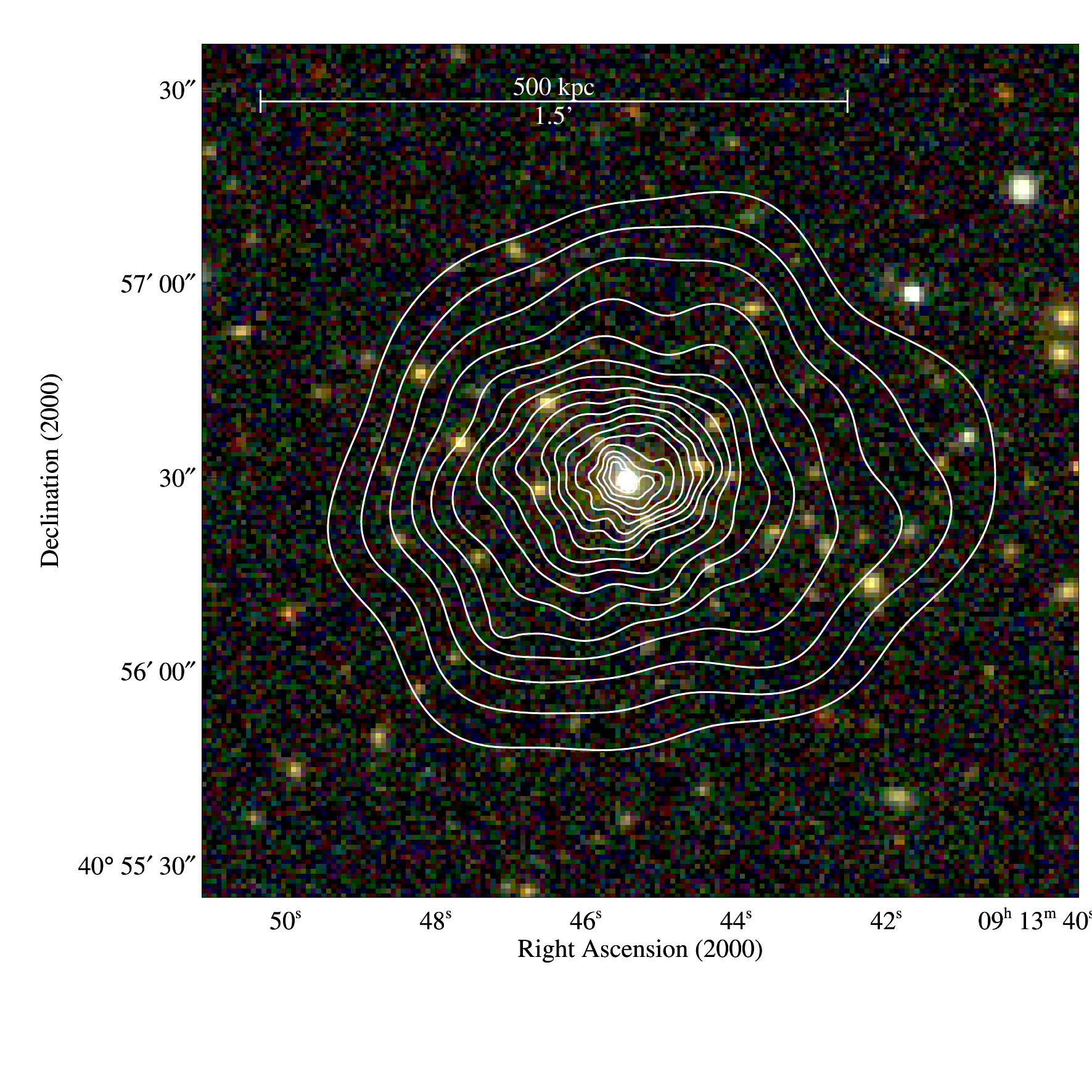}
\includegraphics[width=\overlaysize]{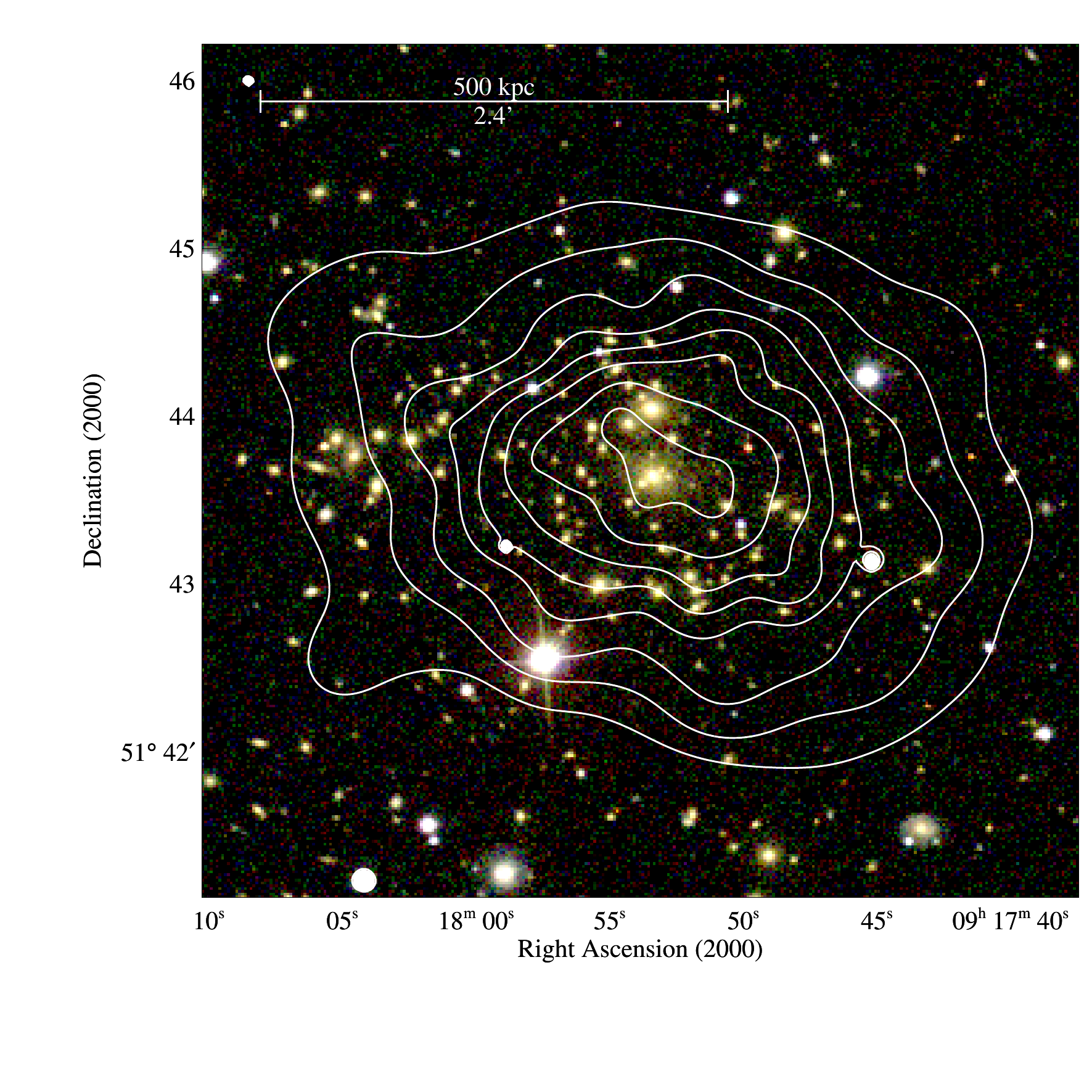}
\includegraphics[width=\overlaysize]{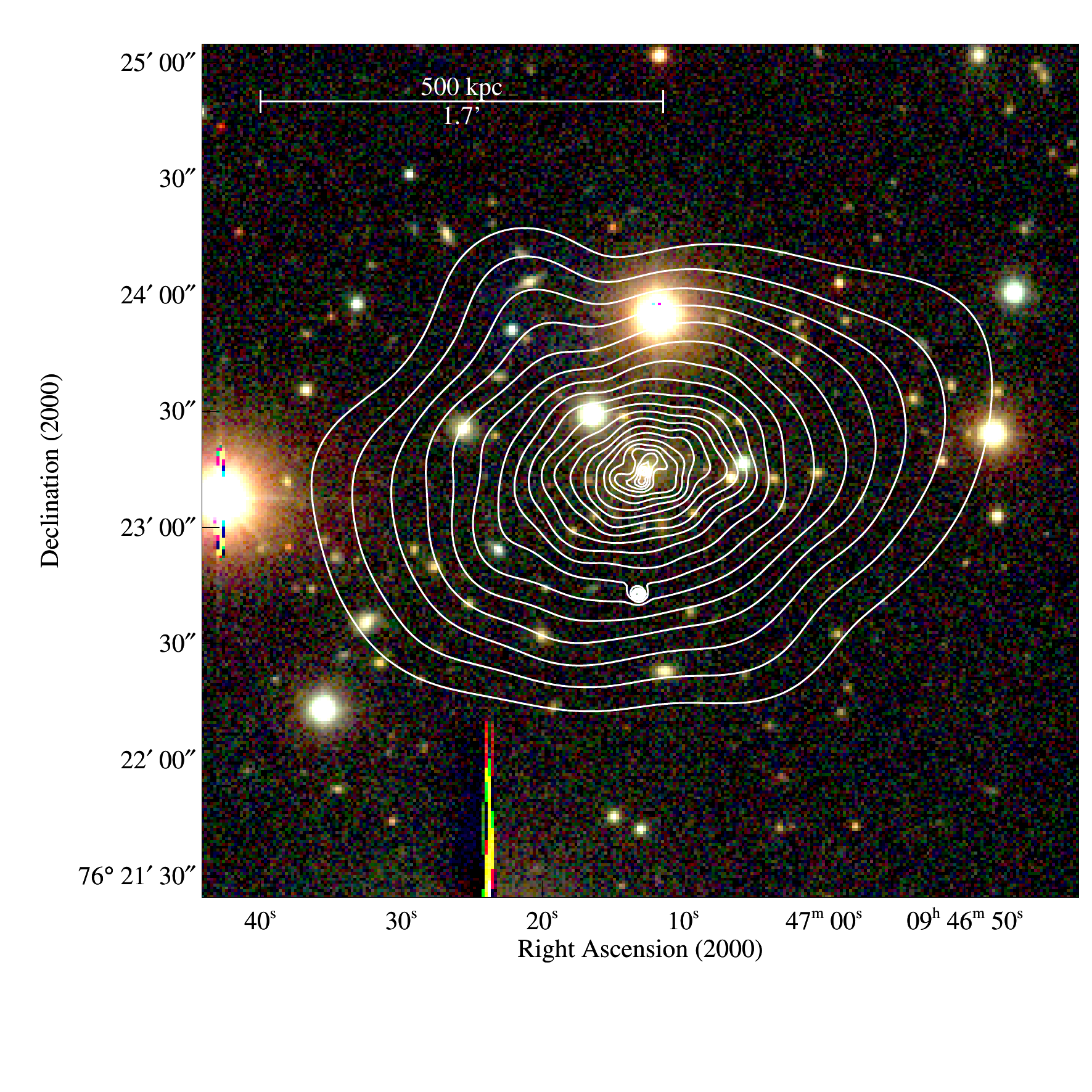}
\includegraphics[width=\overlaysize]{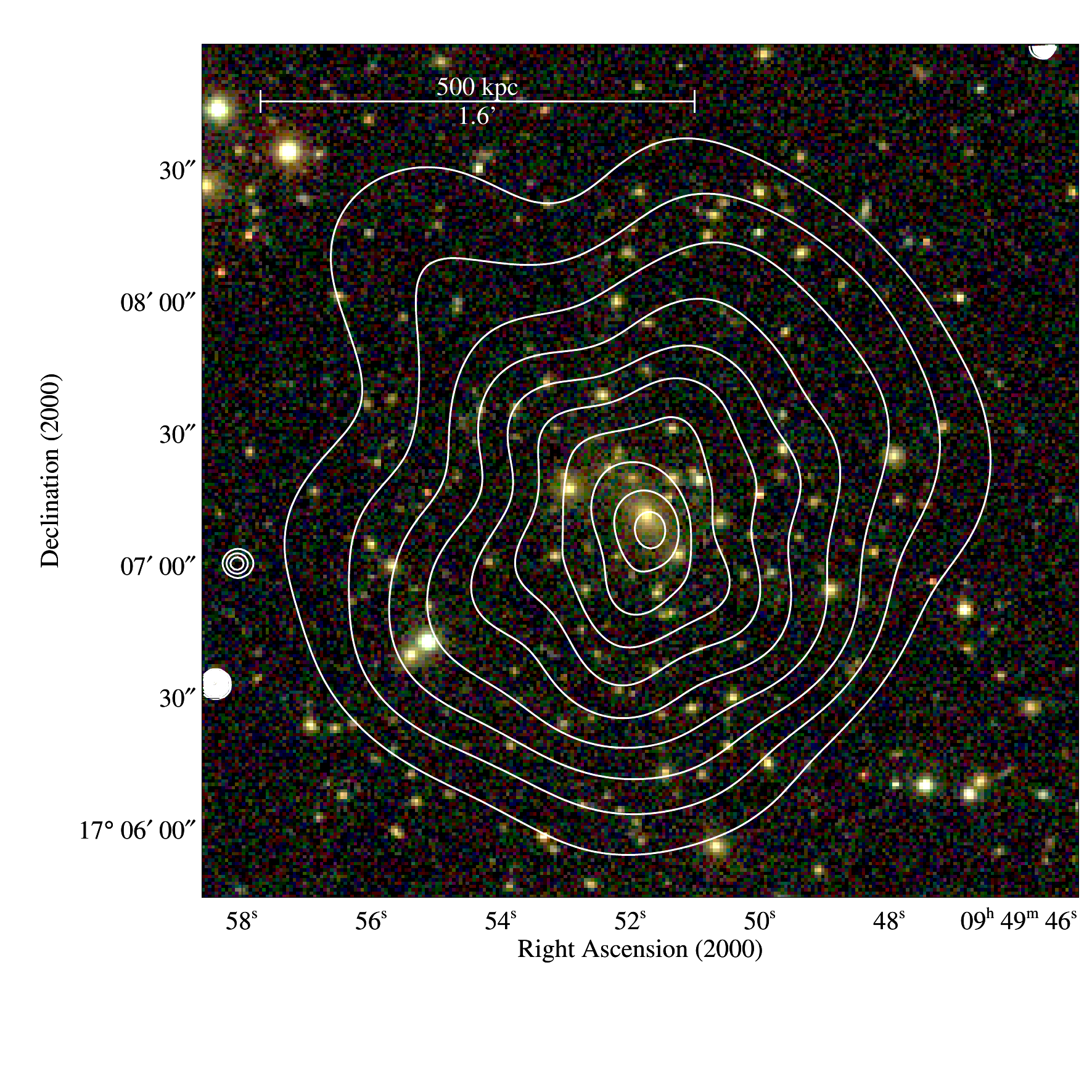}
   \caption{\emph{continued} Shown are (in R.A.\ order) {\it MACS\,J0547.0--3904}, {\it MACS\,J0647.7+7015}, {\it A\,586}, {\it ZwCl\,0735.7+7421}, {\it MACS\,J0744.8+3927}, {\it A\,661}, {\it A\,697}, {\it MACS\,J0911.2+1746}, {\it MACS\,J0913.7+4056}, {\it A\,773}, {\it RBS\,0797}, and {\it ZwCl\,0947.2+1723}.}
   \label{fig:appendix}
\end{figure*}

\setcounter{figure}{0}

\begin{figure*} 
   \centering
\includegraphics[width=\overlaysize]{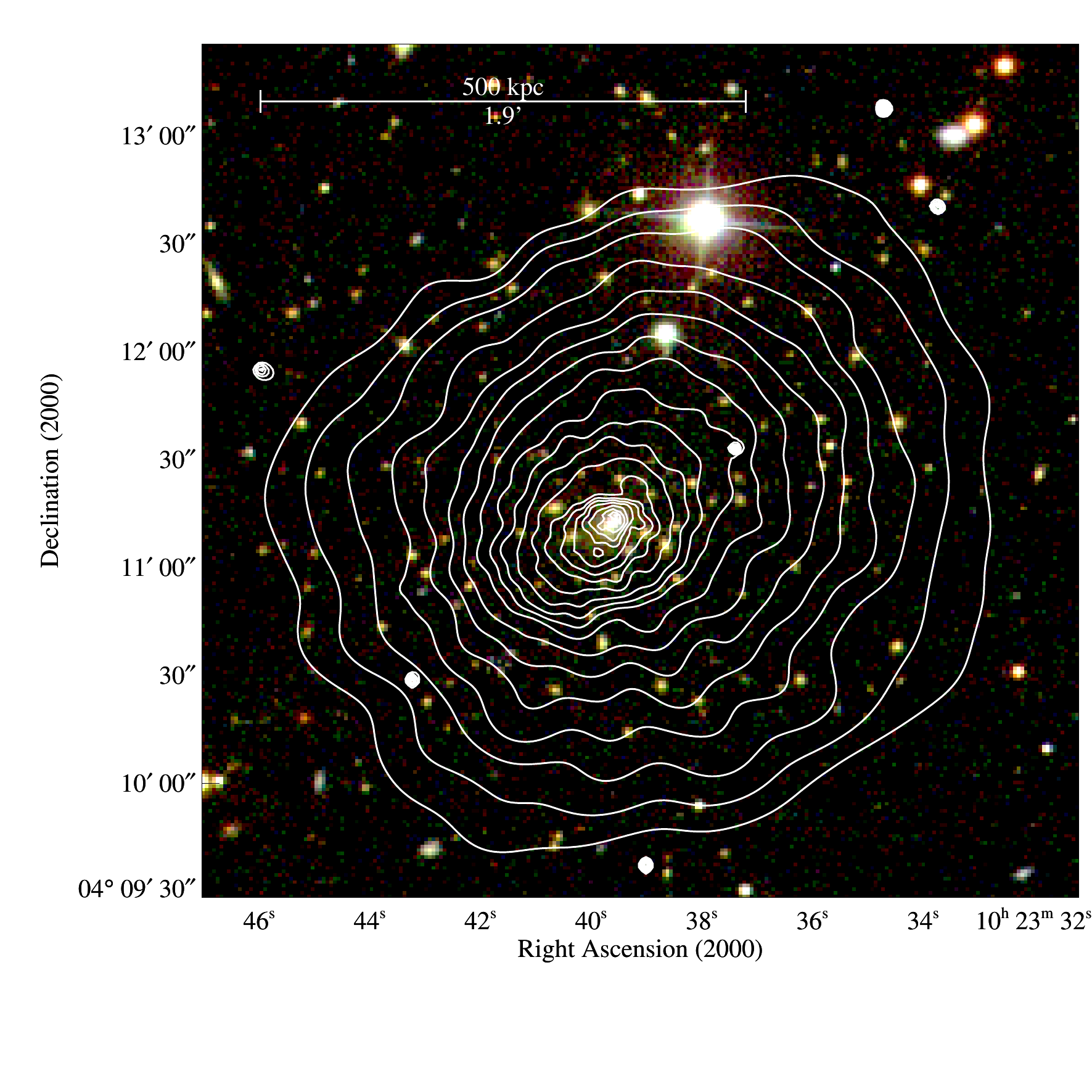}
\includegraphics[width=\overlaysize]{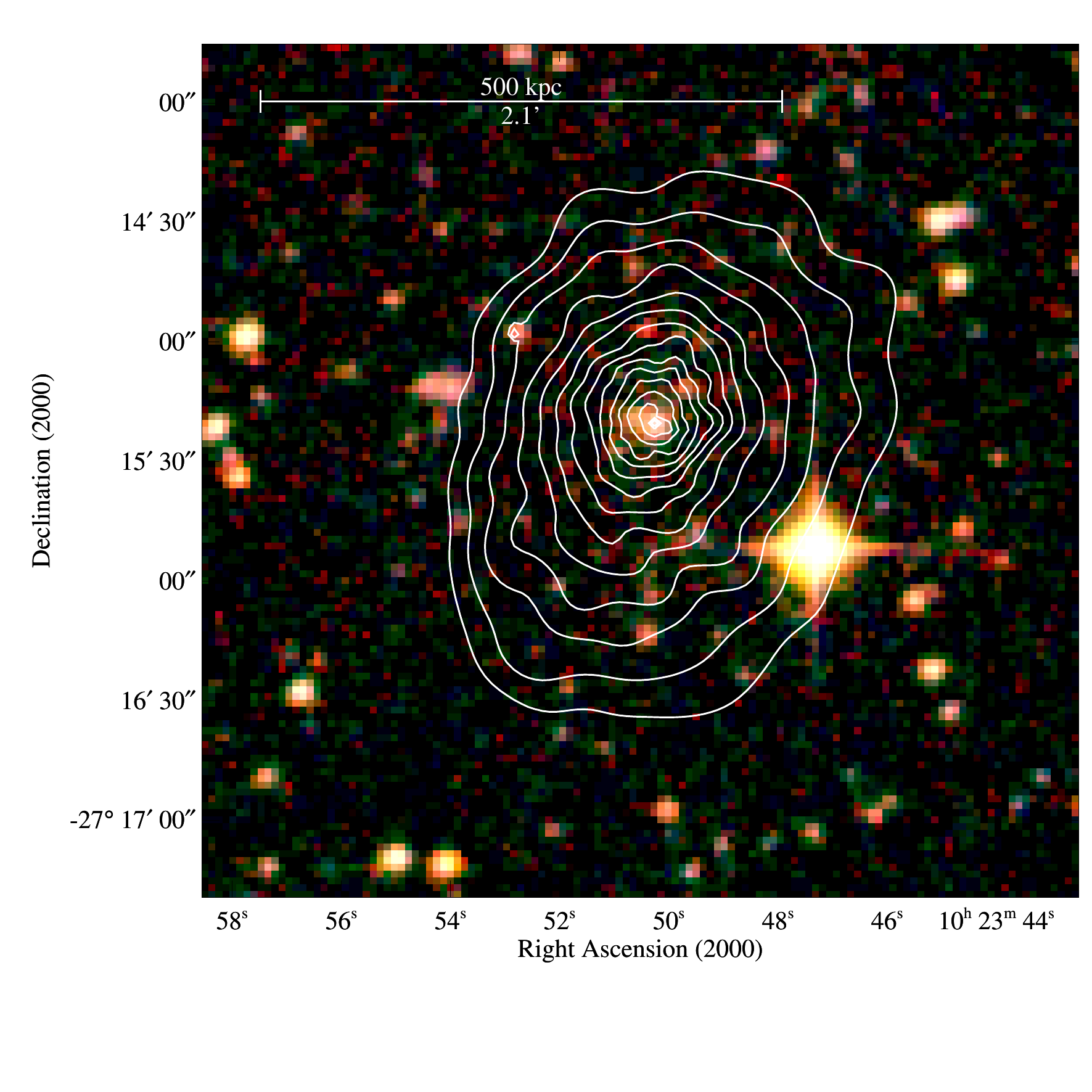}
\includegraphics[width=\overlaysize]{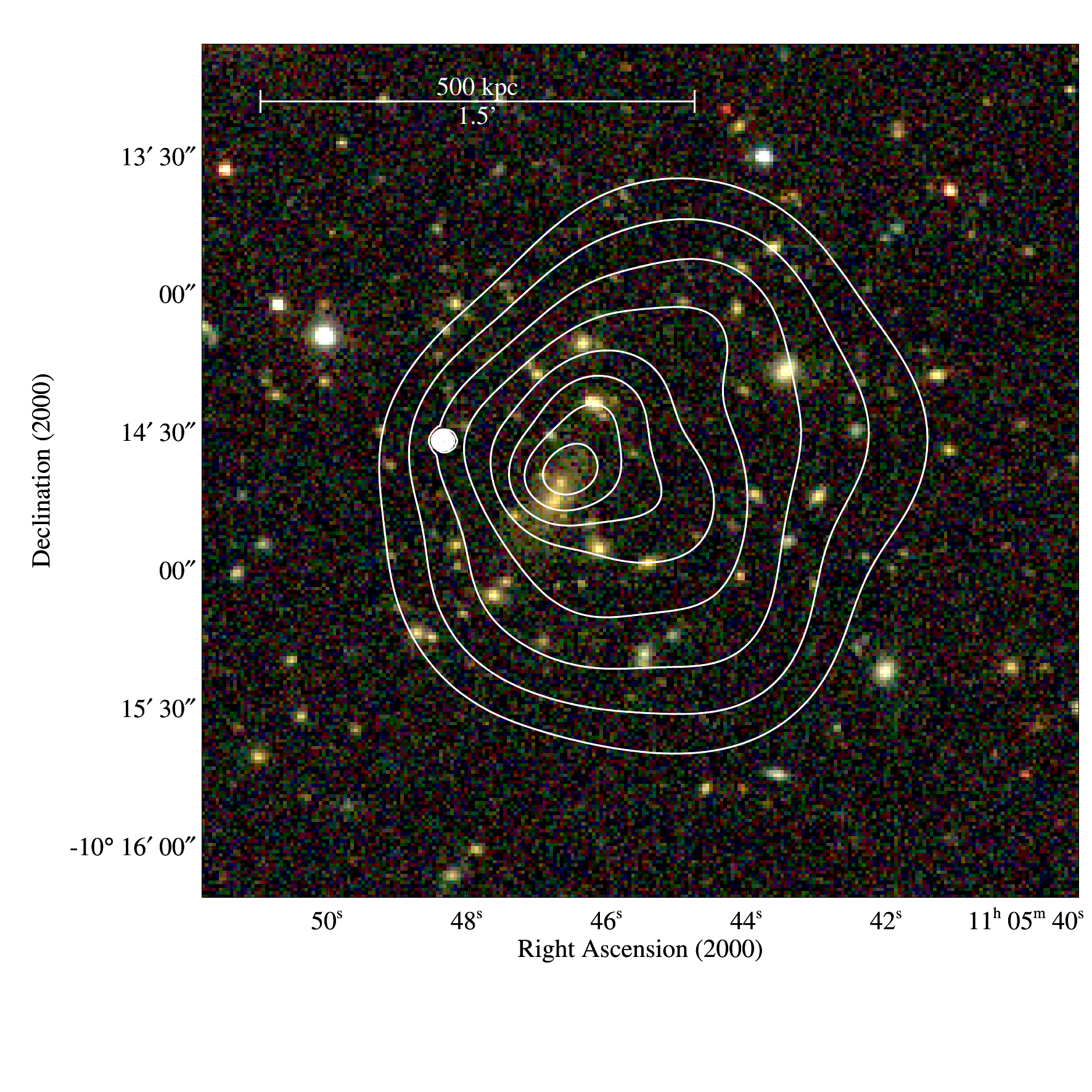}
\includegraphics[width=\overlaysize]{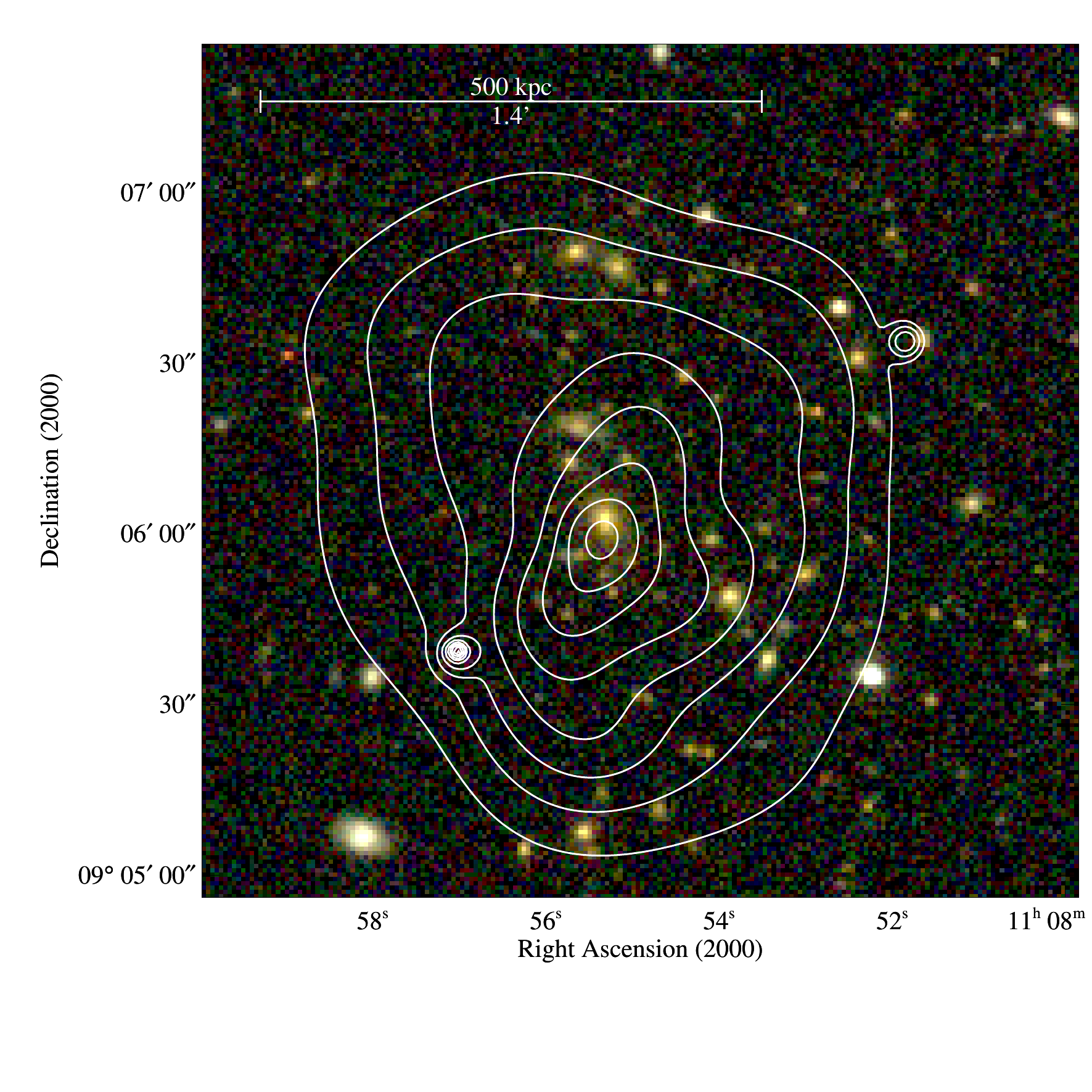}
\includegraphics[width=\overlaysize]{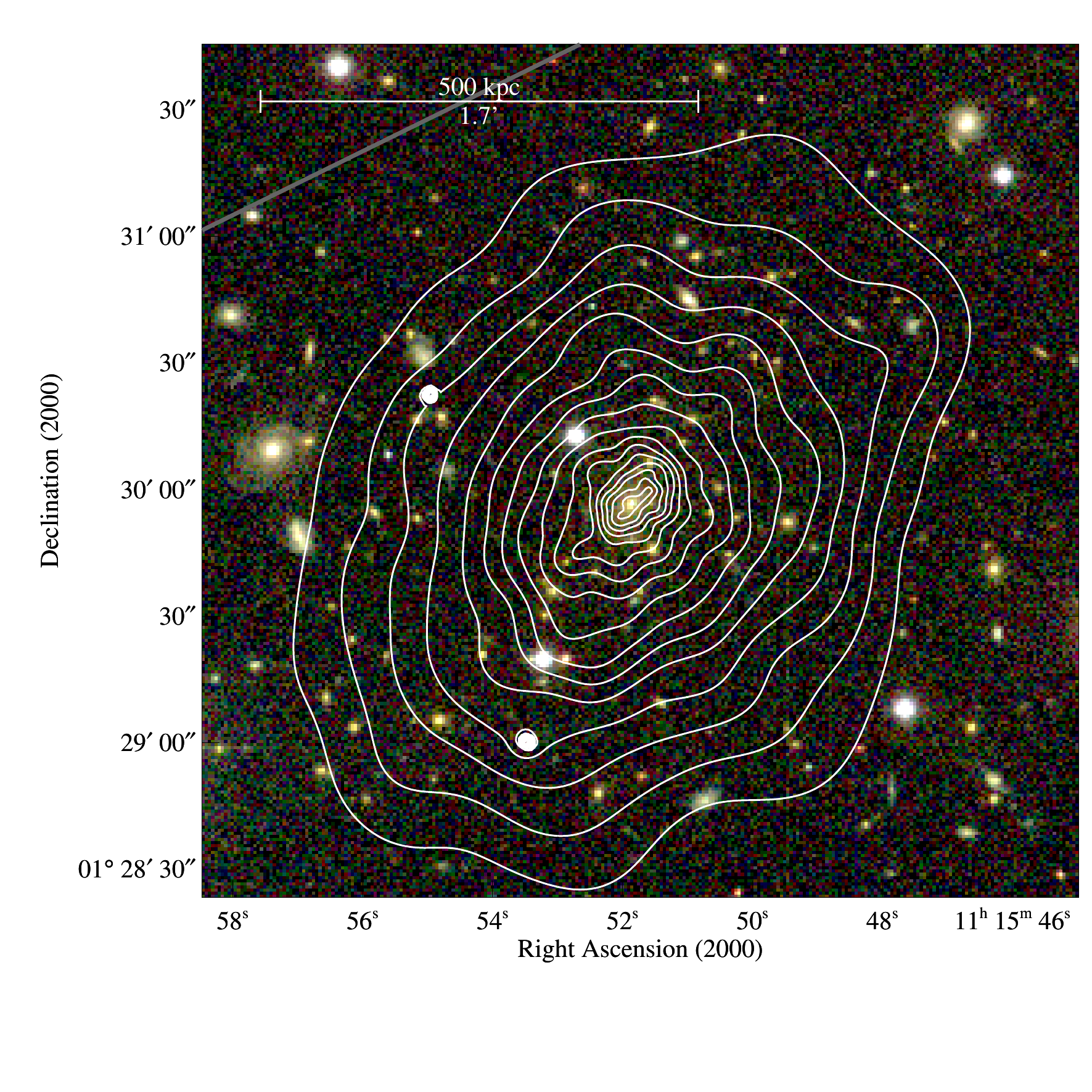}
\includegraphics[width=\overlaysize]{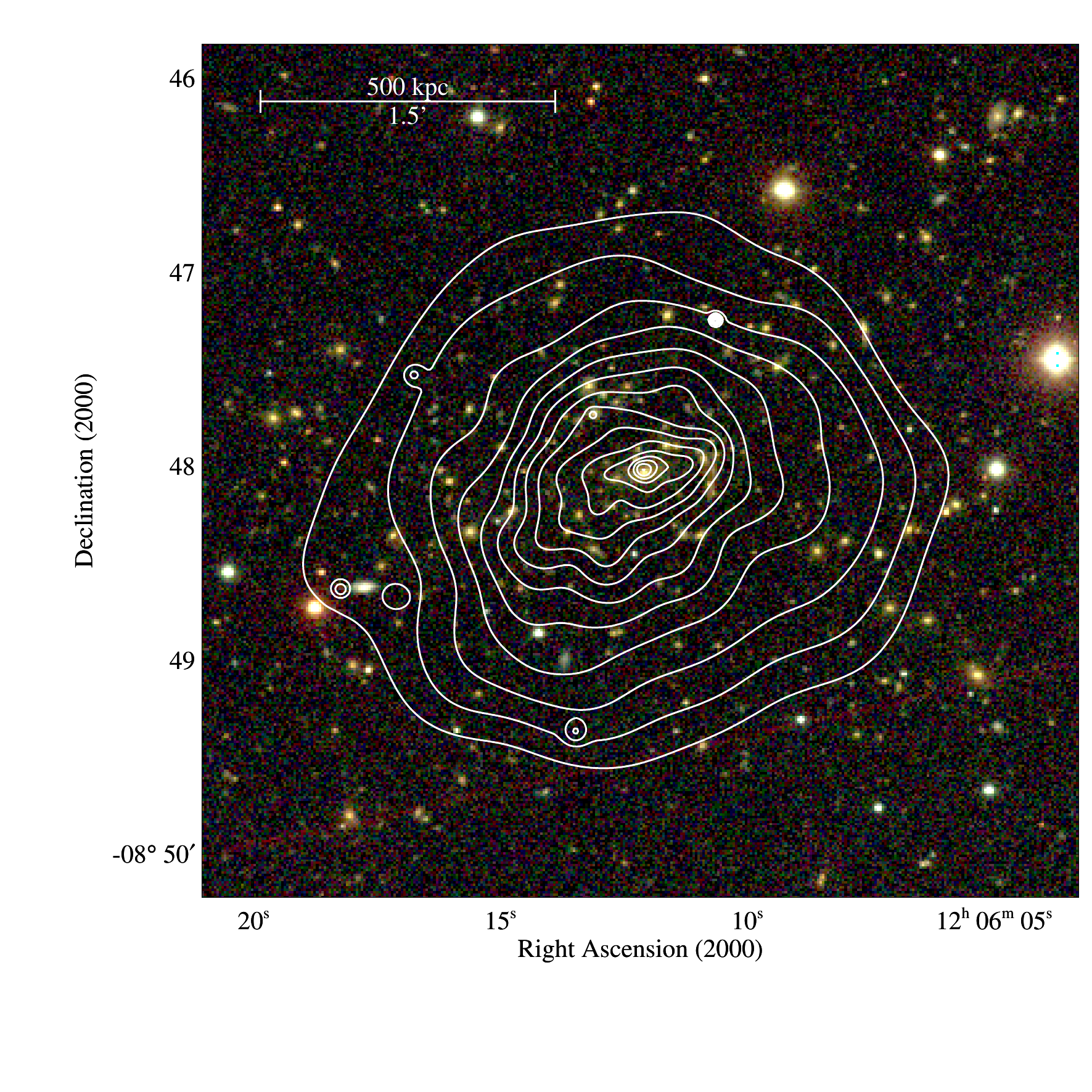}
\includegraphics[width=\overlaysize]{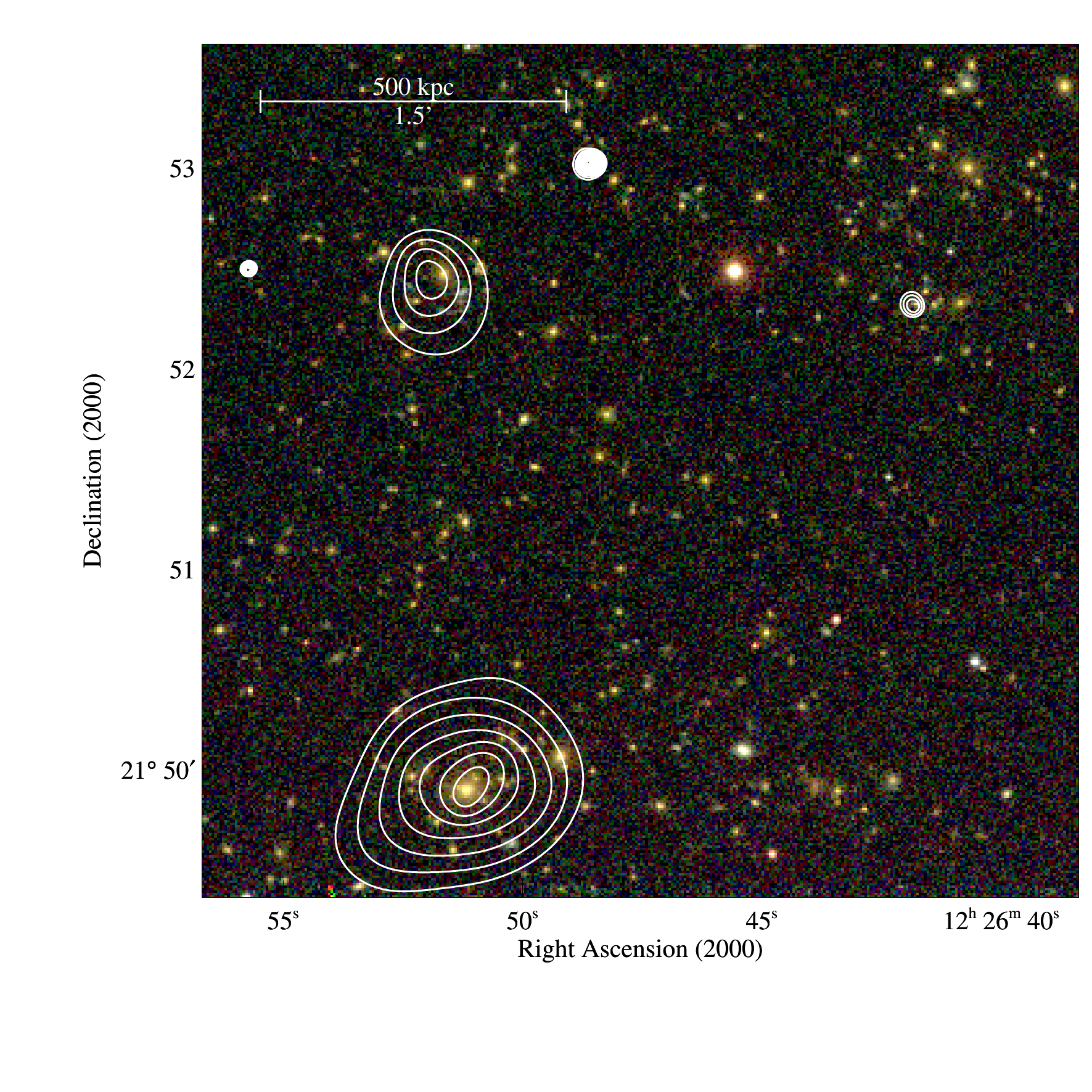}
\includegraphics[width=\overlaysize]{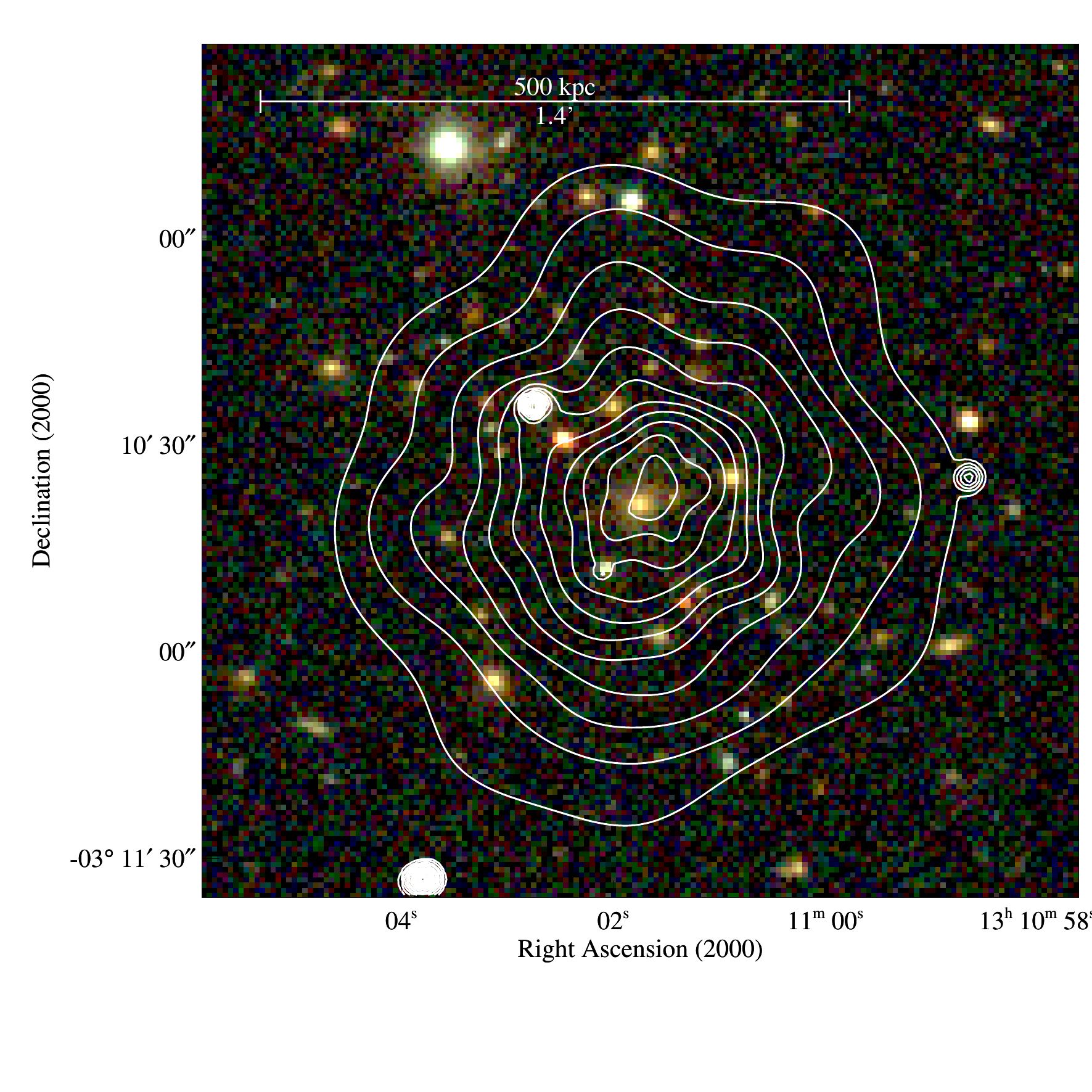}
\includegraphics[width=\overlaysize]{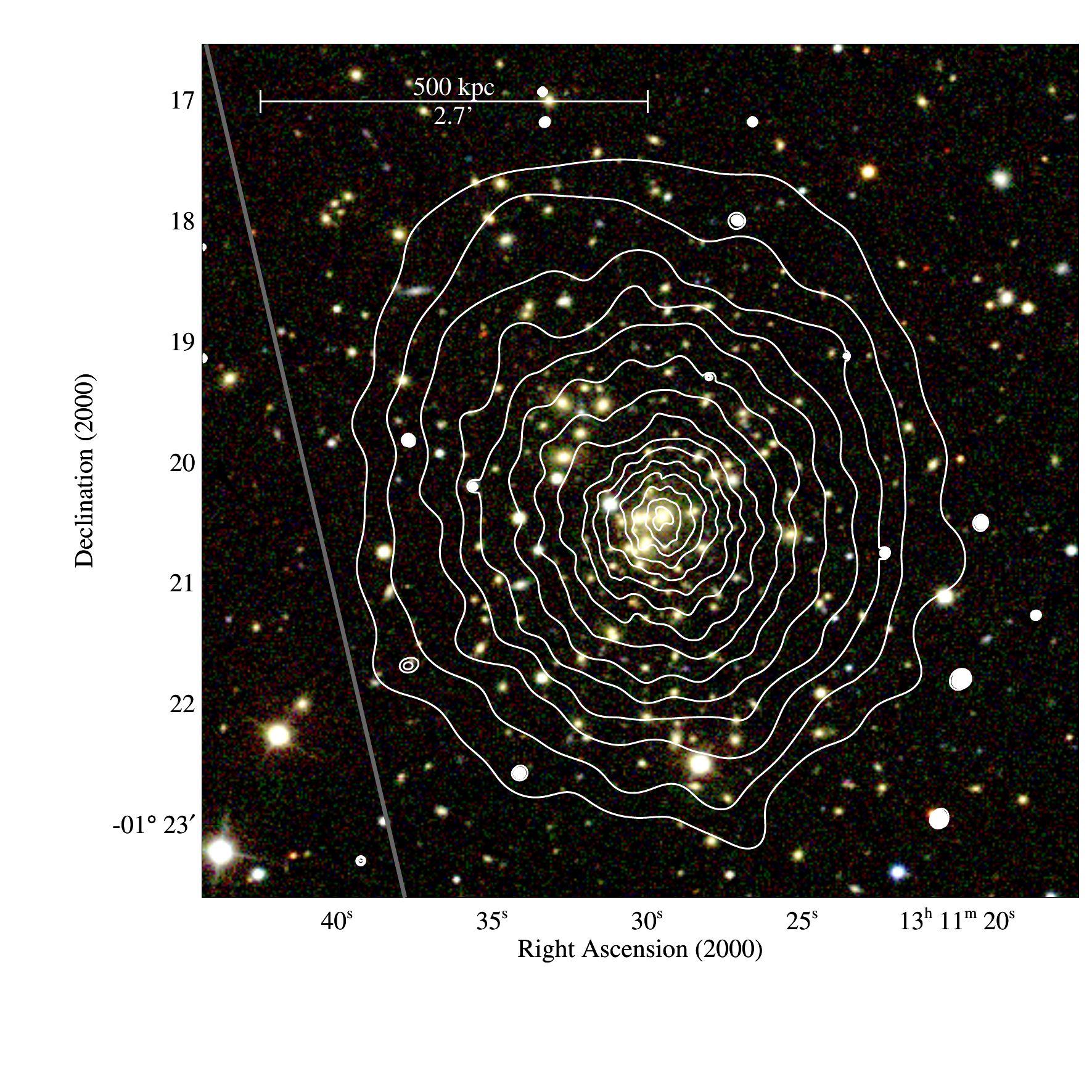}
\includegraphics[width=\overlaysize]{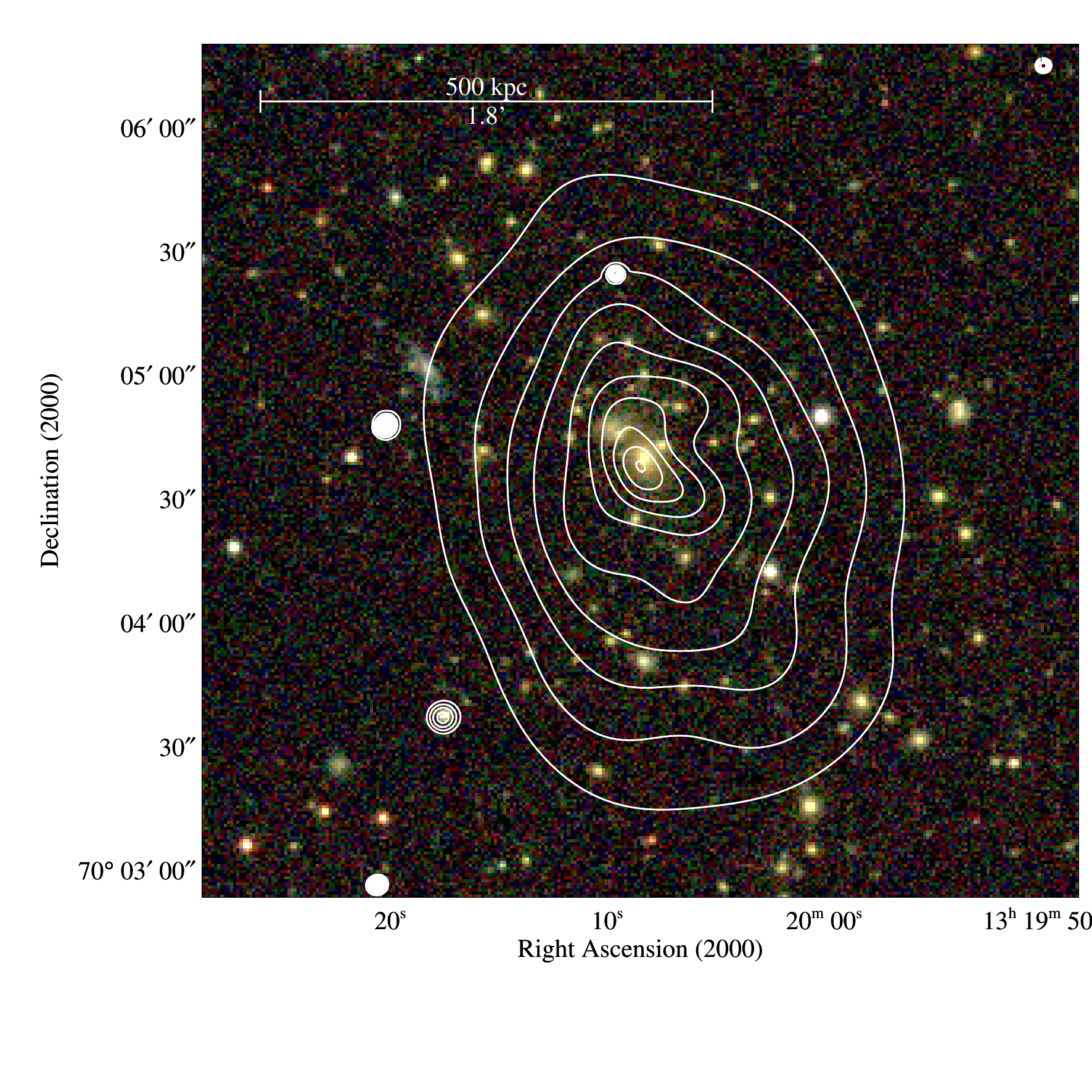}
\includegraphics[width=\overlaysize]{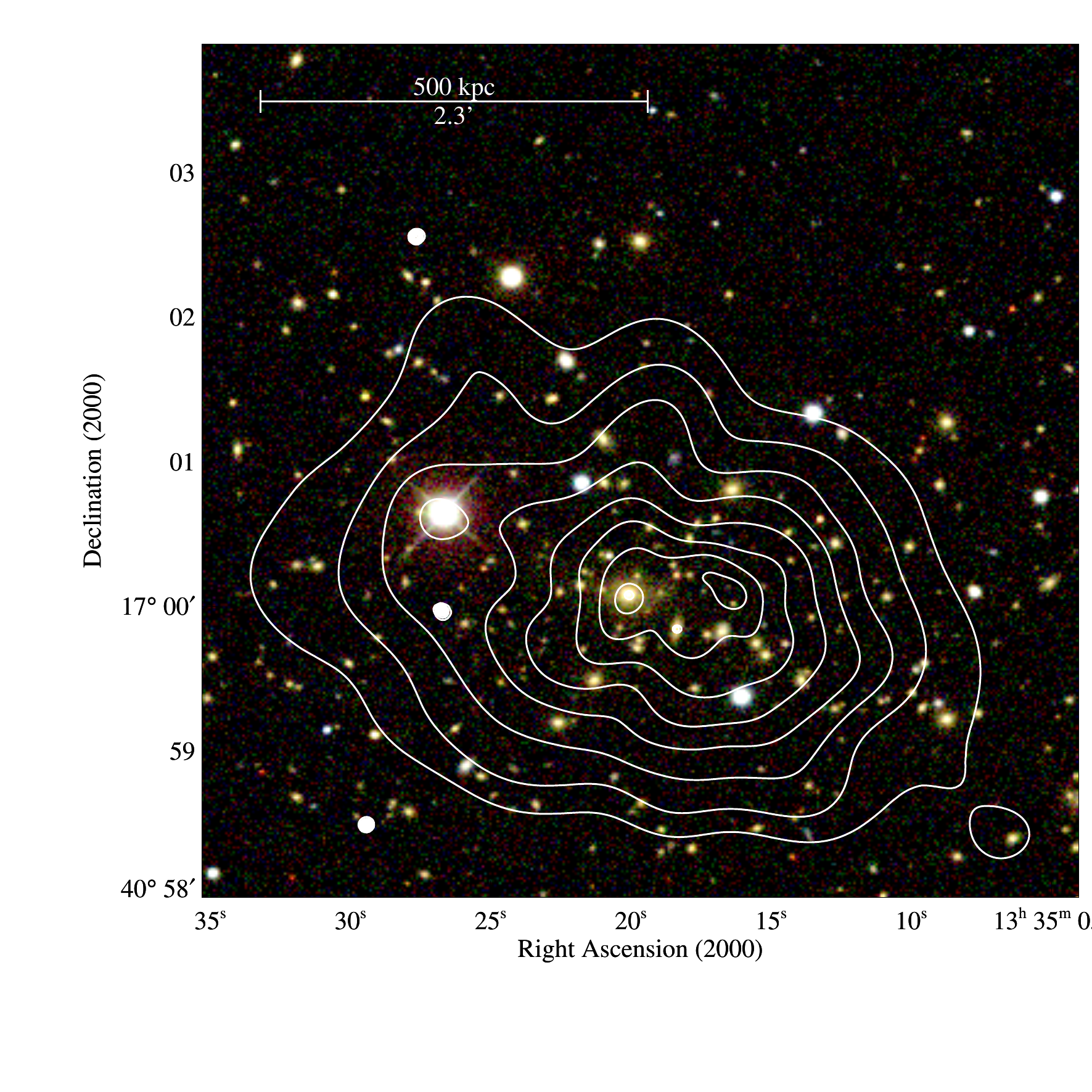}
\includegraphics[width=\overlaysize]{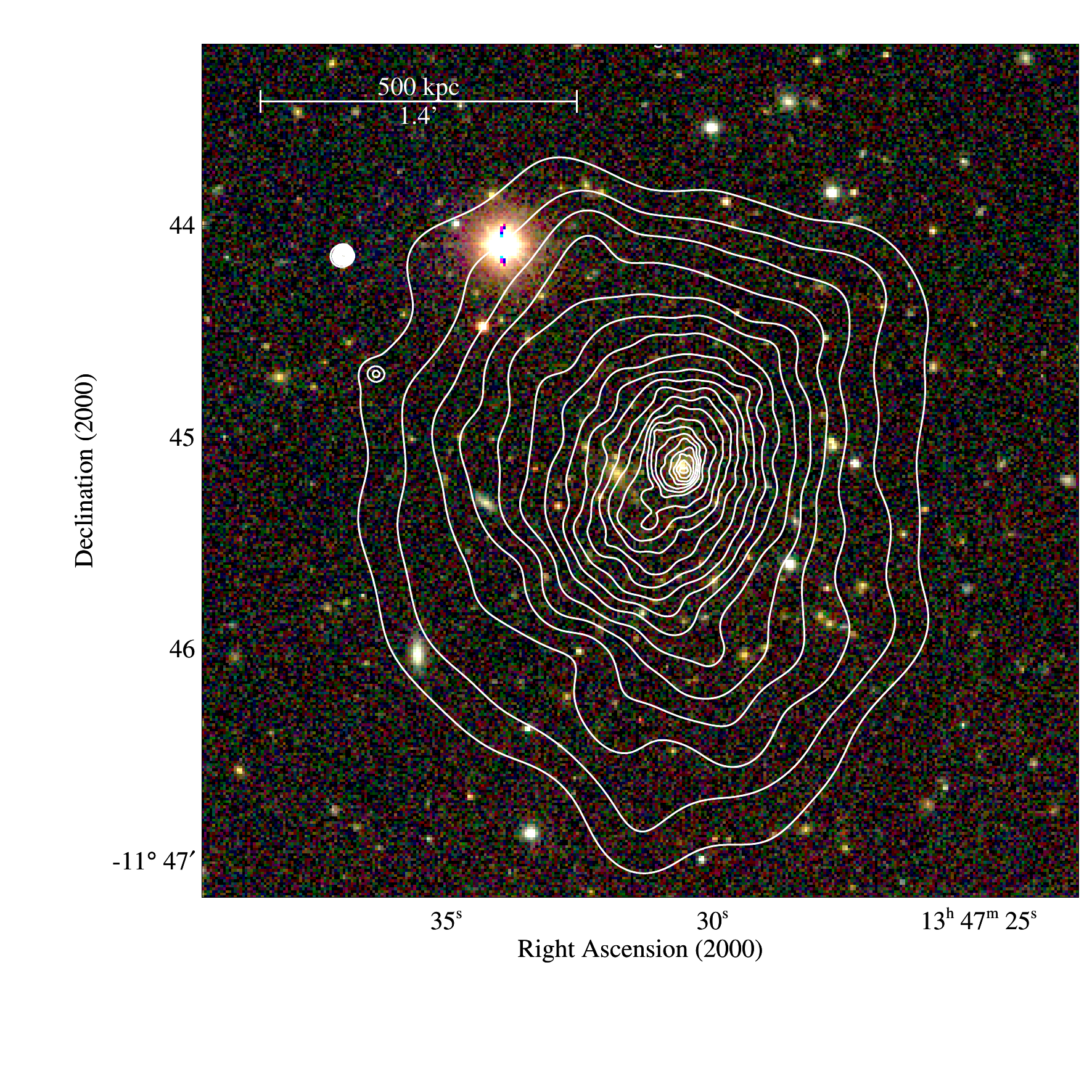}
   \caption{\emph{continued} Shown are (in R.A.\ order) {\it ZwCl\,1021.0+0426}, {\it A\,3444}, {\it MACS\,J1105.7--1014}, {\it MACS\,J1108.8+0906}, {\it MACS\,J1115.8+0129}, {\it MACS\,J1206.2--0847}, {\it MACS\,J1226.8+2153}, {\it MACS\,J1311.0--0310}, {\it A\,1689}, {\it A\,1722}, {\it A\,1763}, and {\it RX\,J1347.5-1145}}
   \label{fig:appendix}
\end{figure*}

\setcounter{figure}{0}

\begin{figure*} 
   \centering
\includegraphics[width=\overlaysize]{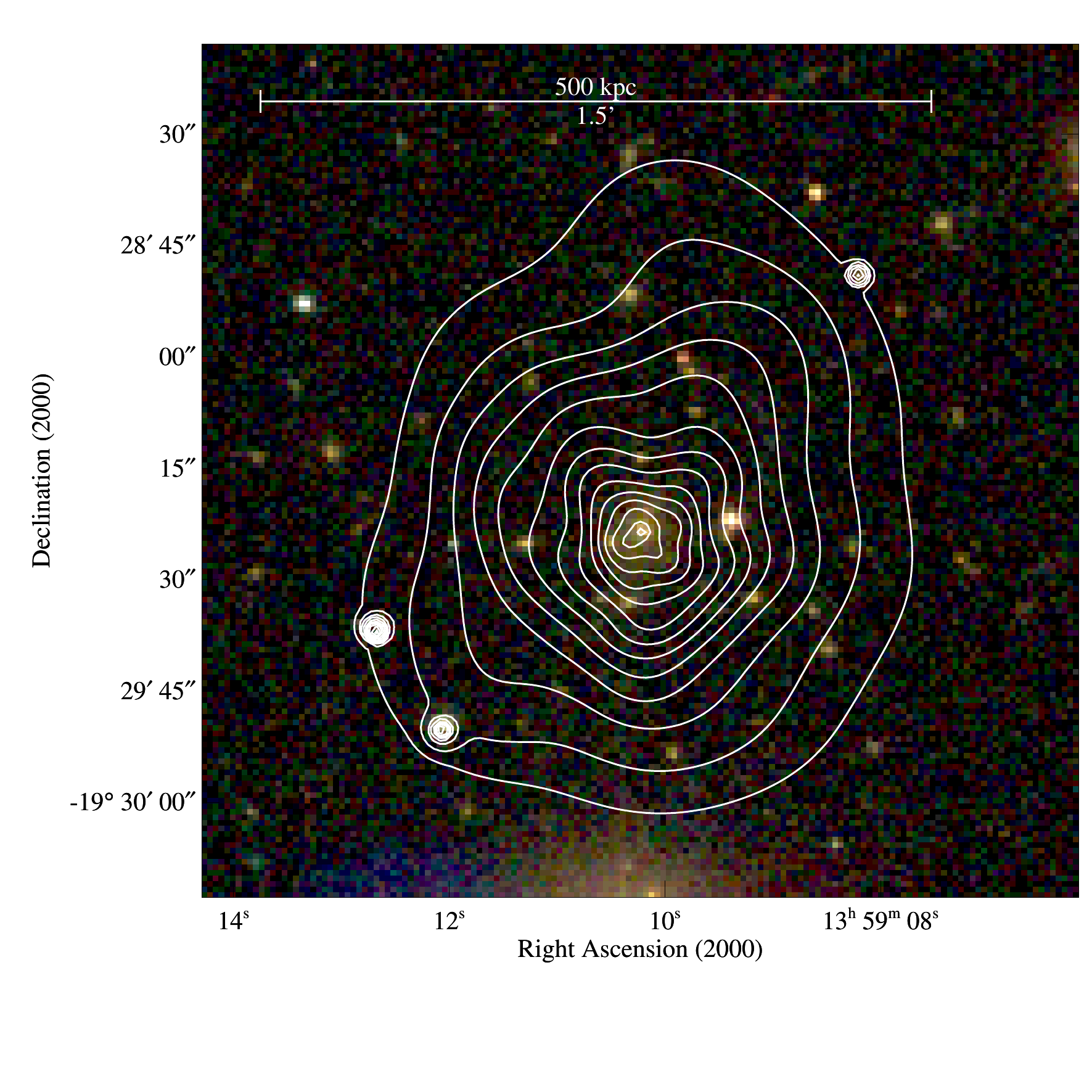}
\includegraphics[width=\overlaysize]{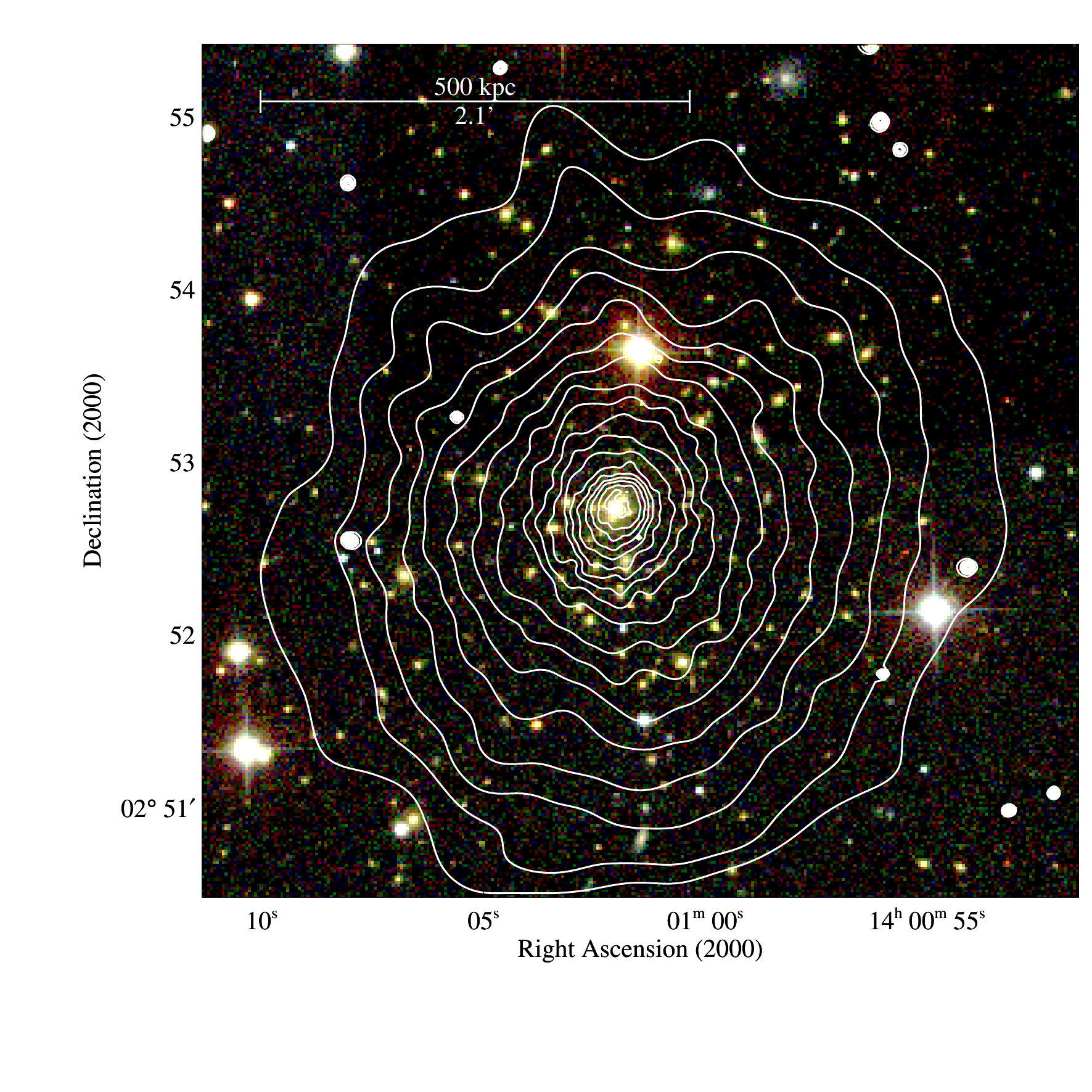}
\includegraphics[width=\overlaysize]{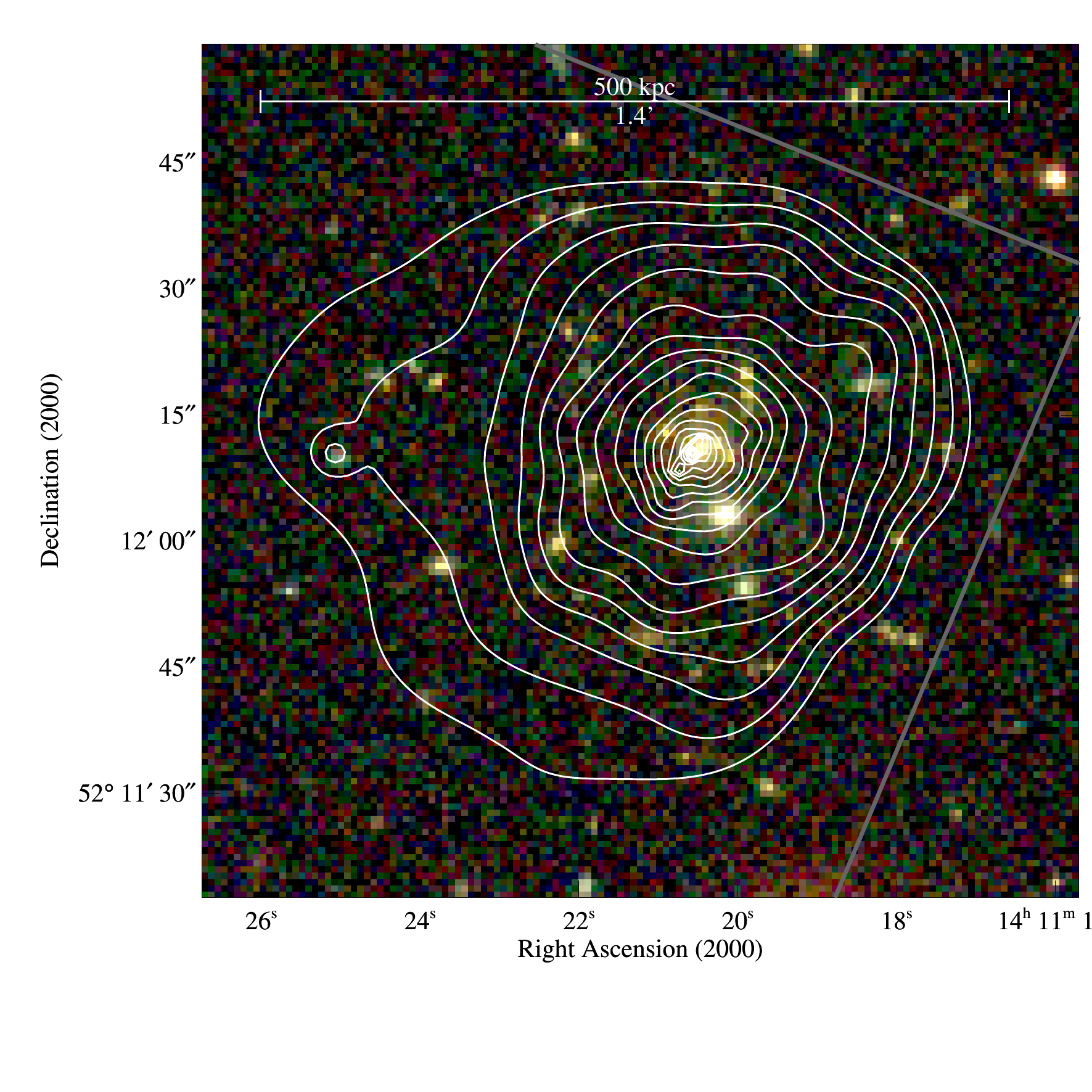}
\includegraphics[width=\overlaysize]{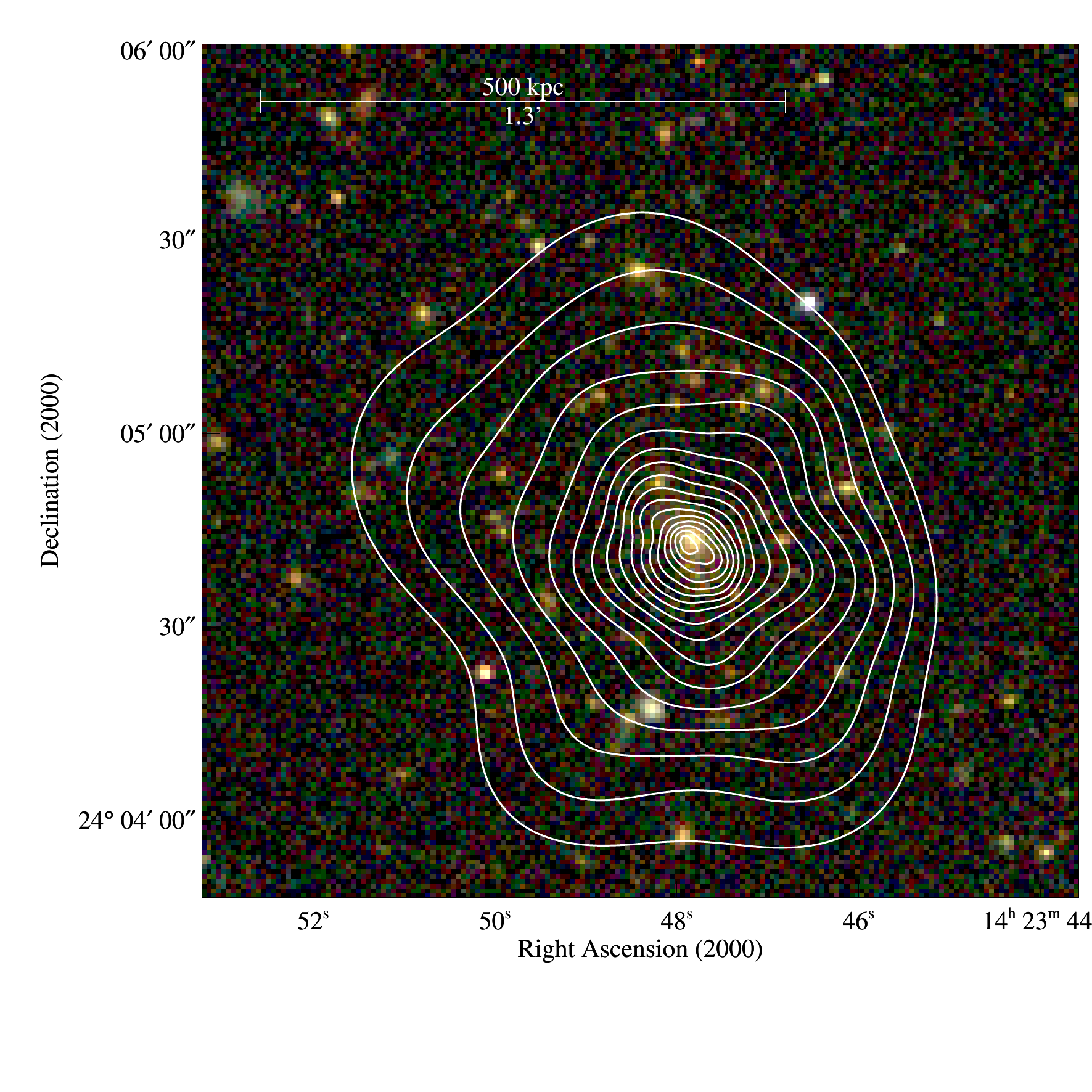}
\includegraphics[width=\overlaysize]{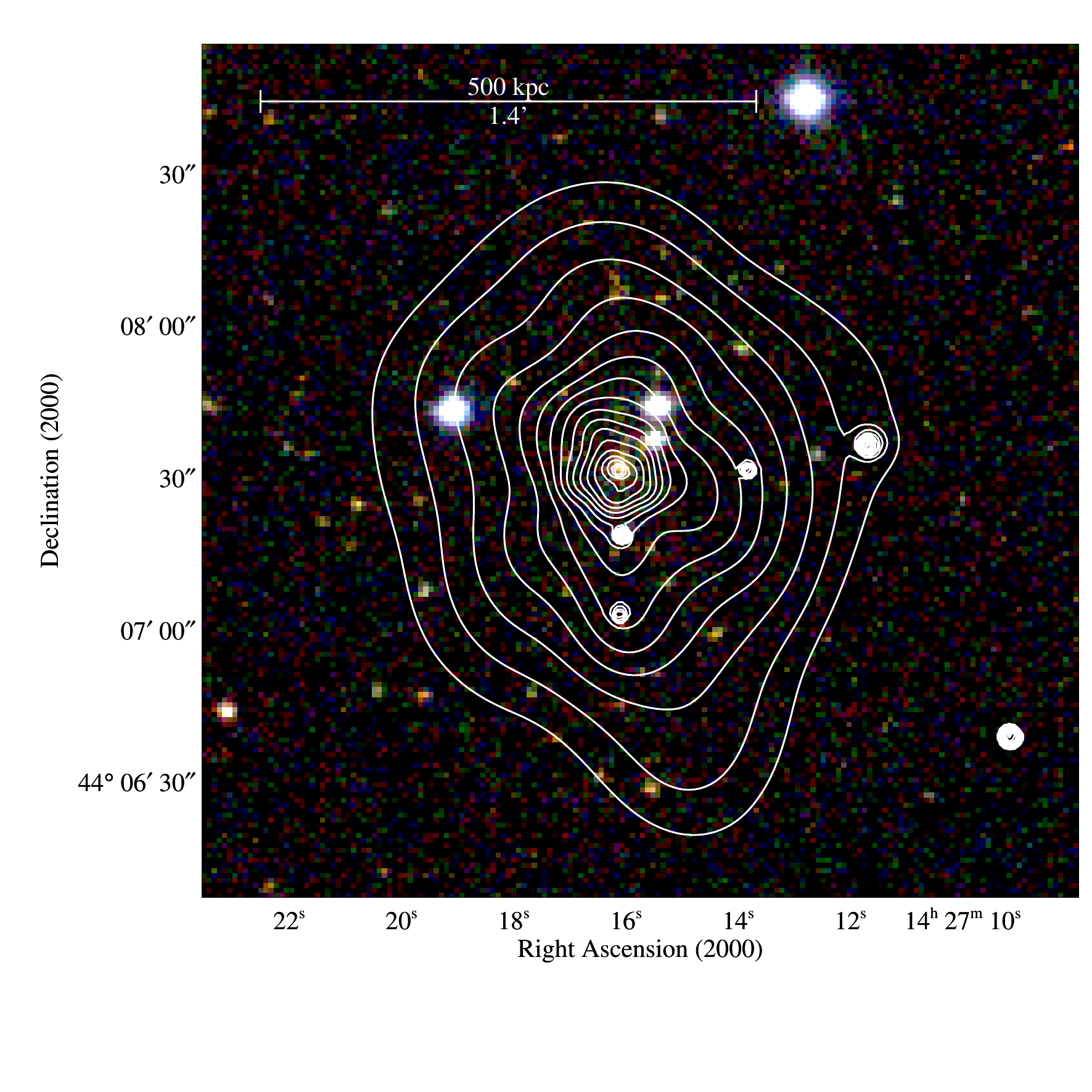}
\includegraphics[width=\overlaysize]{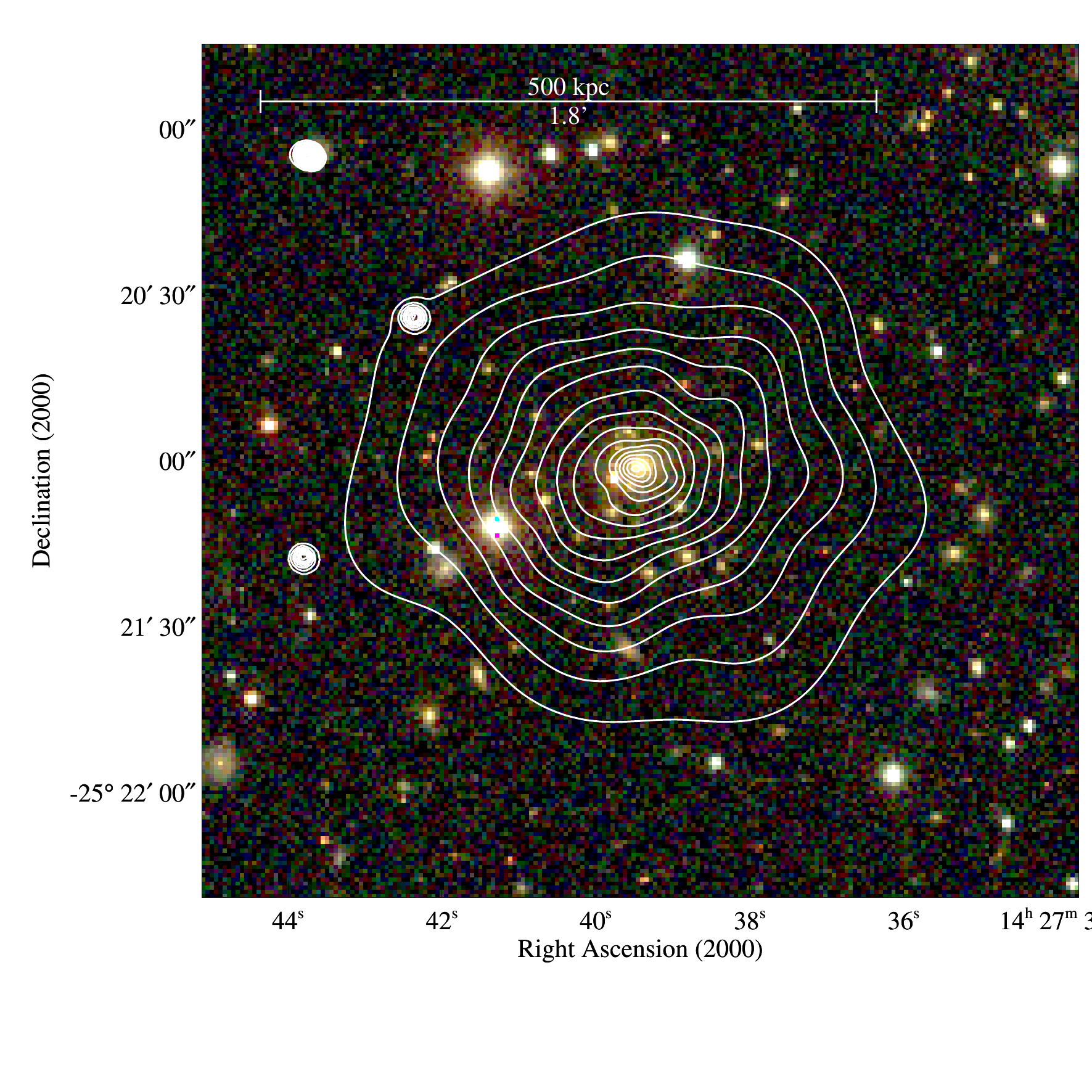}
\includegraphics[width=\overlaysize]{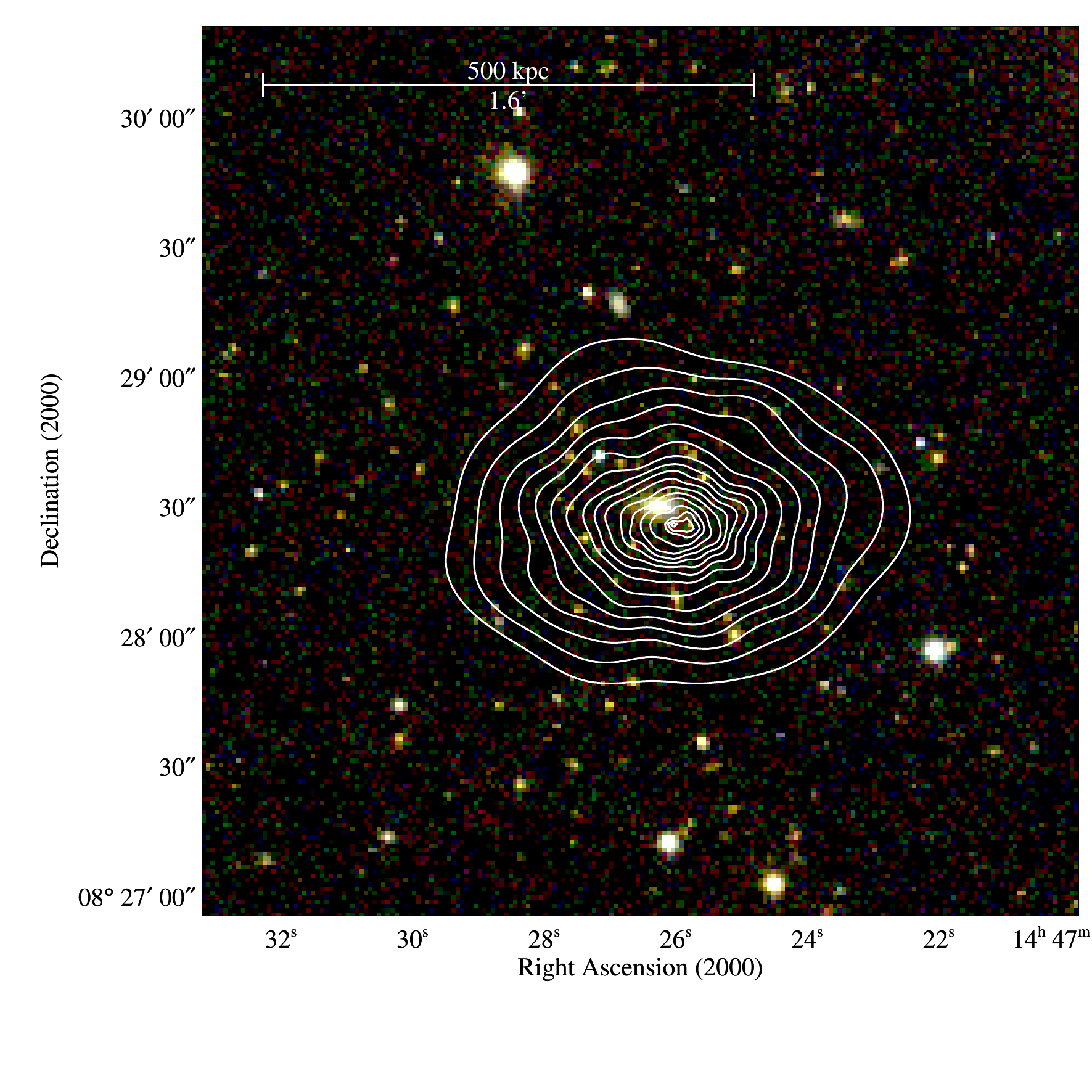}
\includegraphics[width=\overlaysize]{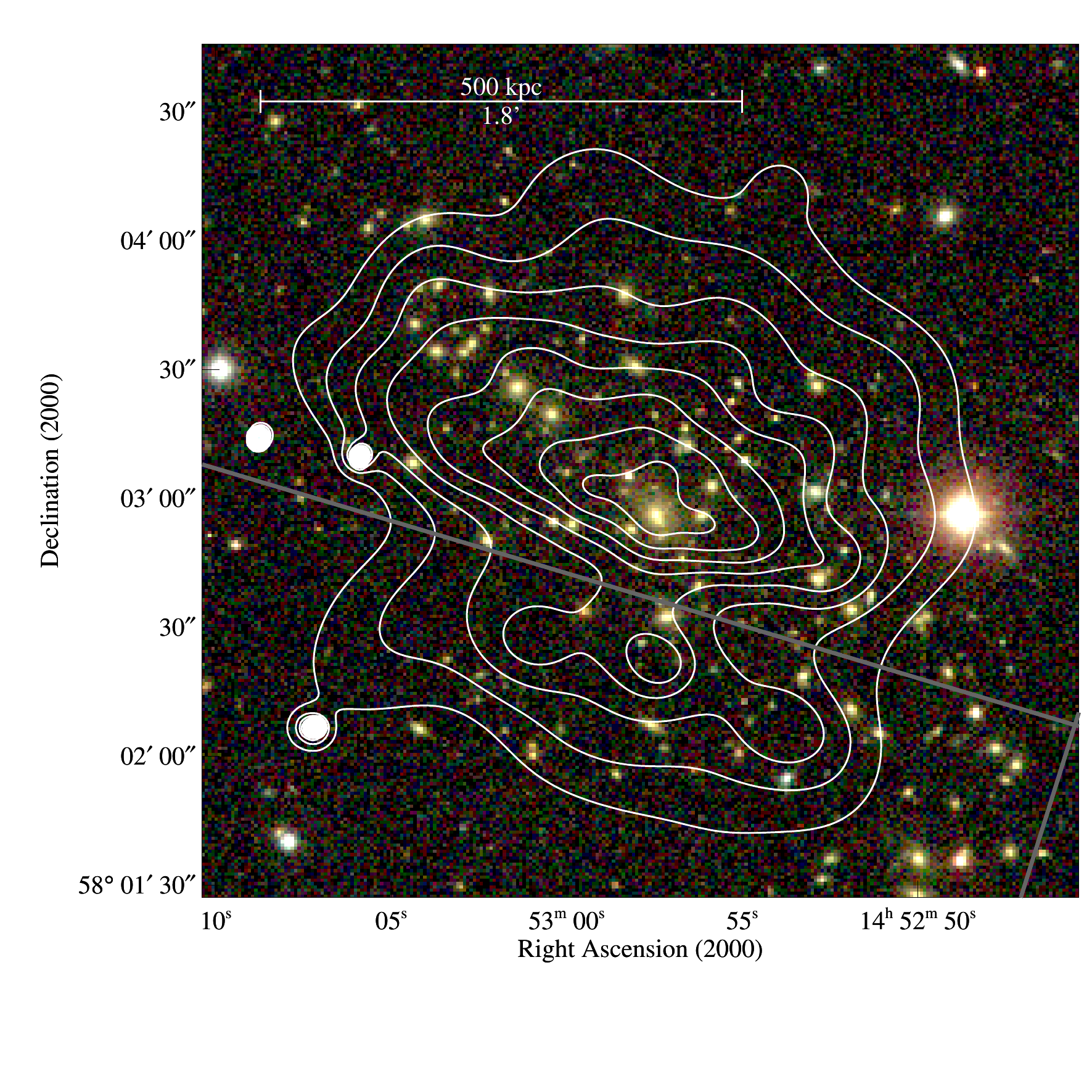}
\includegraphics[width=\overlaysize]{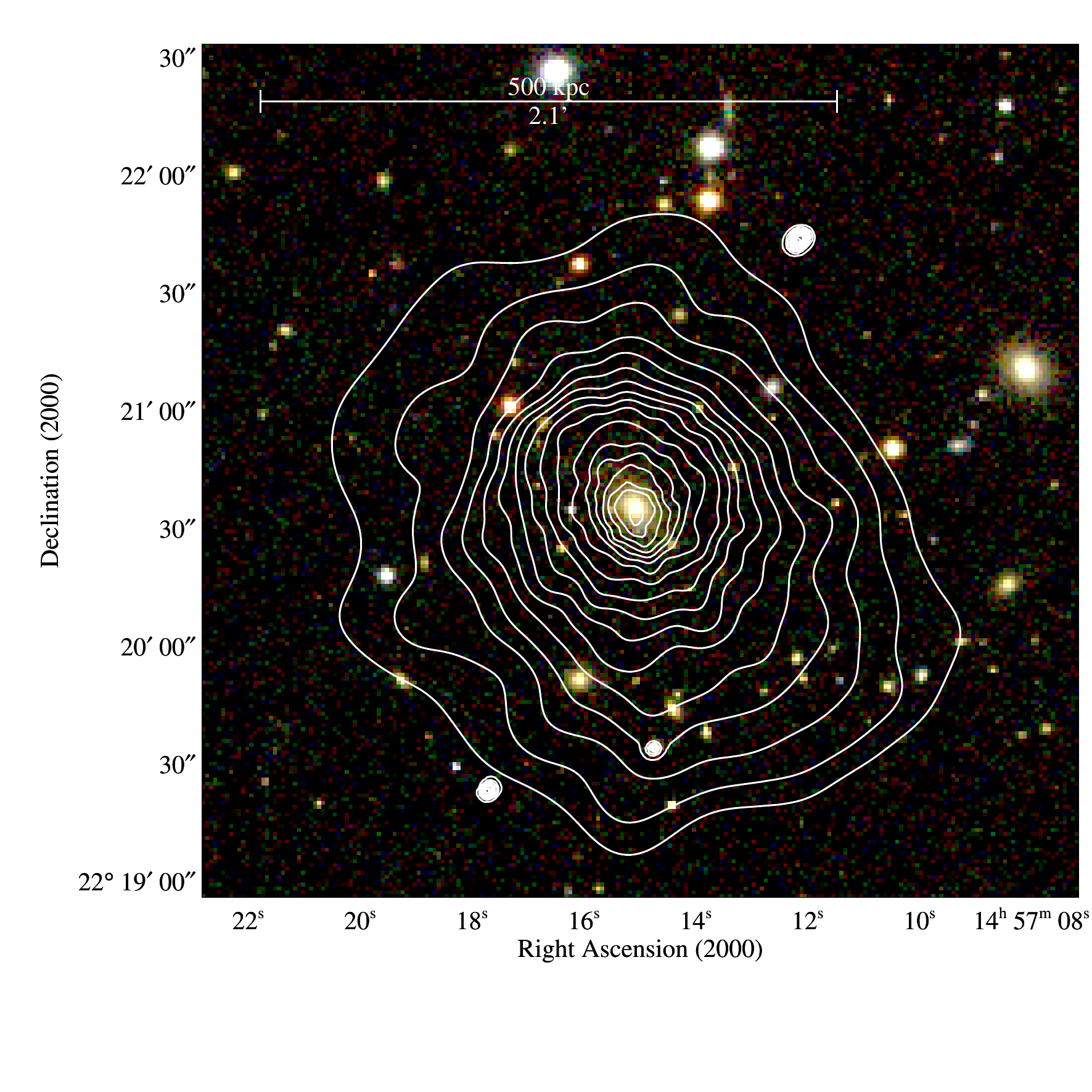}
\includegraphics[width=\overlaysize]{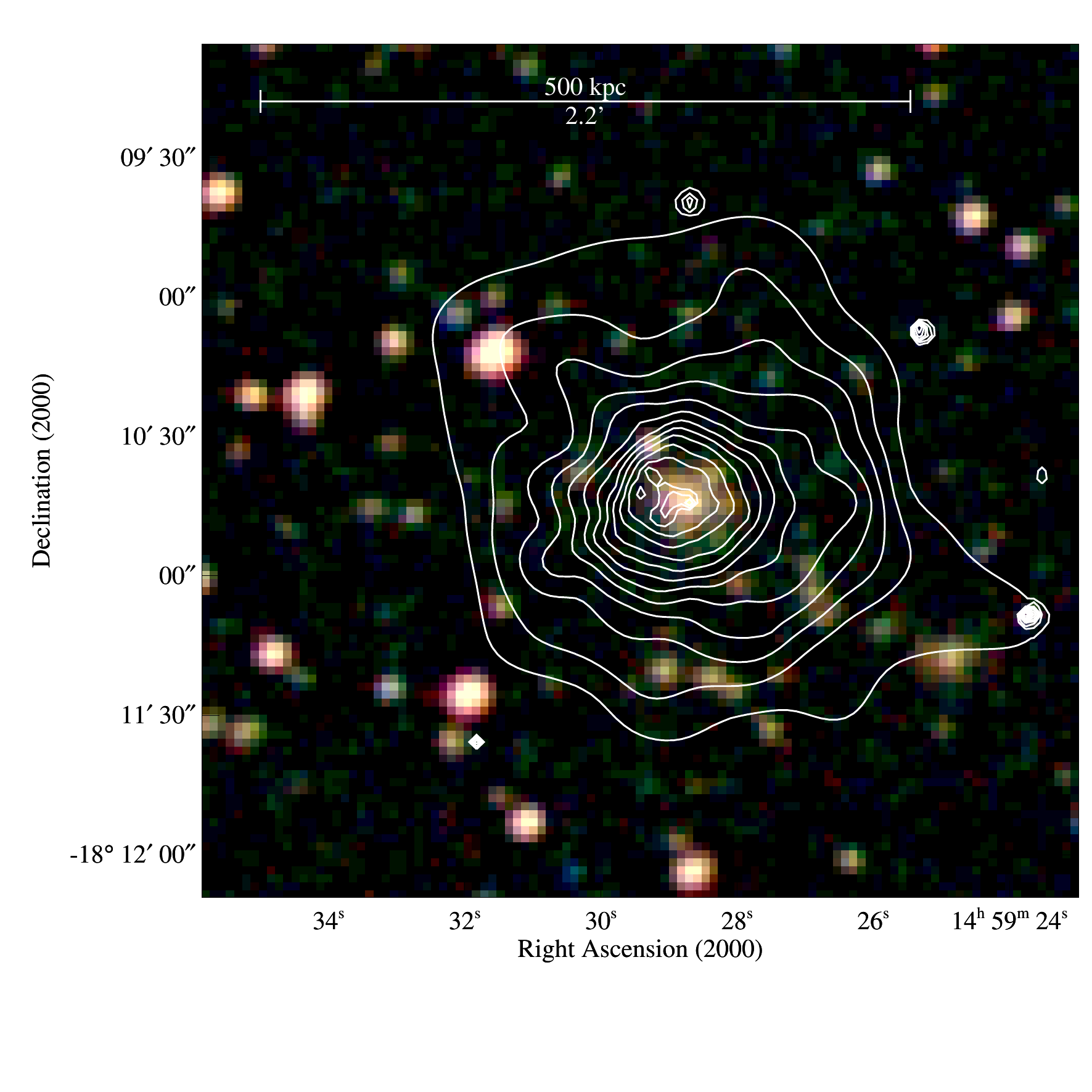}
\includegraphics[width=\overlaysize]{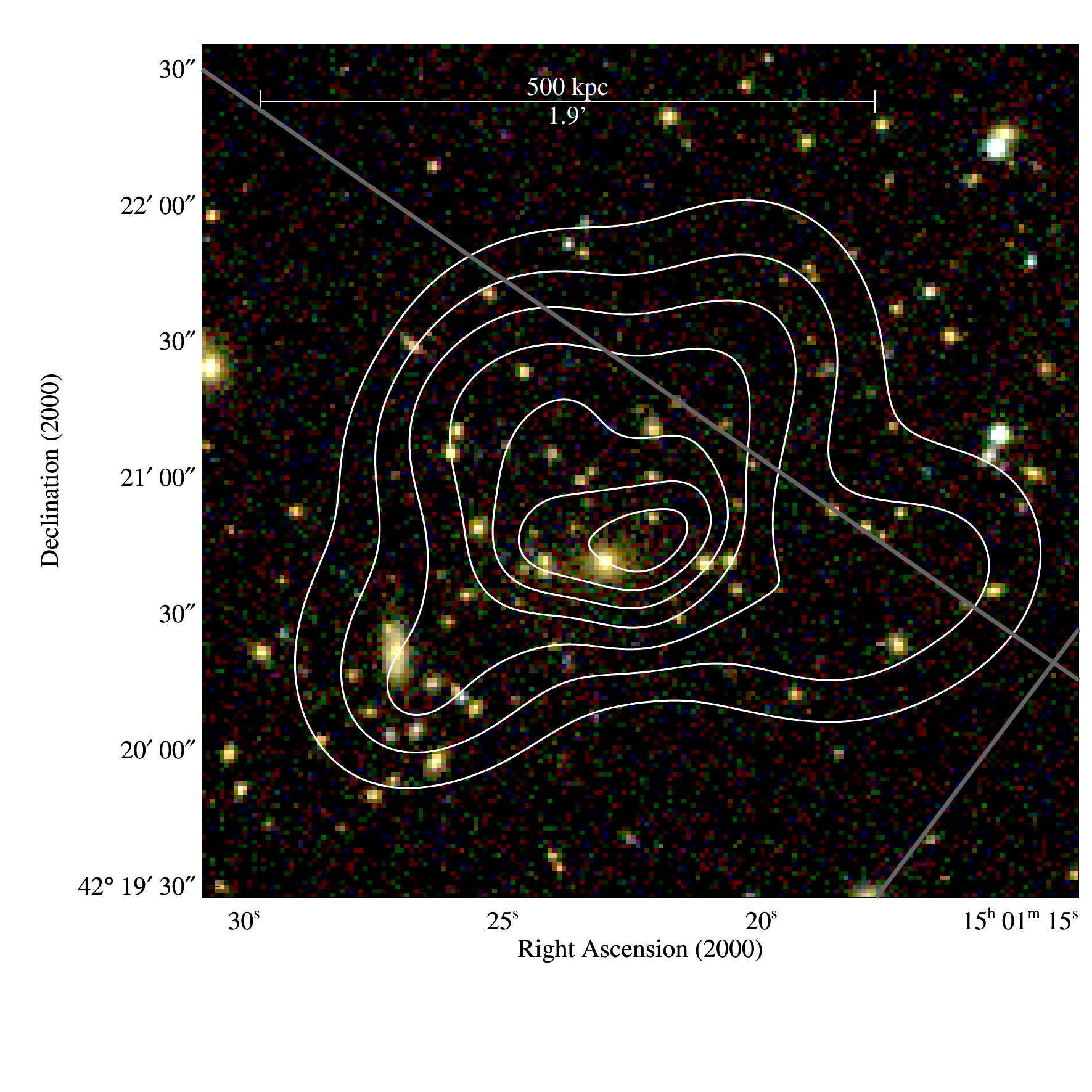}
\includegraphics[width=\overlaysize]{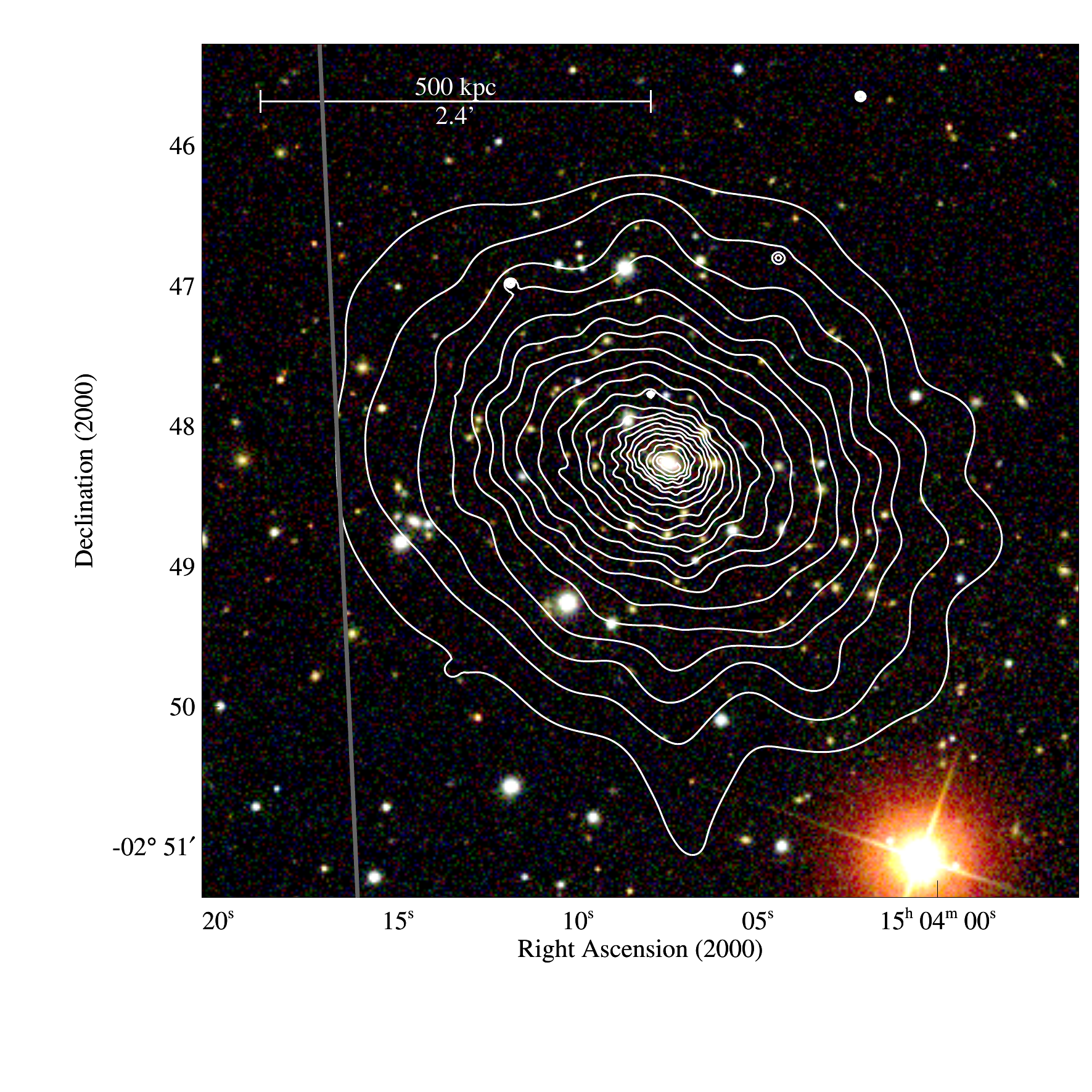}
   \caption{\emph{continued} Shown are (in R.A.\ order) {\it MACS\,J1359.1--1929}, {\it A\,1835}, {\it MACS\,J1411.3+5212}, {\it MACS\,J1423.8+2404}, {\it MACS\,J1427.2+4407}, {\it MACS\,J1427.6--2521}, {\it MACS\,J1447.4+0827}, {\it A\,1995}, {\it ZwCl\,1454.8+2233}, {\it A\,S780}, {\it ZwCl\,1459.4+4240}, and {\it RBS\,1460}}
   \label{fig:appendix}
\end{figure*}

\setcounter{figure}{0}

\begin{figure*} 
   \centering
\includegraphics[width=\overlaysize]{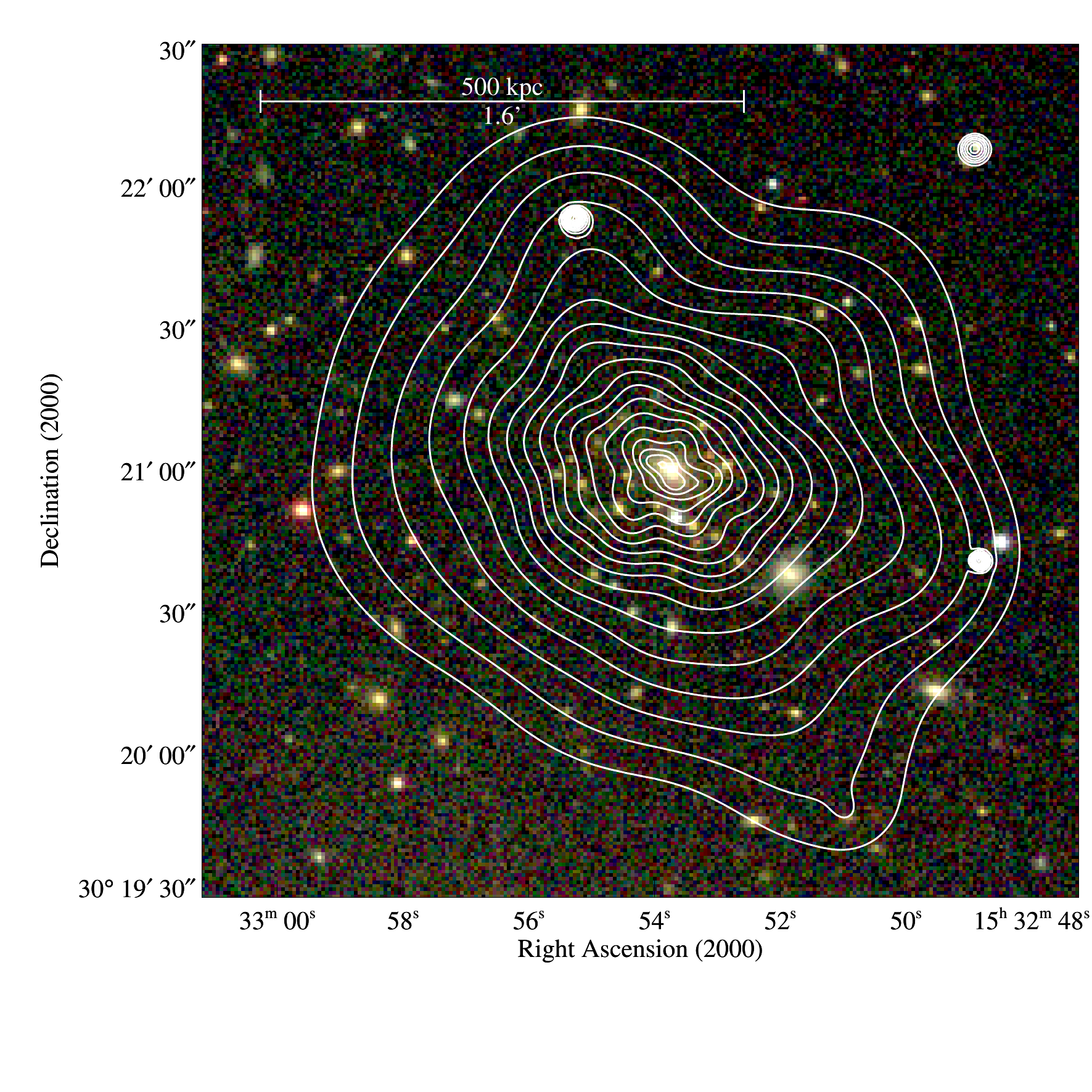}
\includegraphics[width=\overlaysize]{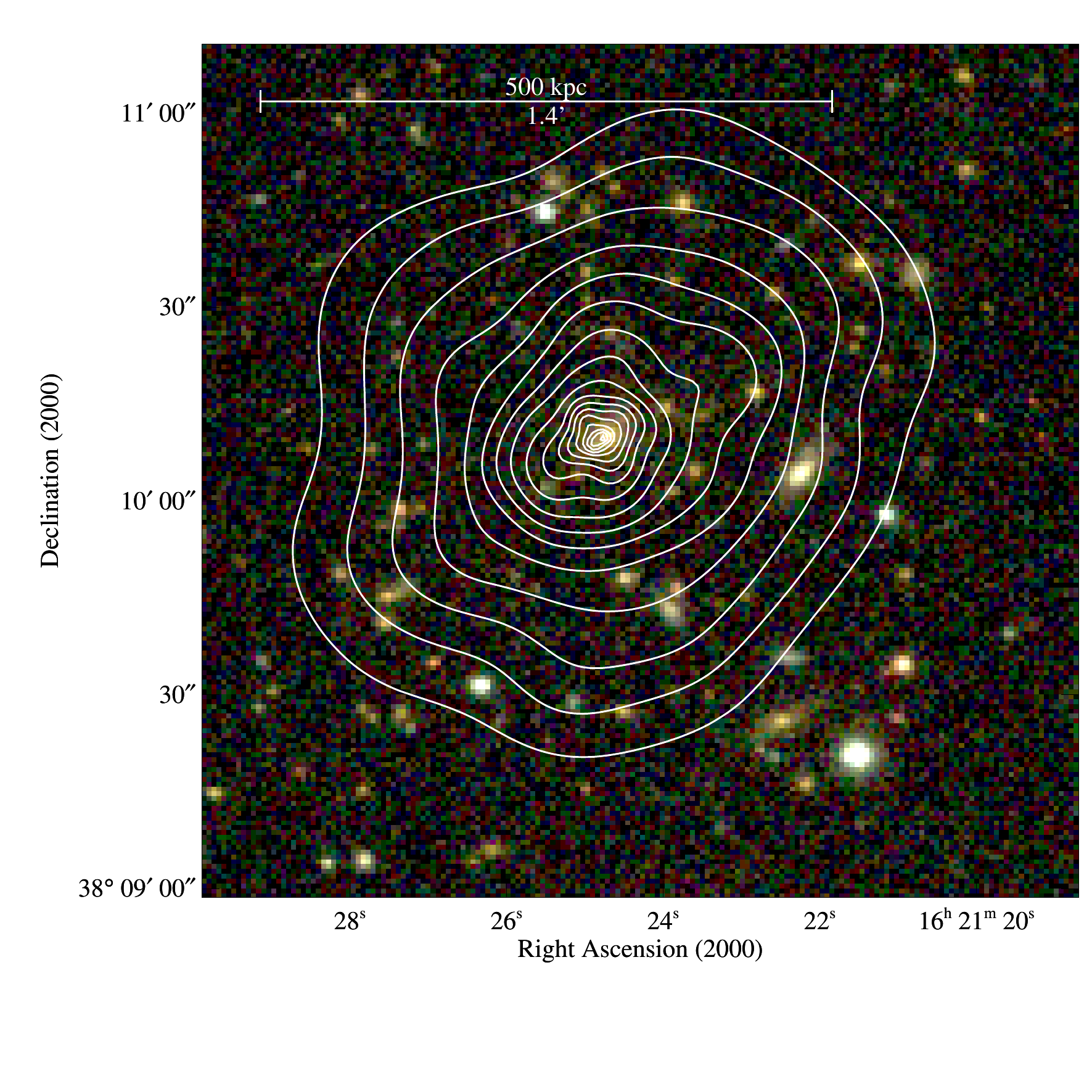}
\includegraphics[width=\overlaysize]{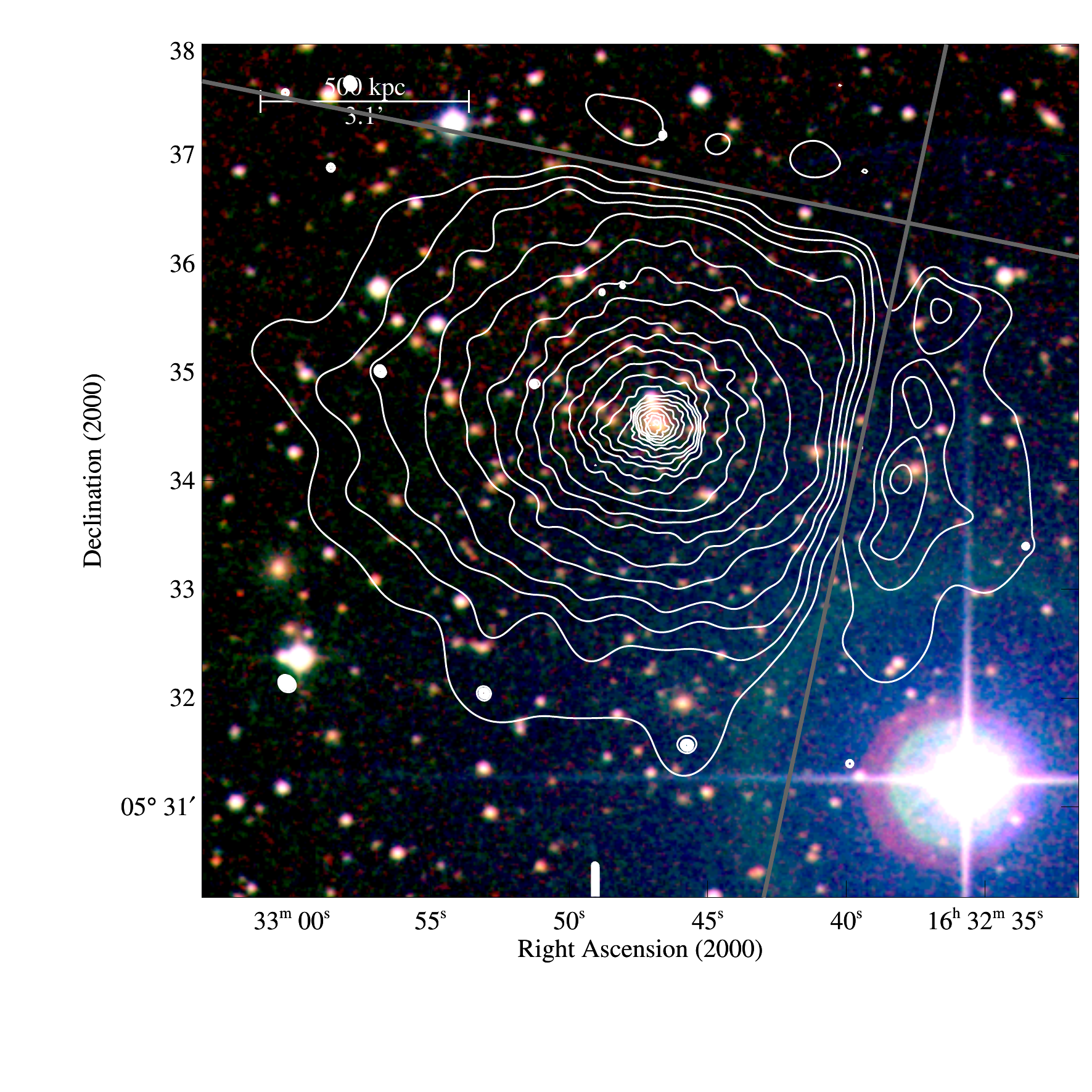}
\includegraphics[width=\overlaysize]{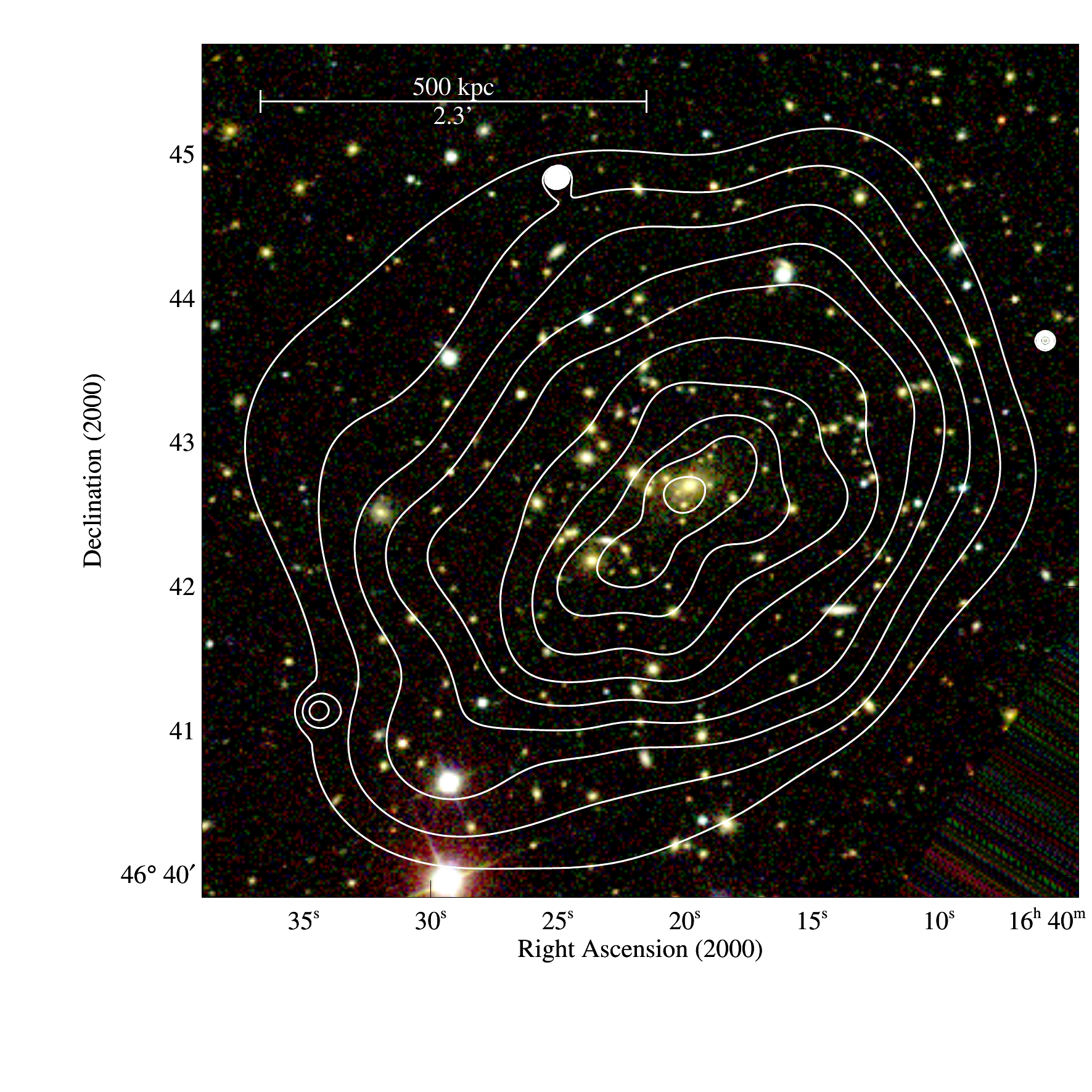}
\includegraphics[width=\overlaysize]{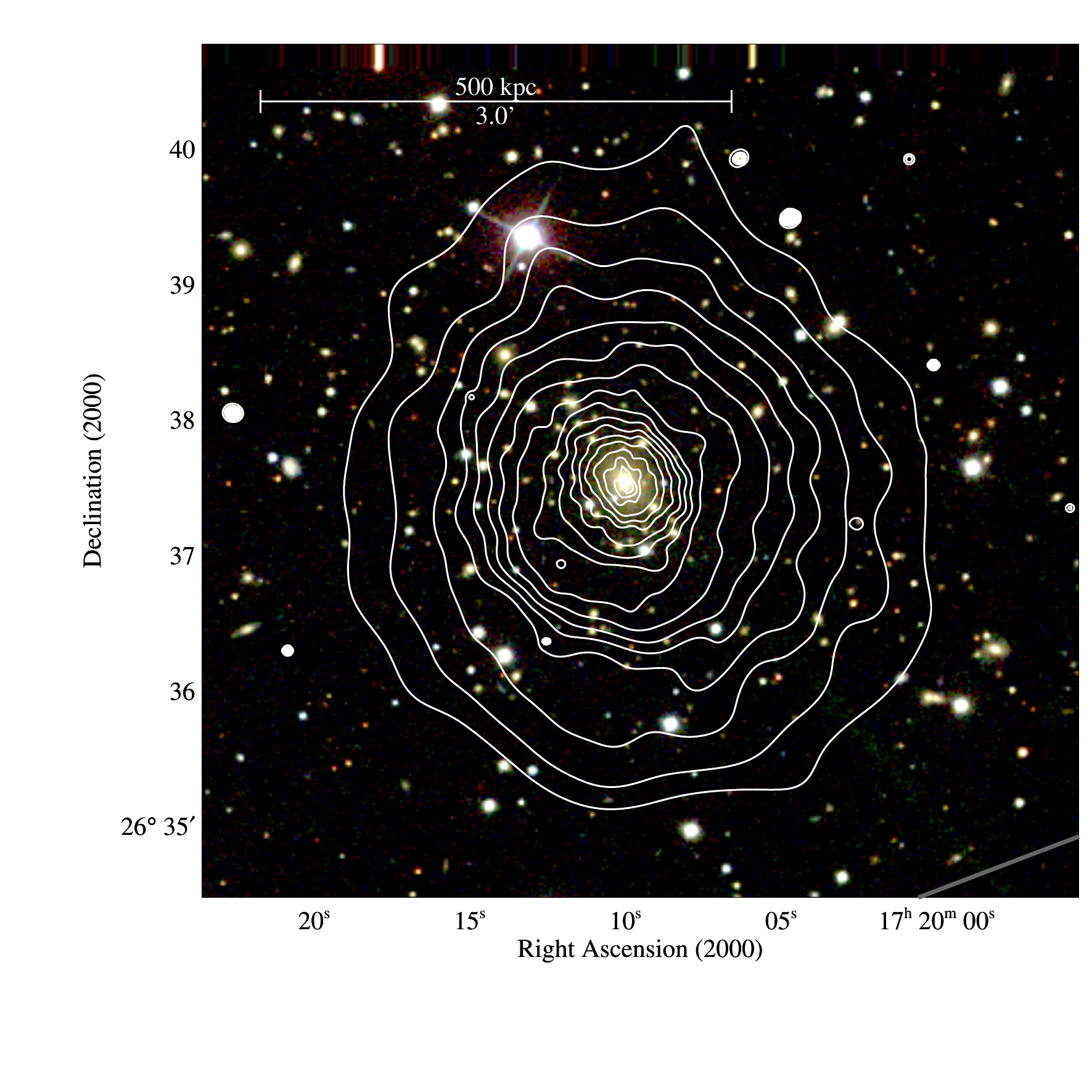}
\includegraphics[width=\overlaysize]{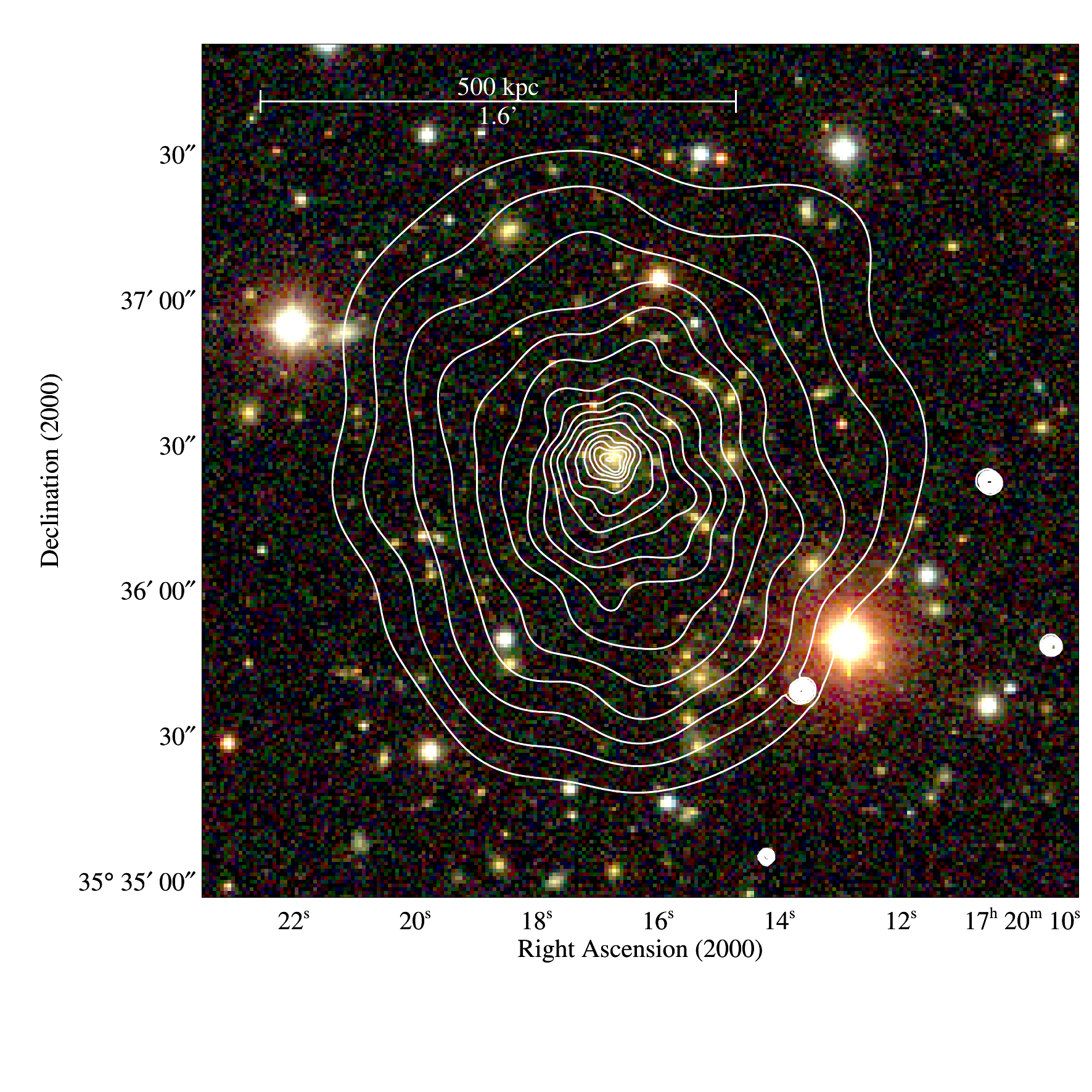}
\includegraphics[width=\overlaysize]{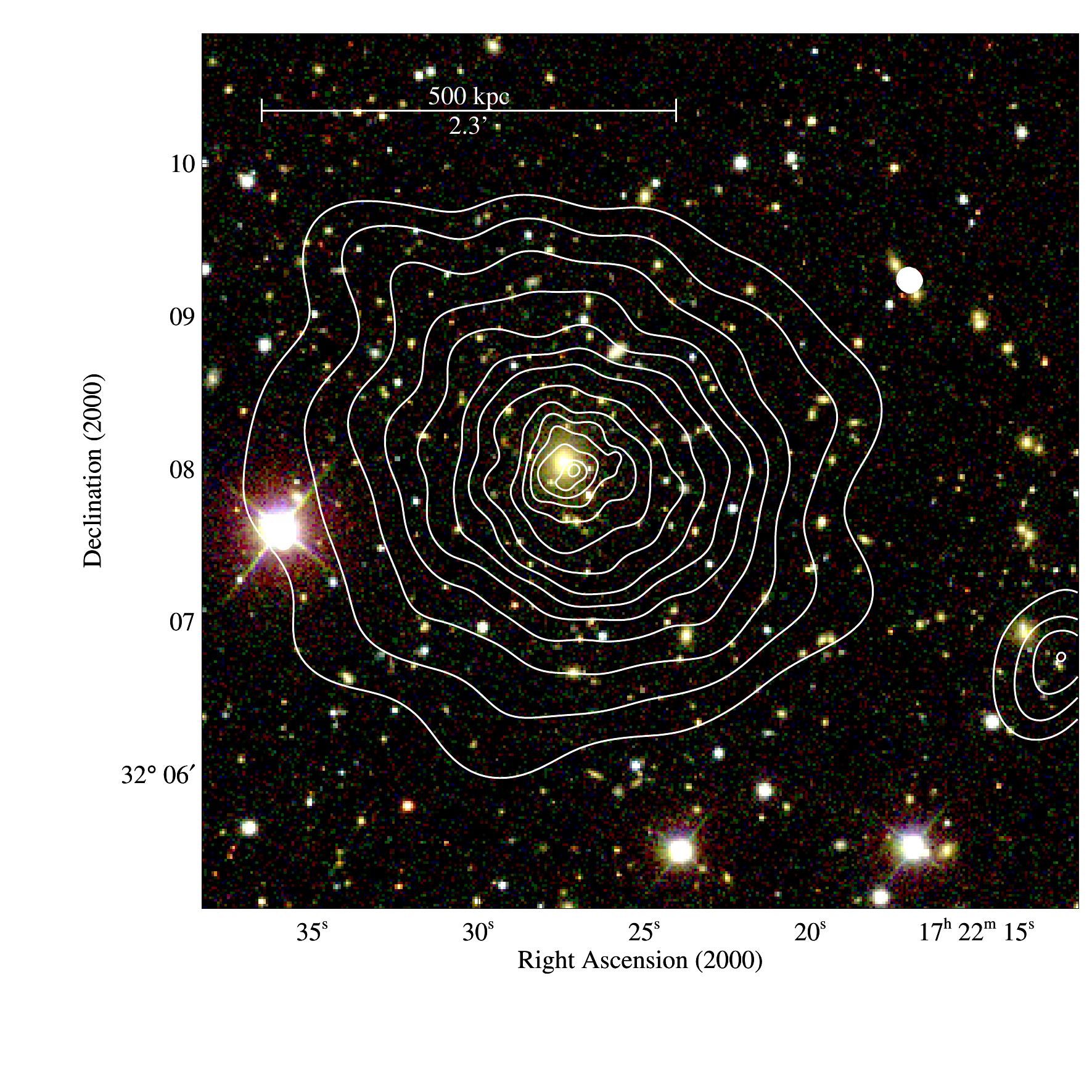}
\includegraphics[width=\overlaysize]{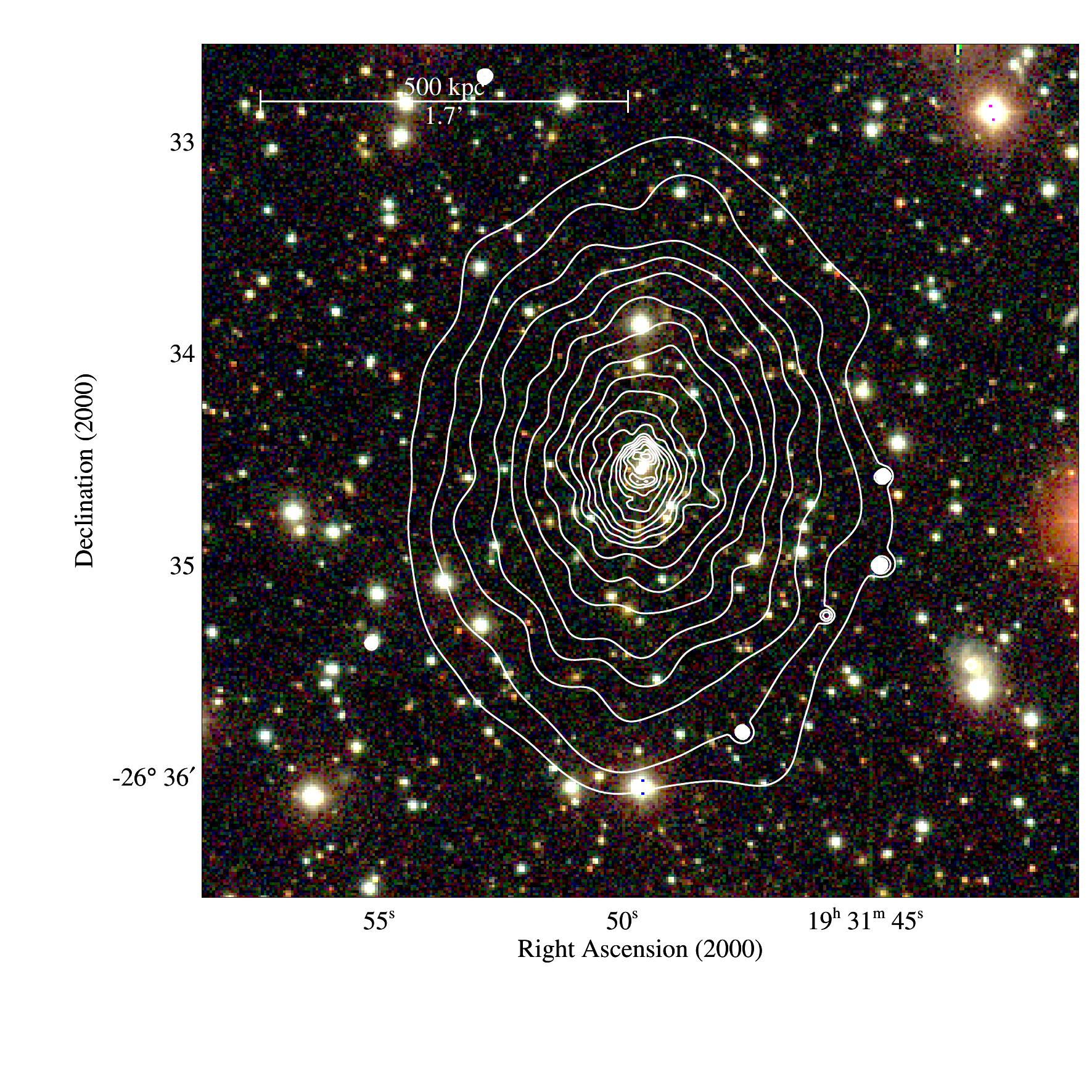}
\includegraphics[width=\overlaysize]{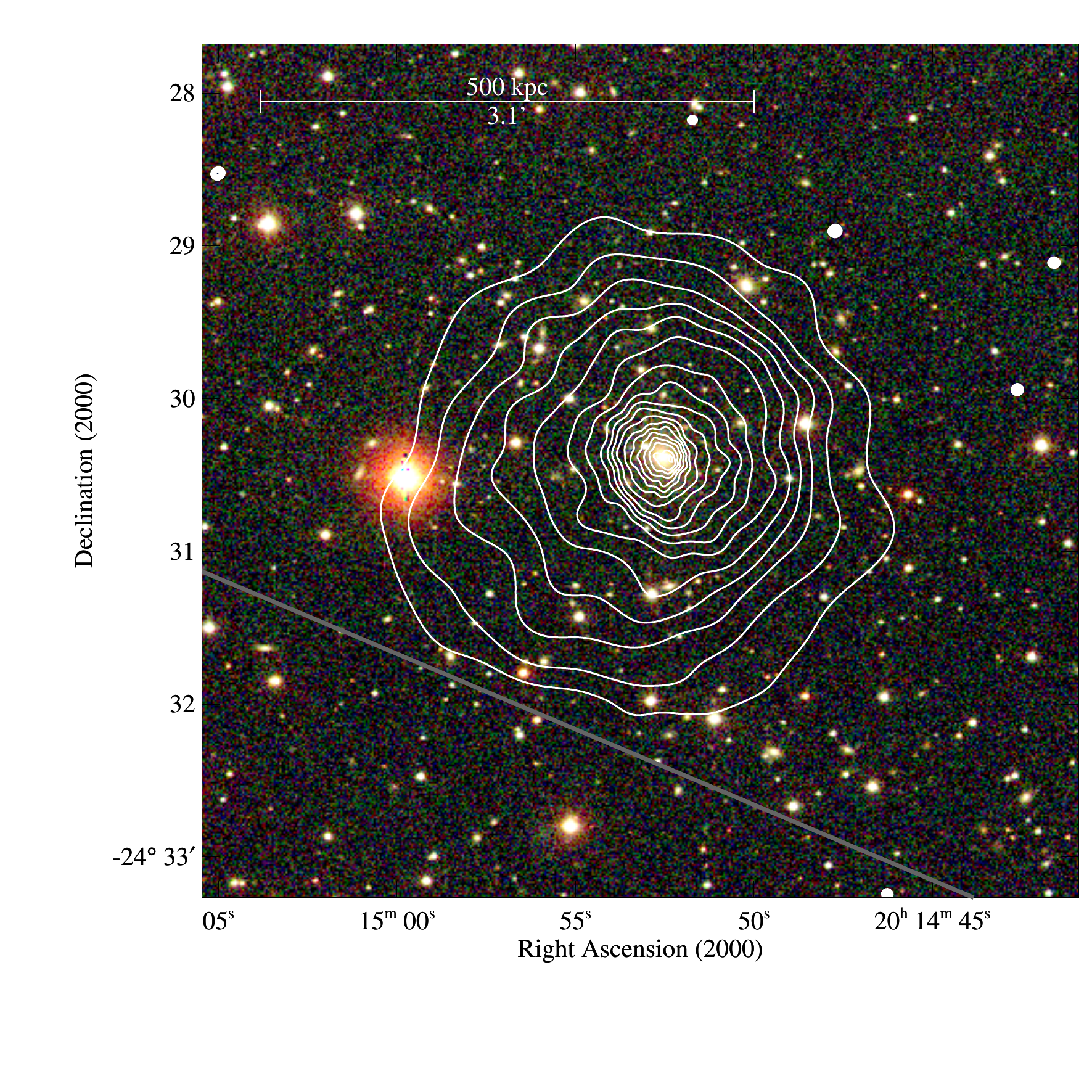}
\includegraphics[width=\overlaysize]{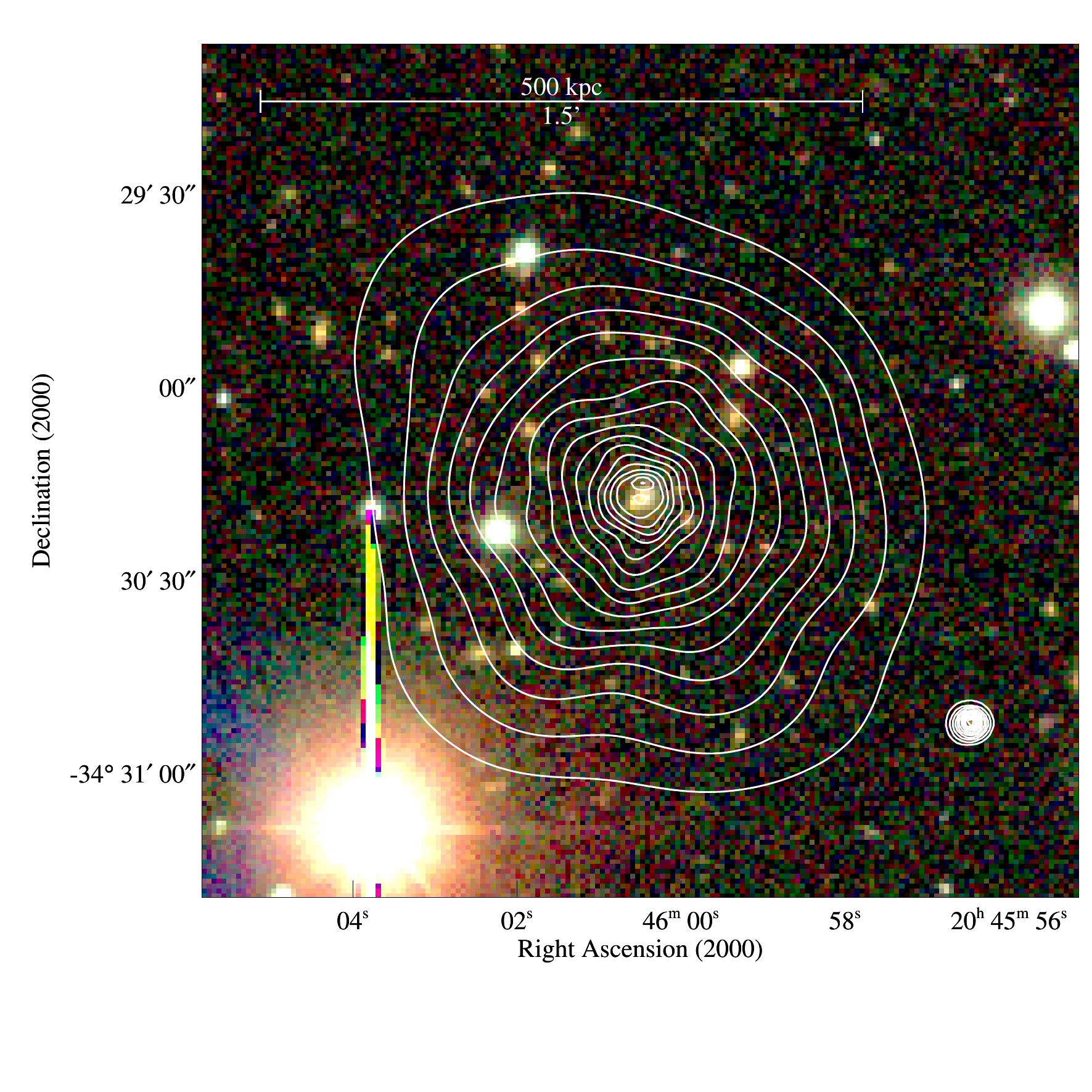}
\includegraphics[width=\overlaysize]{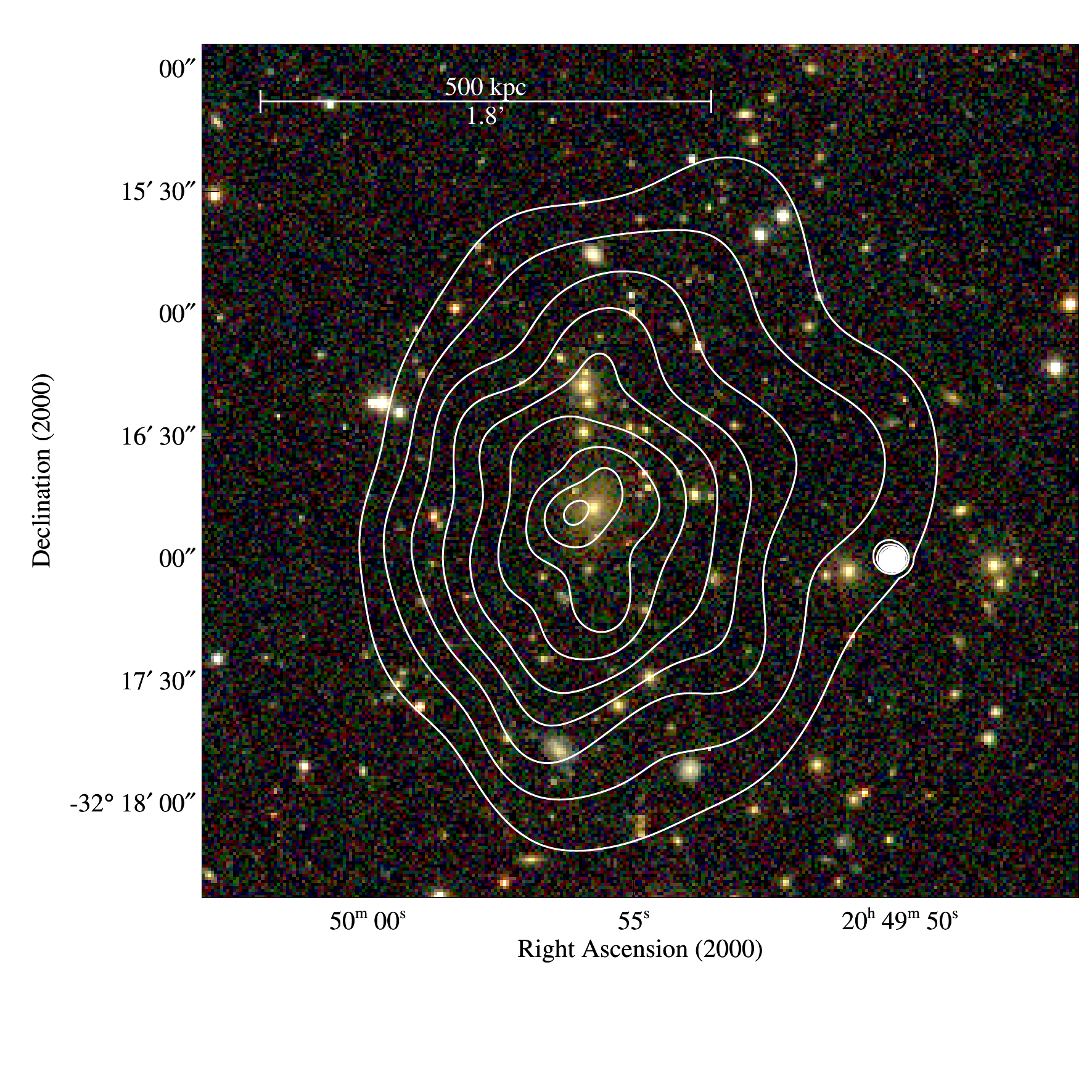}
\includegraphics[width=\overlaysize]{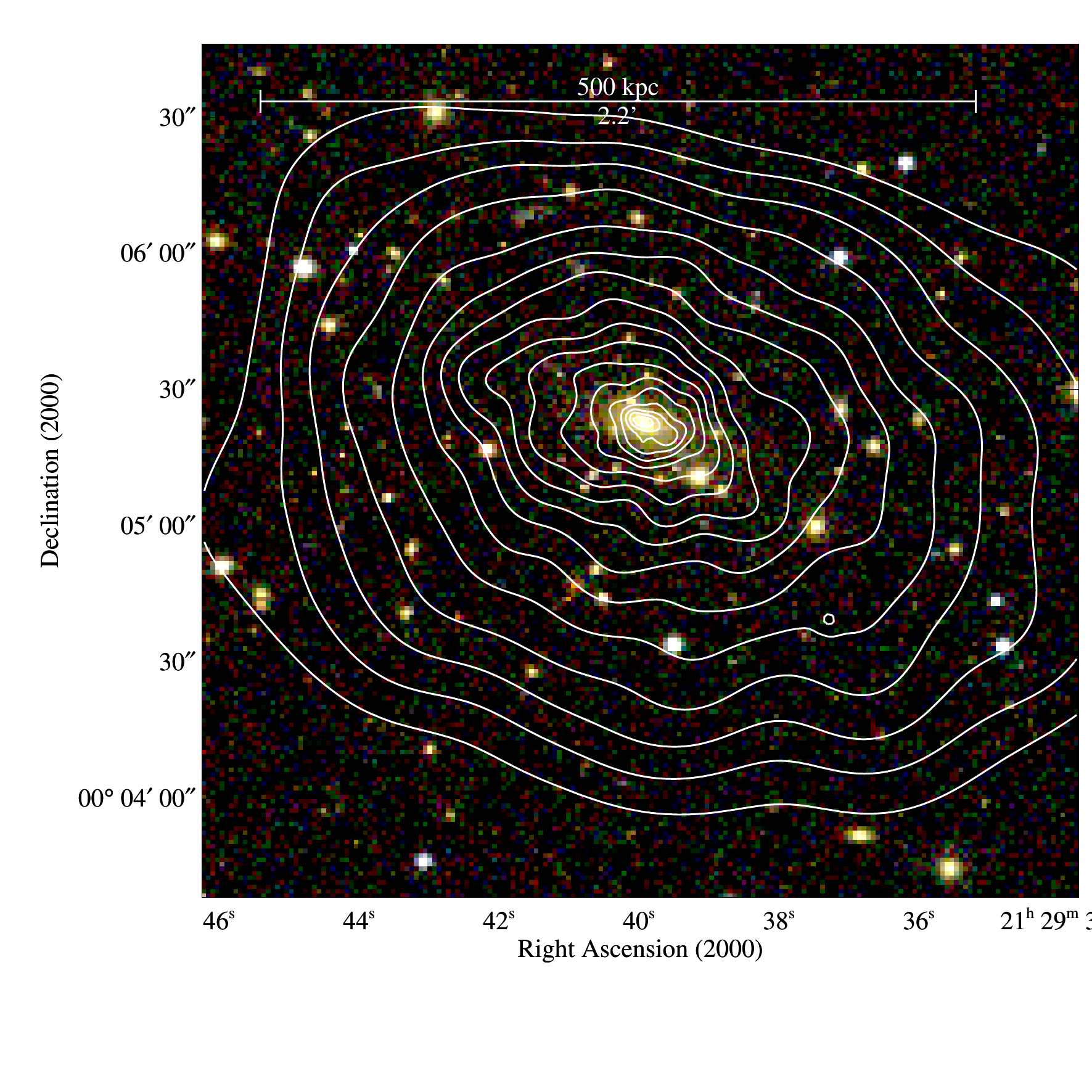}
   \caption{\emph{continued} Shown are (in R.A.\ order) {\it RX\,J1532.9+3021}, {\it MACS\,J1621.3+3810}, {\it A\,2204}, {\it A\,2219}, {\it RXC\,J1720.1+2637}, {\it MACS\,J1720.2+3536}, {\it A\,2261}, {\it MACS\,J1931.8--2634}, {\it RXC\,J2014.8-2430}, {\it MACS\,J2046.0--3430}, {\it MACS\,J2049.9--3217}, and {\it RX\,J2129.6+0005}}
   \label{fig:appendix}
\end{figure*}

\setcounter{figure}{0}

\begin{figure*} 
   \centering
\includegraphics[width=\overlaysize]{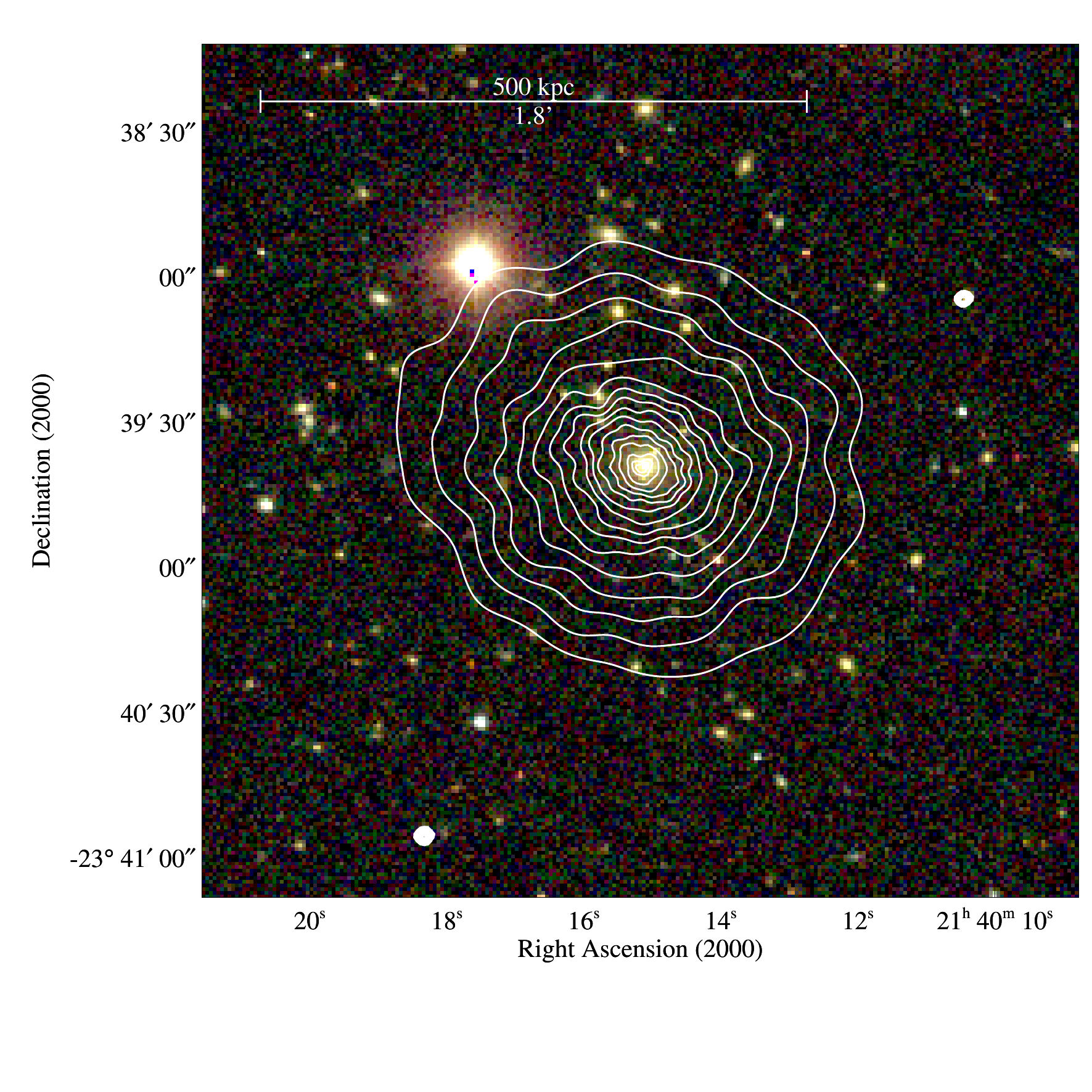}
\includegraphics[width=\overlaysize]{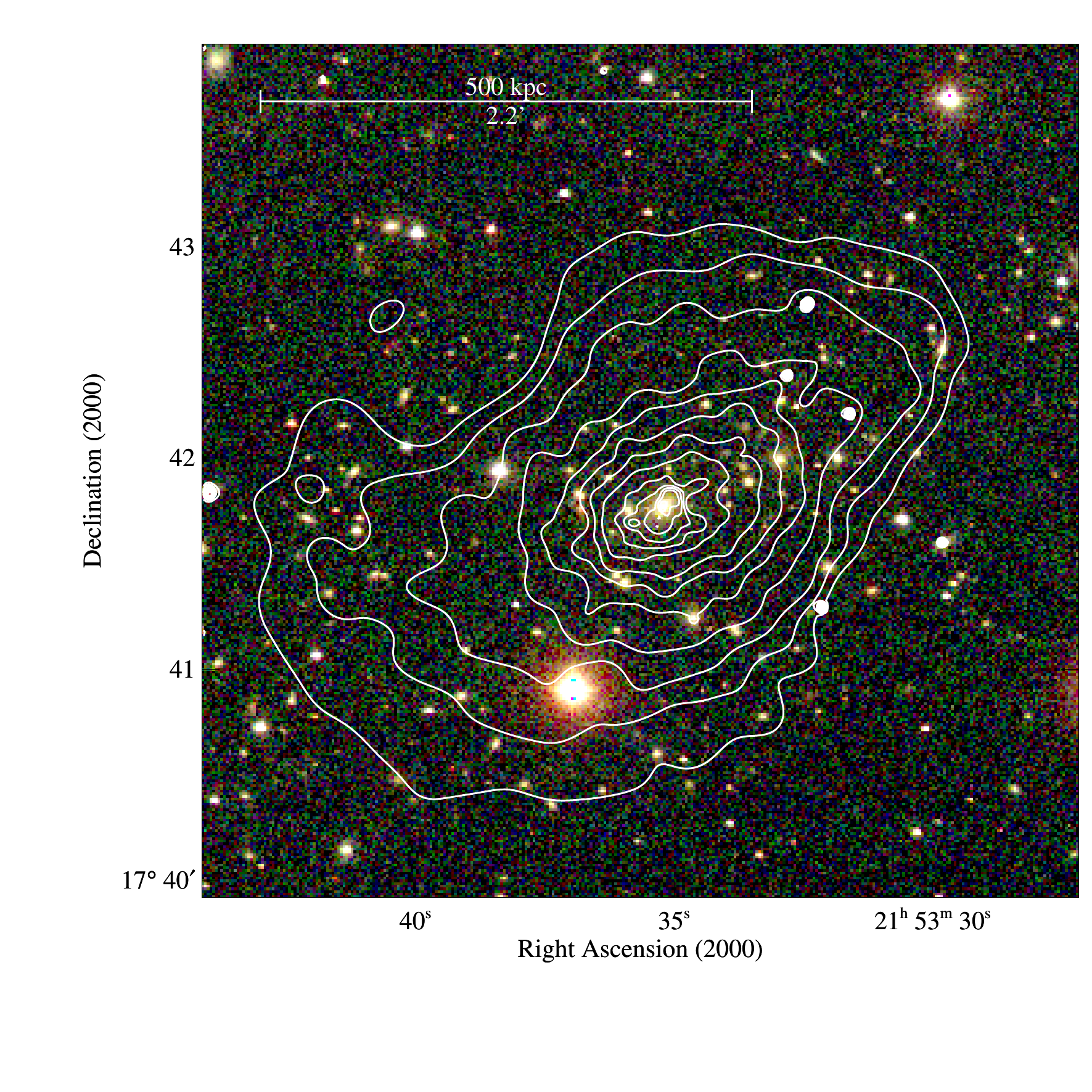}
\includegraphics[width=\overlaysize]{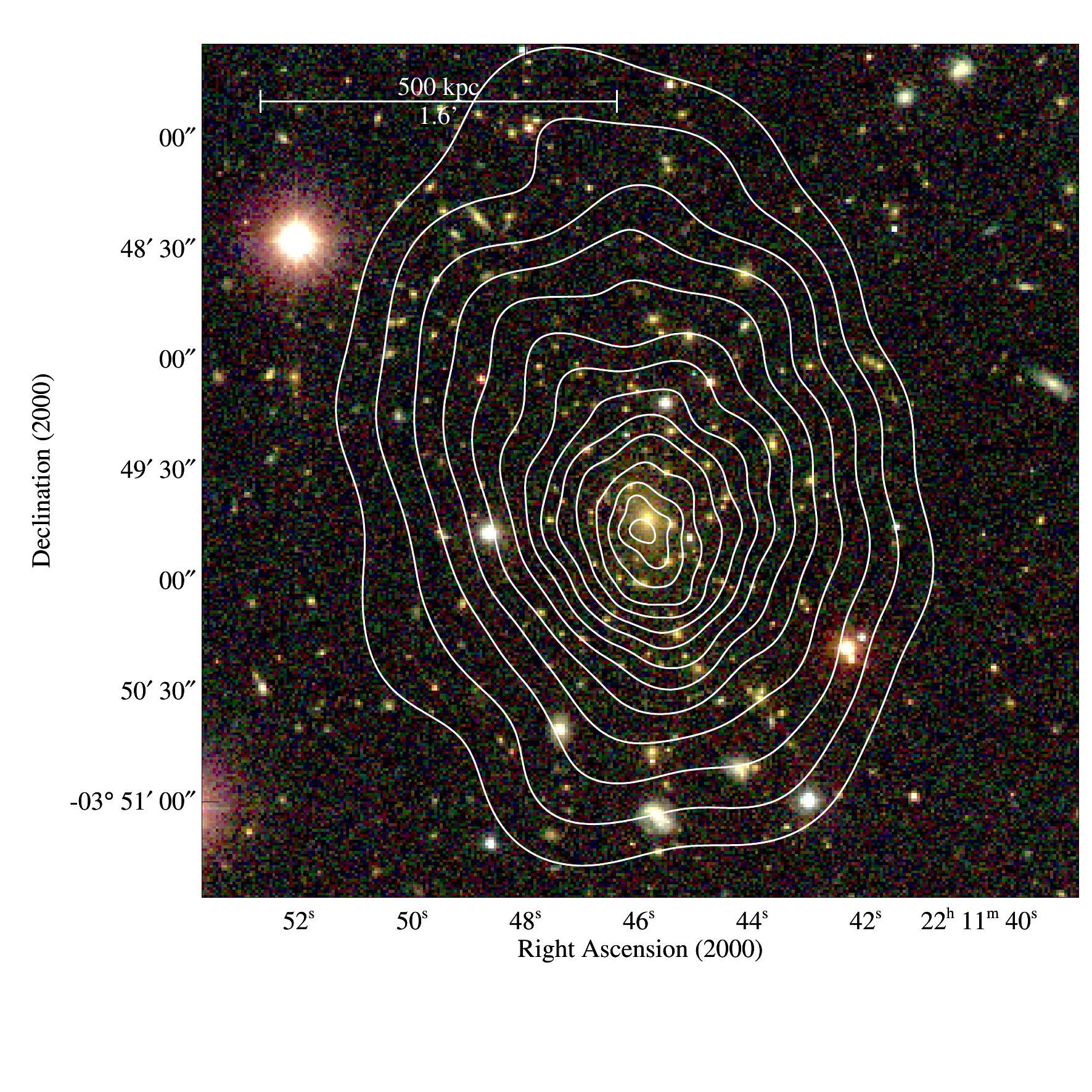}
\includegraphics[width=\overlaysize]{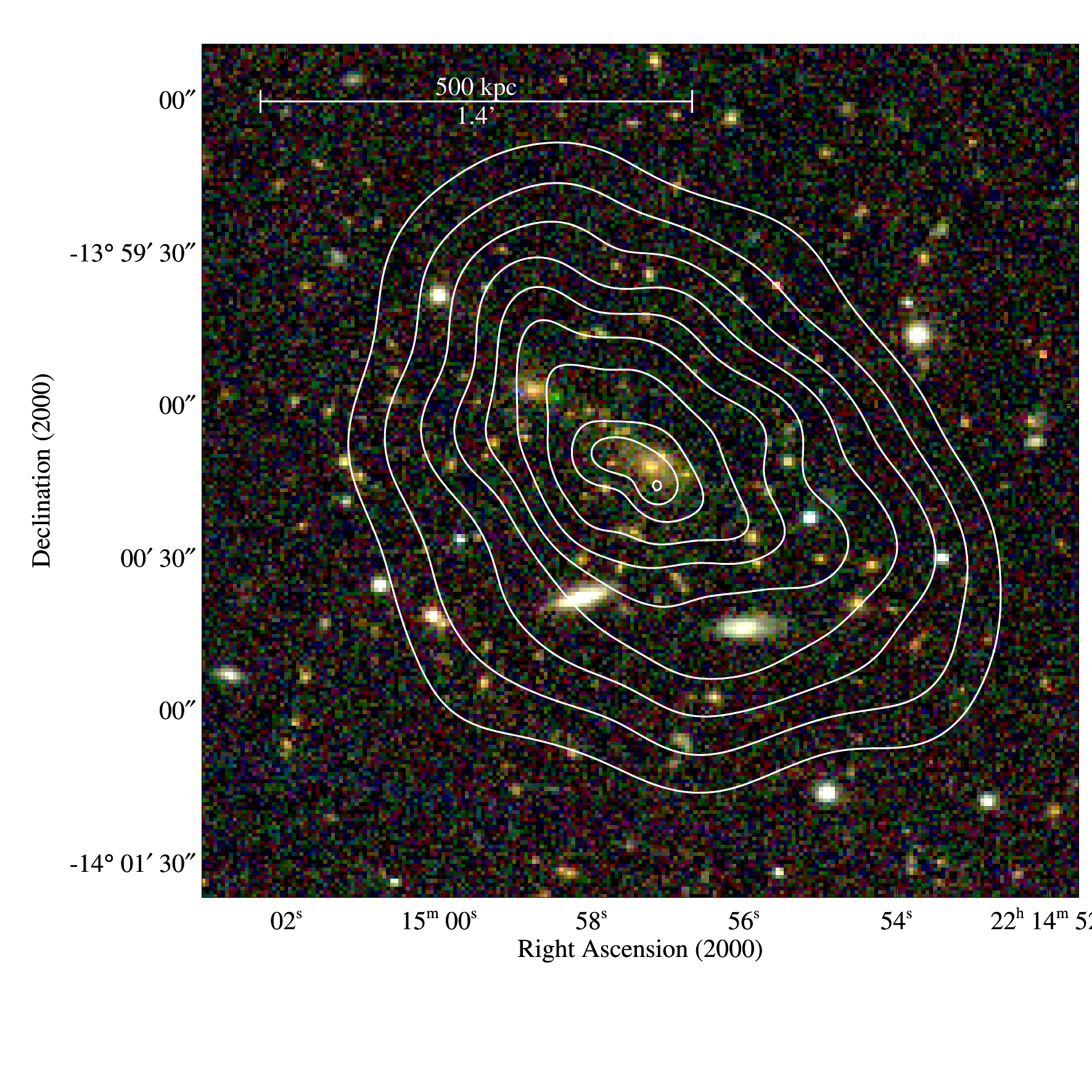}
\includegraphics[width=\overlaysize]{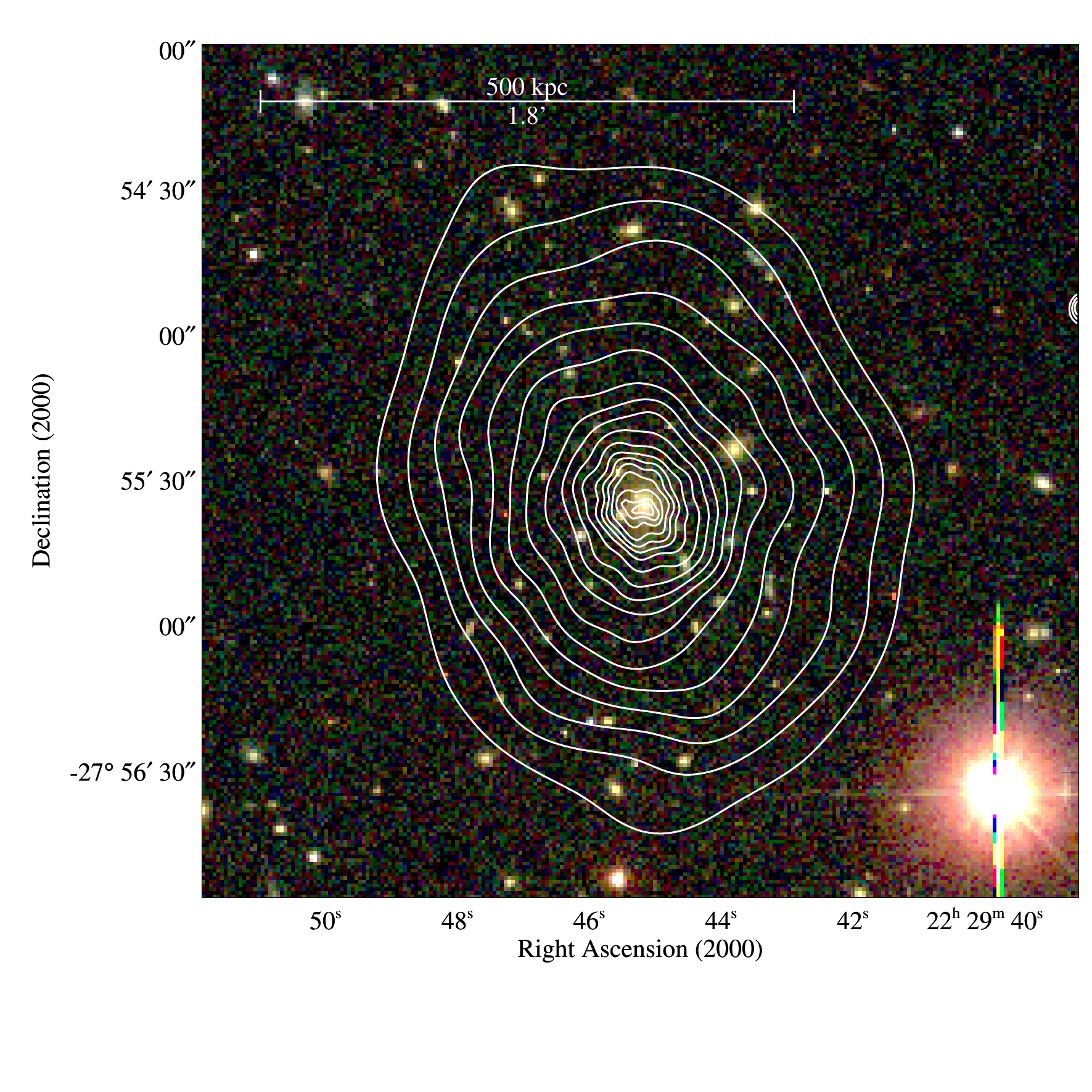}
\includegraphics[width=\overlaysize]{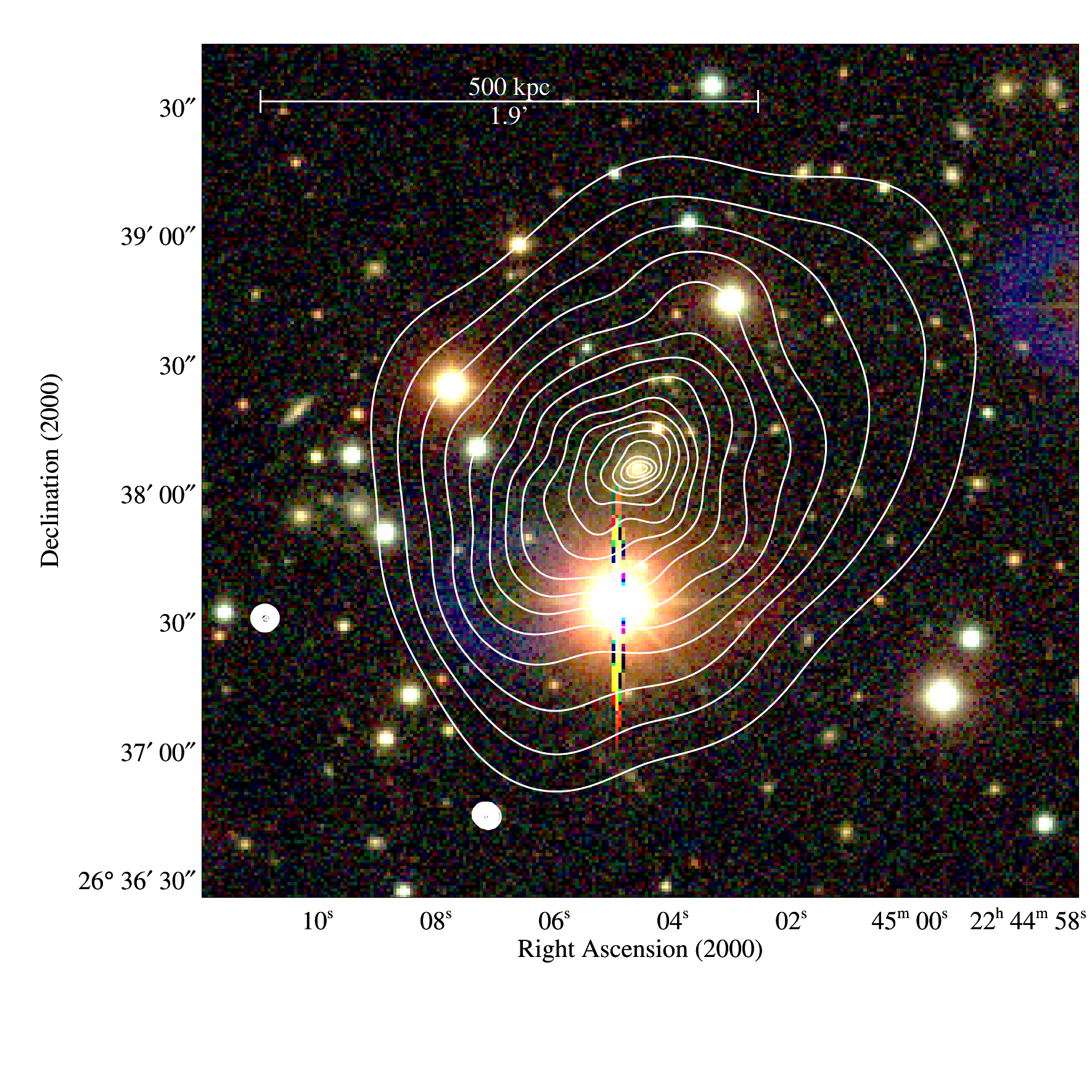}
\includegraphics[width=\overlaysize]{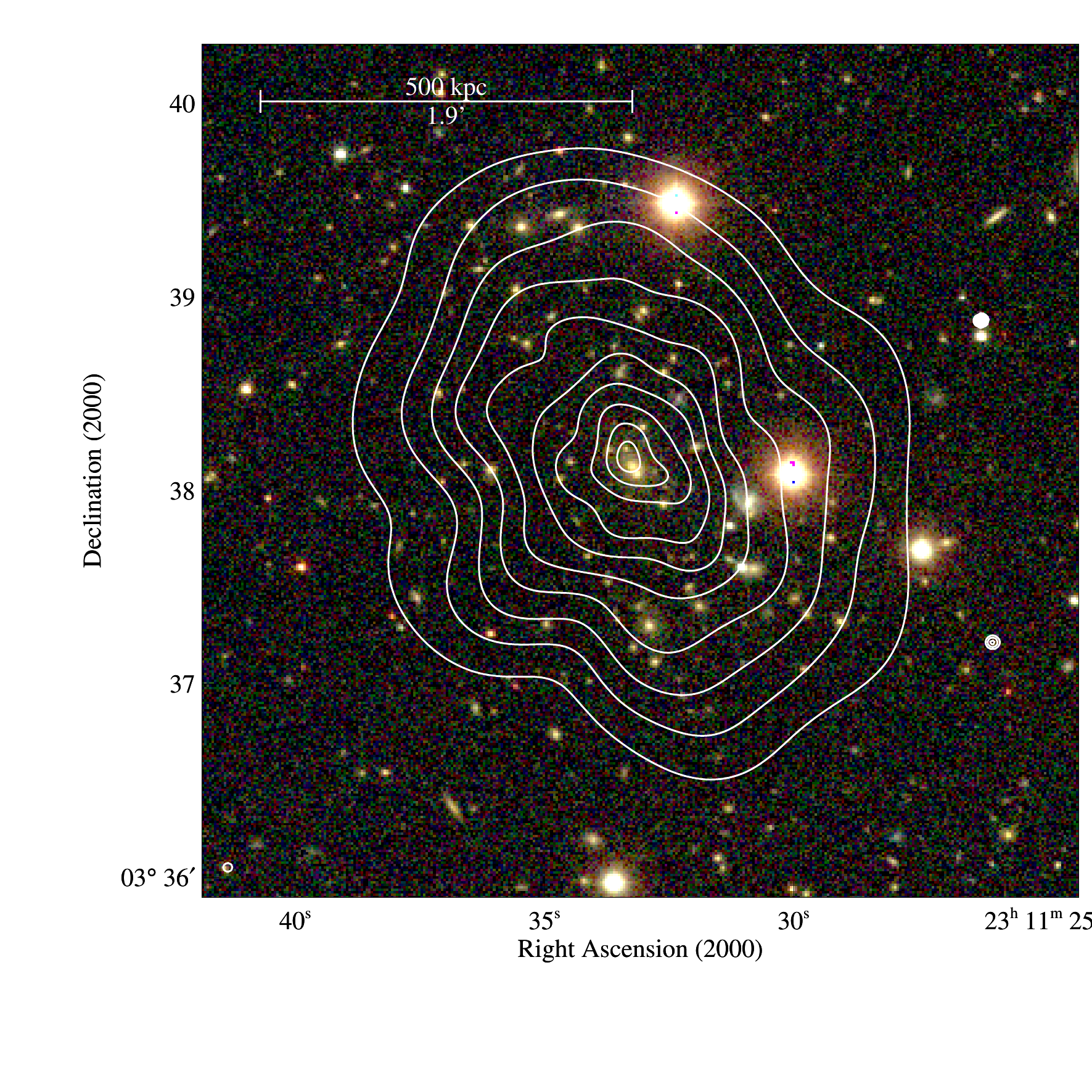}
   \caption{\emph{continued} Shown are (in R.A.\ order) {\it MS\,2137--2353}, {\it A 2390}, {\it MACS\,J2211.7--0349}, {\it MACS\,J2214.9--1359}, {\it MACS\,J2229.7--2755}, {\it MACS\,J2245.0+2637}, and {\it A\,2552}.}
   \label{fig:appendix}
\end{figure*}

\end{document}